\documentclass[a4paper,12pt,openany]{book}
\usepackage{amssymb,epsfig,eucal,amsmath,cite} \psfull

\newlength{\sbilanciamento} 
\newcommand{\ab}{\bar{\al}_\mathrm{s}}
\newcommand{\achi}{\raisebox{1.5pt}{$\chi$}{}}	
\newcommand{\afz}{F}			
\newcommand{\afzt}{\widetilde{F}}		
\newcommand{\al}{\alpha}
\newcommand{\as}{\alpha_{\mathrm{s}}}	
\newcommand{\adp}{{\as\over2\pi}}		
\newcommand{\be}{\beta}
\def\cF{{\mathcal{F}}}			
\def\cFH{\CMcal B}				
\def\cFL{\CMcal A}				
\newcommand{\chil}{\overset{\text{\tiny l}}{\chi}{}}	
\newcommand{\chis}{\overset{\text{\tiny s}}{\chi}{}}	
\newcommand{\chiu}{\overset{\text{\tiny u}}{\chi}{}}	
\newcommand{\ct}{\widetilde{C}}		
\newcommand{\De}{\Delta}
\newcommand{\Dt}{{\mathcal D}_{t}}		
\renewcommand{\d}{\delta}
\newcommand{\dcf}{{\CMcal D}}		
\newcommand{\dcg}{{\cal D}}			
\newcommand{\dd}{\dif\kd}
\newcommand{\de}{\partial}
\newcommand{\dgni}{\CMcal F}		
\newcommand{\dif}{{\rm d}}
\newcommand{\difg}{{\dif\ga\over2\pi\ui}}
\newcommand{\difo}{{\dif\om\over2\pi\ui}}
\newcommand{\dk}{\dif\kk}
\newcommand{\ds}{\displaystyle}
\newcommand{\du}{\dif\ku}
\newcommand{\dug}{\,\raisebox{0.37pt}{:}\hspace{-3.2pt}=}	
\newcommand{\e}{\varepsilon}
\newcommand{\eff}{\text{eff}}			
\newcommand{\esp}[1]{{\rm e}^{#1}}
\newcommand{\F}{{\CMcal F}}			
\newcommand{\FF}{{\mathcal F}}		
\newcommand{\Fp}{\hat{F}{}}			
\newcommand{\f}{{\rm f}}			

\newcommand{\fig}{}
\newcommand{\fin}{\eta}			
\newcommand{\fnp}{\widetilde{f}}		
\newcommand{\fpe}{F}			
\newcommand{\G}{{\CMcal G}}			
\newcommand{\GG}{{\mathcal G}}		
\newcommand{\Ga}{\Gamma}
\newcommand{\GeV}{\text{ GeV}}
\newcommand{\Gl}{\overset{\text{\tiny l}}{\G}{}}	
\newcommand{\Gs}{\overset{\text{\tiny s}}{\G}{}}	
\newcommand{\Gu}{\overset{\text{\tiny u}}{\G}{}}	
\newcommand{\ga}{\gamma}
\newcommand{\gagg}{\gamma^{\pg\pg}}	
\newcommand{\gaggu}{\overset{\mbox{{\tiny (1)}}}{\gamma}{}^{\pg\pg}}
\newcommand{\gagq}{\gamma^{\pg\pq}}	
\newcommand{\gagqu}{\overset{\mbox{{\tiny (1)}}}{\gamma}{}^{\pg\pq}}
\newcommand{\gaqg}{\gamma^{\pq\pg}}	
\newcommand{\gaqgu}{\overset{\mbox{{\tiny (1)}}}{\gamma}{}^{\pq\pg}}
\newcommand{\gaqq}{\gamma^{\pq\pq}}	
\newcommand{\gaqqu}{\overset{\mbox{\text{\tiny (1)}}}{\gamma}{}^{\pq\pq}}
\newcommand{\gb}{\bar{\gamma}}
\renewcommand{\geq}{\geqslant}
\newcommand{\gga}{{\boldsymbol\gamma}}	
\newcommand{\ggau}{\overset{\mbox{\text{\tiny (1)}}}{\gga}{}}
\newcommand{\gl}{\gamma_{\textsc{l}}}		
\newcommand{\gnl}{\ga_{\textsc{nl}}}		
\newcommand{\gnp}{\widetilde{F}}		
\newcommand{\half}{\mbox{\small $\frac{1}{2}$}}
\newcommand{\ho}{\half\om}			
\newcommand{\I}{\mathbb{I}}			
\newcommand{\id}{\boldsymbol{1}}		
\DeclareMathOperator{\im}{\Im\text{m}}		
\newcommand{\imp}{\Longrightarrow}
\newcommand{\intmel}{\int_{{1\over2}-\ui\infty}^{{1\over2}+\ui\infty}}
\newcommand{\K}{{\CMcal K}}			
\newcommand{\KK}{{\mathcal K}}		
\newcommand{\Kl}{\overset{\text{\tiny l}}{\K}}	
\newcommand{\Ks}{\overset{\text{\tiny s}}{\K}}	
\newcommand{\Ku}{\overset{\text{\tiny u}}{\K}}	
\newcommand{\kd}{{\boldsymbol k}_2}
\newcommand{\kk}{{\boldsymbol k}}

\newcommand{\ku}{{\boldsymbol k}_1}
\newcommand{\kw}{{\kappa}}			
\renewcommand{\L}{{\CMcal L}}		
\newcommand{\La}{\Lambda}
\newcommand{\Li}{\text{Li}_{2}}		
\newcommand{\la}{\lambda}
\renewcommand{\leq}{\leqslant}
\newcommand{\lra}{\leftrightarrow}
\newcommand{\M}{{\CMcal M}}		

\newcommand{\MSbar}{\overline{\text{\scriptsize MS}}}	
\renewcommand{\max}{{\text{max}}}
\renewcommand{\min}{{\text{min}}}
\DeclareMathOperator{\Max}{{\rm Max}}
\DeclareMathOperator{\Min}{{\rm Min}}
\renewcommand{\mp}{\mu_{\mathbb{P}}}	
\newcommand{\N}{{\mathbb N}}		
\newcommand{\Nf}{N_{\!f}}			
\newcommand{\ns}{\text{NS}}			
\newcommand{\Om}{\Omega}
\newcommand{\om}{\omega}
\newcommand{\omb}{\bar{\om}}		
\newcommand{\omd}{\Om_2}			
\newcommand{\omu}{\Om_1}			
\newcommand{\op}{\om_{\mathbb{P}}}		
\newcommand{\ord}{{\CMcal O}}		
\renewcommand{\P}{{\mathcal P}}		
\newcommand{\Pu}{\overset{\mbox{{\tiny (1)}}}{P}{}}	
\newcommand{\pa}{\text{\sf a}}		
\newcommand{\pb}{\text{\sf b}}		
\newcommand{\pc}{\text{\sf c}}		

\newcommand{\pf}{\text{\sf f}}		
\newcommand{\pg}{\text{\sf g}}		
\newcommand{\pie}{\pi_\e}			
\newcommand{\pj}{\text{\sf j}}		
\newcommand{\pk}{\text{\sf k}}		
\newcommand{\po}{\epsilon}			
\newcommand{\poc}{\overset{*}{\epsilon}{}}	
\newcommand{\pp}{\hat{p}{}}			
\newcommand{\pq}{\text{\sf q}}		
\newcommand{\ps}{\!\cdot\!}			
\newcommand{\psib}{\bar{\psi}}		

\newcommand{\qq}{{\boldsymbol q}}
\newcommand{\qu}{\qq_1}
\newcommand{\R}{\mathbb{R}}
\DeclareMathOperator{\re}{\Re\text{e}}		

\newcommand{\Si}{\Sigma}
\renewcommand{\Sp}{\text{Sp}}		

\newcommand{\si}{\sigma}
\newcommand{\sip}{\hat{\si}}			
\newcommand{\slarga}{\raisebox{-6mm}{\rule{0mm}{14mm}}}
\newcommand{\Texp}{\mathbb{T}\exp}		
\newcommand{\Th}{\Theta}
\renewcommand{\t}[1]{\overset{#1}{T}{}}		
\newcommand{\tb}{\bar{t}}			
\newcommand{\te}{\theta}
\newcommand{\ts}{\textstyle}
\newcommand{\Up}{\Upsilon}
\newcommand{\ugd}{=\hspace{-3.2pt}\raisebox{0.37pt}{:}\hspace{3pt}} 
\newcommand{\ui}{{\rm i}}
\newcommand{\valutato}{\raisebox{-2mm}{\rule{0.2pt}{6mm}}}
\newcommand{\vm}[1]{\langle#1\rangle}	
\newcommand{\Wp}{\hat{W}{}_{\mu\nu}}		
\newcommand{\xp}{\hat{x}{}}			
\newcommand{\Z}{\mathbb{Z}}		

\newcommand\epj[3]{{\it Eur. Phys. J. }{\bf C #1} (#2) #3}

\newcommand\jhep[3]{{\it JHEP }{\bf #1} (#2) #3}

\newcommand\npb[3]{{\it Nucl. Phys. }{\bf B #1} (#2) #3}

\newcommand\plb[3]{{\it Phys. Lett. }{\bf B #1} (#2) #3}		   
\newcommand\pr[3]{{\it Phys. Rev. }{\bf #1} (#2) #3}

\newcommand\prd[3]{{\it Phys. Rev. }{\bf D #1} (#2) #3}		   
		   
\newcommand\prep[3]{{\it Phys. Rep. }{\bf #1} (#2) #3}

\newcommand\rmp[3]{{\it Rev. Mod. Phys. }{\bf #1} (#2) #3}	   
\newcommand\zpc[3]{{\it Z. Physik }{\bf C #1} (#2) #3}

\newcommand\sjnp[3]{{\it Sov. J. Nucl. Phys. }{\bf #1} (#2) #3}	   
\newcommand\jetp[3]{{\it Sov. Phys. JETP }{\bf #1} (#2) #3}

\newcommand\nc[3]{{\it Nuovo Cim. }{\bf #1} (#2) #3}		   
\newcommand\jetpl[3]{{\it JETP Lett. }{\bf #1} (#2) #3}

\newcommand{\hepph}[1]{{\tt hep-ph/#1}}
\newcommand{\hep}[1]{{\tt hep-ph/#1}}

\newcommand{\hepex}[1]{{\tt hep-ex/#1}}

\newcommand{\didascalia}[1]{\caption{\small\sl{#1}}}
\newcommand{\lab}[1]{\label{#1}}
\newcommand{\labe}[1]{\label{#1}}
\renewcommand{\(}{\big(}
\renewcommand{\)}{\big)}

\input overline
\begin{document}
\begin{titlepage}
\begin{center}
\large UNIVERSIT\`A DEGLI STUDI DI FIRENZE \\ 
Facolt\`a di Scienze Matematiche, Fisiche e Naturali\\
Dipartimento di Fisica\\[2mm]
\normalsize{\sl Corso di Dottorato, XII ciclo}
\vskip 1cm
{\LARGE\sf Tesi di Dottorato in Fisica}
\vskip 1cm
{\huge{\bf Small-$\boldsymbol{x}$ Processes\\in Perturbative\\
Quantum Chromodynamics}
\vskip 17mm   
\it Dimitri Colferai}
\end{center}
\vskip 1cm
\vskip 2cm
\large
\null\hspace{1cm}Supervisore: Prof. M. Ciafaloni\\[5mm]
\vskip 14mm
\begin{center}
{\sl\normalsize Dicembre 1999}
\end{center}
\end{titlepage}
\newpage\null\thispagestyle{empty}

\thispagestyle{empty}
\vskip3cm \null\hfill{\it al mi' babbo Camillo\,...}\hspace{30mm}
\newpage\null\thispagestyle{empty}\newpage

\font\goa=cmfrak scaled 1200
\font\goin=yinit scaled 250
\newfam\gothicfam
\def\gotha{\fam\gothicfam\goa}
\def\gothin{\fam\gothicfam\goin}

\null\vskip 1cm
{\centering\LARGE\bf Ringraziamenti%
\\[16mm]}\thispagestyle{empty}
{\sl
Per me il bello di ultimare un lavoro come questa tesi consiste anche nel comporre la
pagina dei ringraziamenti, e stavolta le persone da ringraziare sono veramente tante.

Comincio da {\sf Marcello}, al quale oso finalmente rivolgermi in tono confidenziale. In 
questi 3 anni ho avuto la fortuna di avere una guida che, accanto a impressionanti doti
professionali, mi ha sempre dimostrato assoluta correttezza e una inattesa
disponibilit\`a, ben oltre il dovuto. Grazie a Marcello (e ai contribuenti) ho anche
avuto l'opportunit\`a di girare un po' per il pianeta, dalla Terrasanta al Nuovo
Mondo. Spesso ho dovuto immolare le pi\`u amate ore del sonno mattutino, ma sono stato
ampiamente ricompensato dal vate dei piccoli $x$.

Ringrazio {\sf Gavin} per le sue pazienti e illuminanti spiegazioni su scale gluoniche,
sigari, anatre, ecc. e soprattutto per i numerosi grafici 
che non ha mai mancato di farmi pervenire alla velocit\`a della luce e che rendono un
po' pi\`u vivace questa tesi.

La realizzazione materiale di questa tesi deve il suo tributo a quell'incredibile
incrocio tra un super-user, madre Teresa di Calcutta e una siepe emisferica noto al
volgo con il nome di {\sf Andrea}, che mi ha anche iniziato al mountain-biking estremo e,
con la super-vespa, aumentato l'apertura alare di almeno 20 cm.

Anche dal lato psico-sentimentale ho avuto molte persone attorno che mi hanno aiutato.
L'insistenza e l'entusiasmo nel portare avanti gli obiettivi prefissati, li ho appresi,
pi\`u che da chiunque altra persona, dalla mia amatissima sorella {\sf Manola}, la
quale, nonostante la lontananza, ho potuto sempre pi\`u apprezzare, ammirare e a volte
anche imitare.

Ringrazio {\sf Linda} per avermi pacificamente consegnato nelle mani di Anna,
dimostrandosi in innumerevoli modi splendida mamma (per me) e suocera (per Anna). 

Dedico questa tesi a {\sf Camillo} per la pazienza, la costanza e altre 
virt\`u che mi ha sempre trasmesso (o per lo meno, ci ha provato):  pi\`u
che per la laurea o per le gare in bici, in questo corso di dottorato mi
sono state indispensabili.

Questa lunga corsa a tappe ha messo a dura prova anche il fisico, e se sono riuscito a
reggere fino alla fine \`e soprattutto per merito delle nonne {\sf Maddalena} e {\sf
Wanda} (in ordine alfabetico) che mi hanno nutrito come un $\dots$ tanto, 
fin dalla mia
infanzia.

Tra i lunghi periodi di studio e lavoro, mi hanno portato ventate di brio
le corse nei boschi
con la moto procuratami dagli zii {\sf Mario} e {\sf Manuela} che
ringrazio anche per le due splendide cuginette. Chi saranno i prossimi?

Sono grato a {\sf Maria Luisa} e al caro {\sf Sergio} per avermi
subito accolto con affetto e per avermi concesso di maritare la 
loro figliuola pi\`u bella, {\gothin A}{\gotha nna}, che con le sue arti domestiche, ciclistiche e
coniugali sa sempre come ricolmarmi di felicit\`a e instillarmi la voglia di fare (e per
questa tesi ne \`e servita molta!)

Ho avuto la fortuna di avere dei compagni/e di dottorato e di ufficio
molto simpatici e amichevoli, con i quali, sia nella superaffollata stanza 001 che
fuori, ho condiviso tante piacevolissime giornate. In particolare mi rivolgo al gi\`a
citato Andrea e a {\sf Riccardo} per aver costantemente inebriato l'aere di
raffinatissimo sigaro cubano, per avermi fatto conoscere di persona i CSI, per il cus
cus speziato, $\dots$

Mi dispiace non aver pi\`u motivi per ringraziare Roperta, Fapia, Paoloz e Tafita con i quali ho
trascorso l'ultima vacanza da scapolo, Stepana e Tiekka coi quali la scappatella a
Imola~'96 \`e diventata pluriennale tradizione fino a Imola~'99, e tutta la schiera di
amici e amiche a cui, per amore della {\it fisica}, ho preferito rinunciare.

Infine, un ``in bocca al lupo'' alla {\sf Martina} con la quale ho potuto condividere la
sensazione di naufrago nell'incantevole mare della QCD.}

\thispagestyle{empty}

\pagenumbering{roman}
\tableofcontents\newpage\null\thispagestyle{empty}\newpage
\pagenumbering{arabic}
\addcontentsline{toc}{chapter}{\hspace{4.7mm} Introduction}
\chapter*{Introduction}

Quantum Chromodynamics (QCD) is a quantum field theory, based on an $SU(N_c)$ non
abelian gauge group, born in order to describe strong interactions. Presently it is a
well defined theory as well as the best candidate nowadays available.

The success of QCD originated from the fact that it provided precisely, for $N_c=3$, the
observed symmetries of strong interaction (such as the statistic of the baryons) and no
more. From a dynamical point of view, its outstanding property of asymptotic freedom ---
due to the non abelian nature of the gauge group --- was able to account for the scaling
properties of cross sections experimentally observed. Furthermore, even if not
rigorously proven, QCD gives strong indications of color confinement, e.g.\ in lattice
simulations.

Asymptotic freedom means that the effective coupling, as defined by the renormalization
group, becomes vanishingly small when large space-like momenta (with respect to the
QCD scale $\La_{\rm QCD}\simeq$ 200 MeV) are transferred. This kind of processes, called
{\em hard} processes, can be investigated by means of perturbative methods.

On the opposite side, soft scattering, hadronization and all long distance effects,
unavoidable in any strong reaction, involve strong coupling features most of which are
far from present computational possibility.

At the basis of several applications of perturbative QCD is the {\em factorization}. For
some measurable quantities, factorization theorems exist which allow the separation of
short distance (perturbative) physics from long distance (non perturbative) physics of
observable hadrons. It should be noted that what can be actually evaluated are not
absolute values of observables corresponding to a given choice of variables, but rather
the evolution of observables in the variables space.

A crucial role in the factorization methods is played by the number and the relative
values of the hard scales involved in the process. In the traditional deeply inelastic
scattering (DIS) and in the old accelerators physics, the virtuality $Q^2$ of the
transferred momentum is the only hard scale (the center of mass energy being of the same
order of magnitude). For this single-scale processes, perturbative QCD predicts the
evolution in $Q^2$ of the relevant quantities as a power series in $\as(Q^2)$. The
natural framework for studying this class of phenomena is the  collinear
factorization in which only the longitudinal (with respect to the incoming hadron)
degrees of freedom of the on-shell partons are present. The transverse degrees of
freedom, peculiar of the interacting theory, give rise to logarithms of $Q^2$. These
large logarithms, which need to be taken into account to all orders in perturbation
theory, are resummed by the DGLAP equations and are responsible for the scaling
violations, i.e., for the deviation from the pure scaling behavior one would obtain by
neglecting the parton-parton interaction.

The coming of high-energy (for the time being) colliders has entered a new era in which
the energy $\sqrt s$ is a scale much harder than the transferred momentum. The
electron-proton collider HERA at DESY is of particular importance, since high energy DIS
has opened the road to new and interesting physics. The large available energy in the
center of mass of the colliding particles $s\simeq(300\GeV)^2$ allows to investigate a
wide kinematic region. It is possible to reach values of $Q^2$ larger than
$10^4\GeV^2$ and very small values of the Bjorken variable $x\sim10^{-5}$. The structure
functions have shown to undergo large scaling violations towards high $Q^2$, especially
at low values of $x$. Besides, a steep rise of the structure functions has been observed
stimulating the interest of a considerable part of both the experimental and the
theoretical community.

In this context, the perturbative QCD description by means of the DGLAP approach reveals
itself very successful, even beyond the expectations, in the sense that starting with
reasonable parametrizations of the parton distribution functions at rather low values of 
$Q_0^2\sim0.35\GeV^2$ --- definitely outside the perturbative domain --- the $Q^2$
evolution of the structure functions is very well described by a next-to-leading order
DGLAP fit.

The basic issue remains to justify or motivate the particular shape of the input parton
densities entering the DGLAP evolution equation. The most pessimistic approach is that
the problem is a non perturbative matter, since the perturbative evolution works even
with initial conditions at very low $Q_0^2$. And actually it could be so.

However one can also argue that we are not allowed to use DGLAP equations outside the
perturbative domain and that it would be preferable to start the evolution at some
higher point of $Q_0^2$ of the order of some GeV$^2$, where the perturbative theory is
presumably trustworthy. In this case the right input functions which are needed to
describe the data present a (small) power-like rise in $1/x$. For instance, the gluon
distribution --- playing a leading role in high energy processes --- has the form
$f^{(\pg)}(x,Q^2)\propto x^{-\la(Q^2)}$ where $\la(Q_0^2\simeq4\GeV^2)\simeq0.17$.

The question then arise: can one justify such a power-like shape for the partonic
densities?  The question reminds us of Regge theory, where it is expected a power-like
growth of the cross sections with $s$ (the latter being proportional to $1/x$). Even if
Regge theory is not based on perturbative physics, nevertheless they should be someway
related.  This hypothesis is confirmed by the fact that, at very low values of $Q^2$,
the $x$-growth exponent of the structure functions reaches the value $\la\simeq0.1$, and
this suggests a smooth junction with the soft pomeron value $\op\simeq0.08$, i.e., the
universal exponent governing the $s$-growth of total cross sections.

It remains to see whether perturbative QCD can explain a small-$x$ rise of the structure 
functions with a power-like law and with a correct exponent in the intermediate
$Q^2\simeq1\div10\GeV^2$ regime.

In connection with that point there are the so-called two-scale processes, such as
$\ga^*\ga^*$ scattering and forward jets, where at the endpoint of the QCD evolution two 
hard scales are present. Here the use of perturbative theory should be more suitable in
order to describe the $Q^2$ and energy behaviour. Preliminary results seem to indicate a 
growth in energy compatible with $s^{\la(Q^2)}$, $\la(Q^2)\simeq0.3$ for $Q^2$ of order
of $3\div40\GeV^2$.

Those high energy-not very large $Q^2$ regimes are referred to as {\em semi-hard}
regimes. Here the perturbative series can be slowly converging or even unstable, because
the higher order terms are accompanied by large $\ln s$ and may be as important as the
first ones. In this case, finite order calculations are no longer reliable and
resummation techniques must be devised in order to take into account all the important
terms.

Different approaches have been adopted in order to study gauge theories at asymptotic
energies, e.g., Regge theory. This ``old'' theoretical problem was for the first
time investigated at a perturbative level in the 70's by the russian school where the
large logarithms of the energy are resummed by means of the BFKL equation, predicting at
low $x$ a power-like growth of the structure functions, but with a too large exponent
$\la(\as)\simeq0.55$ for values of $\as\simeq0.2$ as in the HERA range.

In order to obtain a quantitative agreement with the experimental data, a huge
theoretical effort ('89\,-'98) has been devoted to analyse high-energy QCD beyond the
leading-logarithmic approximation. Last year the next-to-leading logarithmic (NL$x$)
BFKL kernel was found. It immediately appeared that the NL$x$ corrections to the kernel
are quite large and of particular relevance for their negative sign. In particular, the
``pomeron singularity'', which should provide the $s$-growth exponent for high energy
cross sections, undergoes a drastic reduction reaching its maximum value ($\simeq0.11$)
for quite a small value of $\as\simeq0.08$ and in the HERA range it becomes even
negative.

A pathological consequence of the large size of the NL$x$ corrections is that, for
larger values of the coupling, they may even provide negative cross sections in the very
large $s$-limit! The fact is that, for values of $\as\gtrsim0.1$, such corrections are
so large that they cannot be taken literally, suggesting an intrinsic instability of the
perturbative series in the effective parameter $\as\ln1/x$.

A direct analysis shows that responsibles for this instability are the large collinear
contributions to the kernel. The latter are single or double logarithms of the
transverse scales $\kk^2$ and $\kk_0^2$ determining the process. Single logarithms
$(\as\ln\kk^2/\kk_0^2)$ are strictly related to the well known logarithmic corrections
to scaling in the na\"\i ve parton model, providing the scaling violation predicted by
the renormalization group. Double logarithms $(\as\ln^2\kk^2/\kk_0^2)$ are due to the
mismatch between the factorization scale $s_0=\kk\kk_0$ entering the high energy
factorization formula and the Bjorken scale $s_B=\Max(\kk^2,\kk_0^2)$ which is the
relevant scale in the collinear limits $\kk^2\gg\kk_0^2$ and $\kk^2\ll\kk_0^2$.

Since collinear contributions are determined by the renormalization group, it is
mandatory to develop a unified picture where both renormalization group constraints and
small-$x$ features are taken into account.
Single and double $\ln\kk^2/\kk_0^2$ are known to all orders in
$\ln1/x$. Therefore, the first step of the improved small-$x$ formulation consists in
the resummation of the collinear contributions to the BFKL kernel, consistently with the
full L$x$ and NL$x$ expressions. The second step concerns the new method for solving the
improved small-$x$ equation. Starting from the factorization property of the solution in
a perturbative times a non perturbative factor (up to higher twist corrections) the
perturbative part is determined by an expansion with respect to a new parameter $\om$
--- the moment index with respect to $x$. The use of $\om$ instead of $\as$ turns out to
be more convenient in the small-$x$ region where the whole $Q^2$-dependence of
$\as(Q^2)$ is important and has to be taken into account.

The outline of the thesis is the following:
the first chapter introduces the basic objects of our study and recalls the collinear
properties of QCD in the special context of DIS. Scaling violations are explained both
with the operator product expansion analysis and in terms of the renormalization group
improved parton model, in order to identify the results of the latter with the
field-theoretical quantities (e.g.\ the anomalous dimension) of the former.

In Chap.~\ref{c:shp} we concentrate on high energy physics by starting with an overview
of Regge theory in connection with the phenomenological analysis of processes at
asymptotic energies. As far as the perturbative treatment of small-$x$ processes is
concerned, we recall the framework of high energy factorization, its connection to the
collinear one and we show how the resummation of the leading $\ln 1/x$ leads to the BFKL 
equation.

The subsequent chapters contain the author's original contributions, obtained in
collaboration with M.\ Ciafaloni and G.P.\ Salam and published in
Refs.~\cite{CiCo98,CiCo98b,CiCoSa99,CiCoSa99b}.

In Chap.~\ref{c:NLbfkl} we present the generalization of the high energy factorization
formula in NL$x$ approximation. In this context we perform the NL$x$ analysis for both
the impact factors~\cite{CiCo98} and the BFKL kernel. We present the main results as well as
the problems concerned with the large NL$x$ corrections.

With Chap.~\ref{c:rga} we get to the heart of the improving procedure of the NL$x$
approximation. The starting point is the identification of the collinearly-enhanced
physical contributions as the main responsibles of the instability of the BFKL
hierarchy. The collinear singularity resummation procedure devised in
Refs.~\cite{CiCo98b,CiCoSa99} leads to the {\em improved small-$x$ equation} which
incorporates exact leading and NL$x$ BFKL kernels on one hand and renormalization group
constraints in the relevant collinear limits on the other. The basic idea consists in
using a new expansion parameter such that, in the corresponding series, the terms of
order higher than the second are indeed small corrections~\cite{CiCoSa99}.

Besides exhibiting the good qualities of both DGLAP and BFKL formulations, this improved
small-$x$ equation shows several tricky aspects mainly due to the running of the
coupling, the need of regularizing the Landau pole, the diffusion in the infrared
region and the existence of two perturbative exponents for the high-energy
behaviour. For investigating such questions, a simple
model~\cite{CiCoSa99b}, built up with the main ingredients of the RG-improved
formulation, has been introduced and studied in Chap.~\ref{c:cm}. Within this model,
many quantities (among which the full solution of the ensuing equation) can be computed
exactly --- at least numerically --- and the strong coupling physics can be taken into
account so as to see its reflection on the relevant outputs and a clear interpretation of
the small-$x$ growth exponents is at hand. In two-scale processes, this method can find
a direct application, at least in an intermediate small-$x$, large-$Q_i^2$ regime where
perturbative physics appears to dominate. For increasing values of the energy, there are
signals of a new transition mechanism in consequence of which the perturbative features
are lost and ``the pomeron'', i.e.\ the energy growth exponent, belongs to the realm of
non perturbative physics.

The main outcome of the work presented in this thesis is the development of a novel
small-$x$ expansion which turns out to be much more stable than the BFKL method and
which is affected by small next-to-next-to-leading corrections. In particular, the
improved method provides a ``hard pomeron'' index $\om_s(\as)\simeq0.27\div0.32$ for
$\as\simeq0.2\div0.3$ which should be observable in two-scale hard processes.

We are confident that, in two-scale processes, this method can serve to make
reliable predictions, at least in the perturbative regime previously mentioned. Whether
or not it applies to single-scale processes like DIS, remains a question to be
investigated in more detail.

\chapter{Perturbative Treatment Of Hard Processes\labe{c:pthp}}

\section{What's asymptotic freedom?\labe{s:waf}}

The systematic study of strong interactions and the successful description of deeply
inelastic phenomena in terms of free partons inside the hadrons demanded an
asymptotically free theory, i.e., a theory with vanishingly small coupling at short
distances, to be the right and fundamental one.

Non abelian gauge theories are unique among renormalizable theories in 4 space-time
dimensions in having this peculiar feature. The quantum charge giving rise to strong
interaction, called {\em color}, is carried by both the quarks, the fermionic spin $1/2$
matter fields, and the gluons, the bosonic spin 1 gauge fields, mediators of the color
interaction.

Hadronic interactions are effectively described by quantum chromodynamics (QCD) which is
a gauge theory based on a $SU(3)$ group (for a review, see e.g.~\cite{MaPa78} and also
~\cite{Mut87}).  QCD is a conformally invariant theory. Nevertheless, the
renormalization procedure, required in order to remove the divergences of the
perturbative expansion, breaks the conformal invariance and causes the dimensionless
coupling $g$ to depend on dimensional kinematic variables. However, contrary to abelian
theories, the self interaction of gluons ``spread out'' the color charge and an
anti-shielding effect is produced. This is responsible of weakening the color charge at
small distances and presumably to provide the strong coupling regime at large distances
leading to confinement.

In practice, for a process characterized by the scale $Q^2$, the color strength $\as\dug
g^2/4\pi$ among partons must depend on $Q^2$ in a way constrained by the RG equations,
according to which
\begin{equation}\lab{beta}
 \be(\as(Q^2))\dug{\dif\as(Q^2)\over\dif\ln Q^2}=
 -b_0\as^2(Q^2)-b_1\as^3(Q^2)+\cdots\;,
\end{equation}
where the $b_i$'s are pure numbers completely determined by the structure of the gauge
group (and, for $i\geq2$, from the renormalization scheme).

If $b_0>0$ asymptotic freedom holds. In QCD
\begin{equation}\lab{bzero}
 b_0={11N_c-2\Nf\over12\pi}\;,
\end{equation}
where $N_c=3$ is the number of colors and $\Nf=3\div6$ is the number of flavors with
masses less than $Q$, and hence $b_0\simeq0.72\div0.88$.

The solution of Eq.~(\ref{beta}) is known and at lowest order reads
\begin{equation}\lab{alfasQ}
 \as(Q^2)={1\over b_0\ln{Q^2\over\La_{\rm QCD}^2}}\;,
\end{equation}
where $\La_{\rm QCD}$ ($\La$ from now on) is a new mass parameter --- not present in the
original Lagrangian --- introduced by the renormalization procedure. It is apparent
that, for $Q^2\gg\La^2$, $\as\ll1$ so perturbation theory and Eq.~(\ref{alfasQ}) are
reliable, while for $Q^2\lesssim\La^2$ we are in strong coupling regime. The QCD scale
$\La$ has to be determined experimentally. Its value of about 200 MeV sets the lower
bound of the perturbative domain.

\section{Hard and semi-hard processes\labe{s:hshp}}

According to the above considerations, we define {\em hard} processes~\cite{DoDyTr80}
those hadronic processes involving a large momentum transfer $Q^2\gg\La^2$, i.e., $Q^2$
greater than or equal to some GeV$^2$. In this case, perturbative QCD provides
quantitative predictions.

Among hard processes we must distinguish {\em semi-hard} ones~\cite{GrLeRy83} where
various (hard) momentum scales of different orders are present, because the coefficient
of the perturbative series may become large and the latter has to be improved with
resummation techniques of classes of Feynman diagrams.

In this chapter we will consider only hard processes. Semi-hard processes will be
discussed from Chap.~\ref{c:shp}.

\section{Inclusive means perturbative\labe{s:imp}}

As soon as we start a perturbative calculation, we are confronted with divergent
integrals. Besides the ultra-violet (UV) divergences, causing the running of the
coupling constant and, as we shall see, the scaling violations of structure functions
(SF), the perturbative theory suffers also infra-red (IR) divergences, due to the
presence of massless particles\footnote{In hard processes, light quarks can be
considered massless in comparison with the much larger hard scale $Q^2$.}.

IR singularities appear in the calculation of diagrams representing
\begin{itemize}
\item emission of gluons with vanishing 4-momentum ({\em soft} singularities;)
\item emission of massless partons collinearly to the incoming one
({\em collinear} singularities).
\end{itemize}
How to recover finite estimates for cross sections? We have to realize that, from a
physical point of view, it is not possible to distinguish a single quark state from one
in which the quark is accompanied by an arbitrary number of soft\footnote{I.e., in the
limit of vanishing energy.}  or collinear gluons. These degenerate states belong to the
same physical state. When evaluating transition rates, both the initial and final states
should be physical ones. Then, the Kinoshita-Lee-Nauenberg theorem~\cite{Mut87} ---
stating that {\sl in a theory with massless fields, transition rates are free of the IR
(soft and collinear) divergence if the summation over the initial and final degenerate
states is carried out} --- ensures to obtain finite results in {\em completely
inclusive} processes, i.e., processes in which nor quarks neither gluons are registered
in the initial or final states. We can quote, e.g., the total $e^+\,e^-$ annihilation
cross section, the jet cross section at fixed angular and energy resolution, the energy
flow, etc.. These quantities are calculable in perturbative QCD by assuming that the sum
over parton states equals the corresponding hadronic sum, in the same spirit of the
parton model.  In short, hadronization and all long distance effects are unimportant
when neglecting the structure of the hadronic initial and final states.

When one or several partons are identified by measuring their momentum, the hypotheses
of the KLN theorem are no longer fulfilled and the perturbative calculation exhibits
large logarithms due to collinear singularities. However, thanks to the factorization
theorem of collinear singularities (cfr.\ Sec.~\ref{s:ipm}), this logarithms can be
resummed ({\em improved perturbation theory}) and they give rise to non trivial
anomalous dimensions determining scaling violations. We refer to these processes as {\em
inclusive processes with registered partons}. We note that the soft singularities of
virtual corrections and real soft emissions cancel each other because of the inclusive
nature of the process. A few examples: DIS (the registered quark in the initial state
being the active parton), inclusive single particle distribution in $e^+\,e^-$
annihilation (the registered quark being in the final state), jet cross section at fixed
invariant mass of observed particles, etc..

Finally, when one or more hadrons are identified, strong coupling effects become
important. These {\em exclusive} or {\em semi-inclusive} processes are therefore
sensitive to IR physics and require a particular perturbative treatment (see, e.g.,
\cite{BaCiMa83}) in addition to hadronization models.

The aim of this thesis is to test and to improve high-energy perturbative QCD. We will
be concerned mainly with high-energy DIS, double DIS, forward jets and in general with
the class of inclusive processes with registered parton described above.

\section{Deeply inelastic scattering\labe{s:dis}}

In order to introduce one of the main physical processes studied in the field of strong
interactions as well as to present some basic mathematical apparatus relating physical
observables in terms of field-theoretical quantities, we start by specifying DIS.

An effective way to study the structure of the hadrons is to probe them by means of
point-like particles, such as electrons or photons. In DIS of electrons by
protons\footnote{Hereafter we will consider a proton as the target,
but any hadron can be considered as well.}, e.g. in the HERA configuration, a beam of
electrons of momentum $p_1$ collides on a target beam of protons of momentum $p_2$. By
recording only the scattered electron $p_1'$ without care of the hadronic final state,
one obtains a highly inclusive measure. The conventional Lorentz invariant kinematic
variables are defined, in the notations of Fig.~\ref{f:dis}, as follows:
\begin{align*}
 &q&&\hspace{-10mm}\dug\; p_1-p_1'&&
 \text{momentum of the virtual photon probing the proton;}\\
 &s&&\hspace{-10mm}\dug\;(p_1+p_2)^2&&
 \text{center of mass energy squared;}\\
 &Q^2&&\hspace{-10mm}\dug\;-q^2&&\text{virtuality of the photon;}\\
 &W^2&&\hspace{-10mm}\dug\;(p_2+q)^2&&
 \text{invariant mass of the hadronic shower }X;\\
 &\nu&&\hspace{-10mm}\dug\; {p_2\cdot q/ M}&&
 \text{electron transferred energy in the LAB frame;}\\
 &x&&\hspace{-10mm}\dug\; {Q^2/ 2p_2}\cdot q&&\text{Bjorken variable;}\\
 &y&&\hspace{-10mm}\dug\; {p_2\cdot q/ p_1\cdot p_2}&&
 \text{transferred energy fraction in the LAB frame;}
\end{align*}
\begin{figure}[ht!]
\centering\includegraphics[width=11cm,height=67mm]{\fig 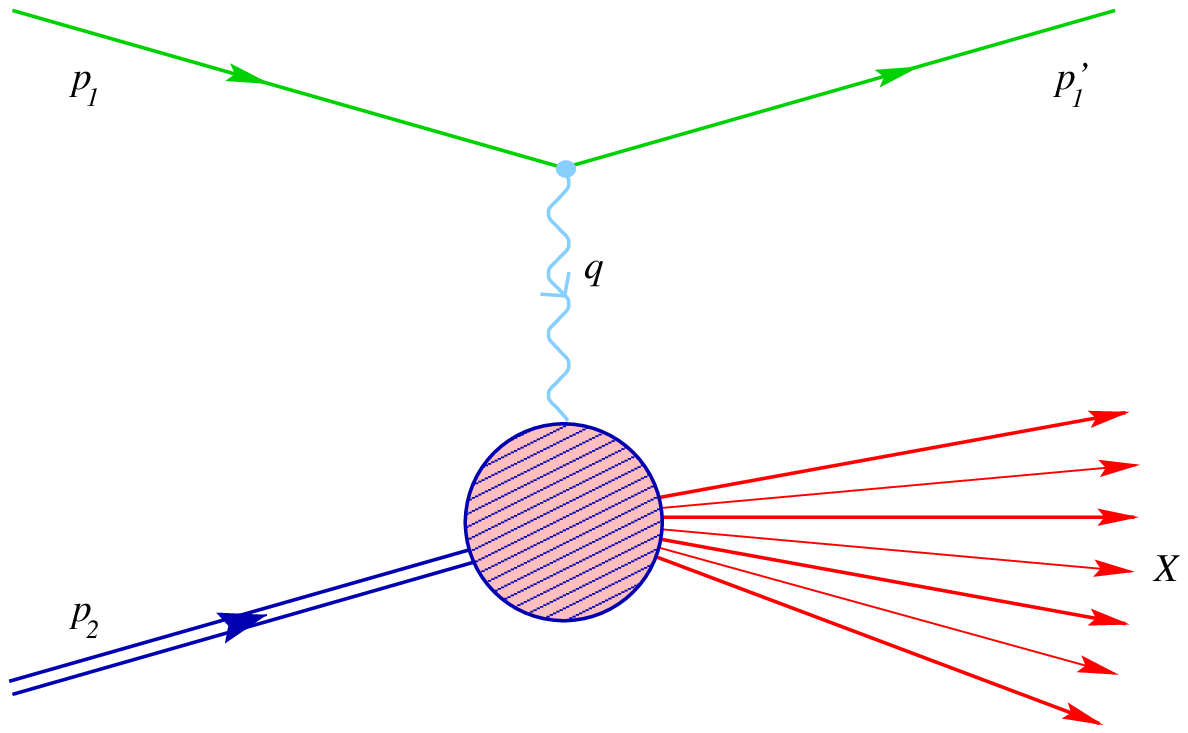}
\didascalia{DIS: $p_1$ ($p_1'$) represents the incoming (scattered) electron, 
	$p_2$ the incoming proton, $q$ the exchanged virtual photon and
	$X$ the hadronic shower.\labe{f:dis}} 
\end{figure}

If the target proton has no spin polarization, the differential cross
section can be parametrized by two\footnote{Here we are concerned only
in the EM part of the neutral current, neglecting $Z^0$ exchanges.}
Lorentz-invariant dimensionless {\em structure functions} according to
App.~\ref{a:sf}\footnote{We have neglected a term of relative order
$M^2/s$ in the coefficient of $F_2$.}:
\begin{equation}\lab{disSF}
 {\dif\si\over\dif x\,\dif Q^2}=
 {2\pi\al^2\over xQ^4}\left\{[1+(1-y)^2]F_2(x,Q^2)-y^2F_L(x,Q^2)\right\}\;.
\end{equation}
What can we say about the structure functions? Even if the final state 
is inclusive as far as the hadrons are concerned and we limit
ourselves to sufficiently high transferred momenta $Q^2$, long
distance effects are important in this kind of reaction. This is due
to the fact that, in the initial state, a particular hadron (the
proton) is recorded by measuring its momentum and therefore the
process is not completely inclusive --- it belongs to the second-type
ones described in the previous section.

Suppose anyhow to undertake the perturbative calculation by assuming the proton composed
of some quarks and gluons. At tree level, the calculation is lead back to the
computation of the electron-quark Born cross section, but as soon as one consider 1-loop
corrections, gluons set in and IR singularities (besides the UV ones) appear giving
infinite cross sections.

Let's stop for a while and try to understand what's happening by looking at scattering
in the center of mass (CM) frame. At high energy $\sqrt{s}\gg M$, the fast motion causes
the proton to be Lorentz contracted in the direction of the collision and its internal
interaction are time-dilated. Thus, the lifetime of any virtual partonic state is
lengthened and the time it takes the electron to traverse the proton is shortened. When
the latter is much shorter than the former, the proton will be in a single virtual state
characterized by a definite number of partons during the entire time the electron takes
to cross it. Since the partons in practice do not interact during this time, each one may
be thought of as carrying a definite fraction $z$ of the proton's momentum%
\footnote{The CM frame is, in this case, a very good approximation of the so called
``infinite momentum frame'' in which the proton is highly boosted so to have only a
(very large) light-cone component of the momentum.}.

Furthermore, if we consider large momentum transfers $Q^2\gg M^2$, then the resolving
power of the virtual photon is much smaller than the transverse size of the proton and
the electron will be able to interact with only a single parton (provided the density of
partons is not too high).

Under these conditions, the electron-parton amplitudes contribute incoherently, hence
the scattering probability on the proton is given by the sum of the probabilities of
scattering on the partons. The latter, in turn, can be thought to be inside the proton
with certain probability distributions.

We have finally the following simple probabilistic interpretation: the electron-proton
cross section is given by the sum over partons of the probabilities for electron-parton
scattering, each of them weighted with the density distribution of that parton to be
found inside the proton. This is the essence of the factorization
theorem of collinear singularities~\cite{CoSoSt89} which, for the structure function,
states
\begin{equation}\lab{fattcoll}
 F_i^{(H)}(x,Q^2)=\sum_{\pa}\int_x^1\dif z\;f^{(\pa/H)}(z,Q^2)\,
 \hat{F}_i^{(\pa)}\({x\over z},\as(Q^2)\)\qquad(i=L,2)\;.
\end{equation}
Here $f^{(\pa/H)}(z,Q^2)\dif z$ is interpreted as the probability%
\footnote{In the literature an alternative notation is also used, differing for a $z$
factor from ours. For instance, the gluon density reads $f^{(\pg)}(z,Q^2)=zg(z,Q^2)$ etc..}%
to find a parton of type $\pa$ (= gluon, $u,\bar{u},d,\bar{d},s,\dots$) in a hadron of
type $H$ carrying a fraction $z$ to $z+\dif z$ of the hadron's momentum ($p_2\equiv p$
in the following), while $\hat{F}_i^{(\pa)}(\xp,\as(Q^2))$ is the partonic structure
function (for the process $e\,\pa\to e\,X$) with respect to the rescaled Bjorken
variable\footnote{The hat $\hat{}\;$ indicates partonic quantities, e.g., $\hat{p}=zp$
is the struck parton momentum.}%
\begin{equation*}
 \xp\dug{Q^2\over2(zp)\cdot q}={x\over z}\;.
\end{equation*}
Note that the partonic distribution functions (PDF) have an explicit
$Q^2$-dependence. In our intuitive picture this is due to the fact
that additional partons --- generated by virtual processes --- can be
resolved by increasing the value of $Q^2$.

\section{The parton model}

If we completely neglect interactions among partons, the parton content of the proton is
fixed, i.e., independent of $Q^2$, since virtual processes are absent. Furthermore, the
virtual photon can be absorbed only elastically by quarks, as depicted in
Fig.~\ref{f:urtoelastico}.
\begin{figure}[ht!]
\centering\includegraphics[height=40mm]{\fig 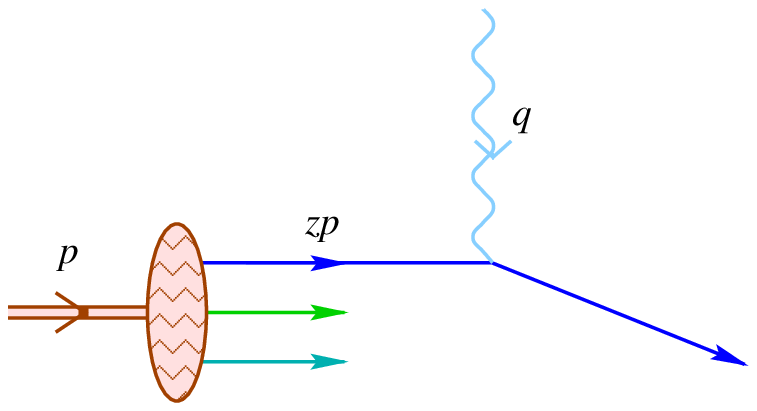}
\didascalia{Elastic scattering of a quark by absorption of a virtual photon.%
\labe{f:urtoelastico}} 
\end{figure}
Since partons are considered massless objects in DIS, it holds
\begin{equation*}
 0=(zp+q)^2=(zp)^2+q^2+2z\,pq=z\,2pq-Q^2
\end{equation*}
or, equivalently
\begin{equation*}
 z={Q^2\over2pq}=x\quad\imp\quad\xp={x\over z}=1\;.
\end{equation*}
This means that the Bjorken variable $x$ coincides with the momentum fraction of the
active quark and that the partonic SF are proportional to $\d(1-\xp)$. The explicit
calculation of the partonic SF yields (omitting the hadron suffix $H$ hereafter)
\begin{subequations}\labe{Fmp}
\begin{align}\lab{F2mp}
 \hat{F}_2^{(\pa)}(\xp,\as=0)&=e_{\pa}^2\d(1-\xp)\;,\\
\lab{FLmp}\hat{F}_L^{(\pa)}(\xp,\as=0)&=0\;,
\end{align}\end{subequations}
$e_{\pa}$ being the EM charge (in unit of the electron charge) of type $\pa$ partons.

The last equation (Callan-Gross relation) follows from the fact that a longitudinally
polarized vector boson cannot be absorbed by a spin-$1/2$ quark (cfr.\ App.~\ref{a:sf}).
However the QCD corrections provide a non-zero coupling between
longitudinal virtual photons and partons, so that $F_L=\ord(\as)$. The SF measurements
show that $F_L\ll F_2$ (see Fig.\ref{f:FL}), confirming the spin $1/2$ property of
quarks.
\begin{figure}[ht!]
\centering\includegraphics[height=75mm]{\fig 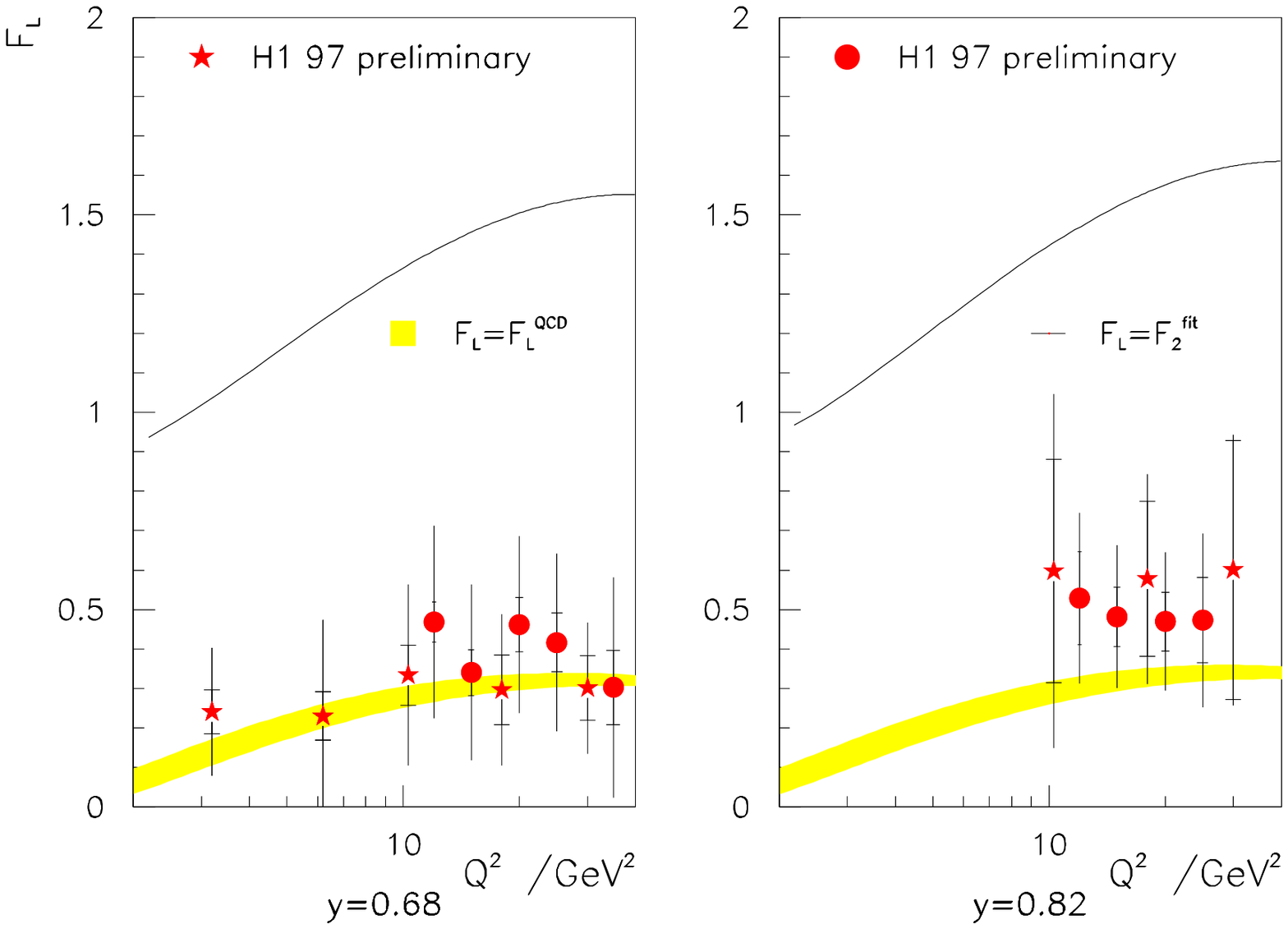}
\didascalia{The longitudinal proton structure function $F_L$ data compared with the $F_2$
structure function fit (the black line).\labe{f:FL}} 
\end{figure}

Because of the $Q^2$-independence of both the PDF and the partonic SF, the SF depend
only on the dimensionless variable $x$, as one would expect from a simple dimensional
analysis. This is the so called {\em scaling} behavior~\cite{Bjo69}. The $Q^2$
dependence of $F_2$ for various values of $x$ is reported in Fig.~\ref{f:F2}.
Approximate scaling is visible, especially around $x\simeq0.2$. Anyway, scaling
violations are also evident, and their origin is the subject of the next sections.
\begin{figure}[ht!]
\centering\includegraphics[height=117mm,width=77mm]{\fig 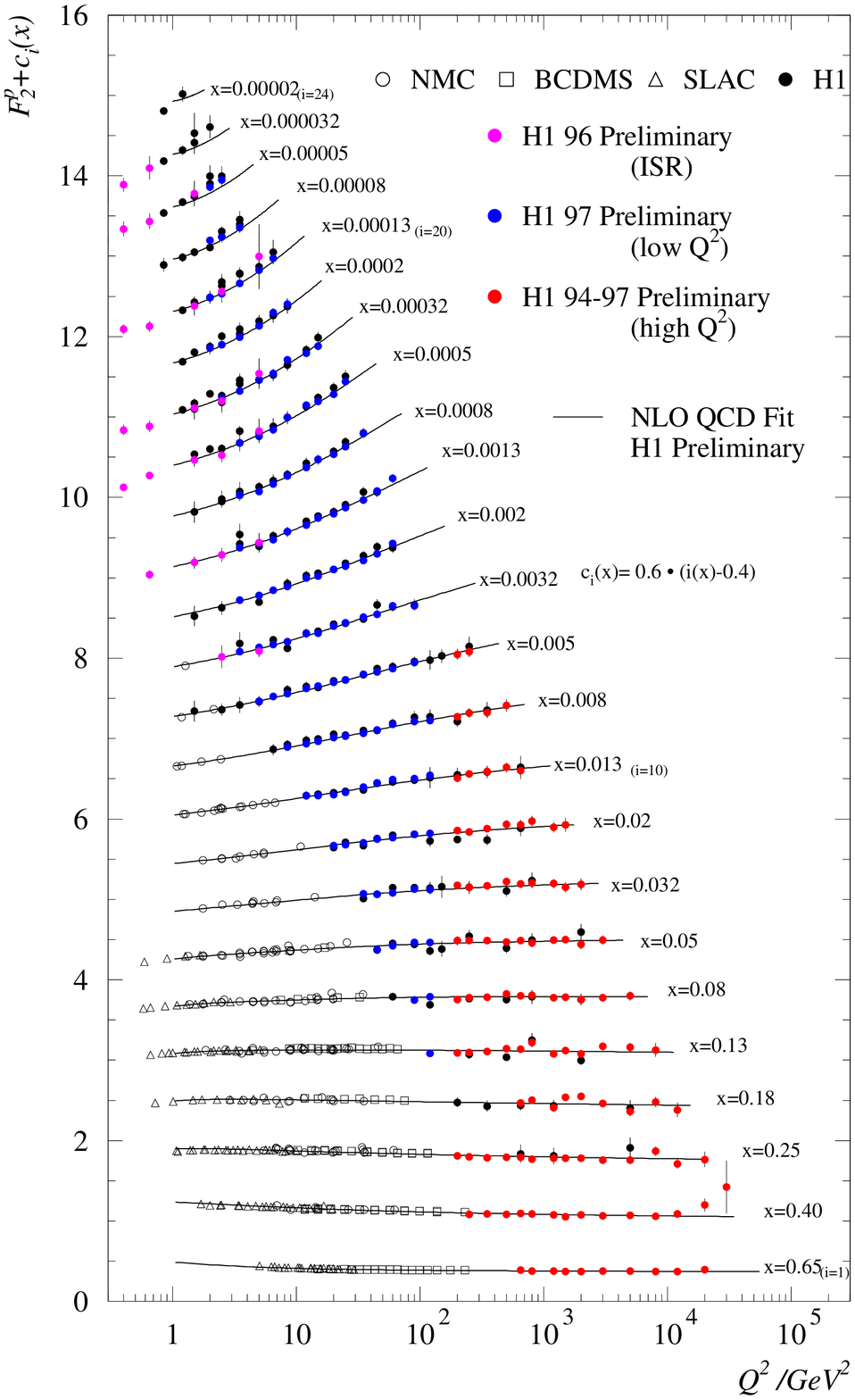}
\didascalia{The proton structure function $F_2$ versus $Q^2$ for various
values of $x$.\labe{f:F2}} 
\end{figure}
\section{Scaling Violations\labe{s:sv}}

Historically, the first approach to next-to-leading order%
\footnote{Next-to-leading in $\as$.}
(NLO) corrections for DIS made use of the OPE. Even if a clear interpretation of the
underlying physics will be available in the parton model picture, this formal approach
will be useful in order (i) to classify the main contributions to scaling violations (by
means of the so called {\em twist} index); (ii) to show explicitly the connection
between anomalous dimensions of certain composite operators and evolution kernels for
partonic distributions; (iii) to provide a first explanation of factorization of
perturbative and non perturbative effects in strong interactions.

\subsection{Operator product expansion (OPE)\labe{ss:ope}}

Starting from the expression (\ref{wcomm}) for the hadronic tensor
\begin{equation}\lab{wcomm1}
 W_{\mu\nu}(p,q)={1\over4\pi}\int\dif^4z\;\esp{\ui qz}\,{1\over2}
 \sum_{\la=\text{spin}}\vm{p,\la|[J_\mu(z),J_\nu(0)]|p,\la}\;,
\end{equation}
we note that the commutator of the EM currents vanishes for $z^2<0$. Furthermore, in the
Bjorken limit $Q^2\to\infty$, $x$ fixed, the dominant domain of integration turns out to
be $0\leq z^2\lesssim\text{const}/Q^2$, i.e., a thin sheet upon the forward light cone.

Notice that the product of two local operators at light-like distances is
singular. Nevertheless, these singularities can be embodied in c-number functions by
mean of the OPE~\cite{Wil69} that near the light-cone takes the form%
\footnote{We omit for simplicity the Lorentz and flavour indices.}
\begin{equation}\lab{jj}
 J(z)J(0)=\sum_{\om=0}^\infty\sum_{\pj} C_\om^{(\pj)}(z^2)\,z^{\mu_1}
 \cdots z^{\mu_\om}\,O_{\mu_1\cdots\mu_\om}^{(\pj)}\;,
\end{equation}
where the c-number {\em coefficient functions} $C_\om^{(\pj)}(z^2)$ are eventually
singular at $z^2=0$ and the {\em composite operators} $O_{\mu_1\cdots\mu_\om}^{(\pj)}$
are well-defined objects, apart from the renormalization infinities. Anyhow, the
singular behaviour in coordinate space is entirely carried by the coefficient functions.

In the Bjorken limit, all the masses can be neglected and the only dimensional variables
in configuration space are the coordinates. A na\"\i ve dimensional analysis of
Eq.~(\ref{jj}) shows that, near the light-cone, the behaviour of the coefficient
functions should be
\begin{equation}\lab{lcb}
 C_\om^{(\pj)}(z^2)\simeq (z^2)^{-3+\half(d_\om^{\pj}-\om)} \quad,\quad z^2\sim0\;,
\end{equation}
where $3=[J]$ and $d_\om^{\pj}=[O_{\mu_1\cdots\mu_\om}^{(\pj)}]$ are the natural
dimensions of the EM current and of the composite operators respectively.

In the non interacting theory the conformal invariance would require such a simple
power-like behaviour. However in the interacting theory the above simple argument does
not hold and in general logarithmic corrections to Eq.~(\ref{lcb}) develop after the
renormalization procedure.

We see from Eq.~(\ref{lcb}) that, the smaller the {\em twist} value $d_\om^{\pj}-\om$
is, the more singular $C_\om^{(\pj)}$ will be, and then the dominant terms as $z^2\to0$
($Q^2\to\infty$) are controlled by the lowest twist operators, higher twists being
suppressed by powers of $Q^2$.

There are two kinds of leading twist operators (twist $=2$) occurring 
in the OPE~(\ref{jj}) for DIS:
\begin{subequations}\labe{leadingtwist}
\begin{align}\lab{Og}
 O_{\mu_1\cdots\mu_\om}^{(\pg)}&=F_{\mu_1\la}\dcg_{\mu_2}
 \cdots\dcg_{\mu_{\om-1}}F_{\mu_\om}{}^\la\;,\\
\lab{Oq}O_{\mu_1\cdots\mu_\om}^{(\pf)}&=\psib{}^{(\pf)}\ga_{\mu_1}
 \dcf_{\mu_2}\cdots\dcf_{\mu_\om}\psi^{(\pf)}
 \quad,\quad(\pf=1,\dots,\Nf)\;,
\end{align}
\end{subequations}
$\Nf$ being the number of (massless) flavours and $\dcg$ ($\dcf$) the covariant
derivative for gluons (quarks). The former are gluonic operators and the latter are
quark operators.

It is to be stressed that the composite operators of Eqs.~(\ref{leadingtwist}) appearing
in the OPE do not depend on the hard probe (photon, weak boson, gluon, etc.) coupled to
the hadronic tensor, and their hadronic matrix elements $A^{(\pj)}_\om$
(cfr.\ Eq.~(\ref{pop})) describe only the partonic content of the initial particle $H$. In this
sense, they are universal, i.e., the same for any hard reaction involving the same
hadron $H$. However, the composite operators describe the long distance properties of
the particles and therefore they cannot be calculable within perturbative QCD and their
ignorance have to be compensated by some modellistic assumption or, better, by
experimental measurements.

The coupling to the probe is accounted for by the coefficient functions, which are
process dependent. They contains the informations of the short distance interactions
between the exchanged boson and the constituents of the hadron as well as of high
transfer momentum processes among partons. The following analysis will show how the
former part of these short distance processes corresponds to the partonic SF and the
latter one can be included in the PDF which acquire a (logarithmic) dependence on the
virtuality $Q^2$ of the probe.
 
What is needed to derive the $Q^2$-dependence of the SF is
\begin{itemize}
\item to relate the objects of the OPE to the SF;
\item to determine the $Q^2$-dependence of the coefficient functions
(the composite operators being independent of $z$ and hence of $Q^2$).
\end{itemize}

The relations between OPE and SF have been derived in \cite{ChHaMu72} and express the
moments (i.e.\ the Mellin transforms) of the SF%
\footnote{You can think $F$ to be any of the SF $F_2$ or $F_L$. Only
the coefficient functions carry the ``$2$'' or ``$L$'' dependence.}
\begin{equation}\lab{mfs}
 F_\om(Q^2)\dug\int_0^1{\dif x\over x}\;x^\om\,F(x,Q^2)
\end{equation}
in terms of the Fourier transforms%
\footnote{Up to a numerical factor $2^m\pi^k\ui^h$.}
of the coefficient functions
\begin{equation}\lab{tffc}
 (Q^2)^\om\int\dif^4z\;\esp{\ui qz}\,z^{\mu_1}\cdots z^{\mu_\om}\,
 C_\om^{(\pj)}(z^2)
 \ugd q^{\mu_1}\cdots q^{\mu_\om}\,\ct_{\om-1}^{(\pj)}(Q^2)
\end{equation}
and of the matrix elements
\begin{equation}\lab{pop}
 {1\over2}\sum_\la\vm{p,\la|O_{\mu_1\cdots\mu_\om}^{(\pj)}|p,\la}
 =A_{\om-1}^{(\pj)}\,p_{\mu_1}\cdots p_{\mu_\om}+\text{terms containing }
 g_{\mu_i\mu_k} 
\end{equation}
in the famous moment sum rules
\begin{equation}\lab{msr}
 F_{\om}(Q^2)=\sum_{\pj} A_\om^{(\pj)}\,C_\om^{(\pj)}(Q^2)\;.
\end{equation}

The determination of the coefficient function dependence on $Q^2$ can
be done by using the RG technique\cite{GrWi74}. The renormalization procedure,
necessary to handle the infinities stemming from the perturbative expansion,
requires the introduction of a mass parameter $\mu$ (the
renormalization scale) not present in the original lagrangian. As
$\mu$ is an artificial parameter, the physical quantities --- in our
case the SF --- do not depend on it. This constraint determine the
$\mu$-dependence of the coefficient functions, of the composite
operators and of the coupling as well.

In addition, the dimensionless coefficient
functions and the coupling can depend on the massive parameter $Q^2$
and $\mu$ only through the ratio $Q^2/\mu^2$. Therefore, their
$Q^2$-dependence is completely calculable within perturbative QCD and, by means
of Eq.~(\ref{msr}), we obtain the $Q^2$ dependence of the SF.

In particular, the RG equation for the coefficient function is
\begin{equation}\lab{rgeq}
 \left[\left(\mu{\de\over\de\mu}+2\be(\as){\de\over\de\as}\right)\d^{\pj\pk}
 +(\gga_\om)^{\pj\pk}\right]\ct_\om^{(\pk)}\(Q^2/\mu^2,\as(\mu^2)\)=0
 \qquad\forall\;\pj\;,
\end{equation}
where the indices $\pj$ and $\pk$ label the different types of operators
in Eqs.~(\ref{leadingtwist}) and $\gga_\om$ is (the opposite of) their
anomalous dimension matrix:
\begin{align}\lab{defanomdim}
 \gga_\om&\dug-\boldsymbol{Z}_\om^{-1}
 \left(\mu{\de\over\de\mu}\boldsymbol{Z}_\om\right)\quad,\quad
 O_{\mu_1\cdots\mu_\om}^{(\pj),\text{bare}}=
 \left[\boldsymbol{Z}_\om\right]^{\pj\pk}
 O_{\mu_1\cdots\mu_\om}^{(\pk),\text{renormalized}}\;,\\
 \lab{sviluppodimanom}\gga_\om&\ugd
 \as\,\ggau_\om+\as^2\,\overset{\mbox{{\tiny (2)}}}{\gga}{}_\om+\cdots\;.
\end{align}

The solution of Eq.~(\ref{rgeq}) is found by introducing the running
coupling function of Eq.~(\ref{beta}) and can be written as a path-ordered exponential
\begin{equation}\lab{solcoef}
 \ct_\om^{(\pj)}(Q^2/\mu^2,\as(\mu^2))=\Texp\left\{\int_{\as(\mu^2)}^{\as(Q^2)}
 {\gga_\om(\al)\over\be(\al)}\;\dif\al\right\}^{\pj\pk}\ct_\om^{(\pk)}\(1,\as(Q^2)\)\;,
\end{equation}
showing that the coefficient functions depend on $Q^2$ only through
the running coupling $\as(Q^2)$.

\subsection{The ``running'' parton distribution functions\labe{ss:rpdf}}

The formal results (\ref{msr}) and (\ref{solcoef}) can be given a
simple and elegant interpretation in terms of PDF which depend upon
$Q^2$: we can write at LO in $\ln Q^2$
\begin{equation}\lab{solcoefLO}
 F_{i,\om}(Q^2)=\sum_{\pj,\pk}A_\om^{(\pj)}\left[
 {\as(\mu^2)\over\as(Q^2)}\right]^{\ggau\!_\om/b_0}_{\;\pj\pk}
 \ct_{i,\om}^{(\pk)}(1,0)\;\times\;\left[1+\ord(1/\ln Q^2)\right]\;.
\end{equation}

First of all, switch off for a minute the strong interaction, so that
$\as(Q^2)=\as(\mu^2)=0$ and
\begin{equation}\lab{foni}
 F_{i,\om}(Q^2)=F_{i,\om}=\sum_{\pj}A_\om^{(\pj)}\,\ct_{i,\om}^{(\pj)}(1,0)\;.
\end{equation}
If we take the Mellin transform of Eq.~(\ref{fattcoll}) by defining
\begin{align}\lab{mfsp}
 \hat{F}_{i,\om}^{(\pa)}\(\as(Q^2)\)&\dug\int_0^1{\dif x\over x}\;x^\om\,
 \hat{F}_i^{(\pa)}\(x,\as(Q^2)\)\;,\\	\lab{mpdf}
f_\om^{(\pa)}(Q^2)&\dug\int_0^1\dif x\;x^\om\,f^{(\pa)}(x,Q^2)\;, 
\end{align}
the moments of the SF turns out to be simply
\begin{equation}\lab{msrLO}
 F_{i,\om}=\sum_{\pa}f_\om^{(\pa)}(Q^2)\,\hat{F}_{i,\om}^{(\pa)}(0)\;.
\end{equation}
In this situation it is natural to identify the short distance and long distance factors as
\begin{align}\lab{idenCF}
 \ct_{i,\om}^{(\pa)}(1,0)&\longleftrightarrow\hat{F}_{i,\om}^{(\pa)}(0)\;,\\
 \lab{idenAf}A_\om^{(\pa)}&\longleftrightarrow f_\om^{(\pa)}(Q^2)=f_\om^{(\pa)}\;.
\end{align}
Equivalently, one can obtain the parton model picture (\ref{fattcoll}) by performing the
inverse Mellin transform of Eq.~(\ref{foni}) by suitably identifying PDF and partonic SF.

We can go further by applying the same reasoning in the interacting case, in which we
use again Eq.~(\ref{idenCF}) --- because at LO the partonic SF are $Q^2$-independent ---
together with the new identification
\begin{equation}\lab{idenAfQ}
 \sum_{\pa}A_\om^{(\pa)}\left[{\as(\mu^2)\over\as(Q^2)}
 \right]^{\ggau\!_\om/b_0}_{\;\pa\pb}\longleftrightarrow f_\om^{(\pb)}(Q^2)\;.
\end{equation}
in such a way that the $Q^2$-dependence is incorporated in the PDF.

At this point we differentiate the above relation with respect to $\ln Q^2$ and, taking
into account Eqs.~(\ref{beta}) and (\ref{alfasQ}), we end up with the evolution equation
for the moments of the PDF
\begin{equation}\lab{EDacc}
 {\dif\over\dif\ln Q^2}f_\om^{(\pa)}(Q^2)=\sum_{\pb}\as(Q^2)
 [\ggau\!_\om]^{\pa\pb}f_\om^{(\pb)}(Q^2)\quad,\quad\pa=\pf,\bar{\pf},\pg\;.
\end{equation}
This set of $2\Nf+1$ coupled differential equations can be diagonalized by analysing the
properties of the PDF under flavour symmetry --- which is an exact symmetry if quark
masses are neglected.

Clearly, the gluon density $f^{(\pg)}$ is invariant under flavour transformations, as
well as the ``modulus'' of the quark vector
\begin{equation}\lab{singPDF}
 f^{(\Si)}(x,Q^2)\dug\sum_{\pf}f^{(\pf)}(x,Q^2)+f^{(\bar{\pf})}(x,Q^2)
\end{equation}
called {\em quark singlet} density.

The non-singlet (NS) components of the quark densities, transforming according to the
adjoint representation of $SU(\Nf)$, are constructed by subtracting from the various
fermionic PDF the common singlet component:
\begin{equation}\lab{nonsingPDF}
 f^{(\ns,\pf)}(x,Q^2)\dug f^{(\pf)}(x,Q^2)-{1\over2\Nf}f^{(\Si)}(x,Q^2)\;.
\end{equation}
Since the NS quark densities carries different quantum numbers (flavours) they don't mix
and renormalize independently. Furthermore, the flavour group commutes with the colour
group, so all NS densities evolve with the same quark-to-quark anomalous dimension
$\ga^{\pq\pq}$
\begin{equation}\lab{EDns}
 {\dif\over\dif\ln Q^2}f_\om^{(\ns,\pf)}(Q^2)=\as(Q^2)\,
 \gaqqu_\om\,f_\om^{(\ns,\pf)}(Q^2)\;.
\end{equation}
Only quark singlet and gluon operators mix in the renormalization and
require a $2\times2$ anomalous dimension matrix such that
\begin{equation}\lab{EDsing}
 {\dif\over\dif\ln Q^2}
 \begin{pmatrix}f_\om^{(\Si)}(Q^2)\\f_\om^{(\pg)}(Q^2)\end{pmatrix}
 =\as(Q^2)
 \begin{pmatrix}\gaqqu_\om&2\Nf\,\gaqgu_\om\\
 \gagqu_\om&\gaggu_\om\end{pmatrix}
 \begin{pmatrix}f_\om^{(\Si)}(Q^2)\\f_\om^{(\pg)}(Q^2)\end{pmatrix}\;.
\end{equation}
Using Eq.~(\ref{mpdf}) and the {\em splitting functions} implicitly
defined by
\begin{equation}\lab{defsplitfun}
 \ga_\om^{\pa\pb}\ugd{\as(Q^2)\over2\pi}\int_0^1\dif z\;z^\om\,P^{\pa\pb}(z)\;,
\end{equation}
we can invert Eqs.~(\ref{EDns}) and (\ref{EDsing}) in $x$ space and reproduce the famous
Dokshitzer-Gribov-Lipatov-Altarelli-Parisi (DGLAP) equations~\cite{DGLAP77}
\begin{subequations}\labe{eqAP}
\begin{align}\lab{eqAPns}
 &{\dif\over\dif\ln Q^2}f^{(\ns,\pf)}(x,Q^2)\hspace{-6mm}&&
 ={\as(Q^2)\over2\pi}\int_x^1\dif z\;
  \Pu^{\pq\pq}\({x\over z}\)\,f^{(\ns,\pf)}(z,Q^2)\;,\\[2mm]	\lab{eqAPsigma}
 &{\dif\over\dif\ln Q^2}f^{(\Si)}(x,Q^2)&&={\as(Q^2)\over2\pi}\int_x^1\dif z\;
  \Pu^{\pq\pq}\({x\over z}\)\,f^{(\Si)}(z,Q^2)
  +2\Nf \Pu^{\pq\pg}\({x\over z}\)\,f^{(\pg)}(z,Q^2)\;,\\[-3mm]  \lab{eqAPglu}
 &{\dif\over\dif\ln Q^2}f^{(\pg)}(x,Q^2)&&={\as(Q^2)\over2\pi}\int_x^1\dif z\;
  \Pu^{\pg\pq}\({x\over z}\)\,f^{(\Si)}(z,Q^2)
  +\Pu^{\pg\pg}\({x\over z}\)\,f^{(\pg)}(z,Q^2)\;.
\end{align}\end{subequations}
It is also possible to diagonalize the anomalous dimension matrix of the singlet sector
by determining the eigenvalues
\begin{equation}\lab{avlsing}
 \ga_\om^\pm={1\over2}\left(\gaqq_\om+\gagg_\om\pm\sqrt{
 (\gaqq_\om-\gagg_\om)^2+8\Nf\gaqg_\om\gagq_\om}\right)
\end{equation}
and the eigenvectors
\begin{equation}\lab{avtsing}
 f_\om^{(\pm)}=f_\om^{(\pg)}+
 {\ga_\om^\pm-\gagg_\om\over2\Nf\gaqg_\om}f_\om^{(\Si)}(Q^2)\;,
\end{equation}
so that
\begin{equation}\lab{EDind}
  {\dif\over\dif\ln Q^2}f_\om^{(\pm)}(Q^2)=\as(Q^2)
 \ga_\om^\pm f_\om^{(\pm)}(Q^2)\;.
\end{equation}
\begin{table}[ht!] \scriptsize\centering
 \begin{tabular}{lll}
 \hline
 \vline~{\pb\,{\tiny$\leftarrow$}\,\pa~}\vline&\raisebox{-3mm}{\rule{0mm}{9mm}}
 \footnotesize{$\Pu^{\pb\pa}(z)$}\hfill\vline&
 \footnotesize{$2\pi\overset{\mbox{{\tiny (1)}}}{\gamma}{}_\om^{\pb\pa}$}\hfill\vline\\
 \hline\hline
\vline ~~\slarga{\pq\hspace{2mm}\pq}\hfill\vline&
 $\ds{C_F\left(\frac{1+z^2}{1-z}\right)_{\!\!{}_+}}$\hfill\vline&
 $\ds{C_F\left[{3\over2}+{1\over(\om+1)(\om+2)}+2\psi(1)-2\psi(\om+2)\right]}
$\quad\vline\\
 \hline
\vline ~~\slarga{\pg\hspace{2mm}\pq}\hfill\vline&
 $\ds{C_F\left[\frac{1+(1-z)^2}{z}\right]}$\hfill\vline&
 $\ds{C_F\,{\om^2+3\om+4\over\om(\om+1)(\om+2)}}$\hfill\vline\\
 \hline
\vline ~~\slarga{\pq\hspace{2mm}\pg}\hfill\vline&
 $\ds{T_R\left[z^2+(1-z)^2\right]}$\hfill\vline&
 $\ds{T_R\,{\om^2+3\om+4\over(\om+1)(\om+2)(\om+3)}}$\hfill\vline\\
 \hline
\vline ~~\slarga{\pg\hspace{2mm}\pg}\hfill\vline&
$\ds{2C_A\left[\frac{1-z}{z}+\frac{z}{(1-z)_{{}_+}}+z(1-z)\right]
	+\d(1-z)}$\hspace{5mm}~&\hspace{-10mm}
$\ds{\left[{11C_A\over6}-{\Nf\over3 N_c^2}\right]}\;$~\hfill\vline\\
\vline\slarga{~}\hfill\vline& &\hspace{-6cm}\hfill
$\ds{2C_A\left[{11\over12}+{1\over\om(\om+1)}+{1\over(\om+2)(\om+3)}
	+\psi(1)-\psi(\om+2)\right]-{\Nf\over3N_c^2}}$\quad\vline\\
 \hline
\vline\slarga&&\hspace{-75mm}$\ds{C_A=N_c\;,\;C_F={N_c^2-1\over2N_c}\;,\;T_R={1\over2}
 \;,\quad\int_x^1f(z)g_+(z)\dif z:=\int_x^1[f(z)-f(1)]g(z)\,\dif z-f(1)\int_0^xg(z)\,\dif z}$
\hfill\vline\\ \hline
\end{tabular}
\didascalia{The 1-loop partonic splitting functions and anomalous dimensions.
\labe{t:AP}}
\end{table}

To summarize the results of the previous sections, we have shown that a certain number
of observables of hard inclusive processes can be expressed as a convolution of process
dependent, perturbative calculable, partonic cross sections and universal, non
perturbative, partonic distribution functions inside hadrons whose dependence on the
virtuality $Q^2$ of the hard probe is however computable within the framework of
perturbative QCD. This means that we are not in such a position to say what the value of
$F_2(x=0.1,Q^2=10\;\text{GeV}^2)$ is, but if we measure with enough accuracy the
$x$-dependence of a sufficient number of structure functions at a certain value of
$Q^2\gg\La^2$, we can predict their behaviour at higher $Q^2$ values. Furthermore, the
consequent knowledge of the PDF allows us to predict absolute values of cross sections
for completely different hadronic processes. This is very important as a test of the
theory as well as experimentally, in order to check normalization on the cross sections,
to estimate the occurrence of yet unobserved phenomena, etc..

\section{The improved parton model\labe{s:ipm}}

In the following, we shall show an alternative approach for dealing with scaling
violations in DIS. This method presents several advantages with respect to the OPE shown
before.

On one hand, there are very few reactions where one can justify the use of the OPE. It
is therefore of great importance to rephrase the physics in the language of a very
general QCD-improved parton model by which we can study a larger variety of phenomena.

On the other hand the improved parton model gives the LO results in a much more
intuitive and economical way and permits, with reasonable efforts, to perform NLO
calculations (e.g., the NLO anomalous dimensions).

In addition, --- and this is the main reason of this lengthy discussion --- the partonic
framework clarifies the connection between general field-theoretical predictions, such
as factorization and OPE, and the resummation of certain classes of Feynman diagrams. A
very similar formalism will be very useful when dealing with high energy reactions,
where Regge theory, BFKL dynamics, high energy factorization and RG constraints meet in
describing semi-hard processes. Also in this case the partonic picture turns out to be
of fundamental importance to extend the calculation to next-to-leading level and to take
into account a whole series of sub-leading contributions to all orders, which is the
central purpose of the present thesis.

\subsection{Next-to-leading order parton model\labe{ss:nlopm}}

Having convinced ourselves of the factorization properties splitting strong
coupling from perturbative physics, we are going to attack the problem from
another point of view. We assume the hadron undergoing DIS to be composed of
several partons $\pa=\pf,\bar{\pf},\pg$ with distribution densities $f^{(\pa)}(\xi)$, $\xi$
being the longitudinal momentum fraction of the parton. We assume also that
parton-photon and parton-parton interactions can be treated perturbatively. The
validity conditions and consequences of these hypotheses are what we wish to
present.

Since among partons only quarks carry EM charge, the partonic interaction at Born level
is represented uniquely by the diagram in Fig.~\ref{f:quafotLO}, where $\pp=\xi p$ is
the incoming quark momentum. The ensuing partonic tensor reads
\begin{subequations}\labe{w0}
\begin{align}
 \Wp^{(\pa,0)}(\pp,q)&=e_{\pa}^2\d(1-\xp)\Ga_{\mu\nu}(\pp,q)\\
 \Ga_{\mu\nu}(\pp,q)&=-{1\over2}\left(g_{\mu\nu}-{q_\mu q_\nu\over q^2}\right)
 +{1\over\pp q}\left(\pp_\mu-{\pp q\over q^2}q_\mu\right)
 \left(\pp_\nu-{\pp q\over q^2}q_\nu\right)\;,
\end{align}\end{subequations}
whose longitudinal and transverse projections give the partonic SF of
Eqs.~(\ref{Fmp}).
\begin{figure}[ht!]
\centering\includegraphics[height=35mm]{\fig 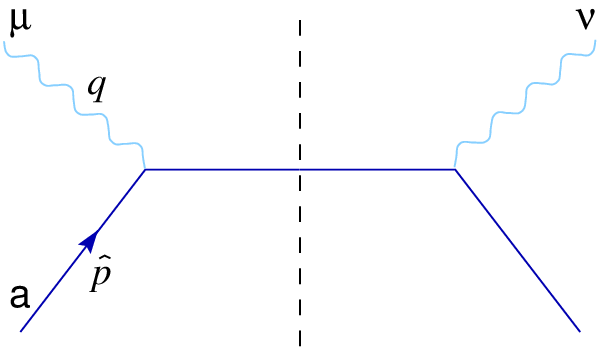}
\didascalia{The diagram representing the partonic tensor $\Wp^{(\pa,0)}$ at Born 
	level. $\pa$ labels the incoming quark flavour an $\pp$ its momentum.%
\labe{f:quafotLO}} 
\end{figure}

At NLO, we should take into account all the insertions of a gluonic line in
Fig.~\ref{f:quafotLO} such as Fig.~\ref{f:fotNLO}{\sl a} and also gluon-photon
coupling via a quark-loop like in Fig.~\ref{f:fotNLO}{\sl b}.
\begin{figure}[ht!]
\centering\includegraphics[width=14cm]{\fig 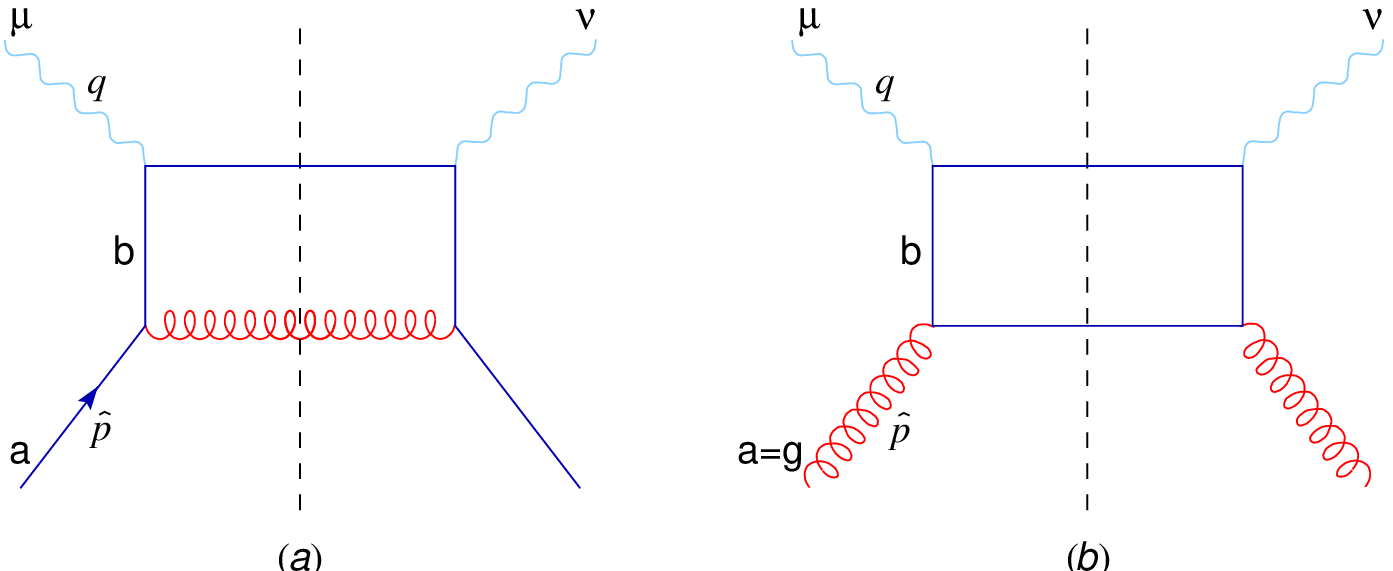}
\didascalia{Examples of 1-loop couplings: quark-photon (a) and gluon-photon (b).
	\labe{f:fotNLO}} 
\end{figure}

Let's start computing the real emission diagram of Fig.~\ref{f:fotNLO}{\sl a} in the
kinematic regime $Q^2\gg M^2$ and $x,1-x$ not too small, where we expect scaling
violations. In this regime, quarks can be considered massless, but in so doing
the integration over the internal momenta $k$ diverges when the transverse
component $\kk$ vanishes, i.e., when the internal quark and the outgoing gluon
are emitted collinearly (collinear singularity). In our notations, all boldface
transverse vectors have to be considered $D-2$ dimensional vectors with euclidean
components, such that $\kk^2>0$.

By means of a suitable regularization, by introducing for instance a (small) cutoff
$\la$, we end up with a finite result which has the interesting feature of containing a
term proportional to $\ln Q^2/\la^2$. The same behaviour is shown by the diagram in
Fig.~\ref{f:fotNLO}{\sl b}.

This logarithmic factor comes from the $\kk$-integration
\begin{equation}\lab{intlead}
 \int_{|\kk|_\min}^{|\kk|_\max}{\dif^2\kk\over\kk^2}\propto\ln{Q^2\over\la^2}
 \quad;\quad\kk_\min^2=\la^2\quad;\quad\kk_\max^2=Q^2{1-\xp\over4\xp}
 =\text{(small const)} Q^2\;.
\end{equation}
It is crucial to note that to the IR logarithmic collinear singularity
corresponds a $\ln Q^2$ term arising from the hard $\kk$ component of the
emitted quark. This is obvious from a simple dimensional analysis, $Q^2$ being
the only dimensional scale which can accompany the cutoff $\la^2$ in the
logarithm. Conversely, to each $\ln Q^2$ is in fact associated a collinear
logarithmic singularity so that one can investigate the structure of the latter
for determining the former. Here resides the intimate connection between
collinear singularities and scaling violations.

The presence of a large $\ln Q^2$ at $\ord(\as)$ should warn us as far as the
convergence of the perturbative series is concerned. In fact, taking for $\as$
its running value $\as(Q^2)\sim1/\ln Q^2$, the NLO correction
$\sim\as(Q^2)\ln Q^2\sim1$ may be large even for small coupling. This is
exactly what happens at higher orders, where in the $\ord(\as^n)$ contribution
we find terms proportional to $\as^n(Q^2)\ln^n(Q^2)\sim1$. Therefore the
relative importance of all the sub-leading corrections may be the same, and in
order to provide quantitative estimates, all of them ought to be taken into
account through a suitable resummation procedure.

The first step of a resummation procedure consists in rearranging the
perturbative series according to a new hierarchy, the first term of which is
formed by the ``largest'' contributions $\sim[\as\ln Q^2]^n:n=0,1,\cdots$
(leading logarithmic (LL) approximation), the second one containing one power of 
$\as$ more than the first one, i.e., $\sim\as[\as\ln Q^2]^n$ (next-to-leading
logarithmic (NLL) approximation), the third one (NNLL) $\sim\as^2[\as\ln Q^2]^n$ 
and so on.

The second step requires the determination of all the LL terms in the
perturbative series by identifying an iterating common structure --- typically
an integral kernel --- representing the building block of the corresponding
Feynman diagrams.

In the third step one expresses the unknown resummed object by means of an
equation --- typically an integro-differential one --- whose basic ingredient is 
the iterating structure previously mentioned.

Finally, one has to solve the resumming equation as best as he or she (or the
computer) can, to check its consistency, to try to fit the data and, possibly, restarting
from the beginning in the next logarithmic approximation.

In this section we are dealing with the real emission, 1-loop correction to SF 
at leading $\ln Q^2$ level. For both \ref{f:fotNLO}{\sl a} and {\sl b}
processes, the logarithmic term stems from the integration over the internal quark
transverse momentum (Eq.~(\ref{intlead})). The link with the collinear
singularity tells us that the coefficient of $\ln Q^2$ is given by the residue
of the $1/\kk^2$ pole.

What about other real emission diagrams? Following the resummation program, we
have to identify the diagrams furnishing the LL terms we are looking for. It
turns out that, adopting axial gauges, i.e., including only physical transverse
polarizations in the gluon propagators, only the diagrams \ref{f:fotNLO}{\sl a}
and {\sl b} contributes in the LL approximation.

It is convenient to introduce the Sudakov parametrization
\begin{equation}\lab{sudakNLO}
 k=z\pp+\bar{z}q'+\kk\quad,\quad q'\dug q+\xp\pp\quad,\quad
 \xp\dug{-q^2\over2\pp\cdot q}\;,
\end{equation}
which are nothing but the components of $k$ with respect to the light-like vectors $\pp$
and $q'$ plus the transverse component. In our notations, $k^2=2z\bar{z}\,\pp\cdot
q'-\kk^2$ and $\kk^2>0$.  For vanishingly small values of $\kk$, the mass-shell
constraints of the outgoing partons force $\bar{z}\simeq0$ and $z\simeq\xp$ so that the
only relevant variable (upon which the residue depends) is the momentum fraction $\xp$
of $k$ with respect to $\pp$.

The residue of the collinear singularities of Figs.~\ref{f:fotNLO}.{\sl a} and
{\sl b} turns out to be respectively%
\footnote{From now on, both for the splitting functions and for the anomalous
dimension coefficients we shall omit the 1-loop suffix $(1)$ which will be
always understood, unless explicitly specified.}
\begin{subequations}
\begin{align}\lab{residuoqq}
 \adp P^{\pb\pf}(z)&=\adp\d^{\pb\pf}P^{\pq\pq}(z)\;,\\ \lab{residuoqg}
 \adp P^{\pb\pg}(z)&=\adp P^{\pq\pg}(z)\;,
\end{align}
\end{subequations}
where $\pq$ denotes a generic quark or antiquark and the quark-to-quark and
gluon-to-quark splitting functions read%
\footnote{The $\pq\pq$ splitting function is not in its full form, having not
yet considered the contribution of virtual corrections to
diagram~\ref{f:quafotLO}.}
\begin{align}\lab{Pqqinc}
 P^{\pq\pq}(z)&=C_F{1+z^2\over1-z}\;,\\ \lab{Pqg}
 P^{\pq\pg}(z)&=T_R[z^2+(1-z)^2]\;.
\end{align}

The partonic tensor $\Wp^{(\pa)}$ at first order in $\as$ assumes then the form
($\pp=\xi p$)
\begin{align}
 \left[\Wp^{(0)}+\as\Wp^{(1)}\right]^{(\pa)}(\pp,q)&=
 e_{\pa}^2\d(1-\xp)\Ga_{\mu\nu}(\pp,q)+\sum_{\pb}e_{\pb}^2\adp
 \ln{Q^2\over\la^2}\,P^{\pb\pa}(\xp)\Ga_{\mu\nu}(\xp\pp,q)\nonumber\\
\lab{w01}&=\sum_{\pb}\int_0^\xi{\dif u\over u}\;\Wp^{(\pb,0)}(up,q)\left[
 \d^{\pb\pa}\d\(1-{u\over\xi}\)+\adp\ln{Q^2\over\la^2}\,P^{\pb\pa}\({u\over\xi}\)
 \right]\;.
\end{align}
It is tempting to suppose that the hadronic tensor $W_{\mu\nu}(p,q)$ could be
obtained by summing up the partonic ones $\Wp^{(\pa)}(\xi p,q)$ weighted with
the parton densities $f^{(\pa)}(\xi)$. Including the $1/\xi$ flux factor
for compensate the fractional momentum of the incoming parton with respect to the
proton, we try writing%
\footnote{The quark absorbing the photon carries a fraction of momentum $x$ with
respect to the proton and thus $x/\xi(\leq1!)$ with respect to the incoming parton.}
\begin{align}\lab{tento}
 W_{\mu\nu}(p,q)&=\sum_{\pa}\int_x^1{\dif\xi\over\xi}\;\Wp^{(\pa)}(\xi p,q)
 f^{(\pa)}(\xi)\\	\lab{trovo}
&=\sum_{\pb}\int_x^1{\dif u\over u}\;\Wp^{(\pb,0)}(up,q)\left[\left(\id+\adp
 \ln{Q^2\over\la^2}\,P\right)\otimes f\right]^{(\pa)}\!\!(u)\;,
\end{align}
where in the last equality we have changed the order of integration
$\xi\leftrightarrow u$ and introduced the shorthand notation for convolutions
\begin{equation}\lab{convol}
 \big[f\otimes g\big](u)\dug\int_u^1{\dif\xi\over\xi}\;f(\xi)g\({u\over\xi}\)\;,
\end{equation}
in which a sum over partonic indices is eventually understood.

A couple of remarks are in order:
\begin{itemize}
\item we are still in presence of the IR singularity $\la\to0$ which prevents us
from interpreting Eq.~(\ref{trovo}) as a reliable expression for $W_{\mu\nu}$;
\item we have used perturbative theory in the (strong coupling) IR region
$k^2\simeq -\kk^2\to0$ without being allowed to. $\as$ is not yet running, but can 
we expect to have obtained the right answer?
\end{itemize}

The way to handle and to eliminate these drawbacks is easily explained. Let
$\mu_F\gg\La$ be a sufficiently hard mass scale, called {\em factorization
scale}. We believe perturbation theory for $|k^2|\simeq\kk^2\geq\mu_F^2$, while
we cannot say anything for $\kk^2<\mu_F^2$. However, the bare parton densities
are completely unknown, so if we insert the non perturbative contribution
$\la^2\leq\kk^2<\mu_F^2$ in $f$ redefining the partonic distributions, we don't
introduce any new uncertainty in Eq.~(\ref{trovo}). Furthermore, in so doing the
new partonic distributions absorb the collinear singularities which do not affect
$W_{\mu\nu}$ anymore. In practice, we write, at $\ord(\as)$ accuracy, 
\begin{equation}\lab{mufdivide}
 \left(\id+\adp\ln{Q^2\over\la^2}\,P\right)\otimes f=
 \left(\id+\adp\ln{Q^2\over\mu_F^2}\,P\right)\otimes
 \left(\id+\adp\ln{\mu_F^2\over\la^2}\,P\right)
\end{equation}
and define
\begin{equation}\lab{effemuf}
 f^{(\pa)}(\xi,\mu_F^2)\dug\left[\left(\id+\adp\ln{\mu_F^2\over\la^2}\,P\right)
 \otimes f\right]^{(\pa)}\!\!(\xi)\;,
\end{equation}
to be considered as an input parameter.

By writing the $W$ tensors in terms of (partonic) structure functions,
Eq.~(\ref{trovo}) yields
\begin{equation}\lab{SFeffemuf}
 F_i(x,Q^2)=\sum_{\pa}\int_x^1\dif z\;\Fp_i^{(\pa)}\({x\over z}\)\left[\left(\id+
 \adp\ln{Q^2\over\mu_F^2}\,P\right)\otimes f_{\mu_F^2}\right]^{(\pa)}\!\!(z)
 \qquad(i=2,L)\;.
\end{equation}
Obviously, the RHS of the above equation cannot depend on the artificial
parameter $\mu_F$.

It is natural to define the $Q^2$-dependent PDF as%
\footnote{To be fair, up to now we can define the ``dressed'' PDF of
Eq.~(\ref{defPDF}) as well as the ``bare'' ones in Eq.~(\ref{effemuf}) only for
quarks ($\pa=\pf,\bar{\pf}$). In the next section we will justify the extension to the
gluonic case we are already using here.}
\begin{equation}\lab{defPDF}
 f^{(\pa)}(z,Q^2)\dug f^{(\pa)}(z,\mu_F^2)+\adp\ln{Q^2\over\mu_F^2}\sum_{\pb}
 \int_z^1{\dif \xi\over \xi}\;P^{\pa\pb}\({z\over \xi}\)f^{(\pb)}(\xi,\mu_F^2)
\end{equation}
ending up with the desired factorization formula
\begin{equation}\lab{fattform}
 F_i(x,Q^2)=\sum_{\pa}\int_x^1\dif z\;f^{(\pa)}(z,Q^2)\,\Fp_i^{(\pa)}\({x\over z}\)\;.
\end{equation}
Our complete ignorance of $f^{(\pa)}(z,\mu_F^2)$ prevents us from determining the PDF
$f^{(\pa)}(z,Q^2)$ at a given value of $Q^2$. Nevertheless, their full $Q^2$-dependence
is contained into the logarithmic factor of Eq.~(\ref{defPDF}). Taking the derivative
with respect to $\ln Q^2$ yields
\begin{align}
 {\dif\over\dif\ln Q^2}f^{(\pa)}(z,Q^2)&=\adp\sum_{\pb}\int_z^1{\dif \xi\over \xi}\;
 P^{\pa\pb}\({z\over \xi}\)\,f^{(\pb)}(\xi,\mu_F^2)\nonumber\\
\lab{eqAPdue}&=\adp\sum_{\pb}\int_z^1{\dif \xi\over \xi}\;
 P^{\pa\pb}\({z\over \xi}\)\,f^{(\pb)}(\xi,Q^2)+\ord(\as^2)\;.
\end{align}
The identification of the splitting functions $P^{\pa\pb}$ in Eqs.~(\ref{eqAP})
and (\ref{eqAPdue}) is straightforward. We are not far from completing the bridge 
connecting OPE and improved parton model.

Before embarking in virtual corrections and higher order diagrams, let's derive
--- for future reference --- the solutions of the AP equations in this
fixed-coupling situation. In moment space Eq.~(\ref{eqAPdue}) reads
\begin{equation}\lab{APdueom}
 {\dif\over\dif\ln Q^2}f^{(\pa)}_\om(Q^2)=\as\sum_{\pb}\ga_\om^{\pa\pb}
 f^{(\pb)}_\om(Q^2)\qquad(\pa=\pf,\bar{\pf},\pg)\;.
\end{equation}
This set of $2\Nf+1$ coupled differential equations can be diagonalized as
usual by introducing the ``non-singlet'' , ``plus'' and ``minus'' components for PDF and
anomalous dimensions $\ga_\om$. The solutions are then readily obtained:
\begin{equation}\lab{solalfafissa}
 f^{(a)}_\om(Q^2)=\left(Q^2\right)^{\as\ga_\om^a}\,C_\om^{(a)}\qquad
 a=\text{NS},+,-\;.
\end{equation}

The 1-loop virtual correction diagrams contributing with
$\ln Q^2$ (Fig.~\ref{f:virtNLO}) are both UV and IR divergent.
\begin{figure}[ht!]
\centering\includegraphics[width=14cm]{\fig 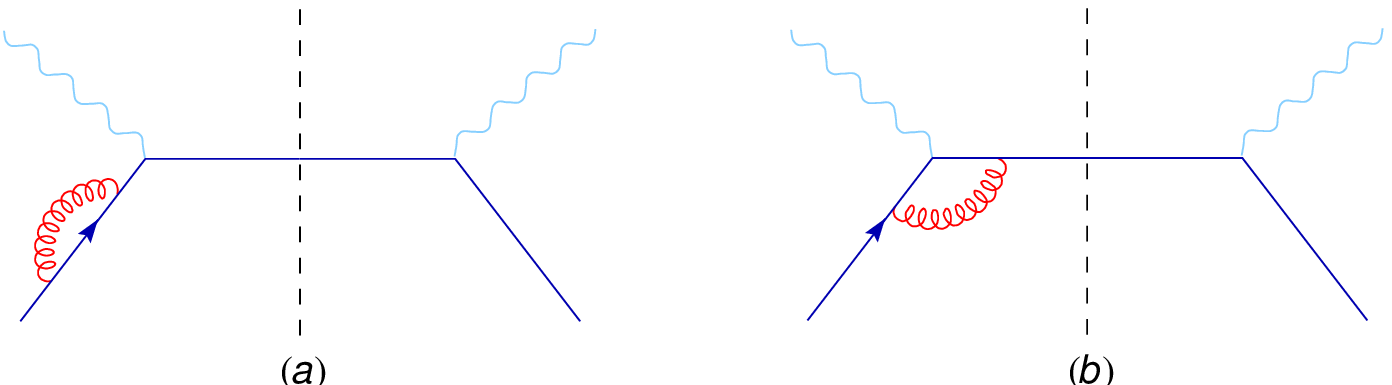}
\didascalia{1-loop virtual diagrams: (a) self-energy and (b) vertex correction.\labe{f:virtNLO}} 
\end{figure}

The UV divergences are regulated with the introduction of the running coupling
constant whose argument has to be of the order of the virtuality of the probing
particle (the photon).

The IR divergences, on the other hand, originate when the incoming quark emits a gluon
with vanishing 4-momentum (soft singularity), i.e., $z=1$, $\bar{z}=0$ and $\kk=0$ in
the quark Sudakov variables, or when the quark emits a collinear gluon. A corresponding
term proportional to $\d(1-z)$ has to be added to the quark-to-quark splitting function
in Eq.~(\ref{Pqqinc})
\begin{equation}\lab{Pqqcom}
 P^{\pq\pq}(z)=C_F\left[{1+z^2\over1-z}-\d(1-z)\int_0^1\dif y\;{1+y^2\over1-y}\right]\;,
\end{equation}
which reproduces the quark-to-quark splitting function in Tab.~\ref{t:AP}.

\subsection{All order resummation of leading logarithms\labe{ss:aorll}}

To complete the resummation program, we should determine the LL contributions to
all order in $\as$. An accurate analysis of higher order diagrams shows that:
\begin{itemize}
\item in axial gauges, the real emission diagrams contributing in LL
approximation are the ladder type ones like in Fig.~\ref{f:diagscala};
\item the phase space region generating large $\ln Q^2$ corresponds, in the
Sudakov variables of the internal particles
\begin{equation}\lab{sudaktutte}
 k_i=z_i\pp+\bar{z}_iq'+\kk_i\;,
\end{equation}
to ordered transverse momenta
\begin{equation}\lab{trasvord}
 Q^2>\kk_1^2>\kk_2^2>\cdots\kk_n^2>\la^2\;,
\end{equation}
in fact, the $j$-th internal parton plays the role of the virtual photon in
regard to the lower part of the diagram, so that the arguments of
Sec.~\ref{ss:nlopm} can be repeated showing that the integration for
$\kk_{j+1}^2>\kk_j^2$ is suppressed;
\item the virtual contributions in LL approximation corresponds to vertex and
self-energy corrections to the ladder diagrams;
\item in addition to $\pg\to\pq$ and $\pq\to\pq$ splitting, we find $\pg\to\pg$
and $\pq\to\pg$ processes involving collinear singularities;
\item the mass-shell constraints of the outgoing partons forces the $z_j$
variables to be ordered along the ladder:
\begin{equation}\lab{longord}
 \xp\simeq z_1<z_2<\cdots<z_n<1\;,
\end{equation}
while the $\bar{z}_i$ components in the ordered transverse momenta region
(\ref{trasvord}) are very small.
\end{itemize}
\begin{figure}[ht!]
\centering\includegraphics[height=7cm]{\fig 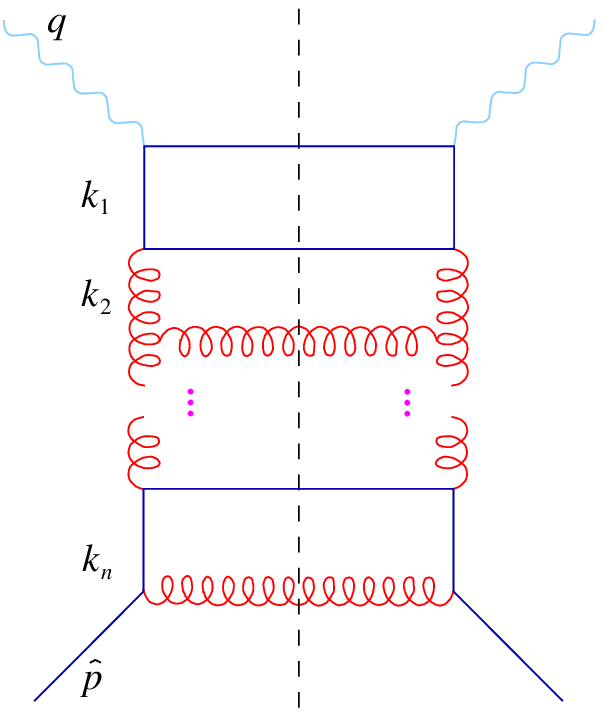}
\didascalia{Ladder-type diagram contributing to DIS in the leading logarithmic
approximation.%
\labe{f:diagscala}} 
\end{figure}

The $n$-loop contribution to the partonic tensor is
\begin{subequations}\labe{winf}
\begin{align}\lab{gammaab}
 \as^n\Wp^{(\pa,n)}(\pp,q)=&\sum_{\pa_1}e_{\pa_1}^2\Ga_{\mu\nu}(\xp\pp,q)\,
 A^{(n)}(Q^2)\,{1\over\xp}B^{(\pa_1\pa,n)}(\xp)\\
\lab{defanq}A^{(n)}(Q^2)\dug&
 \int_{\la^2}^{Q^2}{\dif \kk_1^2\over\kk_1^2}\as(\kk_1^2)
 \int_{\la^2}^{\kk_1^2}{\dif \kk_2^2\over\kk_2^2}\as(\kk_2^2)\cdots
 \int_{\la^2}^{\kk_{n-1}^2}{\dif \kk_n^2\over\kk_n^2}\as(\kk_n^2)\\
 \lab{defbanx}{1\over\xp}B^{(\pa_1\pa,n)}(\xp)\dug&\sum_{\pa_2\cdots\pa_n}
 \int_0^1{\dif z_1\over z_1}\d\(1-{\xp\over z_1}\)
 \left[\prod_{j=2}^n \int_{z_{j-1}}^1{\dif z_j\over z_j}\;
 {1\over2\pi}P^{\pa_{j-1}\pa_j}\({z_{j-1}\over z_j}\)\right]{1\over2\pi}P^{\pa_n\pa}(z_n)
\end{align}
\end{subequations}
By summing over all $n\in\N$ and integrating over $\xi$ with the partonic
densities just like in Eq.~(\ref{tento}) we obtain, for the SF,
\begin{equation}\lab{Ffab}
 F_i(x,Q^2)=\sum_{\pa,\pa_1}\hat{F}_{i,\om}^{(\pa_1)}\int_0^1\dif\xi\;
 \sum_{n=0}^\infty A^{(n)}(Q^2)B^{(\pa_1\pa,n)}\({x\over\xi}\)
 f^{(\pa)}(\xi)\qquad(i=2,L)\;,
\end{equation}
where, from Eqs.~(\ref{Fmp}) and (\ref{mfsp}), $\hat{F}_{2,\om}^{(\pa_1)}=e_{\pa_1}^2$
and $\hat{F}_{L,\om}^{(\pa_1)}=0$.

The integral over transverse momenta can be performed by introducing the variable
\begin{equation}\lab{eLLe}
 L(Q^2)\dug\int_{\la^2}^{Q^2}{\dif k^2\over k^2}\;\as(k^2)
 ={1\over b_0}\ln{\as(\la^2)\over\as(Q^2)}
\end{equation}
and gives
\begin{equation}\lab{anq}
 A^{(n)}(Q^2)={1\over n!}L^n(Q^2)\;.
\end{equation}
By performing a Laplace transformation with respect to L, we obtain
\begin{equation}\lab{anrho}
 A^{(n)}_\rho\dug\int_0^\infty\dif\rho\;A^{(n)}(L)\esp{-\rho L}={1\over\rho^{n+1}}\;.
\end{equation}

The integral over $z_j$ variables is easily evaluated in Mellin space by
defining
\begin{align}
B_\om^{(\pa_1\pa,n)}\dug&\int_0^1\dif z\;z^{\om-1}\,B^{(\pa_1\pa,n)}(z)\nonumber\\
\lab{banom}=&\sum_{\pa_2\cdots\pa_n}\ga^{\pa_1\pa_2}_\om\cdots
 \ga^{\pa_{n-1}\pa_n}_\om\ga^{\pa_n\pa}_\om
 =\big[\gga_\om^n\big]^{\pa_1\pa}\;,
\end{align}
so that
\begin{align}
 F_{i,\om,\rho}=&\sum_{\pa,\pa_1}\hat{F}_{i,\om}^{(\pa_1)}\sum_{n=0}^\infty
 A^{(n)}_\rho B^{(\pa_1\pa,n)}_\om f^{(\pa)}_\om\nonumber\\
 =&\sum_{\pa,\pa_1}\hat{F}_{i,\om}^{(\pa_1)}\sum_{n=0}^\infty{1\over\rho}
 \left({\gga_\om\over\rho}\right)^n_{\pa_1\pa}f^{(\pa)}_\om\nonumber\\
\lab{FrhoomQ}=&\sum_{\pa,\pa_1}\hat{F}_{i,\om}^{(\pa_1)}
 \big[(\rho-\gga_\om)^{-1}\big]^{\pa_1\pa}f^{(\pa)}_\om\;.
\end{align}
In $(\om,\rho)$-space, the above results can be summarized in the following
diagrammatic rules:
\begin{figure}[ht!]
\centering\includegraphics[width=13cm]{\fig 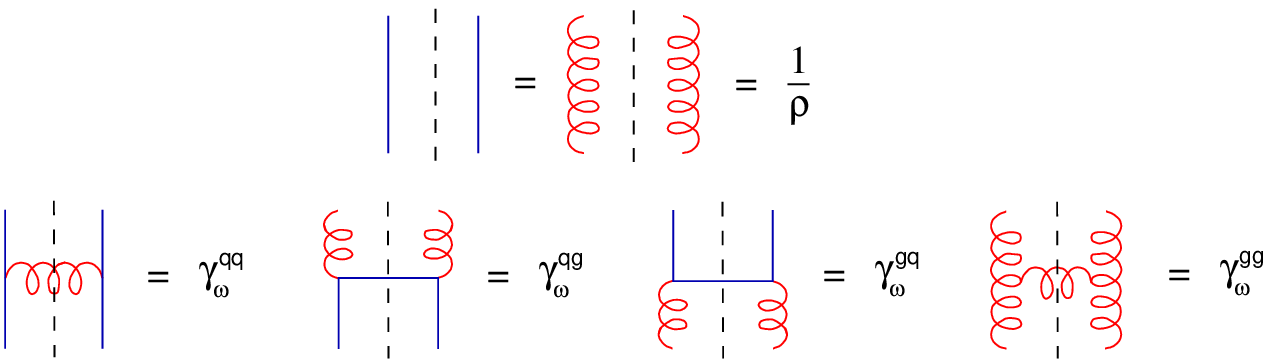}
\end{figure}

By inverting the Laplace transform we obtain
\begin{align}\lab{FiomQ}
 F_{i,\om}(Q^2)\dug\int_{c-\ui\infty}^{c+\ui\infty}{\dif\rho\over2\pi\ui}\;
 \esp{\rho L}\,F_{i,\om,\rho}
 =\sum_{\pa,\pa_1}\hat{F}_{i,\om}^{(\pa_1)}\exp\{\gga_\om L\}^{\pa_1\pa}f^{(\pa)}_\om\;.
\end{align}
By splitting the exponential factor
\begin{equation}\lab{dividoexp}
 \exp\{\gga L\}=\left[{\as(\la^2)\over\as(Q^2)}\right]^{\gga/b_0}
 =\left[{\as(\mu_F^2)\over\as(Q^2)}\right]^{\gga/b_0}
 \left[{\as(\la^2)\over\as(\mu_F^2)}\right]^{\gga/b_0}
\end{equation}
and redefining the bare parton densities
\begin{equation}\lab{PDFnude}
 f_\om^{(\pa)}(\mu_F^2)\dug\sum_{\pb}\left[{\as(\la^2)\over\as(\mu_F^2)}
 \right]^{\gga/b_0}_{\pa\pb}f_\om^{(\pb)}\;,
\end{equation}
we get an expression for the SF free of singularities which is the $\om$-space
analogue of Eq.~(\ref{SFeffemuf}). The moments of the $Q^2$-dependent PDF are
defined by
\begin{equation}\lab{defPDFom}
 f_\om^{(\pa)}(Q^2)\dug\sum_{\pb}\left[{\as(\mu_F^2)\over\as(Q^2)}
 \right]^{\gga/b_0}_{\pa\pb}f_\om^{(\pb)}(\mu_F^2)\;,
\end{equation}
in terms of which we finally obtain
\begin{equation}\lab{fattformom}
 F_{i,\om}(Q^2)=\sum_{\pa}\hat{F}_{i,\om}^{(\pa)}\,f_\om^{(\pa)}(Q^2)\;,
\end{equation}
i.e., in $x$-space, Eq.~(\ref{fattform}).

The evolution equation for the PDF, obtained by differentiating
Eq.~(\ref{defPDFom}) with respect to $\ln Q^2$ reads
\begin{equation}\lab{eqAPtre}
 {\dif\over\dif\ln Q^2}f^{(\pa)}_\om(Q^2)=\as(Q^2)\sum_{\pb}\ga_\om^{\pa\pb}
 f^{(\pb)}_\om(Q^2)\qquad(\pa=\pf,\bar{\pf},\pg)\;.
\end{equation}

The main differences of the resummed formula (\ref{eqAPtre}) with respect to the
1-loop one (\ref{eqAPdue}) are that (i) it has been obtained without
approximations apart from the LL restrictions, (ii) the coupling is running and (iii) the
evolution equation is extended to the gluon densities where $\gagq$ and $\gagg$
anomalous dimensions are present.
\chapter{Small-$\boldsymbol x$ hard processes\labe{c:shp}}

The coming of high energy colliders is making it possible to investigate strong
interaction processes in very peculiar kinematic regimes, where the CM energy is much
larger both than the hadronic masses and of the transferred momenta.

From a theoretical point of view, this new class of phenomena is of great importance by
testing the asymptotic behaviour of cross sections in the high energy limit on one side
and to check QCD theory in a much wider kinematic domain on the other side.

From the phenomenological point of view, the discovery of a marked rise of structure
functions at HERA~\cite{Aid96} has increased the interest of physicists for
understanding the high energy features of QCD. What is actually observed
(Fig.~\ref{f:crescita}) is a surprising growth of the $F_2$ structure function towards
low values of $x$. Note that $1/x$ is nearly proportional to the CM energy squared of
the virtual photon-proton system
\begin{equation*}
 W^2\simeq{Q^2\over x}(1-x)\;\to\;{Q^2\over x}\qquad(x\ll1)\;.
\end{equation*}
The growth is particularly marked for large values of $Q^2$, but it is also evident down 
to very low values (Fig.~\ref{f:crescitapiccoliQ}).
\begin{figure}[ht!]
\centering
\includegraphics[height=130mm]{\fig 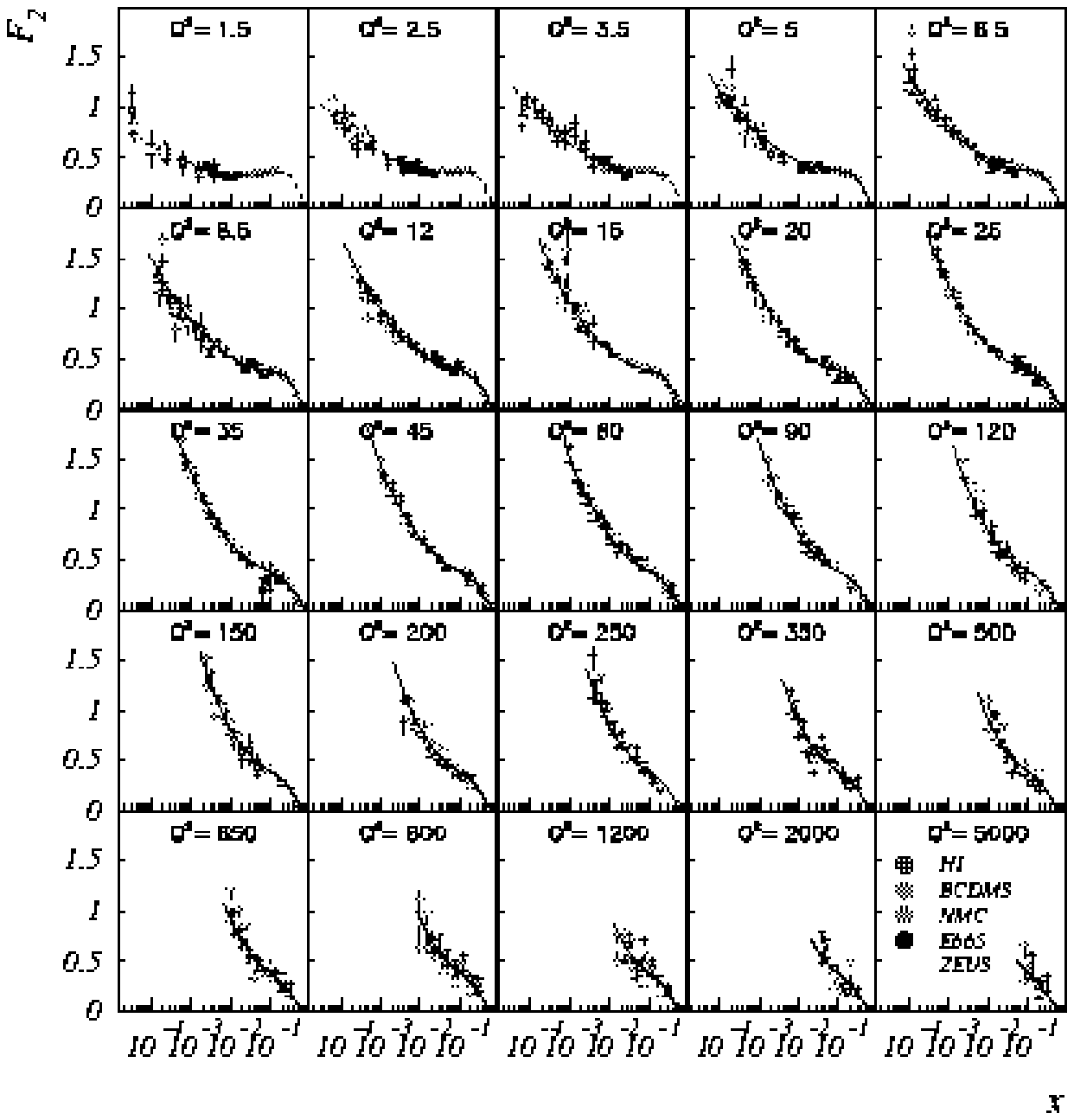}
\didascalia{$x$-dependence of the $F_2$ structure function at different
values of $Q^2$, showing the steep rise towards small-$x$ for all values of
$Q^2$.\labe{f:crescita}} 
\end{figure}
\begin{figure}[ht!]
\centering
\includegraphics[height=100mm]{\fig 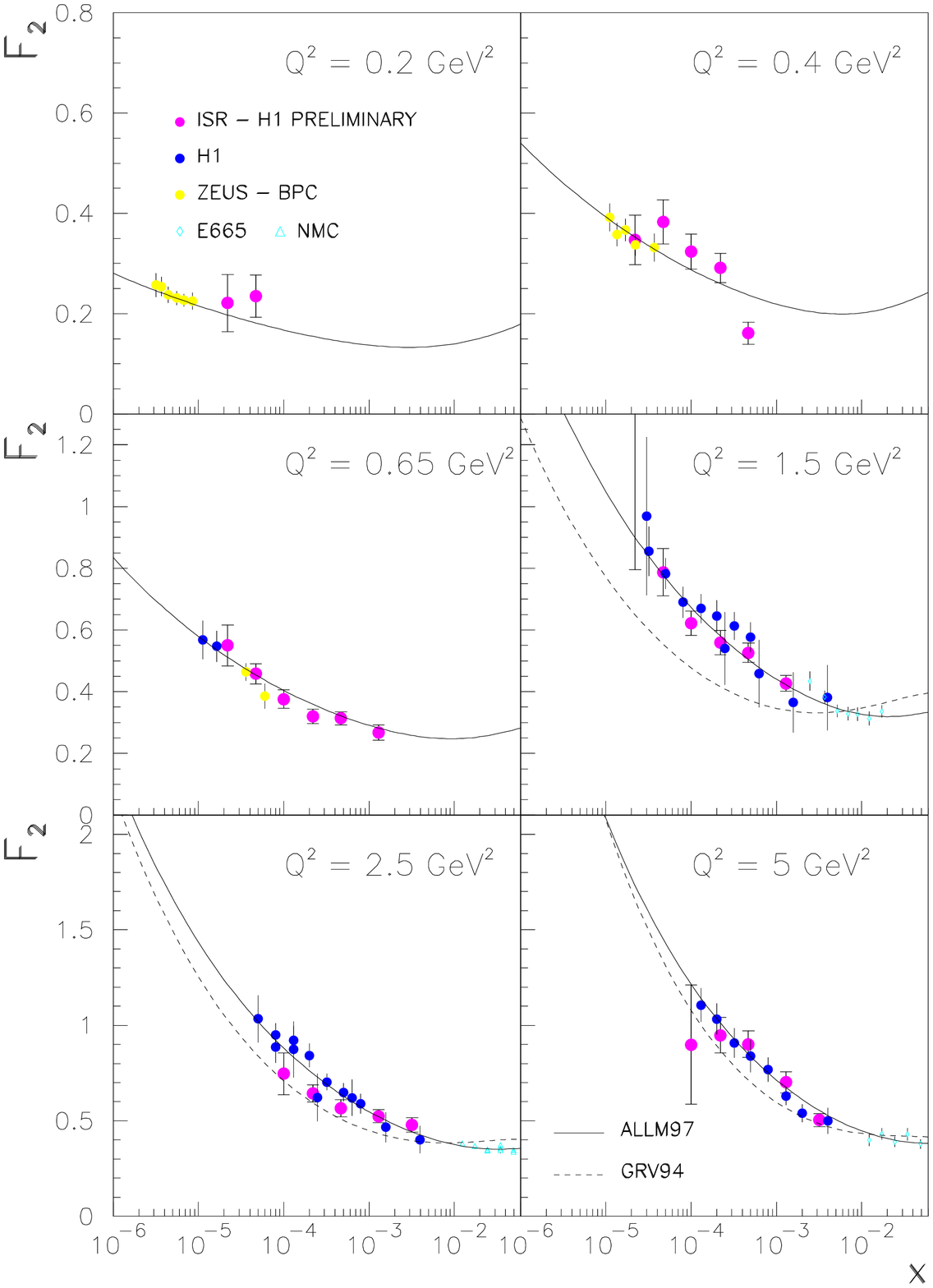}
\didascalia{Small-$x$ rise of $F_2$ for very low values of $Q^2$.\labe{f:crescitapiccoliQ}} 
\end{figure}

In this chapter, after a brief introduction to Regge theory, we expose the general
framework of the perturbative treatment of high energy inclusive strong interactions.

\section{Regge behaviour\labe{s:rb}}

Before a fundamental theory for strong interactions --- such as QCD --- was established,
the study of scattering of hadronic particles relied basically on rather general
assumptions on the scattering matrix, such as Lorentz invariance, crossing symmetry,
unitarity, causality, analyticity, asymptotic states etc.. The subsequent development
led essentially to what is known as Regge theory~\cite{CoSq68}, which determines the
asymptotic behaviour of cross sections in the high energy limit regardless the strength
of the coupling, i.e., independently of perturbation theory.

In this section we outline some important results of Regge theory which will be
useful for our study on semi-hard processes. Keeping in mind our interest in
high-energy inclusive reactions, unitarity allows us to relate the total cross
section for the scattering of two initial particles ``1'' and ``2'' to the
elastic scattering amplitude of the same particles. More precisely, by denoting
with $\M^{\pa_1^{}\pa_1'}_{\pa_2^{}\pa_2'}(s,t)$ the scattering amplitude for the
$s$-channel ($s>0,-s\leq t\leq0$) process
\begin{equation*}
 (p_1,\pa_1)+(p_2,\pa_2)\to(p_1',\pa_1')+(p_2',\pa_2')\quad;
 \quad s\dug(p_1+p_2)\;,\;t\dug(p_1-p_1')\;,
\end{equation*}
($\pa_j$ labelling the discrete quantum numbers) the optical theorem states
that\footnote{All masses are supposed to be much smaller than the CM energy
$\sqrt{s}$ and are thus neglected.}
\begin{equation}\lab{teorott}
 \si_{\text{tot}}^{\pa_1,\pa_2}(p_1,p_2)(s)=
 {1\over s}\im\M^{\pa_1\pa_1}_{\pa_2\pa_2}(s,t=0)\;.
\end{equation}
Consider on the contrary the $t$-channel ($t>0,-t\leq s\leq0$) process
\begin{equation*}
 (p_1,\pa_1)+(-p_1',\bar{\pa}{}_1')\to(-p_2,\bar{\pa}_2)+(p_2',\pa_2')\;,
\end{equation*}
where the antiparticle $\bar{\pa}_2$ of ``2'' is now outgoing and the
antiparticle $\bar{\pa}{}_1'$ of ``1$^\prime\,$'' is incoming. By exploiting the crossing
symmetry and the analyticity properties of the amplitude, we can say that the
corresponding amplitude is given by the same function
$\M^{\pa_1\pa_1'}_{\pa_2\pa_2'}(s,t)$, or better, by its analytic continuation
from the $s$-channel to the $t$-channel region.

In the $t$-channel CM frame ($\vec{p}_1-\vec{p}_1{}'=0$), the scattering angle
$\te$ between $\vec{p}_1$ and $-\vec{p}_2$ is given by
\begin{equation}\lab{costeta}
 \cos\te=1+{2s\over t}\quad;\quad-1\leq\cos\te\leq1\;.
\end{equation}
By expressing the scattering amplitude by means of $(\cos\te,t)$ instead of
$(s,t)$, it is easy to perform the expansion in series of Legendre functions
\begin{equation}\lab{legendre}
 \M(s,t)=M(\cos\te,t)=16\pi\sum_{l=0}^\infty(2l+1)M_l(t)P_l(\cos\te)
\end{equation}
which is called the partial wave series for $\M$.

We know from quantum mechanics that $P_l(\cos\te)$ is the emission pattern of a state
whose total angular momentum (in the CM system) is $l$. Hence we interpret the {\em
partial wave amplitude} $M_l(t)$ as the contribution to the total amplitude
corresponding to the $l^{\text{th}}$ angular momentum component.

Coming back to the $s$-channel domain, $|\cos\te|>1$ loses its physical
meaning. Nonetheless, thanks to analyticity, the contribution from the $l^{\text{th}}$
partial wave amplitude is still
\begin{equation*}
 16\pi(2l+1)M_l(t)P_l\(1+{2s\over t}\)\;.
\end{equation*}
In the high energy limit $s\gg|t|$ it holds
\begin{equation}\lab{limpl}
 P_l\(1+{2s\over t}\)\to{\Ga(2l+1)\over 2^l\Ga^2(l+1)}\left({s\over t}\right)^l
\end{equation}
which indicates a power-like behaviour in $s$ for the elastic scattering amplitude
whose power equals the angular momentum of the state mediating the
interaction. Moreover, the leading contribution to the amplitude will be given by the 
highest spin intermediate state allowed in the $t$-channel process.

As far as the differential cross section is concerned, we have ($Q^2=-t$)
\begin{equation}\lab{2lm2}
 {\dif\si\over\dif Q^2}={1\over16\pi}{|\M(s,-Q^2)|^2\over s^2}\sim s^{2l-2}\;.
\end{equation}
At this point we can anticipate that, in QCD, the leading contributions to cross
sections will be given by gluon exchanges in the $t$-channel. In fact, $s$ and $u$
channel exchanges are suppressed by the large denominators $1/s$ and
$1/u$ of the virtual particle propagators with respect to $t$-channel ones
which involve $1/t$ ($s\simeq|u|\gg|t|$); the highest spin field in the QCD
Lagrangian is the spin 1 gluon which, according to Eq.~(\ref{2lm2}) is responsible of 
nearly constant differential cross sections.

What we have sketched are nothing but rough estimates, which give us an idea about
the ``order of magnitude'' of the phenomena we are investigating. It is possible to
push further the partial wave analysis in a beautiful mathematical description
where the amplitudes $M_l(t)$ are analytically continued to complex values of the
angular momentum variable $l$.

The complex functions $l\mapsto M_l(t)$ may present poles or branch cuts whose
positions $\al_i(t)$ are smooth functions of $t$. Having in mind the $t$-channel
process ($t>0$), we expect the amplitude to have a pole corresponding to the exchange 
of a physical particle of spin $J$ and mass $m$, thus $\exists i:\al_i(m^2)=J$. It
turns out that the functions $t\mapsto\al_i(t)$, which are called {\em Regge
trajectories}, assume physical (semi)integer values when $t$ equals the squared mass
of a particle or of a resonance. For each set of quantum numbers (except spin!) there 
is a definite Regge trajectory along which hadronic particles carrying those quantum
numbers lie.

However also non-physical value of $\al_i(t)$ are important: each Regge
trajectory contributes to the scattering amplitude in the $s$-channel. The kind of
contribution depends on the particular nature of the singularity associated to the
trajectory, e.g., 
\begin{equation}\lab{poliregge}
 M_l(t)\sim{1\over[l-\al(t)]^{\be(t)}}\;\longleftrightarrow\;\M(s,t)\sim
s^{\al(t)}[\ln s]^{\be(t)-1}\;.
\end{equation}
as if a particle of ``effective spin'' $\al(t)$ were exchanged.
We are not surprised to encounter logarithmic corrections to the na\"\i ve power-like 
estimates: these are normal features in interacting theories.

According to the optical theorem, we argue the total cross section to be expressed
--- apart from logarithmic corrections --- as a sum of powers of $s$ involving the
intercepts of the various Regge trajectories:
\begin{equation}\lab{sigmaregge}
 \si_{\text{tot}}(s)=\sum_i c_i\,s^{\al_i(0)-1}\;\overset{s\to\infty}{\sim}\;
 c_M s^{\al_M(0)-1}\;,
\end{equation}
where $\al_M(0)$ is the largest among the intercepts.

A careful analysis carried out from Froissart, based on general grounds and which
assumes a small range force (as the strong interaction is) shows that unitarity
limits the asymptotic behaviour of the scattering amplitude, setting an upper bound
on the largest $s$-growth exponent: $\al_M(0)-1\leq0$, i.e., the total cross section of
any reaction cannot grow faster than some power of $\ln s$.

In this connection the phenomenological analysis seems to present a sort of puzzle: a
Regge inspired fit like Eq.~(\ref{sigmaregge}) with two terms ($i=1,2$), as performed by
Donnachie and Landshoff~\cite{DoLa92}, describe very well the experimental data, with
the same Reggeon intercepts $\al_{\R}(0)-1=-0.45$ and $\al_{\mathbb{P}}(0)-1=0.08$ for
different processes (see Fig.~\ref{f:DLfit}). However, the latter clearly violates the
Froissart bound. The answer to this apparent paradox has to be found in the kinematic
domain so far investigated which is evidently outside the regime where unitarity effects
become important.
\begin{figure}[ht!]
\vskip4mm
\centering
\includegraphics[height=57mm]{\fig 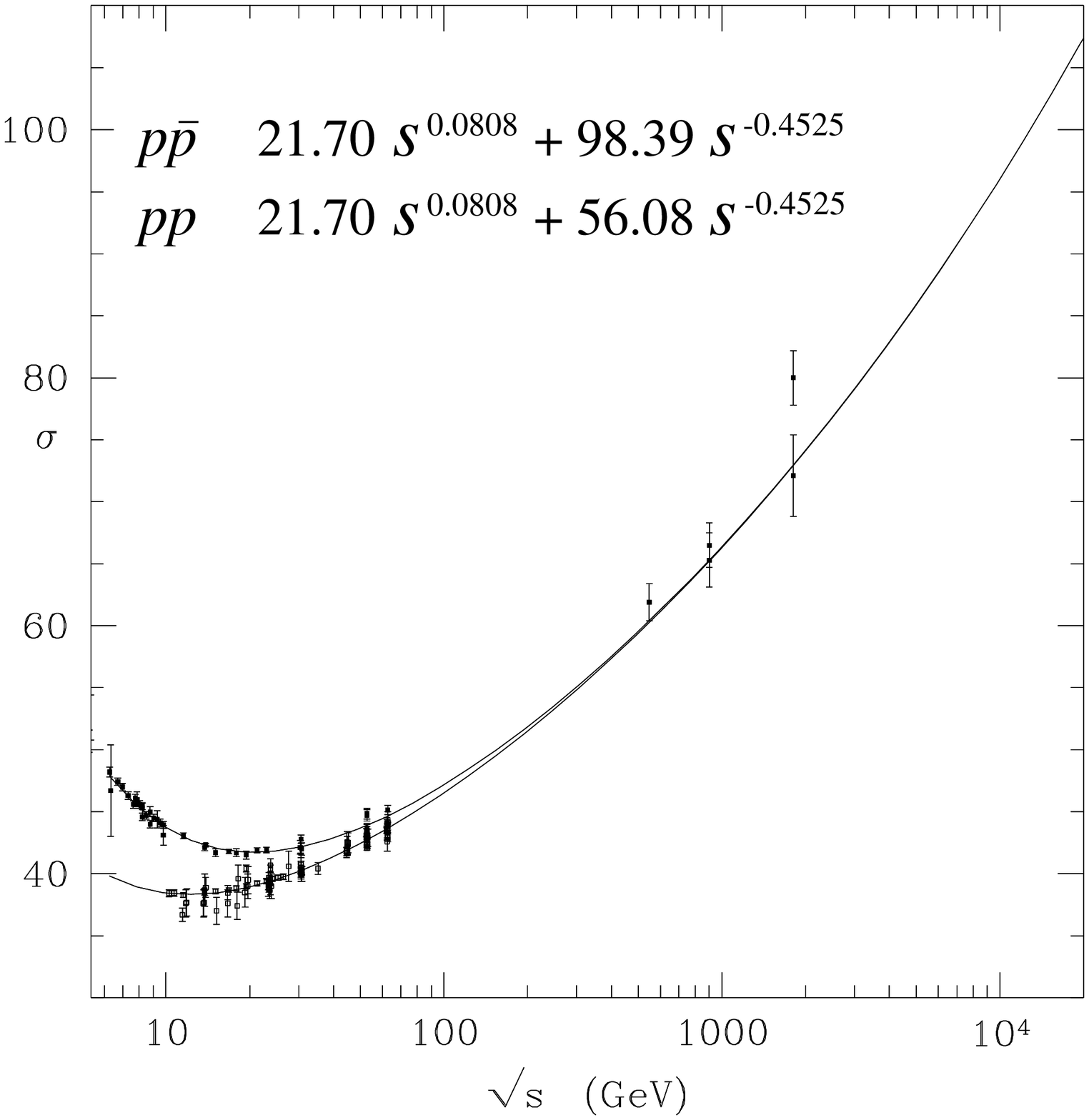}\hspace{13mm}
\includegraphics[height=57mm]{\fig 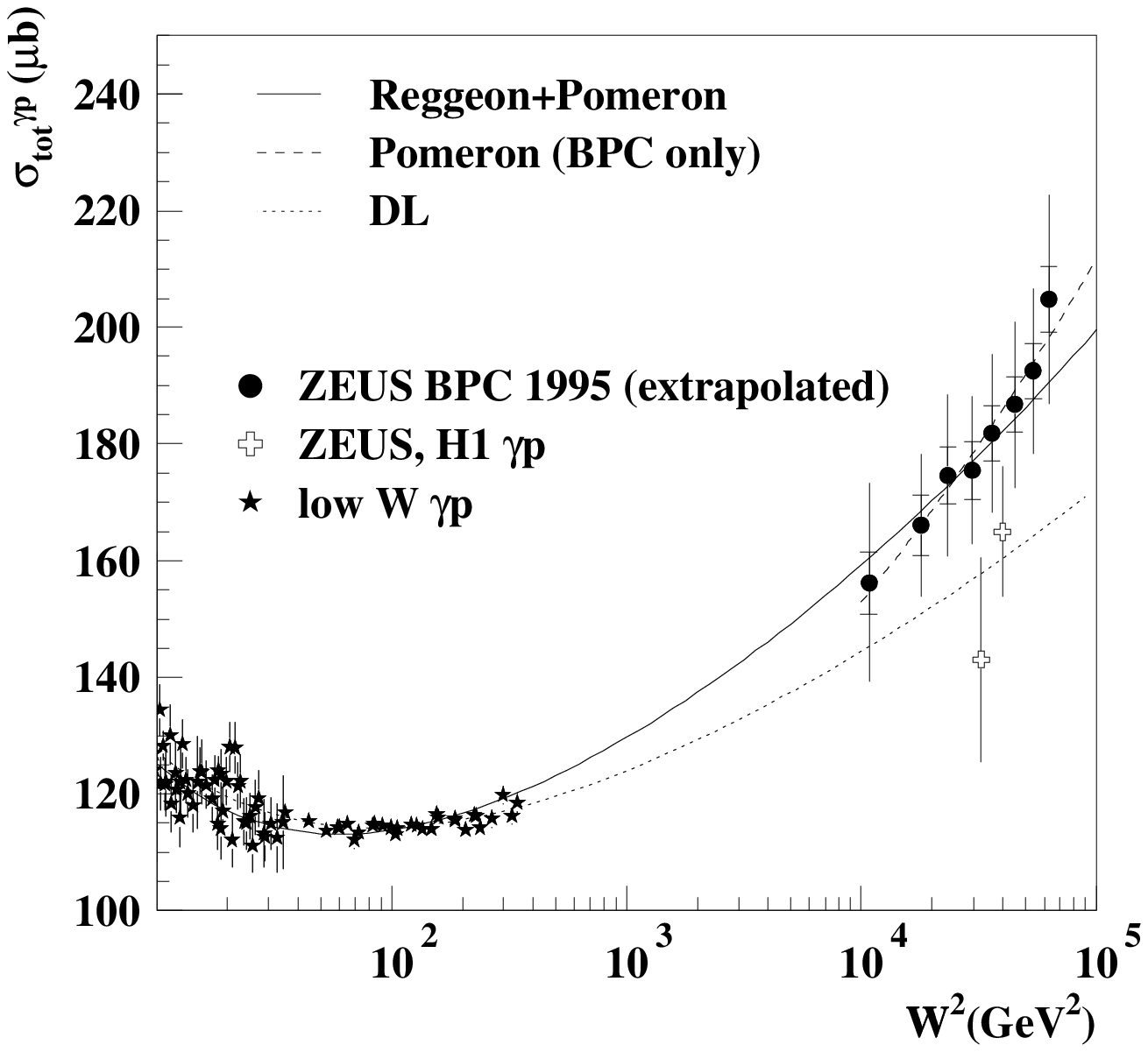}\vspace{8mm}
\didascalia{The Donnachie-Landshoff fit to ({\sl a}) proton-proton and
antiproton-proton and ({\sl b}) to photon-proton total cross section.%
\labe{f:DLfit}} 
\end{figure}

Anyway, there is evidence of a dominance of Regge-particles in the high energy region so
far investigated. The lower Regge trajectory $\al_{\R}$ corresponds to the family of
particles including the $\rho$, $\om$, etc.\ mesons, and has odd parity and charge
conjugation. This is the reason of the different coefficient $c_{\R}$ in the $pp$ and
$p\bar{p}$ cross sections.

The upper trajectory does not distinguish charge conjugated particles --- it has the
quantum number of the vacuum --- and does not correspond to any particle till now
observed. This elusive particle has been given the name of {\em pomeron} and is the
responsible of the high energy behaviour of cross sections, at least before the
unitarity regime, where pomeron self-interactions are no longer negligible.

Is perturbative QCD able to describe the pomeron? and, in general, the high energy
processes? These are the questions we wish to give an answer.

\section{Perturbative analysis of DIS in leading $\boldsymbol{\ln1/x}$\\ approximation
\labe{s:padla}}

We are ready to start the analysis of DIS in the high energy limit $s\gg Q^2\gg\La^2$,
i.e., from Eq.~(\ref{qxys}), $xy\ll1$. In particular, we are interested in the kind of
process where $x\ll1$ and $y\ll1$.

The large virtuality $Q^2\gg\La^2$ of the virtual photon should justify the use of
perturbation theory, regardless of the value of $s$. However, in this regime where there 
are two very different hard scales, the QCD perturbative expansion is affected by large
coefficients which, to order $\as^n$ and for inclusive observables, are
$\ord[(\ln W^2/Q^2)^m]\sim\ln^m1/x:m\leq n$. We have to face again a resummation
problem, this time with respect to the effective expansion parameter $\as\ln1/x$ which
can be of the order of unity for small values of $x$ even if $\as\ll1$. The leading
$\ln1/x$ (L$x$) approximation consists in taking into account all the perturbative terms 
of order $[\as\ln1/x]^n$, the next-to-leading $\ln1/x$ (NL$x$) considers all
contributions like $\as[\as\ln1/x]^n$ and so on.

In order to resum the leading logarithms of $x$ a technique is usually adopted which is
in certain aspects similar to the one employed in the resummation of the leading $\ln
Q^2$. First of all, the collinear factorization formula for SF has to be replaced with a
corresponding high energy ($\kk$-dependent) factorization~\cite{CaCiHa90}. The latter
should take into account the larger phase space available and should generalize the
former.  The second step is to calculate the evolution (in $x$-space) of the ``parton
density'' factor which obey an equation obtained by Balitski\u{\i}, Fadin, Kuraev and
Lipatov (BFKL)~\cite{BFKL76}.  By performing the transverse-space integration of the
parton densities with the partonic cross section, which can be explicitly evaluated, one
obtains a factorized expression for the high energy (small-$x$) SF where both partonic
SF and PDF are resummed.

\subsection{High energy factorization\labe{ss:hef}}

As we noticed in Sec.~\ref{s:rb} and we shall show in more detail in Sec.~\ref{s:rlxla},
the terms contributing to the L$x$ approximation arise from gluon exchanges in the
$t$-channel. Since gluons can couple to the (virtual) photon only via quarks, we are led
to consider diagrams where single gluon (Fig.~\ref{f:fqgF}{\sl a}) and multiple gluon
(Fig.~\ref{f:fqgF}{\sl b}) exchanges are possible.
\begin{figure}[ht!]
\centering
\includegraphics[width=14cm]{\fig 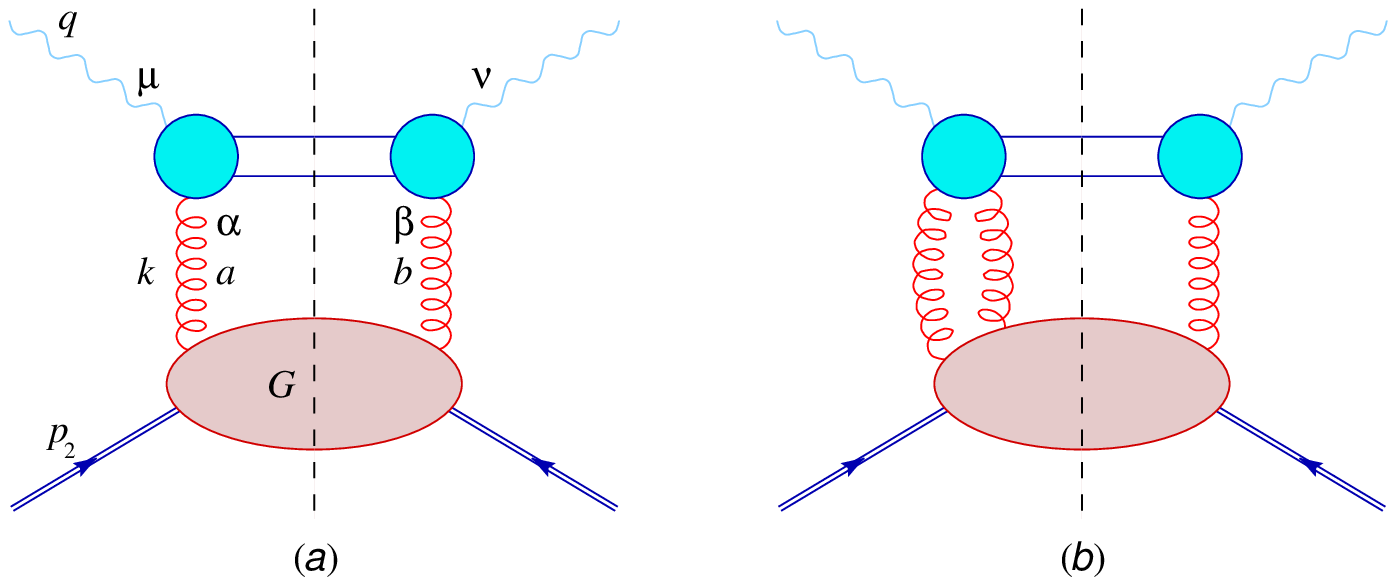}
\didascalia{Single (a) and multiple (b) gluon exchanges in the high energy limit
of photon-proton total cross section.%
\labe{f:fqgF}} 
\end{figure}

If we work in axial gauges, the
multiple gluon exchange diagrams are suppressed by powers of $\as(Q^2)$, because the
gluons' endpoints have a very large relative velocity, the upper one being nearly
collinear to $p_1$ and the lower one nearly collinear to $p_2$ (see Eq.~(\ref{sudakqk})).

Single $t$-channel gluon exchanges in L$x$ approximation are then factorizable as we
will briefly explain in the following. According to Eq.~(\ref{d:tensadr}), the
contribution to the hadronic tensor from Fig.~\ref{f:fqgF}{\sl a} can be written
\begin{equation}\lab{WAG}
 W^{\mu\nu}(p_2,q)={1\over4\pi}\int{\dif^4 k\over(2\pi)^4}\;\overset{ab}{A}{}^{\mu\nu}
 _{\al\be}(q,k)\,\overset{ab}{G}{}^{\al\be}(p_2,k)
\end{equation}
(summation over colour indices $a,b$ is understood) where
\begin{equation}\lab{d:Aabmn}
 \overset{ab}{A}{}^{\mu\nu}_{\al\be}(q,k)\dug\int\dif\Phi(p_3,p_4)\;
 \overset{\;a}{\M}{}^\mu_\al(q,k,p_3,p_4)\overset{\;b}{\M}{}^\nu_\be(q,k,p_3,p_4)^*
\end{equation} 
denotes the lowest order $\pq\bar{\pq}$ contribution to the $\ga^*\pg^*\to\ga^*\pg^*$
absorptive part (Fig.~\ref{f:Aabmn}) integrated over the final particle phase space and
$\overset{ab}{G}{}^{\al\be}(p_2,k)$ is the full $p\pg^*\to p\pg$ absorptive part,
including the gluon propagator.
\begin{figure}[ht!]
\centering
\raisebox{16mm}{$\overset{\;a}{\M}{}^\mu_\al\quad=\quad$}
\includegraphics[width=13cm]{\fig 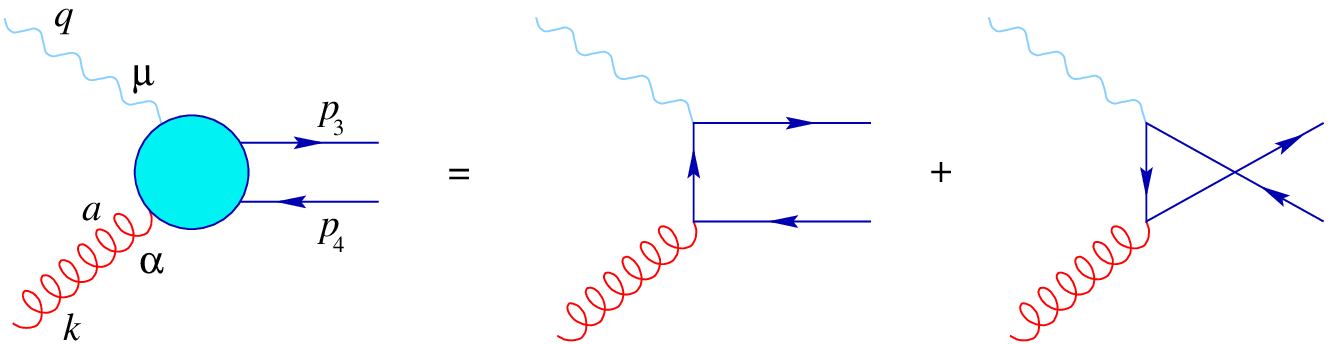}
\didascalia{Feynman diagrams for photon-gluon coupling at lowest order in $\as$.%
\labe{f:Aabmn}} 
\end{figure}

In the high energy, small-$y$ regime, the differential cross section is essentially
given by the $F_2$ structure function, and can be obtained by the eikonal approximation
at the electron-photon vertex. The eikonal approximation can be applied when the
components of the transferred momentum $q$ are much smaller than that of the external
momenta $p_1$ and $p_1'$; it consists in the replacement of the helicity conserving
vertex $\bar{u}_\si(p_1')\ga^\mu u_\si(p_1)\simeq2p_1^\mu$ and neglecting the
non conserving one, i.e., to approximate the leptonic tensor (\ref{d:tenslep})
\begin{equation}\lab{apptenslep}
 L_{\mu\nu}(p_1,q)\simeq2p_{1\mu}2p_{1\nu}\;.
\end{equation}
After contraction with the hadronic tensor we obtain
\begin{equation}\lab{Weik}
 2p_{1\mu}2p_{1\nu}\,W^{\mu\nu}=-Q^2F_1+2s{(1+y/2)^2\over y}F_2\simeq{2s\over y}F_2
 \qquad(y\ll1)
\end{equation}
and hence, using Eq.~(\ref{qxys}),
\begin{equation}\lab{F2AG}
 F_2\simeq2x\({y\over Q}p_{1\mu}\)\({y\over Q}p_{1\nu}\)
 \int{\dif^4 k\over(2\pi)^4}\;\overset{ab}{A}{}^{\mu\nu}_{\al\be}(q,k)\,
 {1\over4\pi}\overset{ab}{G}{}^{\al\be}(p_2,k)
\end{equation}
By adopting the Sudakov parametrization
\begin{subequations}\labe{sudakqk}
\begin{align}
 q&=yp_1+\bar{y}p_2+\qq\\
 k&=\bar{z}p_1+zp_2+\kk
\end{align}
\end{subequations}
the high energy limit constrains $\bar{y}\ll y\ll1$ and it turns out that the dominant
integration region in the $k$ variable is the one with fixed $\kk^2$, $\bar{z}\ll z\ll1$
and $|k^2|\simeq\kk^2=\ord(yzs)=\ord({Q^2})$.

The next step is to show that only a single gluon polarization contributes in the
diagram of Fig.~\ref{f:fqgF}{\sl a}. Since the amplitude
$\overset{\;a}{\M}{}^\mu_\al\ugd\M^\mu_\al\t{a}$ involves a
colourless object (the photon), the tensor
$\overset{ab}{A}{}^{\mu\nu}_{\al\be}\ugd A^{\mu\nu}_{\al\be}\d^{ab}$ satisfies the Ward
identities of electrodynamics
\begin{equation}\lab{Ward}
 k^\al A^{\mu\nu}_{\al\be}=k^\be A^{\mu\nu}_{\al\be}=0
\end{equation}
and thus the following decomposition holds:
\begin{subequations}\labe{decompA}
\begin{align}
 A_{\al\be}\dug&\({y\over Q}p_{1\mu}\)\({y\over Q}p_{1\nu}\)
 A^{\mu\nu}_{\al\be}(q,k)\\
 =&A_1\left({k_\al k_\be\over k^2}-g_{\al\be}\right)-A_2{1\over k^2}
 \left(k_\al-{k^2\over qk}q_\al\right)\left(k_\be-{k^2\over qk}q_\be\right)
\end{align}
\end{subequations}
where the dimensionless Lorentz invariant functions $A_i$ depends on $Q^2$, $\kk^2$,
$2qk\simeq yzs$, $2(yp_1)k=yzs$ or, equivalently, on the rescaled variables $x/z$ and
$Q^2/\kk^2$.

In the kinematic regime mentioned above, the amplitudes $A_1$ and $A_2$ satisfy
\begin{equation*}
 A_1\simeq A_2\simeq\ord\({x\over z}\)
\end{equation*}
for fixed values of $\kk^2/zs$ (they are exactly equal when $k^2=0$ in order to cancel
the spurious pole at $k^2=0$ in Eq.~(\ref{decompA})). This is due to the fact that the
corresponding diagrams Fig.~\ref{f:Aabmn} involve quark (spin $1/2$) exchanges,
vanishing like $s^{-1}\sim x$ for increasing energy (see Eq.~(\ref{2lm2})). Therefore,
the dominant contribution in Eq.~(\ref{F2AG}) is obtained for fixed values of
$x/z=\ord(1)$ and hence $z\to0$. On the other hand, when $z\to0$, $k^2\simeq-\kk^2$
 and $\kk^2/zs$ is fixed, the polarization tensor in front of $A_2$ can be replaced with
\begin{equation}\lab{poldom}
 4\;{\kk^2\over zs}\,{p_{1\al}p_{1\be}\over s}\,{1\over z}\left[1+\ord\(z,{|\kk|\over s}\)
 \right]\;,
\end{equation}
which clearly shows an $1/z$ enhancement factor with respect to the $A_1$ one and a
``selection'' of the gluon polarization along $p_2$.

By using Eqs.~(\ref{decompA}) and (\ref{poldom}), Eq.~(\ref{F2AG}) can be rewritten as
\begin{equation}\lab{F2ppA2G}
 F_2=2x\int_0^1{\dif z\over z}\int\dif^2\kk\int{\dif k^2\over2(2\pi)^4}\;
 {4\kk^2\over z^2s^2}p_{1\al}p_{1\be}\,A_2\,\d^{ab}{\overset{ab}{G}{}^{\al\be}\over4\pi}\;.
\end{equation}
By defining the {\em off-shell partonic cross section}
\begin{equation}\lab{d:sigma2}
 \hat{\si}_2\({x\over z},{Q^2\over \kk^2}\)\dug{2x\over z}
 A_2\({x\over z},{Q^2\over \kk^2}\)\simeq{2xz\over\kk^2}p_2^\al p_2^\be A_{\al\be}
\end{equation}
and the {\em unintegrated gluon density}
\begin{equation}\lab{d:dgni}
 \dgni(z,\kk)\dug\int{\dif k^2\over(2\pi)^5}\;{\kk^2\over zs^2}p_{1\al}p_{1\be}
 \sum_a \overset{aa}{G}{}^{\al\be}(p_2,k)
\end{equation}
the high energy, $\kk$-dependent, factorization formula is established~\cite{CaCiHa90}:
\begin{equation}\lab{ffae}
 F_2(x,Q^2)=\int_0^1{\dif z\over z}\int\dif^2\kk\;
 \hat{\si}_2\({x\over z},{Q^2\over\kk^2}\)\dgni(z,\kk)\;.
\end{equation}
Eq.~(\ref{d:sigma2}) shows that the hard partonic cross section $\hat{\si}_2$ is given
by the diagram in Fig.~(\ref{f:Aabmn}) where the soft ($z\to0$) off-shell gluon is
coupled to an external fast quark (or gluon) with an eikonal vertex $p_2^\al$. It is
important to note that $\hat{\si}_2$ is a gauge invariant object, despite the
off-shellness of the ``incoming'' virtual gluon $k$. The reason for this is that the
eikonal coupling induces physical polarization for the incoming gluon.

The formula (\ref{ffae}) can be written in diagonal form in Mellin space. With the
definitions of Eqs.~(\ref{mellinomga}) we get
\begin{equation}\lab{ffaeom}
 F_{2,\om}(Q^2)=\int\dif^2\kk\;\sip_\om\({Q^2\over\kk^2}\)\,\F_\om(\kk)
\end{equation}
or, equivalently,
\begin{equation}\lab{ffaega}
 F_{2,\om}(\ga)=\sip_\om(\ga)\,\F_\om(\ga)\;.
\end{equation}

\subsection{Relation with the collinear factorization\labe{ss:rwcf}}

The high energy factorization formula (\ref{ffae}) is a generalization of the collinear
one (\ref{fattcoll}) which holds not only for $s\sim Q^2\gg\La^2$ but also in the
semi-hard regime $s\gg Q^2\gg \La^2$. This can be easily seen if we remember that, for
$s\sim Q^2$, the leading contributions to the cross section arise from emission of
partons whose transverse momenta are strongly ordered. In particular, the integration
over $\kk$ is important only for $\kk^2\ll Q^2$, where the hard cross section assume the 
typical form of collinear emission
\begin{equation}\lab{sigcoll}
 \sip\(\xp,{Q^2\over\kk^2}\)\simeq\sum_{\pa\in\{\text{partons}\}}\hspace{-4mm}
 e_{\pa}^2\,{\adp}\ln{Q^2\over\kk^2}\,\xp P^{\pa\pg}(\xp)\qquad(\kk^2\ll Q^2)\;.
\end{equation}
In $\om$-space this means that
\begin{equation}\lab{F2aecoll}
 F_{2,\om}(Q^2)=(\sum_{\pa}e_{\pa}^2)\int^{Q^2}\dif^2\kk\;\as\gaqg_\om
 \ln{Q^2\over\kk^2}\F_\om(\kk)\;.
\end{equation}
Since ${1\over2\Nf}\sum_{\pa}e_{\pa}^2$ is just the partonic structure function for the
quark singlet density, we identify
\begin{equation}\lab{singcoll}
 f_\om^{(\Si)}(Q^2)=\int^{Q^2}\dif^2\kk\;\as\,2\Nf\gaqg_\om
 \ln{Q^2\over\kk^2}\F_\om(\kk)\;.
\end{equation}
Taking the derivative with respect to $\ln Q^2$ yields
\begin{align}\lab{evolsing}
 {\dif\over\dif\ln Q^2}f_\om^{(\Si)}(Q^2)=&\as\,2\Nf\gaqg_\om\int^{Q^2}\dif^2\kk\;
 \F_\om(\kk)\\ \nonumber
 =&\as\,2\Nf\gaqg_\om f_\om^{(\pg)}(Q^2)
\end{align}
once we have made the further identification
\begin{equation}\lab{d:DGI}
 f_\om^{(\pg)}(Q^2)=\int^{Q^2}\dif^2\kk\;\F_\om(\kk)\;.
\end{equation}
In conclusion, $\F_\om(\kk)$ represents the unintegrated gluon density in $\kk$-space
which is related to the usual gluon PDF by Eq.~(\ref{d:DGI}). The former gives the most
important contribution to the quark singlet density in the framework of high energy
factorization.

However, when $s\gg Q^2$ we cannot exploit the collinear picture to describe the $Q^2$
evolution of the singlet density, since $\sip$ does not necessarily contain a leading
$\ln Q^2$. Rather, $\sip$ has to be considered the $\ord(\as)$ coefficient of the gluon
density contributing to $F_2$. The $\kk$-convolution of $\sip$ with the unintegrated
gluon density provides the resummation of the large $\ln1/x$ in the coefficient
function, as we shall see in Sec.~\ref{ss:rcf}.

\section{Resummation of $\boldsymbol{\ln1/x}$ in the leading approximation\labe{s:rlxla}}

The high energy factorization formula (\ref{ffae}) shows that the leading high energy
contribution to semi-hard DIS is governed by the unintegrated gluon density for small
values of the gluon momentum fraction $z$. As one can easily realize, it is not possible 
to determine perturbatively the gluon density, since long distance effects are
unavoidable when dealing with hadrons. Nevertheless, in a particular regime (to be
specified soon), the perturbative analysis allows us to extract information about the
dependence of the unintegrated gluon density on the ``hard'' variables $z$ and $\kk$
involved in the hard gluon-photon vertex. This is done by means of an evolution equation 
in $z$-space obtained more than twenty years ago by the russian school.

In this section we will go over the road that led to the BFKL
equation~\cite{BFKL76}. The latter resums all the leading logarithmic ($\ln s$)
coefficients of the $\as$ perturbative series for inclusive semi-hard processes. The
following presentation will be useful in order to identify the basic physical quantities
(in connection with the DGLAP ones) and to fix the framework for the NL$x$ formulation
of the BFKL equation (Cap.~\ref{c:NLbfkl}) and for its subsequent improvement
(Chap.~\ref{c:rga}).

Following the idea of factorization of non perturbative effects, we can guess that the
probability to find a gluon with momentum fraction $z$ with respect to the parent proton 
and transverse momentum $\kk$ should be given by a convolution of bare partonic
densities $f^{(\pa)}$ with a perturbative calculable function $\G^{(\pa)}$ which
embodies all kinds of intermediate processes between the bare parton $\pa$ and the hard
gluon. At high energies, only gluon exchanges in the $t$-channel contribute in the L$x$
approximation. Therefore we assume a sort of factorized expression for the unintegrated
gluon density of the form
\begin{equation}\lab{fattDGNI}
 \F_\om(\kk)\simeq\int\dif^2\kk_0\;f_\om^{(\pg)}(\kk_0)\,\G_\om(\kk,\kk_0)
\end{equation}
which stresses again the relevance of the transverse degrees of freedom. Here
$f_\om^{(\pg)}(\kk_0)$ denotes the Mellin transform of the probability density of
finding a gluon with transverse momentum $\kk_0$ inside the proton; it is to be
considered as an unknown input function to be eventually specified in particular
phenomenological applications.

The function $\G_\om(\kk,\kk_0)$ is the basic object we wish
to investigate; it will be extensively studied throughout this thesis.

\subsection{Leading $\boldsymbol{\ln1/x}$ BFKL equation\labe{ss:lxBFKLe}}

The function $\G_\om(\kk,\kk_0)$ represents the set of gluon exchanges contributing in
L$x$ approximation to high energy scattering of two strong interacting objects. Also
high energy scattering of colourless particles is governed by gluon exchanges: in this
case the coupling to gluons occurs via quark or EW boson intermediate states.

In order to determine $\G_\om$, the simplest situation we can consider is high energy
scattering of two partons $\pa$ and $\pb$.
\begin{figure}[ht!]
\centering
\includegraphics[height=39mm]{\fig 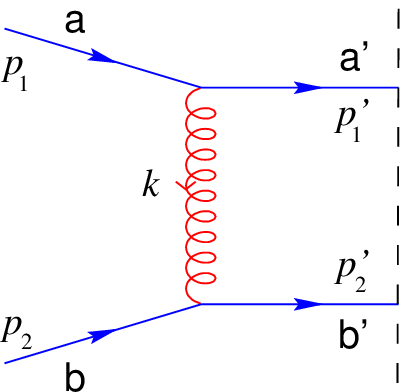}
\didascalia{The lowest order Feynman diagram contributing to leading L$x$ order.%
\labe{f:born}} 
\end{figure}
At Born level, the cross section is dominated by one gluon exchange in the $t$-channel
(Fig.~\ref{f:born}) corresponding to the amplitude 
\begin{equation}\lab{Aborn}
 \M_{2\to2}^{(0)}=-\ui\,2s\,\Ga_{\pa'\pa}^r\,{1\over t}\,\Ga_{\pb'\pb}^r\quad,\quad
 s=(p_1+p_2)^2\quad,\quad t=(p_1-p_1')^2
\end{equation}
where summation over repeated colour indices is understood and the parton-gluon vertices 
are
\begin{subequations}\labe{PGvert}
\begin{align}\lab{PGvertg}
 \Ga_{\pa'\pa}^r\dug g\,\d_{\la_{\pa'}\la_{\pa}}\,
 f^{a'ra}\qquad(\pa=\pg=\text{gluon})\;,\\    \lab{PGvertq}
 \Ga_{\pa'\pa}^r\dug-\ui g\,\d_{\la_{\pa'}\la_{\pa}}\,\t{r}_{\rm a'a}
 \qquad(\pa=\pq=\text{quark})\;.
\end{align}
\end{subequations}
Since gluon exchanges do not affect the flavour of the quarks and the
polarization/\-helicity $\la_{\pa'}=\la_{\pa}$ is conserved, only colour degrees of
freedom and kinematics have to be taken into account. Note also that the longitudinal
variables corresponding to the Sudakov decomposition $k=zp_1+\bar{z}p_2+\kk$ are small
in the high energy regime: $z=|\bar{z}|=|t|/s\ll1$.
\begin{figure}[ht!]
\centering
\includegraphics[width=143mm]{\fig 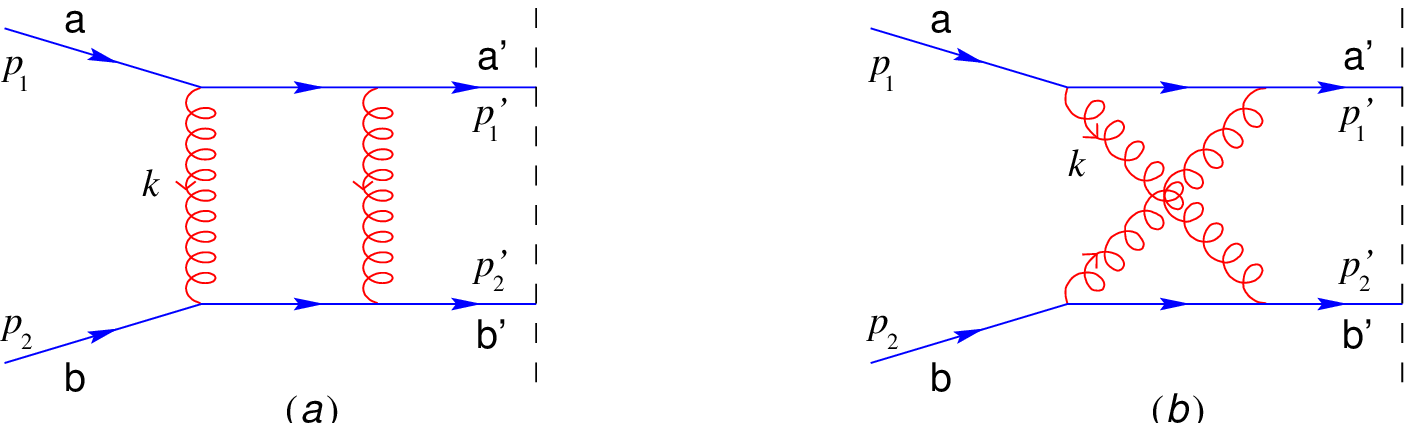}
\didascalia{The two particle final states virtual correction diagrams contributing at 
leading L$x$ order. Iteration of gluon exchange in the $s$-channel ({\sl a}) and in the
$u$-channel ({\sl b}).%
\labe{f:boxcross}}
\end{figure}

The 1-loop virtual corrections to the elastic amplitude give rise to a $\ln s$
coefficient. In covariant gauges, only the ``box'' and ``crossed'' diagrams of
Fig.~\ref{f:boxcross} contributes with logarithms of $s$. These amplitudes
can be projected in the subspaces of the irreducible representations of
$8\otimes8$ where the couple of exchanged gluon lives. In other words, a given amplitude
can be considered as a sum of terms representing the exchange in the $t$-channel 
of colour structures belonging to the various $8\otimes8$ irreducible representations.
When summing the amplitudes $\M^{(1,s)}_{2\to2}$ and $\M^{(1,u)}_{2\to2}$ of
Fig.~\ref{f:boxcross}{\sl a} and {\sl b} respectively, which kinematically
corresponds to the replacement $s\leftrightarrow u$, only the octet representation 
of the antisymmetric part $(8\otimes8)_A=1\oplus8\oplus10\oplus\overline{10}$
interferes constructively. For this reason the colour factors can be well represented by
the parton-gluon vertices of Eqs.~(\ref{PGvert}).

The octet part of $\M^{(1,s)}_{2\to2}$ reads
\begin{equation}\lab{M1s8}
 \M^{(1,s)}_{2\to2}\cong g^2s^2N_c\,\Ga_{\pa'\pa}^r\,I(s,t)\,\Ga_{\pb'\pb}^r
\end{equation}
where the loop integral $I$ yields%
\footnote{The 1-loop Regge-gluon trajectory $\ab\Om^{(0)}$ is mostly known in the
literature as $\om^{(1)}$. We prefer to adopt a different notation in order to avoid
confusion with the $\om$ variable and because of its connection with the leading BFKL
kernel $K^{(0)}$. In this chapter the suffix $^{(0)}$ will be understood.}
\begin{subequations}\labe{loopint}
\begin{align}
 I(s,t)\simeq&-{\ui\over(2\pi)^2\,s\,t}\ln\(-{s\over s_0}\)\Om^{(0)}(-t)\\ \lab{interc}
 \Om^{(0)}(\kk^2)\dug&-{\kk^2\over4\pi}\int{\dif\qq\over\qq^2(\kk-\qq)^2}
\end{align}
\end{subequations}
The logarithmic factor stems from integration over the longitudinal variables of $k$.
In the massless theory the {\em scale of the energy} $s_0$ is of the order of the
transverse momentum squared of the outgoing particles. In the L$x$ approximation,
however, the choice of $s_0$ is irrelevant, because with an other choice $s'_0$ the
difference $\ln s/s_0-\ln s/s_0'=\ln s'_0/s_0$ is independent on $s$ and therefore
subleading.

The coefficient $\Om(-t)$ of the logarithmic term is an IR divergent
two-dimensional integral. The IR divergence will cancel when considering real emission
correction to the Born amplitude (see App.~\ref{a:avlbfkl}).

We note that among the subleading 1-loop corrections we have neglected there are the
self-energy and vertex-correction diagrams which determine the running of the coupling
constant. Accordingly, in a L$x$ treatment of the radiative corrections, $\as$ must be
regarded as fixed.

By collecting Eqs.~(\ref{Aborn}), (\ref{M1s8}) and (\ref{loopint}) the full 1-loop
expression for the elastic amplitude results
\begin{equation}\lab{Aelas1loop}
 [\M^{(0)}+\M^{(1)}]_{2\to2}=-\ui\,2s\,\Ga_{\pa'\pa}^r\,{1+\ab\Om(-t)\ln(s/s_0)\over t}\,
 \Ga_{\pb'\pb}^r\;.
\end{equation}
Eq.~(\ref{Aelas1loop}) is intriguing, since the logarithmic correction looks like the
$\ab$ truncation of $(s/s_0)^{\ab\Om(-t)}$ which would correspond to Regge behaviour.
The 2-loop correction has been computed~\cite{BFKL76} and confirms this assumption. In
conclusion, we make the ansatz that, in L$x$ approximation, the amplitude for elastic
scattering with the gluon quantum numbers in the $t$-channel is
\begin{equation}\lab{gluonregge}
 \M_{2\to2}=-\ui\,2s\,\Ga_{\pa'\pa}^r\,{(s/s_0)^{\ab\Om(-t)}\over t}\,
 \Ga_{\pb'\pb}^r
\end{equation}
which states the {\em reggeization of the gluon}.

Upon colour averaging and by using the phase space measure for two particle final states 
(including the flux factor $(2s)^{-1}$)
\begin{equation}\lab{sfdue}	       
 \dif\phi^{(2)}=\frac{1}{(4\pi)^{2}}\,\frac{1}{s^2}\,\dif\kk\;,
\end{equation}			       
Eq.~(\ref{gluonregge}) yields the two-particle final state contribution to the
differential cross section
\begin{equation}\lab{sigma2}
  \dif\si_{\pa\pb\to2}
 =h_{\pa}^{(0)}(\kk)\left({s\over s_0}\right)^{\ab2\Om(-t)}h_{\pb}^{(0)}(\kk)\,\dif\kk
\end{equation}
in terms of the Born impact factors
\begin{equation}\lab{hzero}       
 h_{\pa}^{(0)}(\kk)\dug{2\as C_{\pa}\over\sqrt{N_c^2-1}}\,{1\over\kk^2}
 \quad;\quad C_{\pg}=C_A\quad;\quad C_{\pq}=C_F\;.
\end{equation}
\begin{figure}[ht!]
\centering
\includegraphics[width=16cm]{\fig 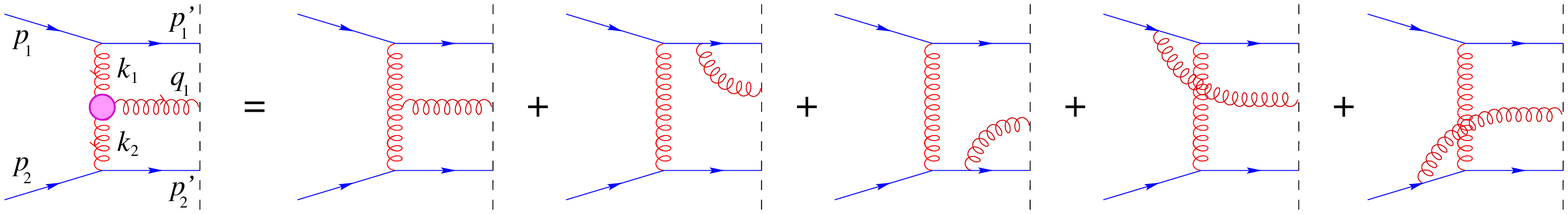}
\put(-154,10){\footnotesize{$\Up$}}
\didascalia{The Lipatov effective vertex for the emission of a real gluon: the blob
represents gluon emission both from the exchanged gluon and from the external particles.%
\labe{f:lipatov}}
\end{figure}

Let's now consider the 1-loop order real emission corrections. At L$x$ level, the
high-energy kinematics favours the emission of an additional gluon (Fig.~\ref{f:lipatov}) of
momentum $q_1$ and polarization $\po$ in the central region
$q_1^+\sim q_1^-\sim|\qq_1|$~. The corresponding amplitude is
\begin{align}\lab{Magb}
 \M^{(0)}_{2\to3}=&-\ui\,2s\,\Ga_{\pa'\pa}^{r_1}\,{1\over t_1}\Up^c_{r_1r_2}
 {1\over t_2}\,\Ga_{\pb'\pb}^{r_2}\;,\\
 &\Up^c_{r_1r_2}(k_1,k_2)\dug\,\ui g\,f^{r_1c\,r_2}\,\poc_\mu(q_1)J^\mu(k_1,k_2)\;,
\end{align}
where $q_1^{\mu}=k_1^{\mu}-k_2^{\mu}$, $t_i=k_i^2\simeq-\kk_i^2$ and
\begin{equation}\lab{verlip}
 J^{\mu}(k_1,k_2)=-(\ku+\kd)^{\mu}+p_1^{\mu}\left({2p_2k_1\over s}+{\ku^2\over p_1k_2}
 \right)+p_2^{\mu}\left({2p_1k_2\over s}+{\kd^2\over p_2k_1}\right)
\end{equation}
is the Lipatov effective vertex.

The 1-loop virtual corrections contributing to the three-particle final states in L$x$ 
approximation can be computed, and their final effect --- according to gluon
reggeization --- consists simply in the replacement of the propagators
\begin{equation}\lab{propregge}
 {1\over t_i}\;\longrightarrow\;{1\over t_i}\left({s_i\over s_0}\right)^{\ab\Om(-t_i)}\quad,
 \quad s_i=(q_i+q_{i-1})^2\qquad(i=1,2)\;.
\end{equation}
in the amplitude (\ref{Magb}).
			
After polarization and colour averaging, by using the three-body phase space measure
($z_1$ being the momentum fraction of $k_1$ with respect to $p_1$)
\begin{equation}\lab{sftre}
 \dif\phi^{(3)}=\frac{1}{\pi(4\pi)^{4}}\,\frac{\dif z_1\,\du\,\dd}{s^2 z_1}
\end{equation}
and the properties of the $J$ vertex
\begin{equation}\lab{prop}
 q^\mu J_\mu(k_1,k_2)=0\quad\imp\quad\sum_\la\poc^\mu_\la\po^\nu_\la J_\mu J_\nu=
 -J^2=4{\ku^2\kd^2\over \qq^2}
\end{equation}
we obtain the three-particle final state contribution to the differential cross section
\begin{equation}\lab{sigma3}
 \dif\si_{\pa\pb\to3}=h_{\pa}(\ku)\left({s_1\over s_0}\right)^{\ab2\Om(-t_1)}
 \frac{\ab}{\pi\qq^2}\left({s_2\over s_0}\right)^{\ab2\Om(-t_2)}h_{\pb}(\kd)\,
 {\dif z_1\over z_1}\,\dif\ku\,\dif\kd
\end{equation}			       
whose leading logarithmic character follows from the $z_1$ integration with infrared
boundary $z_1>s_0/s$ (pure phase space would yield
$\frac{\qq^2}{s}<z_1<1-\frac{\ku^2}{s}$).
					
To summarize the results till now obtained
\begin{itemize}
\item the huge part of the cross section arise from $t$-channel small virtualities gluon
exchanges;
\item the longitudinal phase space grows logarithmically with $s$ and the differential
cross section is almost independent of the longitudinal variables (apart in the
proximity of their upper boundary\footnote{I.e., for $z_1\to1$ where the longitudinal
components of $q_1$ get large.}) whose integration yields just the $\ln s$ term;
\item most of the longitudinal phase space corresponds to small values of the $z_i$
Sudakov variables;
\item the mass-shell conditions of the outgoing particles causes the transverse momenta
of the exchanged (reggeized) gluons to be the only relevant degrees of freedom for the
amplitudes, furthermore $k_i^2\simeq-\kk_i^2$.
\end{itemize}
Going to higher order in $\ab$ is then straightforward.
\begin{figure}[t]
\centering
\includegraphics[height=6cm]{\fig 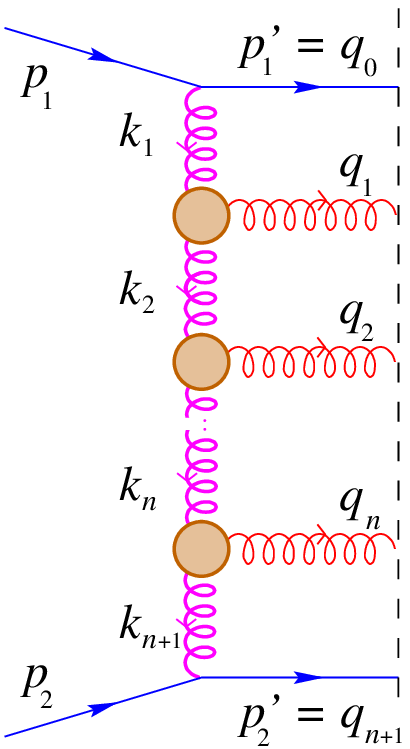}
\didascalia{The $n+2$ final particle gluon ladder contributing at L$x$ to high energy
scattering. The thick wavy line represents reggeized gluon exchanges.%
\labe{f:scalagluoni}}
\end{figure}

The scattering amplitude with $2+n$ particles in the final state receives its L$x$
contribution in the phase space region where the outgoing particles are strongly ordered
in rapidity. At tree level, this corresponds to (half) a gluon ladder
(Fig.~\ref{f:scalagluoni}) in {\em multi-Regge kinematics} (MRK):
\begin{subequations}\labe{mrk}
\begin{align}
 &k_i=z_ip_1+\bar{z}_ip_2+\kk_i\qquad(i=1,\dots,n+1)\\
 &1\equiv z_0\gg z_1\gg\cdots\gg z_{n+1}={-t_{n+1}\over s}\\
 &{t_1\over s}=\bar{z}_1\ll\bar{z}_2\ll\cdots\ll\bar{z}_{n+1}\ll1 
\end{align}
\end{subequations}
where the squared transferred momenta $t_i=k_i^2\simeq-\kk_i^2$ are of the same order
and much smaller than $s$. Because of the mass-shell constraints and total momentum
conservation, we can take $\{z_i:i=1,\dots,n\}$ and $\{\kk_j:j=1,\cdots,n+1\}$ as
independent variables; accordingly the phase space measure reads
\begin{equation}\lab{sf2n}
 \dif\phi^{(2+n)}={1\over s^2\,(4\pi)^{2n+2}\pi^n}
 \prod_{i=1}^{n}{\dif z_i\over z_i}\prod_{j=1}^{n+1}\dif\kk_j\;.
\end{equation}

The real emission amplitude is nothing but the generalization of Eq.~(\ref{Magb}), where 
outgoing gluons are coupled to internal lines by the effective vertex
(\ref{verlip}). Finally, the virtual corrections provides the reggeization of the
exchanged gluons with the replacement of the propagators as in Eq.~(\ref{propregge}),
yielding~\cite{BFKL76,Lip89}
\begin{equation}\lab{a2n}
 \M_{2\to2+n}=-\ui\,2s\,\Ga_{\pa'\pa}^{r_1}\,{(s_1/s_0)^{\ab\Om_1}\over t_1}
 \Up^{c_1}_{r_1r_2}{(s_2/s_0)^{\ab\Om_2}\over t_2}\cdots\Up^{c_n}_{r_nr_{n+1}}
 {(s_{n+1}/s_0)^{\ab\Om_{n+1}}\over t_{n+1}}\,\Ga^{r_{n+1}}_{\pb{}'\pb}
\end{equation}
with the shorthand notation $\Om_i\equiv\Om(-t_i)$.

Having so obtained the $2+n$ particle differential cross section
\begin{equation}\lab{sez2n}
 \dif\si_{\pa\pb\to2+n}=h_{\pa}(\ku)h_{\pb}(\kk_{n+1})\left({\ab\over\pi}\right)^n
 {(s_1/s_0)^{\ab2\Om_1}\cdots(s_{n+1}/s_0)^{\ab2\Om_{n+1}}\over\qu^2\cdots\qq_n^2}
 \prod_{i=1}^{n}{\dif z_i\over z_i}\prod_{j=1}^{n+1}\dif\kk_j
\end{equation}
there remains to sum up over all values of $n$ from 0 to infinity. This is easily done
in Mellin space with respect to the $\om$ variable conjugated to $s$ by defining
\begin{equation}\lab{d:sigom}
 \si^\om_{\pa\pb}\dug\int_{s_0}^\infty{\dif s\over s}\;\left(s\over s_0\right)^{-\om}
 \sum_{n=0}^\infty\si_{\pa\pb\to2+n}(s)\;.
\end{equation}
By using the multi-Regge kinematic relations
\begin{equation}\lab{mrkrel}
 s_i\simeq{z_{i-1}\over z_i}\qq_i^2\quad(i=1,\dots,n+1)\quad,\quad
 z_{n+1}\simeq{\qq_{n+1}^2\over s}\quad,\quad
 s={s_1\cdots s_{n+1}\over \qq_1^2\cdots\qq_n^2}\;,
\end{equation}
we can perform the change of variables $\{s,z_i\}\to\{s_i\}$ such that
\begin{equation}\lab{dazasi}
 {\dif s\over s}{\dif z_1\over z_1}\dots{\dif z_n\over z_n}=
 {\dif s_1\over s_1}\dots{\dif s_{n+1}\over s_{n+1}}\;
\end{equation}
and, after integration over the sub-energies $s_i\in[s_0,\infty[$, we obtain
\begin{align}\lab{difsigom}
 {\dif\si_{\pa\pb}^\om\over\dif\kk\dif\kk_0}&=h_{\pa}(\kk)h_{\pb}(\kk_0)\\
 &\quad\times\sum_{n=0}^\infty\int\dif\kd\cdots\dif\kk_n\;{1\over\om-\ab2\Om_1}
 \;{\ab\over\pi\qu^2}\;{1\over\om-\ab2\Om_2}\cdots{\ab\over\pi\qq_n^2}\;
 {1\over\om-\ab2\Om_{n+1}}\;,\nonumber
\end{align}
where $\kk=-\ku$ and $\kk_0=\kk_{n+1}$ are the transverse momenta of the scattered
partons $\pa'$ and $\pb'$. 
The sum on the RHS is just the series expansion of the operator\footnote{%
We ask the reader to forgive us for the ``na\"\i ve'' notation!}
\begin{equation}\lab{d:GGF}
 \G_\om={1\over\om-\K^{(\text{V})}-\K^{(\text{R})}}=
 {1\over\om-\K^{(\text{V})}}+{1\over\om-\K^{(\text{V})}}\K^{(\text{R})}
 {1\over\om-\K^{(\text{V})}}+\cdots
\end{equation}
where the real and virtual BFKL kernels are
\begin{subequations}\labe{d:nucleiBFKL}
\begin{align}
 \K^{(\text{R})}(\kk,\kk')&\dug{\ab\over\pi\,(\kk-\kk')^2}\\
 \K^{(\text{V})}(\kk,\kk')&\dug\ab2\Om(\kk^2)\d^2(\kk-\kk')\;.
\end{align}
\end{subequations}

It is evident that the distribution $\G_\om(\kk,\kk_0)$, determining the cross section
apart from the impact factors $h_{\pa}$ and $h_{\pb}$, represents all exchanges and
emissions stemming from the two gluons that couples directly to the incoming
partons. Therefore, we identify $\G_\om(\kk,\kk_0)$ as the function introduced in
Eq.~(\ref{fattDGNI}) and representing the evolution of the unintegrated
gluon density at high energy. 

Finally, we can invert the Mellin transform (\ref{d:sigom}) to $s$-space and represent
the L$x$ partonic differential cross section as
\begin{equation}\lab{fattLx}
 {\dif\si_{\pa\pb}(s)\over\dif\kk\,\dif\kk_0}=\int\difo\;\left({s\over s_0}\right)^\om
 h_{\pa}(\kk)\G_\om(\kk,\kk_0)h_{\pb}(\kk_0)\;.
\end{equation}
The parton-parton cross section just analized, due to the chromatic nature of the
scattering particles, suffers severe Coulomb singularities, since the impact factors
$h(\kk)\sim(\kk^2)^{-1}$ prevent the definition of a total cross section for coloured
objects.

On the other hand, by high energy factorization, Eq.~(\ref{fattLx}) is valid as well for 
scattering of colourless particles, provided the corresponding impact factors are
employed. For instance, in the case of virtual photons, the ensuing impact factors ---
depending on the virtuality $Q^2$ of the photon --- assume the form
\begin{equation}\lab{fiQ}
 h_{\pa}^{(0)}=h_{\pa,\om}^{(0)}(Q,\kk)={1\over Q^2}f_\om^{\pa}\({Q^2\over\kk^2}\)
\end{equation}
which is non singular for $\kk^2\to0$ and hence provide, upon $\kk$ and $\kk_0$
integration in Eq.~(\ref{fattLx}), a finite total cross section (see Sec.~\ref{ss:cif}).

The identity
\begin{equation}\lab{idenGGF}
 [\om-(\K^{(\text{R})}+\K^{(\text{V})})]\G_\om=\id
\end{equation}
suggests that $\G_\om$ is the solution of the Green function equation
\begin{equation}\lab{eqBFKL}
 \om\G_\om(\kk,\kk_0)=\d^2(\kk-\kk_0)+\int\dif\kk'\;\K(\kk,\kk')\G_\om(\kk',\kk_0)
\end{equation}
where $\K=\K^{(\text{R})}+\K^{(\text{V})}$ is the L$x$ BFKL kernel.

The {\em gluon Green's function} (GGF) $\G_\om$ can be determined once a complete set of 
eigenfunctions for the integral operator $\K$ is found. In App.~\ref{a:irtms} it is
shown that the functions
\begin{equation}\lab{afzK0}
 f_{\ga,m}(\kk)=(\kk^2)^{\ga-1}\;{\esp{\ui m\phi}\over\sqrt\pi}\qquad,\quad
 \ga=\half+\ui\nu\quad,\quad\nu\in\R\quad,\quad m\in\Z
\end{equation}
form a complete set of eigenfunctions for $\K$ whose eigenvalue is $\ab\chi(\ga,m)$ where
\begin{equation}\lab{avlK0}
 \chi(\ga,m)=2\psi(1)-\psi\(\ga+{|m|\over2}\)-\psi\(1-\ga+{|m|\over2}\)\;.
\end{equation}
The finite expression of $\chi(\ga,m)$ confirms the cancellation of the IR divergencies of the real and virtual kernels.

The GGF can be expressed by the spectral representation
\begin{equation}\lab{rapspeG}
 \G_\om(\kk,\kk_0)=\sum_{m\in\Z}\intmel\difg\;{f_{\ga,m}(\kk)f^*_{\ga,m}(\kk_0)\over
 \om-\ab\chi(\ga,m)}
\end{equation}
which, by deforming the contour of integration to the left for $\kk^2>\kk_0^2$, can be
evaluated by the residue method.

\subsection{Resummation of the anomalous dimension\labe{ss:rad}}

The full determination of the L$x$ GGF permits the high energy resummation of both the
gluon anomalous dimension~\cite{BFKL76} and of the coefficient function~\cite{CaCiHa90}
(at L$x$ level).

First of all, let's note that the unintegrated gluon density (\ref{fattDGNI}) satisfies
the BFKL equation
\begin{equation}\lab{Fbfkl}
 \om\F_\om(\kk)=\om\F_\om^{(0)}(\kk)+\int\dif\kk\;\K(\kk,\kk_0)\,\F_\om^{(0)}(\kk_0)
\end{equation}
where an $\om$ has been factorized in the inhomogeneous term
$\om\F_\om^{(0)}(\kk)=f_{\om}^{(\pg)}(\kk)$ for later convenience.

If we adopt an azimuthally symmetric input function $\F_\om^{(0)}$, then only the $m=0$
subset of eigenfunctions $f_{\ga,m}$ contributes to the expansion of $\F_\om^{(0)}$ and
$\F_\om$. In this case the spectral representation corresponds to the Mellin
transformation with respect to $\kk^2$ in $\ga$-space (App.~\ref{a:mellin}), where
Eq.~(\ref{Fbfkl}) reduces to the simple algebraic equation
\begin{equation}\lab{FconF0}
 \F_\om(\ga)={\F_\om^{(0)}(\ga)\over1-{\ab\over\om}\chi(\ga)}\;.
\end{equation}
By inverting Eq.~(\ref{ffaega}) in favour of the $Q^2$ variable, we get
\begin{equation}\lab{intF2ga}
 F_{2,\om}(Q^2)=\intmel\difg\;\left({Q^2\over\La^2}\right)^\ga{\F_\om^{(0)}(\ga)
 \sip_\om(\ga)\over1-{\ab\over\om}\chi(\ga)}\;.
\end{equation}
Since we are interested to $Q^2\gg\La^2$, we displace the contour of integration to the
left so as to enclose all the singularities whose real part is less than $1/2$. Provided 
a smooth input function $\F_\om^{(0)}$ has been adopted, the singularities in the
integrand of Eq.~(\ref{intF2ga}) can arise either from the hard cross section
$\sip_\om(\ga)$ or from a vanishing denominator.

$\sip_\om(\ga)$ has poles both at $\ga=0$ --- corresponding to its collinear behaviour
for $Q^2\gg\kk^2$ --- and $\ga=1$ ($Q^2\ll\kk^2$)~\cite{CaCiHa90}. We explicitly show
the former by writing
\begin{equation}\lab{d:acca}
 \sip_\om(\ga)\ugd{1\over\ga}h_\om(\ga)\;.
\end{equation} 

In the denominator of Eq.~(\ref{intF2ga}) we have zeroes corresponding to spectral
points of $\G_\om$, i.e., when $\om=\ab\chi(\ga)$. The symmetry, hermiticity and
boundedness properties of $\K$ causes $\chi(\ga)$ to assume real values only for
$\ga-1/2\in\I$ (the imaginary axis) where $\chi(\ga)\leq\chi_m\dug\chi(1/2)=4\ln2$ and
for $\ga\in\R$.

If $\om>\ab\chi_m$ then the rightmost singularity $\gl$ in the half plane $\re\ga<1/2$
--- governing the high-$Q^2$ behaviour of $F_2$ --- is a simple pole determined by the
conditions
\begin{equation}\lab{d:gl}
 \om=\ab\chi(\gl)\quad,\quad0<\gl<{1\over2}\quad,\quad\gl=\gl\({\ab\over\om}\)\;.
\end{equation}
The other singularities to the left of $\gl$ give terms which are suppressed as (almost
integer) powers of $Q^2$ (higher twist) and will not be considered in the
following%
\footnote{The eigenvalue functions $\chi(\ga,m)$ corresponding to conformal spin
$m\neq0$, which has been neglected in the present analysis, would contribute with
subleading poles. In fact, the position of the rightmost pole of $\chi(\ga,m)$ in the
half-plane $\re\ga<1/2$ is $\ga=-|m|/2$}.
Anyhow, they may be important in moderate and intermediate-$Q^2$ regimes.

By applying the residue theorem to Eq.~(\ref{intF2ga}) yields
\begin{equation}\lab{F2dimanom}
 F_{2,\om}(Q^2)=\left({Q^2\over\La^2}\right)^{\gl}{\F_\om^{(0)}(\gl)\over\gl{\ab\over\om}
 |\chi'(\gl)|}h_\om(\gl)\;.
\end{equation}
In order to extract the anomalous dimension, we note that in the $\as\to0$ limit
\begin{equation}\lab{limitialfa0}
 \gl{\ab\over\om}|\chi'(\gl)|\to1\quad,\quad\F_\om^{(0)}(\gl)\to\F_\om^{(0)}(0)
 \quad,\quad h_\om(\gl)\to h_\om(0)
\end{equation}
and both $F_\om^{(0)}(0)$ and $h_\om(0)$ are finite%
\footnote{Even if $h_\om(\ga)$ is singular for $\ga\to0$ in massless quark emission
processes, nevertheless it is $\ord(\as)$, and in the $\as\to0$ limit the singularity
arising from the implicit $\as$ dependence of $\gl$ is cancelled by the explicit $\as$
dependence of $h_\om$.}.

By comparing Eq.~(\ref{F2dimanom}) with Eq.~(\ref{solalfafissa}), which gives --- up to a 
constant coefficient function factor --- the $Q^2$ dependence of the SF in the
fixed-$\as$ formulation, we identify $\gl(\ab/\om)$ as the L$x$ gluon anomalous
dimension. $\gl$ resums all the leading $\ab/\om$ terms of the gluon anomalous dimension
\begin{equation}\lab{seriegl}
 \gl\({\ab\over\om}\)={\ab\over\om}\sum_{n=0}^\infty c_n\left({\ab\over\om}\right)^n=
 {\ab\over\om}+2\zeta(3)\left({\ab\over\om}\right)^4
 +2\zeta(5)\left({\ab\over\om}\right)^6+\ord\({\ab\over\om}\)^7
\end{equation}
or, in $x$-space, the $\ab\ln1/x$ terms of the gluon splitting function
\begin{equation}\lab{seriePgg}
 \adp P^{\pg\pg}(x)={\ab\over x}\sum_{n=0}^\infty c_n\left(\ab\ln{1\over x}\right)^n\;.
\end{equation}

Outside the anomalous dimension regime, i.e.\ for $\om<\ab\chi_m$, the function $\gl$
develop a branch-cut singularity at $\ab/\om=1/\chi_m$ corresponding to local
non-invertibility of the eigenvalue function: $\chi'(\gl=1/2)=0$.

\subsection{Resummation of the coefficient function\labe{ss:rcf}}

By applying the same complex integration techniques to the integrated gluon density
whose Mellin transform is
\begin{equation}\lab{melDGI}
 \int{\dif Q^2\over Q^2}\;\left({Q^2\over\La^2}\right)^{-\ga}
 \int^{Q^2}\dif\kk\,\F_\om(\kk)={\F_\om(\ga)\over\ga}
 ={1\over\ga}\,{\F_\om^{(0)}(\ga)\,\over1-{\ab\over\om}\chi(\ga)}\;,
\end{equation}
it is easy to see that, in the anomalous dimension regime $\om>\ab\chi_m$,
\begin{equation}\lab{gludimanom}
 f_\om^{(\pg)}(Q^2)=\left({Q^2\over\La^2}\right)^{\gl}{\F_\om^{(0)}(\gl)\over\gl
 {\ab\over\om}|\chi'(\gl)|}\;,
\end{equation}
The expression (\ref{F2dimanom}) is then consistent with the standard QCD factorization
theorem in which the remaining factor $h_\om\(\gl(\ab/\om)\)$ represents the L$x$ gluon
coefficient function~\cite{CaCiHa90} for $F_2$. The resummed effect is incorporated
through the $\ab/\om$ dependence of $\gl$ of Eq.~(\ref{seriegl}) and the
$\ga$-dependence of $h_\om$.

\section{High energy behaviour of the structure functions\labe{s:hebSF}}

The inverse Mellin transformation in $\om$-space of Eq.~(\ref{F2dimanom}) allows us to
determine the small-$x$ behaviour of $F_2$ according to the BFKL resummation of the
leading $\ln x$.

Regarded as a function of $\om$, $F_{2,\om}(Q^2)$ in Eq.~(\ref{F2dimanom}) presents its
rightmost singularity at
\begin{equation}\lab{d:pomLx}
 \om=\op(\ab)\dug\ab\chi_m=\ab\,4\ln2
\end{equation}
both for the presence of the branch-cut of $\gl(\ab/\om)$ and for the vanishing of
$\chi'(\gl)$ in the denominator. By assuming $F^{(0)}$ and $h$ not too strong dependent
on $\om$ and $\ga$ around $\op$ and $\gl=1/2$ respectively, we have
\begin{equation}\lab{F2intdelta}
 F_{2,\om}(Q^2)\simeq{H_{\op}x^{-\op}\over a\chi''_m}\left({Q^2\over\La^2}\right)^{1/2}
 \int_{-\ui\infty}^{\ui\infty}{\dif\De\over2\pi\ui}\;{x^{-\De}\esp{-at\De^{1/2}}\over
 \De^{1/2}}
\end{equation}
where $\De=\om-\op$, $\chi''_m=\chi''(1/2)=28\zeta(3)$,
$H_{\op}=4\F^{(0)}_{\op}(1/2)h_{\op}(1/2)$ and
\begin{equation}\lab{apprgl}
 a=\sqrt{2\over\ab\chi''_m}\quad,\quad\gl\({\ab\over\om}\)\simeq\half-a\De^{1/2}
 \quad,\quad t=\ln{Q^2\over\La^2}\;.
\end{equation}
The integration in the complex $\De$-plane can be performed explicitly and gives
\begin{equation}\lab{F2pom}
 F_{2,\om}(Q^2)\simeq H_{\op}\sqrt{\ab\over2\pi\chi''_m}\,{x^{-\op}\over\sqrt{\ln{1/x}}}
 \left({Q^2\over\La^2}\right)^{1/2}\exp\left\{-{\ln^2Q^2/\La^2\over2\ab\chi''_m\ln1/x}
 \right\}\;.
\end{equation}
Several facts can be learned from the last expression:
\begin{itemize}
\item at high energy the leading anomalous dimension $\gl$ saturates at the value $1/2$;
\item correspondingly the coefficient function $h_\om(\gl)$ undergoes a large
enhancement with respect to its ``collinear'' value where $\gl\simeq0$ since
$h_\om(1/2)$ is usually much larger than $h_\om(0)$ because of the singularity of
$h_\om$ at $\ga=1$;
\item apart from small logarithmic terms, the $F_2$ structure function --- and hence the 
total cross section --- presents a power-like rise towards small values of $x$.
\end{itemize}
As explained in Sec.~\ref{s:rb}, the last point is in disagreement with the general
results of Regge theory: the Froissart bound $\si_{\text{tot}}\leq\text{const}(\ln s)^2$
is violated. The reason is that the $s$-channel unitarity constraints for scattering
amplitudes are not fulfilled in L$x$ approximation. Nevertheless, a power-like growth in
$x$, or equivalently, in $s$, has been observed for the structure functions. The
agreement of the L$x$ prediction (\ref{F2pom}) is only qualitatively (power-like
growth), the small-$x$ exponent $\op=\ab4\ln2$ being too large to fit the data, but the
order of magnitude is OK. $\as$ is not yet running, NL$x$ corrections have to be taken
into account...

Does that mean that we are on the right track?

\chapter{High energy processes in next-to-leading
$\boldsymbol{\ln1/x}$ approximation\labe{c:NLbfkl}}

The analysis of high energy scattering in QCD by means of the high energy factorization
and of the BFKL approach has clarified the mechanisms of logarithmic enhancement of the
cross sections and has provided a powerful tool for the resummation of the large
coefficients of the perturbative series in the L$x$ approximation.  Nevertheless, we
have remarked that the BFKL equation, at L$x$ accuracy, is not satisfactory both from
the theoretical and from the phenomenological point of view.

Concerning the last point, the small-$x$ rise of $F_2$ shown in Figs.~\ref{f:crescita}
and \ref{f:crescitapiccoliQ} is compatible with a power-like growth
\begin{equation}\lab{lambdaQ}
 F_2(x,Q^2)\simeq C(Q^2)x^{-\la(Q^2)}\;.
\end{equation}
Figs.~\ref{f:esponcresc} show the measured value of $\la(Q^2)$. Note that $\la(Q^2)$ is
larger than the pomeron intercept $\op\simeq0.08$ but the former joins smootly the
latter for $Q^2<1\GeV^2$.
\begin{figure}[ht!]
\centering
\includegraphics[width=86mm,height=49mm]{\fig 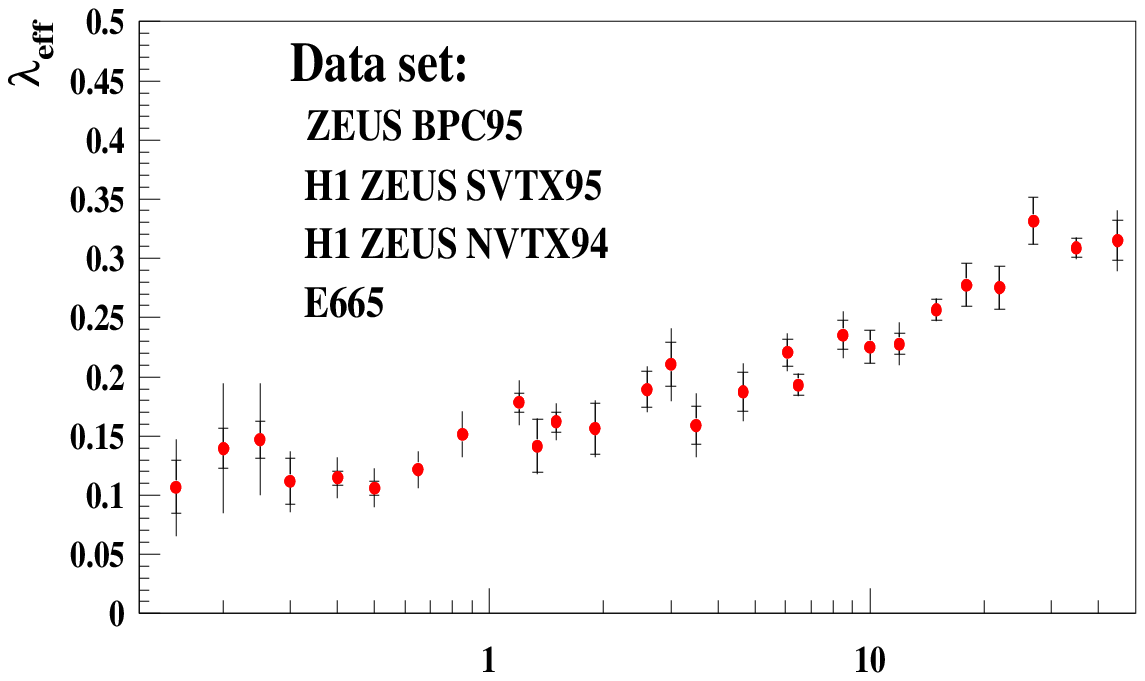}\hspace{4mm}
\includegraphics[width=68mm]{\fig 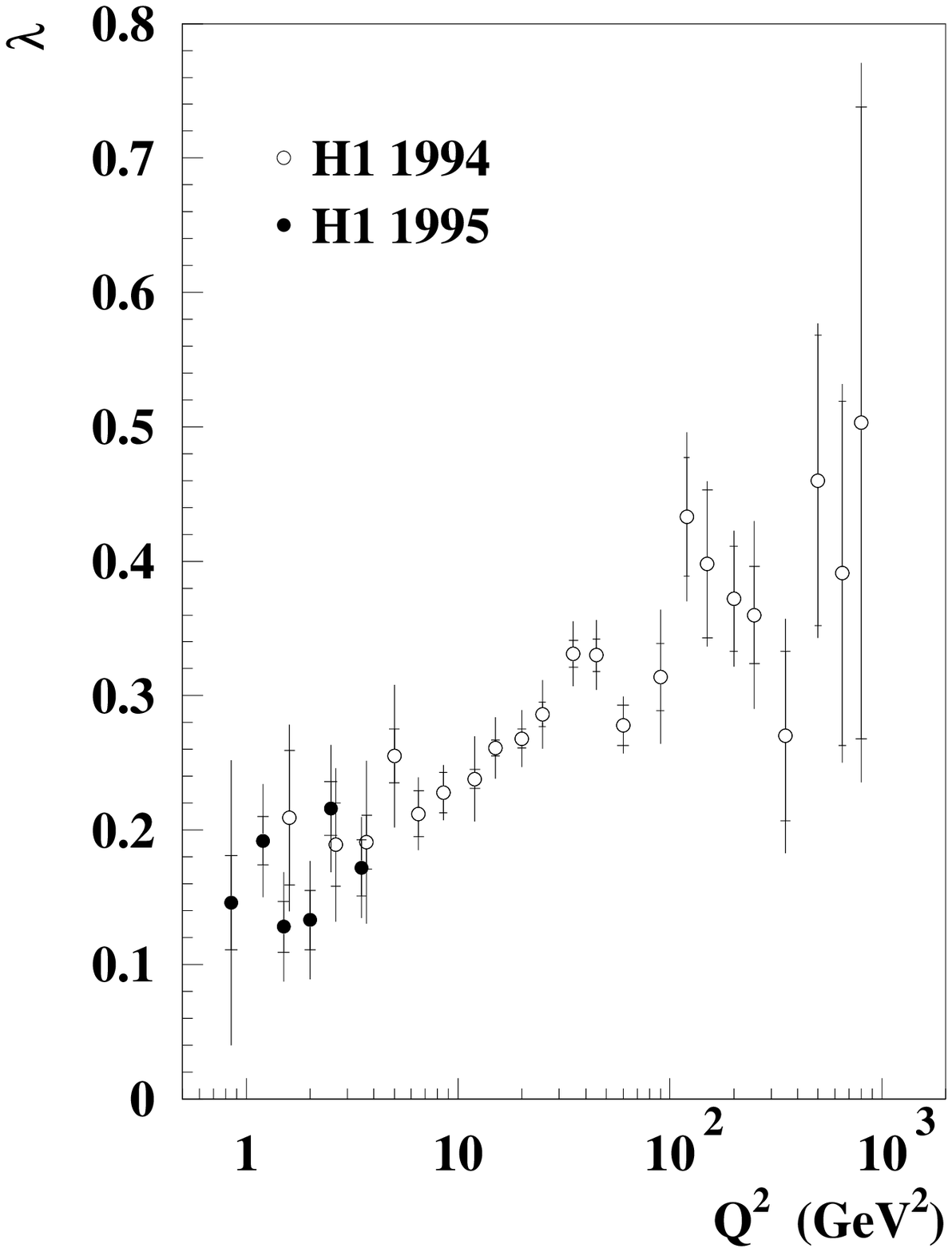}\\[7mm]
\didascalia{Small-$x$ growth exponent for the $F_2$ structure function versus $Q^2$.
\labe{f:esponcresc}} 
\end{figure}

The small-$x$ growth exponent appear to be much smaller than the L$x$ BFKL result
$\op\(\ab(Q^2)\)\simeq2.77\ab(Q^2)$ (where the running coupling replaces the fixed one).
In fact, for a typical value of $Q^2\simeq30\GeV^2$ so that $\ab\simeq0.2$, the
experimental exponent is $\la(Q^2)\simeq0.3$, definitely smaller than
$\op(0.2)\simeq0.55$.

Furthermore, $\la(Q^2)$ increases with $Q^2$, while $\op\(\ab(Q^2)\)$ decreases. The
DGLAP equations, on the contrary, provide an accurate description of the $Q^2$ rise of
$\la(Q^2)$.

The rise of $\la(Q^2)$ is a consequence of scaling violations, and one can roughly
understand the underlying mechanism by observing that
\begin{equation}\lab{segnodimanom}
 \begin{matrix}\hspace{6.8mm}\gaqq_\om+\gagq_\om\\2\Nf\gaqg_\om+\gagg\end{matrix}
 \quad\left\{\begin{matrix}\geq0\text{ for }\om\leq1\\ \leq0\text{ for }\om\geq1
 \end{matrix}\right.\;,
\end{equation}
so that the $Q^2$-evolution enrich the low-$x$ ($\om\to0$) content of the PDF at the
expenses of their large-$x$ ($\om\to+\infty$) one.

Indeed the HERA small-$x$ $F_2$ data are very well described by NLL DGLAP fits (see
Figs.~\ref{f:crescita},\ref{f:crescitapiccoliQ}), provided suitable parton distributions
are taken at the beginning of the $Q^2$-evolution. In particular, nearly constant (in
$x$) PDF at very small $Q_0^2\simeq0.35\GeV^2$ are able to take into account the
observed behaviour for the SF~\cite{GlReVo95}. As an example, consider the singlet
density --- governing the high energy regime --- $f^{(+)}(x,t)$, $t\dug\ln Q^2/\La^2$,
whose small-$x$ behaviour is controlled by the singular part of the gluon anomalous
dimension
\begin{equation}\lab{dimdom}
 \ga^+_\om\simeq{\ab\over\om}={1\over b\om t}\;.
\end{equation}
The DGLAP equation, for a flat starting density
\begin{subequations}\labe{eqdglap}
\begin{align}
 {\dif\over\dif t}f_\om^{(+)}(t)&={1\over b\om t}f_\om^{(+)}(t)\;,\\ \lab{condiniz}
 f^{(+)}(x,t_0)&=\text{const}\;,
\end{align}
\end{subequations}
has the simple solution
\begin{equation}\lab{solsemp}
 f_\om^{(+)}(t)=\text{const}\times\exp\left\{{1\over b\om}\ln{t\over t_0}\right\}\;.
\end{equation}
The $x$-distribution is then given by
\begin{equation}\lab{xdistr}
 f^{(+)}(x,t)=\int\difo\;x^\om\,f_\om^{(+)}(t)
\end{equation}
which, by a saddle point evaluation, gives the asymptotic solution
\begin{equation}\lab{solasint}
 f^{(+)}(x,t)\simeq\text{const}\;{(\ln t/t_0)^{1/4}\over(\ln1/x)^{3/4}}\exp\left\{
 \sqrt{{4\over b}\ln{t\over t_0}\ln{1\over x}}\right\}\;.
\end{equation}
Eq.~(\ref{solasint}) shows essentially the $t$-evolution obeyed by the SF. The
$x$-dependence is however incomplete, because of the crude
approximation~(\ref{condiniz}).

Since the DGLAP approach is based on a purely perturbative ground, we may wonder whether
we are allowed to start the evolution at so a small value of $Q_0^2$ as a fraction of
GeV$^2$. The same evolution for $F_2$ is obtained with power-like behaved input
PDF~\cite{MaRoSt95}
\begin{equation}\lab{inputpert}
 f(x,Q^2)\propto x^{-0.17}\quad\text{for}\quad Q_0^2\simeq4\GeV^2\;.
\end{equation}
This is a reasonable starting point for apply perturbation theory.

Can perturbative QCD explain the power-like shape (\ref{inputpert}) for the PDF? In
principle it should, since $\as(4\GeV^2)\simeq0.3$ should be enough small. Of course,
the high energy regime involved at small-$x$ requires the resummation of the logarithms
of $1/x$ so that BFKL appears to be the most natural approach. On the other hand, we
should not expect BFKL to give a reasonable answer at higher values of $Q^2$, since it
ignores the $\ln Q^2$ resummation.

The natural field of applicability of BFKL evolution is the class of processes with two
hard scales. We can mention $\ga^*\ga^*$ scattering~\cite{BrHaSo97}, which occurs, e.g., in double DIS
(Fig.~\ref{f:2scale}{\sl a}), the two hard scales being the (large) virtualities of the
photons. Another two-scales process being studied concerns the so called forward jet
(Fig.~\ref{f:2scale}{\sl b}), where the two hard scales are the photon virtuality and
the jet transverse energy. By selecting large rapidity differences between the two
observed objects we demand a large BFKL evolution, since we allow strong ordering in
longitudinal space along the ladder. On the contrary, we require the two extreme
transverse scales to be as close as possible, so as to suppress DGLAP evolution.
\begin{figure}[ht!]
\centering
\includegraphics[width=135mm]{\fig 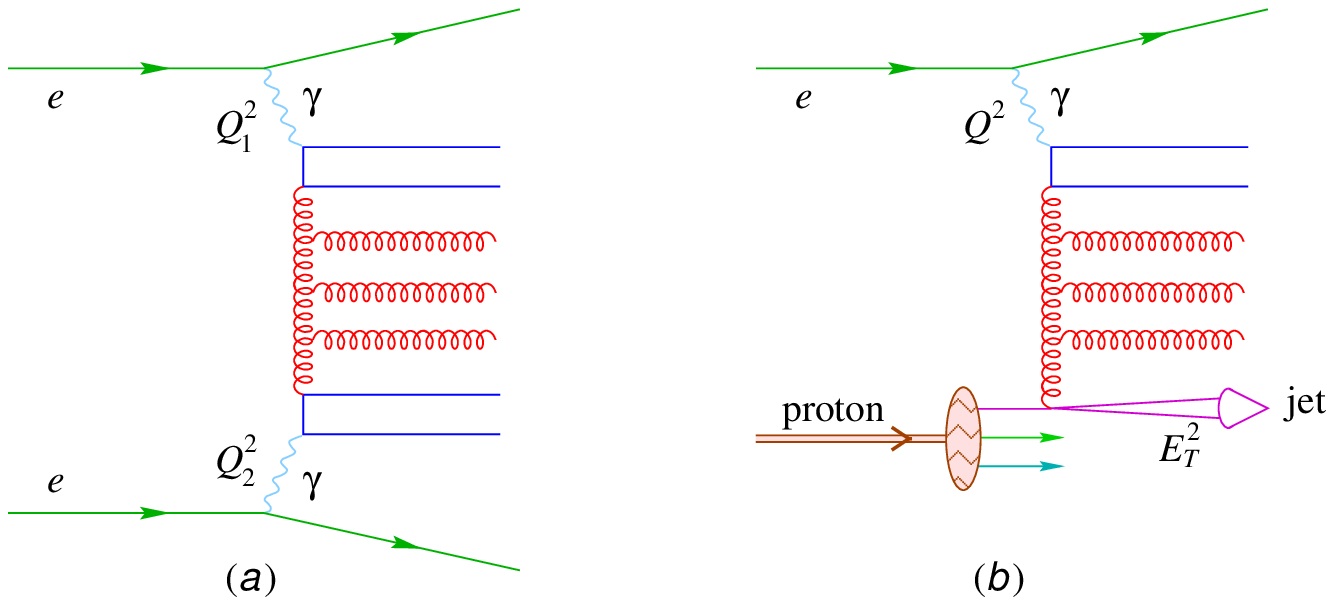}
\didascalia{Examples of two-scale processes: (a) double DIS: two highly virtual photons
couples to a gluon ladder by quark loops; (b) a high $E_T$ jet is emitted in the forward 
region with respect to the incoming proton.
\labe{f:2scale}} 
\end{figure}

While $\ga^*\ga^*$ measurements are not entering the small-$x$ domain yet
\cite{ex9907016}, the forward jet analysis is affected by rather large
uncertainties mainly concerning hadronization phenomena and limited statistics, giving
consistent but not very significant agreement with BFKL resummation\cite{KwMaOu99}.
\vskip2mm

We will not deal with unitarity corrections, which are hopefully not too important for
the perturbative indices that we shall determine. On the other hand, their study for
the full "pomeron" behaviour is outside the scope of the present thesis, and we refer
to the available literature on the subject~\cite{unitarieta}.

On the contrary, we are going to extensively study the
next-to-leading $\ln 1/x$ (NL$x$) corrections to the BFKL equation and to the resummed
anomalous dimensions in semi-hard processes with one or more hard scales.

In this chapter we present a formula which generalizes at NL$x$ level the factorization
properties already determined at L$x$ accuracy by Eq.~(\ref{fattLx}) and we compute on
one hand the NL$x$ BFKL kernel~\cite{FaLi98,CaCi98}, which is supposed to be independent
of the external probes, and on the other hand the impact factors~\cite{Cia98,CiCo98},
which characterize the probe.

The separation of impact factors and kernel is made on the basis of high-energy
$\kk$-dependent factorization (Sec.~\ref{ss:hef}) which is therefore to be extended at
NL$x$ level~\cite{Cia98,CaCi98}.

Particular care has to be taken in the use of a proper factorization scheme, because the
separation of the cross section in impact factor contributions and gluon Green's
function suffers of some ambiguity, analogous to the one between coefficient functions
and anomalous dimensions in collinear factorization. In fact, the subtraction of the
leading $\ln s$ terms involves a prescription which is not unique, not only for
choosing the scale of $s$, which is undetermined in L$x$ approximation, but
also for the form of $\G_{\om}$ at finite energies. Therefore, some probe-independent
NL$x$ terms can be attributed to either the impact factors or to the Green's function,
depending on the scheme being adopted.

First of all, we shall introduce a definition of impact factors on the basis of a proper scale
choice in such a way to prevent the presence of spurious infrared divergences. The
definition, carried out for partonic particles in a first time, will be generalized to
the case of colourless probes, which is the most important one for applications to DIS
and to heavy quark processes.

Secondly, we shall proceed to the determination of the NL$x$ part of the BFKL kernel,
discussing at length the relevance of the choice of the energy-scale $s_0$.

Finally, we shall extract the physical quantity, i.e., the resummed anomalous dimension and
the hard pomeron intercept stemming from the NL calculation, by discussing also the running
coupling features. By pointing out the large size of the NL$x$ corrections, we will
prepare the ground for the next chapter, where a further improvement of the BFKL
equation will cure a lot of pathologies of the $\ln1/x$ hierarchy.

\section{Next-to-leading high energy factorization\labe{s:nlhef}}

In the calculation of the L$x$ cross section, two basic properties have led us to the 
factorized form (\ref{fattLx}):
\begin{itemize}
\item the dominance, in multi Regge kinematics (MRK), of the amplitude with the gluon
quantum numbers in the $t$-channel (only for them there is no cancellation between $s$
and $u$ channel contributions);
\item the gluon reggeization, which determines in each $t$-channel gluon exchange the
correction to the propagator (\ref{propregge}) coming from all order leading virtual
contributions.
\end{itemize}
The above properties holds also in the NL$x$ approximation, as the explicit calculations
show.

In conclusion, we will assume the factorization formula (\ref{fattLx}) to be valid also
at NL$x$ level, with appropriate NL$x$ impact factors and GGF. This assumption has to be 
checked ``a posteriori'', and we anticipate that this is actually the case, even though
the GGF has not a simple resolvent structure as in Eq.~(\ref{d:GGF}).

\section{Impact factors\labe{s:IF}}

Among the ambiguities concerning the definition of the NL$x$ impact factors and kernel,
the most important is due to the dependence of some NL$x$ features on the determination
of the energy-scale $s_0$ in the $\ln s$ dependence of the cross section.

In hard processes like DIS, the scale of $s$ is taken to be $Q^2$, the virtuality of the
photon or EW boson involved. Thus the SF are basically dependent on the Bjorken scaling
variable $x\simeq Q^2/s$, with scaling violations induced by $\as(Q^2)$. Similar
considerations can be made for double DIS~\cite{BrHaSo97} of quarkonium production,
where two hard scale are present.

On the other hand, the computation of cross sections in process where the probe doesn't
couple directly to the gluons involves diagrams with at least one loop more than the
ones with partonic probes. For this reason, the high energy cluster expansion (see
Sec.~\ref{ss:cluster}) needed for the definition of the NL$x$ kernel has been mostly
investigated in the case of parton-parton scattering, in which no physical hard scale is
present.

While the total cross section of this process has severe power-like Coulomb
singularities, it is hoped that by fixing the transverse momenta $\kk$ and $\kk_0$ of
the fragmentation jets (corresponding to the virtualities of the exchanged gluons,
Fig.~\ref{f:diagfatt}, one is able to define a two-scale hard process. A similar but not
identical procedure was devised by Mueller and Navelet \cite{MuNa87} in hadron-hadron
scattering (cfr.\ Sec.~\ref{s:cffp}).

We then consider high energy scattering of two partons $\pa,\pb=$ quark~($\pq$) or
gluon~($\pg$) with momenta $p_1$ and $p_2$ and we factorize it in impact factors and GGF 
contribution as in Fig.~\ref{f:diagfatt}.
\begin{figure}[ht!]
\centering
\includegraphics[height=52mm]{\fig 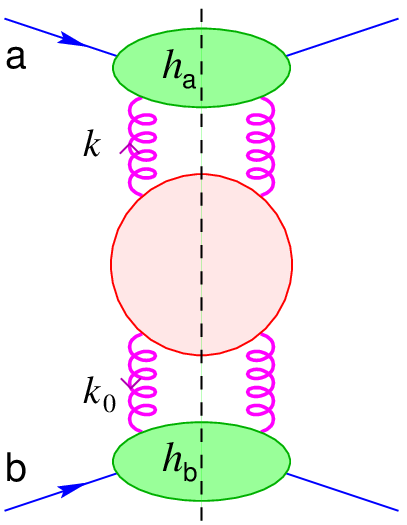}\put(-25,25){$\G_\om$}
\didascalia{High energy factorization diagram for parton-parton scattering in dijet
	production.\labe{f:diagfatt}} 
\end{figure}

Formally, we can write the colour and polarization averaged differential cross section
\begin{equation}\lab{fatt1loop}
 {\dif\si_{\pa\pb}\over\dif\kk\,\dif\kk_0}=\int\difo\;\left({s\over s_0(\kk,\kk_0)}\right)
 ^\om h_{\pa}(\kk)\,\G_\om(\kk,\kk_0)\,h_{\pb}(\kk_0)\;.
\end{equation}
By assuming the GGF to be the resolvent of a NL$x$ kernel $\G_\om=[\om-\K]^{-1}$ and 
by expanding in $\ab$ the impact factors and the kernel
\begin{align}\lab{esph}
 h(\kk)&=h^{(0)}(\kk)+\ab h^{(1)}(\kk)+\cdots\\ \lab{espK}
 \K(\kk,\kk')&=\ab K^{(0)}(\kk,\kk')+\ab^2 K^{(1)}(\kk,\kk')+\cdots\;,
\end{align}
we obtain, after $\om$-integration, the 1-loop expansion of (\ref{fatt1loop})
\begin{equation}\lab{espfatt}
 {\dif\si_{\pa\pb}\over\dif\kk\,\dif\kk_0}=h_{\pa}^{(0)}h_{\pb}^{(0)}+\ab\left[h_{\pa}^{(0)}
 K^{(0)}h_{\pb}^{(0)}\ln{s\over s_0}+h_{\pa}^{(1)}h_{\pb}^{(0)}+h_{\pa}^{(0)}h_{\pb}^{(1)}
 \right]+\cdots\;.
\end{equation}
It is easy to recognize, in the second term of the above expansion, the leading $\ln s$
term whose coefficient provides the L$x$ BFKL kernel, whilst the constant (in $s$) ones
yields the impact factor corrections. However, the ambiguity in the energy scale $s_0$ affects the
definition of the impact factor correction, and a ``scale fixing'' prescription has to be
introduced.

We define the impact factors by the following factorization procedure~\cite{Cia98,CiCo98}:
\begin{itemize}
\item[(i)] subtract the leading term with a reference scale, consistently with the IR
and collinear properties of the process;
\item[(ii)] interpret the remaining constant at single-$\kk$ factorization level as the
1-loop correction $h^{(1)}$.
\end{itemize}

The definition of the impact factor correction $h^{(1)}$ can be accomplished by the 1-loop
calculation of the differential cross section, which involves the known \cite{PPRcorr}
particle-particle-reggeon (PPR) vertex at virtual level (the reggeon trajectory
correction giving a L$x$ contribution) and require an accurate treatment of the squared
matrix element for one particle production in the fragmentation regions%
\footnote{The fragmentation region of the incoming particle $\pa$ is the phase space
region corresponding to outgoing particles being almost collinear to $\pa$. In the CM
frame, this corresponds to large and positive rapidities with respect to $\pa$.}
of the incoming particles~\cite{Cia98,CiCo98}.

Because of the UV singularities stemming from the loop integrals in the virtual
correction diagrams, we adopt dimensional regularization in $D=4+2\e$ space-time
dimensions. In this way we regularize at the same time the IR singularities when
considering real and virtual emission separately. Definitions and details are deferred
to App.~\ref{a:regdim}.

\subsection{Quark impact factor\labe{ss:qif}}

Let's start by considering quark-quark scattering in which an additional gluon $\pg$ of
momentum $q$ is produced. According to the Sudakov parametrization
\begin{subequations}\labe{sudak12}
\begin{align}
 k_1&=z_1p_1+\bar{z}_1p_2+\ku\qquad(\ku=-\kk)&\bar{z}_1&=-{\ku^2\over(1-z_1)s}\;,\\
 k_2&=z_2p_1+\bar{z}_2p_2+\kd\qquad(\kd=\kk_0)&z_2&=-{\kd^2\over(1-\bar{z}_2)s}\;,
\end{align}
\end{subequations}
we define the ``rapidity'' of the gluon%
\footnote{The usual rapidity variable would be $y=\ln\((z_1+z_2)\sqrt{s}/|\qq|\)$ which
reduces to the the definition of the text for $z_2\ll z_1$, i.e., in the central region
and in the fragmentation region of $\pa$.}
$y\dug\ln(z_1\sqrt{s}/|\qq|)$ which can assume the values $y\in[-Y,Y]$ where
$Y\dug\ln(\sqrt{s}/|\qq|)$.

In the central region, corresponding to $|y|\ll Y$, the squared matrix element is
correctly obtained by the amplitude
\begin{equation}\lab{sezcentraleQ}
 {\dif\si_{\pq\pq\to\pq\pg\pq}^{(\text{L}x)}\over\dif z_1\,\dif\ku\dif\kd}=
 h_{\pq}^{(0)}(\ku)\,{1\over z_1}\,
 {\ab\over\pie\qq^2}\,h_{\pq}^{(0)}(\kd)\quad,\quad\pie\dug\pi^{1+\e}\Ga(1-\e)\mu^{2\e}
\end{equation}
which is nothing but Eq.~(\ref{sigma3}) without reggeization of the gluon.

Outside the central region ($|y|\gg1$), Eq.~(\ref{sezcentraleQ}) does not hold at NL$x$
accuracy. The real emission amplitude, assuming the gluon in the $\pa$ quark
fragmentation region, can be computed by means of high energy factorization
\begin{equation}\lab{M1232}
 \overline{|\M_{\pq\pq\to\pq\pg\pq}|^2}={1\over N_c^2-1}g_\e^2C_F
 {1\over(k_2^2)^2}\sum_{c,d,\la}|\poc_{(\la)}^\mu\,2p_2^\nu\,A_{\mu\nu}^{cd}|^2\;,
\end{equation}
i.e., by coupling the fragmentation tensor $A_{\mu\nu}^{cd}$ of Fig.~\ref{f:QGfrag} to the
$\pb$ quark with an eikonal vertex, because of the strong ordering between the
sub-energies $s_1\dug(p'_1+q)^2\ll s_2\dug(q+p_2')^2$.
\begin{figure}[ht!]
\centering
\includegraphics[width=147mm]{\fig 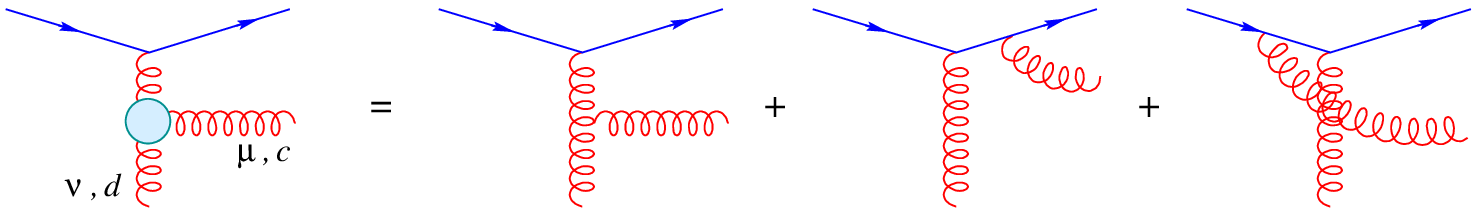}
\didascalia{Quark-to-gluon fragmentation tensor $A_{\mu\nu}^{cd}$.\labe{f:QGfrag}} 
\end{figure}

By using the three-body phase space at NL$x$ accuracy (\ref{sftreE}) we get
\begin{align}
 \frac{\dif\sigma_{\pq\pq\to\pq\pg\pq}^{(\text{NL}x)}}{\dif z_1\,\du\,\dd}=\;&
 h_{\pq}^{(0)}(\ku)\,h_{\pq}^{(0)}(\kd)\,\frac{\P_{\pg\pq}(z_1,\e)}{\pie}\times\nonumber\\
\lab{sezframmQ}&\times\left(\frac{C_A\as}{\pi}\,
 \frac{(1-z_1)\qq\ps(\qq-z_1\kd)}{\qq^2\,(\qq-z_1\kd)^2}+
 \frac{C_F\as}{\pi}\,\frac{z_1^2\ku^2}{\qq^2\,(\qq-z_1\kd)^2}\right)
\end{align}
where
\begin{equation}\lab{pgq}
 \P_{\pg\pq}(z,\e)=\frac{1}{2z}\left[1+(1-z)^2+\e z^2\right]
\end{equation}
is related to the quark-to-gluon splitting function. 

We can easily check that $\dif\sigma_{\pq\pq\to\pq\pg\pq}^{(\text{NL})}$ matches
$\dif\sigma^{(\text{L})}_{\pq\pg\pq}$ (\ref{sezcentraleQ})
in the central region $z_1\ll1$ --- the $C_F$ term being suppressed by a factor $z_1^2$ ---
so that we can take it as the right differential cross section in the
whole positive $q^{\mu}$ rapidity range
\begin{equation}\lab{rapidposit}
 y>0\quad\iff\quad z_1>\sqrt{\qq^2\over s}\;.
\end{equation}
In the remaining half phase-space $y<0$ the cross section has the
same expression provided we exchange
$\ku\leftrightarrow\kd\;,\;z_1\leftrightarrow z_2$.

It is worth noting that the splitting function (\ref{pgq}) is factored
out in Eq.~(\ref{sezframmQ}) even outside the collinear regions
$\qq^2\ll\ku^2\simeq\kd^2$ and $\ku^2\ll\qq^2\simeq\kd^2$ thus
suggesting a smooth extrapolation between collinear and Regge regions
\cite{CCFM88}.

The cross section in Eq.~(\ref{sezframmQ}) contains two colour factors which have a simple
interpretation, depending on the collinear singularities involved. The $C_F$ term, with
singularities at $\qq^2=0$ ($(\qq-z_1\kd)^2=0$), comes from the Sudakov jet region, in
which the emitted gluon is collinear to the incoming (outgoing) quark.

On the other hand, the $C_A$ term is not really singular at either $\qq^2=0$ or
$(\qq-z_1\kd)^2=0$, except for $z_1\simeq0$, which correspond to the central region. It
comes from the ``coherent'' region in which the gluon is emitted at angles which are
large with respect to the $\widehat{\vec{p}_1\vec{p}_1{}'}$ angle, and is thus sensitive
to the total $\pq\pq'$ charge $C_A$. This is the region we are interested in, which is
relevant for the energy scale, because it tells us how the leading matrix element, valid
in the central region, is cut-off in the fragmentation region.

We can simply realize the angular ordering phenomenon in the $C_A$ part of the real
emission contribution to the cross section (\ref{sezframmQ}) upon azimuthal averaging in 
$\ku$ at fixed $\qq$, obtaining
\begin{equation}\lab{medazim}
  \left\langle\frac{\dif\sigma_{\pq\pq\to\pq\pg\pq}}{\dif z_1\,\du\,\dd}\right\rangle=\;
 h_{\pq}^{(0)}(\ku)\,h_{\pq}^{(0)}(\kd)\,
 \frac{\P_{\pg\pq}(z_1,\e)}{\pie\qq^2}\frac{C_A\as}{\pi}\,\Th(|\qq|(1-z_1)-|\ku| z_1)
\end{equation}
The latter is the coherence effect prescription of Ref.~\cite{CCFM88} in which the polar
angles of the emitted gluon and recoiling quark obey
\begin{equation}\lab{ordang}
 {|\qq|\over z_1}>{|\ku|\over1-z_1}\quad\iff\quad\te_q>\te_{p_1'}\;.
\end{equation}

Roughly speaking, for $\qq^2\gtrsim\ku^2$ the half phase space $\sqrt{\qq^2/s}<z_1<1$
is available in full, whilst, for $\ku^2\ll\qq^2$ it is dynamically restricted to
$\sqrt{\qq^2/s}<z_1<|\qq|/(|\ku|+|\qq|)\simeq|\qq|/|\ku|$. With the parallel
consideration in the other half phase space, the longitudinal integration of
(\ref{medazim}) provides the logarithmic factor
\begin{equation}\lab{fatlog}
 \int_{q/\sqrt{s}}^1\dif z_1\;\P_{\pg\pq}(z_1,\e)\Th(|\qq|(1-z_1)-|\ku|z_1)+\{1\lra2\}\simeq
 \ln{s\over\Max(\qq,\ku)\Max(\qq,\kd)}
\end{equation}
suggesting the natural scale of $s$ to be $s_0=s_M\dug\Max(\qq,\ku)\Max(\qq,\kd)$ and
hence the subtraction to be adopted in order to isolate the 1-loop impact factors.

More precisely, we define as leading real emission term of the differential cross
section in the half phase space $y>0$ the quantity
\begin{equation}\lab{d:termdom}
  h_{\pa}^{(0)}(\ku)\,h_{\pb}^{(0)}(\kd)\,{\Th(|\qq|-z_1|\ku|)\over z_1}\,{\as C_A\over\pi}
 {1\over\pie\qq^2}\qquad\(z_1>\sqrt{\qq^2\over s}\)
\end{equation}
which we subtract from Eq.~(\ref{sezframmQ}) and interpret as the real emission
contribution to the $\ab h_{\pa}^{(1)}(\ku)h_{\pb}^{(0)}(\kd)$ term in the expansion
(\ref{fatt1loop}). In this way we have fulfilled the first prescription for the
definition of the impact factors, because the ($C_A$ part of the) subtracted term is
finite in the $\qq\to0$ limit even for $z=0$ and therefore, according to the second
prescription, after integration over all but one transverse variable, it defines an
impact factor correction with the right collinear singularities. Explicitly, after
subtracting (\ref{d:termdom}) from (\ref{sezframmQ}), dropping the ``lower'' impact
factor $h_{\pq}^{(0)}(\kd)$, and integrating over $z_1>\sqrt{\qq^2/s}$ and $\ku$ at
fixed $\kd$ we get
\begin{subequations}\labe{h1re}
\begin{align}
 h_{\pq}^{(1)}(\kd)\Big|^{\text{real}}_{C_A}&\simeq\int\du
 \;h_{\pq}^{(0)}(\ku){1\over\pie\qq^2}\nonumber\\
 &\quad\times\int_0^1\dif z_1\;
 \P_{\pg\pq}(z_1,\e){(1-z_1)\qq\ps(\qq-z_1\kd)\over(\qq-z_1\kd)^2}-
 {\Th(|\qq|-z_1|\ku|)\over z_1}\nonumber\\ \lab{h1reca}
 &=\left(-{3\over4\e}-2\psi'(1)-{1\over4}\right)\left({\kd^2\over\mu^2}\right)^\e\,
 h_{\pq}^{(0)}(\kd)
\end{align}
for the $C_A$ part of real emission and
\begin{align}
 h_{\pq}^{(1)}(\kd)\Big|^{\text{real}}_{C_F}&\simeq{C_F\over C_A}\int\du
 \;h_{\pq}^{(0)}(\ku){1\over\pie\qq^2}\int_0^1\dif z_1\;
 \P_{\pg\pq}(z_1,\e){z_1^2\ku^2\over(\qq-z_1\kd)^2}\nonumber\\ \lab{h1recf}
 &=\ab{C_F\over C_A}\left({1\over\e^2}-{3\over2\e}+4-\psi'(1)\right)
 \left({\kd^2\over\mu^2}\right)^\e\,h_{\pq}^{(0)}(\kd)
\end{align}
\end{subequations}
for the $C_F$ part. Note that the lower bound of the $z$-integrations has been pushed
down to zero since, in the $s\to\infty$ limit, the error is negligible. The real
emission correction to the lower impact factor are provided by the left half phase space 
contribution $y<0$.

The ${1/\e}$ pole in Eq.~(\ref{h1recf}) comes as no surprise, because it
corresponds to the collinear divergence at $\ku^2=0$ due to $\pq\to\pg$ transition of
the initial massless quark. This is what we intended with ``consistent with the collinear
properties''. A different subtraction could have produced a double pole in $\e$ which is 
not consistent with the collinear singularity.

As far as the $C_F$ term is concerned, we do have a double pole in $\e$. This is not a
problem, because, as we shall see soon, the whole $C_F$ contribution is cancelled by
virtual corrections.

The correction to the cross section due to virtual emission, including
subleading effects, for general parton-parton scattering, can be
extracted from the amplitude of Ref.~\cite{PPRcorr} 
\begin{equation}\lab{corrvirt}
 \M^{(0+1,\text{NL})}_{\pa\pb\to\pa\pb}=-i\,2s\,\Ga_{aa'}^c
 \Big(1+\ab\Pi^{(+)}_{\pa\pa}\Big)\frac{1}{t}
 \left[1+\Om^{(0)}(-t)\ln\frac{s}{-t}\right]\Big(1+\ab\Pi^{(+)}_{\pb\pb}\Big)
 \Ga_{bb'}^c
\end{equation}
where $\Ga_{aa'}^c$ is the particle-gluon vertex of Eqs.~(\ref{PGvert}). The $\ln s$
term provides the 1-loop Regge gluon trajectory (Eq.~(\ref{interc})) and the finite
terms $\Pi^{(+)}(\kk^2)$ define the PPR vertex correction in the helicity conserving channel.

In the quark case
\begin{equation} \lab{corrqq}
 \Pi^{(+)}_{\pq\pq}={1\over2}\left[-\frac{11}{12\e}+\left(\frac{85}{36}+\frac{\pi^2}{4}
 \right)+\frac{\Nf}{N_c}\left(\frac{1}{6\e}-\frac{5}{18}\right)-\frac{C_F}{N_c}\left(
 \frac{1}{\e^2}-\frac{3}{2\e}+4-\psi'(1)\right)\right]\!\!\left({\kk^2\over\mu^2}\right)^{\!\e}\,.
\end{equation}
The virtual contribution to the 1-loop cross section is given by the interference
between (\ref{corrvirt}) and the Born amplitude (\ref{Aborn}) together with the two-body 
phase space measure (\ref{sfdueE}) yielding
\begin{equation}\lab{sig1loopvirtQ}
 {\dif\si^{(1)}_{\pq\pq\to\pq\pq}\over\dif\ku\,\dif\kd}=\ab h_{\pq}^{(0)}(\ku)
 h_{\pq}^{(0)}(\kd)\,\left[2\Om^{(0)}(\ku^2)\ln{s\over\ku^2}
 +4\Pi_{\pq\pq}^{(+)}(\ku^2)\right]\d(\ku-\kd)\;.
\end{equation}
The first term in the above equation is the virtual part of the leading BFKL kernel. The 
second one provides the virtual contribution to the last two terms in
Eq.~(\ref{fatt1loop}), so that
\begin{subequations}\labe{h1vi}
\begin{align}
 h_{\pq}^{(1)}(\kd)\Big|^{\text{virt}}_{C_A}&=\left[-\frac{11}{12\e}+\left(\frac{85}{36}
 +\frac{\pi^2}{4}\right)+\frac{\Nf}{N_c}\left(\frac{1}{6\e}-\frac{5}{18}\right)\right]
 \left({\kd^2\over\mu^2}\right)^\e h_{\pq}^{(0)}(\kd)\\
 h_{\pq}^{(1)}(\kd)\Big|^{\text{virt}}_{C_F}&={C_F\over C_A}\left(-{1\over\e^2}
 +{3\over2\e}-4+\psi'(1)\right)\left({\kd^2\over\mu^2}\right)^\e h_{\pq}^{(0)}(\kd)
\end{align}
\end{subequations}
When adding the RHS of Eqs.~(\ref{h1re}) and (\ref{h1vi}), one can immediately see the
cancellation of the $C_F$ terms and obtain the complete 1-loop impact factor
correction~\cite{Cia98}
\begin{equation}\lab{hunoq}
 h_{\pq}^{(1)}(\kk)=\left[-{1\over\e}\left(\frac{11N_c-2\Nf}{12N_c}\right)-{1\over2\e}
 \left(\frac{3}{2}-\frac{1}{2}\e\right)+{1\over2}\left(\frac{67}{18}-\frac{\pi^2}{6}
 -\frac{5\Nf}{9N_c}\right)\right]\left({\kk^2\over\mu^2}\right)^\e h_{\pq}^{(0)}(\kk)
\end{equation}

\subsection{Gluon impact factor\labe{ss:gif}}

We consider now gluon-gluon scattering. In this case we have to
distinguish two different final states for three-particle production.
In addition to the gluon $\pg'$ there may be:
\begin{itemize}
 \item[{\it i})] two gluons;
 \item[{\it ii})] a quark-antiquark pair.
\end{itemize}

\subsubsection{$\pg\pg\to\pg\pg\pg$\labe{sss:ggggg}}

\begin{figure}[ht!]
\centering
\includegraphics[width=147mm]{\fig 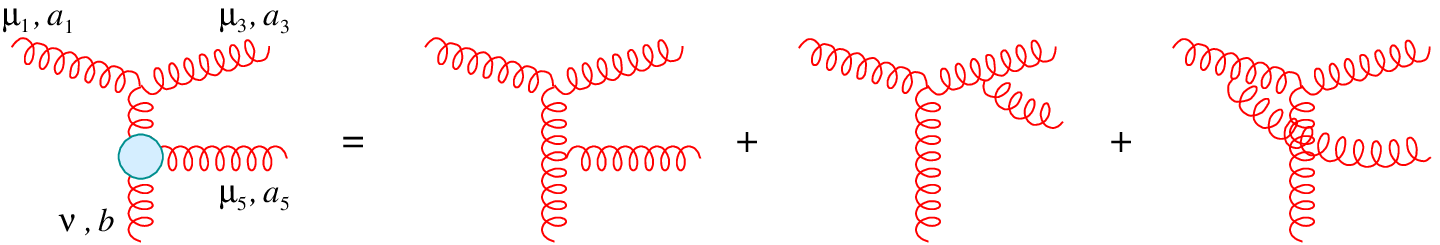}
\didascalia{Gluon-to-gluon fragmentation tensor
	$A_{\mu_1\mu_3\mu_5\nu}^{a_1a_3a_5b}$.\labe{f:GGfrag}} 
\end{figure}
Assume that the additional gluon is emitted in the gluon {\sf a} fragmentation
region. By using again high energy factorization, the scattering amplitude in the
Feynman gauge is given by
\begin{equation}
 \M^{(1)}_{\pg\pg\to\pg\pg\pg}=
 \po_1^{\mu_1}\,\poc\,_3^{\mu_3}\,\poc\,_5^{\mu_5}
 \;A_{\mu_1\mu_3\mu_5\,\nu}^{a_1a_3a_5\,b}\,\frac{1}{k_2^2}\;
 g\,2p_2^{\nu}\,\t{b}_{\pb}\,\delta_{\la_4\la_2} \label{aggg}
\end{equation}
where the amplitude $A$, corresponding to the diagrams of
Fig.~\ref{f:GGfrag} was found in Ref.~\cite{FaLi89} to be
\begin{align}
 A_{\mu_1\mu_3\mu_5\,\nu}^{a_1a_3a_5\,b}\,2p_2^{\nu}&=
 4g^2\,g_{\mu_1\mu_3}\Big[\f^{a_1a_5e}\f^{ea_3b}D_{\mu_5}(-p_1,p_3,p_5)+
 \f^{a_3a_5e}\f^{ea_1b}D_{\mu_5}(p_3,-p_1,p_5)\Big]+\nonumber \\
 &\qquad+\,\begin{picture}(23,9)(0,5)
               \put(0,4.5){$\Bigg($}\put(19,4.5){$\Bigg)$}
               \put(4,10){$p_3\leftrightarrow p_5$}
               \put(3.7,5){$\mu_3\leftrightarrow\mu_5$}
               \put(4,0){$a_3\leftrightarrow a_5$}
           \end{picture}
        +\,\begin{picture}(23,9)(0,5)
               \put(0,4.5){$\Bigg($}\put(21,4.5){$\Bigg)$}
               \put(3,10){$-p_1\leftrightarrow p_5$}
               \put(6,5){$\mu_1\leftrightarrow\mu_5$}
               \put(6.3,0){$a_1\leftrightarrow a_5$}
           \end{picture} \quad, \label{amumu}
\end{align}
where the current
\begin{equation}\lab{d:Dmuxyz}
 D^{\mu}(x,y,z)=\frac{1}{x\ps y}\left[\left(y\ps z-p_2\ps p_4\,
 \frac{p_2\ps y}{p_2\ps z}\right)p_2^{\mu}+\frac{p_2\ps y}{x\ps z}
 (y\ps z-p_2\ps p_4)x^{\mu}+(x\ps p_2)y^{\mu}\right]
\end{equation}
is a function of the momenta $x^{\alpha},y^{\alpha},z^{\alpha}$, with
$x\ps y=x^{\alpha}y_{\alpha}$.

The squared helicity amplitudes corresponding to Eq.~(\ref{amumu}) were given in
Ref.~\cite{Del96}. Since we work in $D=4+2\e$ dimensions, we perform explicitly the
polarization sum on Eq.~(\ref{aggg}). The averaged squared matrix element that we
obtain is actually independent of the space-time dimensionality (i.e.\
$\e$-independent) so that we finally get
\begin{align}\lab{sggg}
 \frac{\dif\sigma_{\pg\pg\pg}}{\dif z_1\,\du\,\dd}=\;&h_{\pg}^{(0)}(\ku)\,
 h_{\pb}^{(0)}(\kd)\,\frac{\P_{\pg\pg}(z_1)}{\pie}\times\nonumber\\
 &\times\frac{C_A\as}{\pi}\,\frac{z_1^2\ku^2+(1-z_1)^2\qq^2+z_1(1-z_1)\ku\ps\qq}{%
 \qq^2\;(z_1\ku+(1-z_1)\qq)^2}\;.
\end{align}
This expression explicitly exhibits the symmetry in the exchange
$-\ku\leftrightarrow\qq\;;\;z_1\leftrightarrow 1-z_1$ due to the identity
of the gluons emitted in the fragmentation region of gluon {\sf a},
but will be used for the softer gluon only $(z_1<\frac{1}{2})$, in
order to avoid double counting.

The function
\begin{equation}
 \P_{\pg\pg}(z_1)=\P_{\pg\pg}(1-z_1)=\frac{1+z_1^4+(1-z_1)^4}{2z_1(1-z_1)}
 \label{pgg}
\end{equation}
is related to the gluon to gluon splitting function, and is factored out
in Eq.~(\ref{sggg}) in the whole fragmentation region,
as for quark scattering. In this case also the cross section matches
the leading expression (\ref{sezcentraleQ})
in the central region $z_1\ll1$, and thus is taken to be valid in
the half phase space (\ref{rapidposit}) in which two of the three gluons
have positive rapidity.

\subsubsection{$\pg\pg\to\pq\bar{\pq}\pg$\labe{sss:ggqqg}}

The main contribution to the cross section from this kind of final
states is reached when the fermion pair belongs to the same
fragmentation vertex, as in Fig.~\ref{diagqqg}{\sl a}.
Other  graphs like Fig.~\ref{diagqqg}{\sl b}
are suppressed by a factor $|t|/s$ because of fermionic exchange.
\begin{figure}[h]
\vskip4mm
\centering
\includegraphics[width=80mm]{\fig 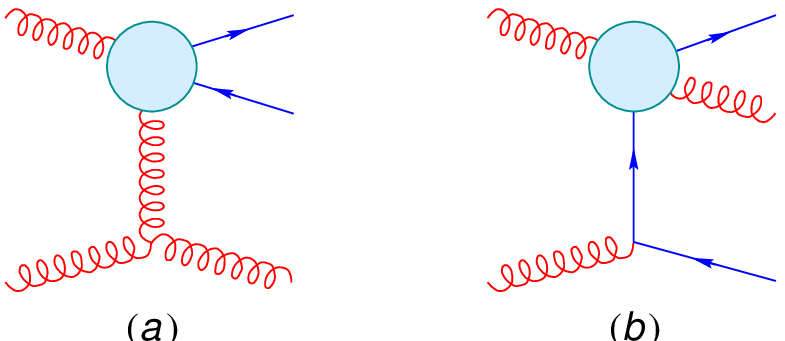}
\didascalia{Quark-antiquark emission amplitudes:
   (a) The pair is emitted in the fragmentation region of the upper incoming gluon;
   (b) This kind of diagram is suppressed by a factor $|t|/s$.\labe{diagqqg}}
\end{figure}

Assuming $\pq\bar{\pq}$ being emitted in the fragmentation region of gluon $\pa$, and
labelling $\pq$ with ``3'' and $\bar{\pq}$ with ``5'' ($p_5=q$), the corresponding
amplitude is
\begin{equation}\lab{aqqg}
 \M_{\pq\overline{\pq}\pg}=\po^{\mu}\,
 \overline{u}_3 B_{\mu\nu}^{ab}v_5\,\frac{1}{k_2^2}\;g\,2p_2^{\mu}\,
 \t{b}_{\pq}\,\delta_{\la_4\la_2}
\end{equation}
where $B_{\mu\nu}^{ab}$ is the sum of the diagrams depicted in
Fig.~\ref{f:GQfrag}.
\begin{figure}[ht!]
\centering
\includegraphics[width=134mm]{\fig 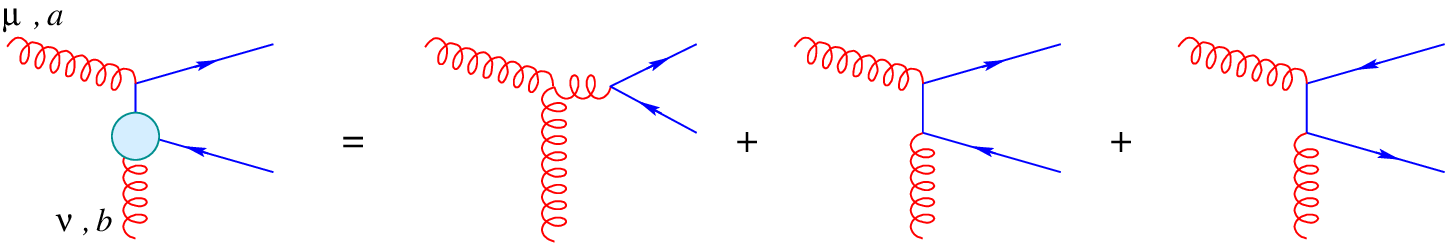}
\didascalia{Gluon to quark fragmentation tensor $B_{\mu\nu}^{ab}$.\labe{f:GQfrag}}
\end{figure}

This calculation is very similar to the $\pq\pq\to\pq\pg\pq$ cross section, and yields
\begin{align}\label{sqqg}
  \frac{\dif\sigma_{\pq\bar{\pq}\pg}}{\dif z_1\,\du\,\dd}
 &=\Nf\,h_{\pg}^{(0)}(\kd)\,\frac{\as N_{\e}}{\ku^2\mu^{2\e}}\,
  \frac{\P_{\pq\pg}(z_1,\e)}{2\pie}\\ \nonumber
 &\quad\times\left[\frac{C_A\as}{\pi}\,
  \frac{-z_1(1-z_1)\ku\ps\qq}{\qq^2\,(z_1\ku+(1-z_1)\qq)^2}
  +\frac{C_F\as}{\pi}\,\frac{1}{\qq^2}\right]
\end{align}
in which the function
\begin{equation}
 \P_{\pq\pg}(z_1,\e)=1-\frac{2z_1(1-z_1)}{1+\e}\quad, \label{pqg}
\end{equation}
related to the gluon to quark splitting function, is regular in
$z_1\in[0,1]$. Hence these final states do not produce any $\ln s$
term, and need no subtraction.

The virtual correction to the amplitude, including NL$x$ effects, is given by an
expression analogue to Eq.~(\ref{corrvirt}) where $\Pi^{(+)}_{\pa\pa}(\kk^2)$ is
replaced by the gluon-gluon-reggeon vertex correction~\cite{PPRcorr} in the helicity
conserving channel
\begin{equation}\lab{corrgg}
 \Pi^{(+)}_{\pg\pg}={1\over2}\left[-\frac{1}{\e^2}+\frac{11}{12\e}-\left(\frac{67}{36}
 -\frac{\pi^2}{4}+\psi'(1)\right)-\frac{\Nf}{N_c}\left(\frac{1}{6\e}-\frac{5}{18}-\psi'(1)
 \right)\right]\left({\kk^2\over\mu^2}\right)^\e\;.
\end{equation}
Now, with the result of experience of Sec.~\ref{ss:qif}, we can obtain the gluon impact
factor in a more direct way: let's introduce the {\em effective fragmentation vertex}
$F_{\pa}(z_1,\ku,\qq)$ of the incoming particle $\pa$ defined in such a way that the
1-loop differential high energy cross section for $\pa\pb$ scattering with possible
emission of a new particle $\pc$ with positive rapidity with respect to $\pa$ is given
by $F_{\pa}$ times the exchanged Regge-gluon propagator and the full impact factor of
particle $\pb$. In practice the effective fragmentation vertex represents the cross
section of $\pa$ with a Regge-gluon, including PPR virtual corrections and one-particle
emission in NL$x$ approximation, as in Fig.~\ref{f:efffragvert}.
\begin{figure}[ht!]
\centering
\includegraphics[width=149mm]{\fig 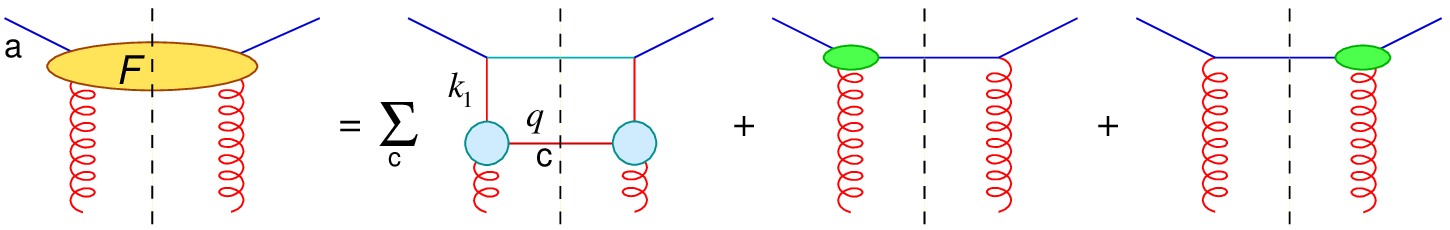}
\didascalia{Effective parton Regge-gluon fragmentation vertex $F_{\pa}$. In the RHS
	the circular blobs denotes the fragmentation tensors of Figs.~\ref{f:QGfrag},
	\ref{f:GGfrag} and \ref{f:GQfrag}; the elliptic blobs corresponds to the PPR
	vertices $\Pi^{(+)}$.\labe{f:efffragvert}}
\end{figure}

In order to extract the 1-loop correction to the impact factor, we have to eliminate the
$\ln s$ term by subtracting the leading fragmentation vertex --- i.e.,
Eq.~(\ref{d:termdom}) without the second impact factor --- containing the $1/z_1$
singularity coming from the splitting functions $\P_{\pg\pa}$:
\begin{equation}\lab{d:h1}
 \ab h_{\pa}^{(1)}(\kd)=\int\du\int_0^1\dif z_1\;F_{\pa}(z_1,\ku,\qq)-\ab h_{\pa}^{(0)}(\ku)\,
 {\Th(q-z_1 k_1)\over z_1}\,\frac{1}{\pie\qq^2}
\end{equation}

From Eqs.~(\ref{sggg}), (\ref{sqqg}) and (\ref{corrgg}), we get the gluon fragmentation vertex
\begin{align}\lab{verframG}
 F_{\pg}&(z_1,\ku,\qq)=h_{\pg}^{(0)}(\ku)\bigg\{\d(\qq)\delta(1-z_1)\ab
 2\Pi_{\pg\pg}^{(+)}(\ku^2)\\
 &+\Nf\frac{\as}{\pi}\,\frac{\P_{\pq\pg}(z_1,\e)}{2\pie}
 \left[\frac{-z_1(1-z_1)\ku\ps\qq}{\qq^2\,(z_1\ku+(1-z_1)\qq)^2}
 +\frac{C_F}{C_A}\,\frac{1}{\qq^2}\right]\nonumber\\
 &+\frac{\ab}{\pie}\,\Th(\frac{1}{2}-z_1)\P_{\pg\pg}(z_1)\,
 \frac{z_1^2\ku^2+(1-z_1)^2\qq^2+z_1(1-z_1)\ku\ps\qq}{%
 \qq^2\;(z_1\ku+(1-z_1))\qq^2}\bigg\}\;.
\end{align}

Having identified $z_1$ as the momentum fraction of the softer gluon in
$\pg\pg\to\pg\pg\pg$ scattering, to avoid double counting due to final gluons identity,
we restrict the naive half phase space to $z_1\in[\frac{q}{\sqrt{s}},\frac{1}{2}]$. In
so doing, the 1-loop correction to the gluon impact factor is given by~\cite{CiCo98}
\begin{align}\lab{hunog}
 h_{\pg}^{(1)}(\kk)&=\left[-{1\over\e}\left(\frac{11N_c-2\Nf}{12N_c}\right)
 +{1\over2N_c}{1\over\e}\left(-{11N_c+(2+\e)\Nf\over6}\right)\right.\\ \nonumber
 &\quad\left.+{1\over2N_c}{C_F\over C_A}{2\Nf T_R\over\e}\left(\frac{2}{3}
 +\frac{1}{3}\e\right)+{1\over2}\left(\frac{67}{18}-\frac{\pi^2}{6}-\frac{5\Nf}{9N_c}\right)
 \right]\left({\kk^2\over\mu^2}\right)^\e h_{\pg}^{(0)}(\kk)\;.
\end{align}

\subsection{Colourless impact factors\labe{ss:cif}}

It is possible to define impact factors for colourless sources --- as in double-DIS
processes --- in a way analogous to the one adopted for partons. In this case, there is
an additional dependence on the $\om$ variable and on the external hard scale(s) so
that, at leading level, the impact factors
\begin{equation}\lab{hbianco}
 h_{\pa}^{(0)}=h_{\pa,\om}^{(0)}(Q_1,\ku)=\frac{1}{Q_1^2}\,f^{\pa}_{\om}
 \left(\frac{Q_1^2}{\ku^2}\right)
\end{equation}
yield a nontrivial $\ku$-dependence, which has been explicitly computed in several
pro\-ces\-ses~\cite{CaCi95}.  In Eq.~(\ref{hbianco}) the function $f_{\pa}$ has the role
of setting $k_1$ of order $Q_1$ in the total cross section, and has no $\ku^2=0$
singularity at all, because there is no initial colour charge.

Therefore, by translating angular ordering into the $\om$-space
threshold factor $(\qq/\ku)^{\om}$, the analogue of Eq.~(\ref{d:h1}) becomes
\begin{align}\lab{d:h1bianco}
 \ab h_{\pa,\om}^{(1)}(Q_1,\kd)&=\!\int\!\!\du\!\int_0^1\!\!\dif z_1\,
 F_{\pa,\om}(Q_1,z_1,\ku,\qq)
 -\ab h_{\pa,\om}^{(0)}(Q_1,\ku)\,\frac{1}{\om\pie\qq^2}\left(\frac{\qq}{\ku}
 \right)^{\!\om\,\Th(|\ku|-|\qq|)}.
\end{align}
In this case, no $\qq=0$ nor $\ku=0$ singularities are expected in
$F_{\pa}$, so that it is again important that the leading term be
subtracted out with the angular ordering constraint.

We thus conclude that the definition of the impact factors in Eqs.~(\ref{d:h1}) and
(\ref{d:h1bianco}), defines a self-consistent $\kk$-factorization scheme of
Eq.~(\ref{fatt1loop}) for both coloured and colourless sources.

\section{Collinear factorization and finite parts}\label{s:cffp}

We want now to analyze the structure of the quark and gluon impact
factors derived in Secs.~(\ref{ss:qif}) and (\ref{ss:gif}). The
explicit expressions reported in Eqs.~(\ref{hunoq}) and (\ref{hunog})
show the presence of
\begin{itemize}
\item a $\be$-function coefficient $\ds{\frac{\pi}{N_c}\,b_0=
      \frac{11N_c-2\Nf}{12N_c}}$;
\item the finite part $\tilde{P}_{\pb\pa}(\om=0)$ in the $\om$ expansion
of the Mellin transform of the $\sf a\to b=q,g$ splitting functions
for the incoming parton {\sf a} under consideration, as depicted in
Tab.~(\ref{tabella});
\item a common and $\e$-finite factor
$\ds{\frac{\K}{N_c}\dug\left(\frac{67}{18}-\frac{\pi^2}{6}-\frac{5\Nf}{9N_c}\right)}$.
\end{itemize}
\newcommand{\vlinei}{\hspace{-2.1mm}\vline\hspace{1mm}}
\newcommand{\vlinem}{\hspace{-2.5mm}\vline\hspace{2.5mm}}
\newcommand{\vlinef}{\hspace{1.5mm}\vline\hspace{-2.2mm}\null}
\begin{table}[h!]
 \begin{tabular}{lll}%
 \hline%
\vlinei$\pc$\hspace{1pt}\raisebox{1pt}{\tiny$\leftarrow$}\hspace{1pt}$\pa$&
\vlinem$P_{\pc\pa}(z,\e)$\raisebox{-3mm}{\rule{0mm}{8.9mm}}\hfill&\vlinem
 $\int_0^1 z^{\om}P_{\pc\pa}(z,\e)\,\dif z$\hfill\vlinef\\
 \hline\hline%
\vlinei\slarga$\,\pq\hspace{2mm}\pq$&\vlinem
\footnotesize$\ds{C_F\left[\left(\frac{1+z^2}{1-z}\right)_+
		-\e(1-z)+\frac{\e}{2}\,\delta(1-z)\right]}$&\vlinem
\footnotesize $ 0+\ord(\om)$\hfill\vlinef\\
 \hline%
\vlinei\slarga$\,\pg\hspace{2mm}\pq$&\vlinem
\footnotesize $\ds{C_F\left[\frac{1+(1-z)^2}{z}+\e z\right]}$&\vlinem
\footnotesize $\ds{\frac{2C_F}{\om}-C_F\left(\frac{3}{2}
	  -\frac{1}{2}\e\right)+\ord(\om)}$\hfill\vlinef\\
 \hline%
\vlinei\slarga$\,\pq\hspace{2mm}\pg$&\vlinem
\footnotesize $\ds{T_R\left[1-\frac{2z(1-z)}{1+\e}\right]}$&\vlinem
\footnotesize $\ds{T_R\left(\frac{2}{3}+\frac{1}{3}\,\e\right)
		      +\ord(\om)}$\hfill\vlinef\\
 \hline%
\vlinei\slarga$\,\pg\hspace{2mm}\pg$&\vlinem
\footnotesize$\ds{2N_c\left[\frac{1-z}{z}+\frac{z}{(1-z)_+}
		      +z(1-z)\right]+\frac{11N_c-2\Nf}{6}\delta(1-z)}$\hfill&\vlinem
\footnotesize$\ds{\frac{2N_c}{\om}-\frac{11N_c+(2+\e)\Nf}{6}
		     +\ord(\om)}$\vlinef\\
 \hline%
\end{tabular}
\caption{\it Partonic splitting functions and Mellin transforms in
	  $\om$-space.}
\label{tabella}
\end{table} 
This suggests writing the complete one-loop impact factors
$h_{\pa}(\kk)=h^{(0)}_{\pa}(\kk)+\ab h^{(1)}_{\pa}(\kk)$ for both $\pa=\pq,\pg$ in the
following manner:
\begin{align}\lab{hak}
 h_{\pa}(\kk)&=h^{(0)}_{\pa}(\kk)+\ab\left[-{\pi\over N_c}{b_0\over\e}h^{(0)}_{\pa}(\kk)
 +{1\over2N_c}\sum_{\pc}h_{\pc}^{(0)}(\kk){\tilde{P}_{\pc\pa}\over\e}
 +{1\over2N_c}h^{(0)}_{\pa}(\kk)\,\K\right]\left({\kk^2\over\mu^2}\right)^\e\\
 &=\left[1-{b_0\over\e}\as\left({\kk^2\over\mu^2}\right)^\e\right]\left\{h^{(0)}_{\pa}(\kk)
 +{\as\over2\pi}\left[h^{(0)}_{\pa}(\kk)\,\K+\sum_{\pc}h_{\pc}^{(0)}(\kk){\tilde{P}_{\pc\pa}
 \over\e}\left({\kk^2\over\mu^2}\right)^\e\right]\right\}\;.\nonumber
\end{align}
The factor in front of the last expression provides the
renormalization of the coupling constant, and by introducing  the
one-loop running coupling
\begin{equation}
 \as(\kk^2)=\as\left[ 1-b_0\,\frac{\as}{\e}
 \left(\frac{\kk^2}{\mu^2}\right)^{\e}\right]\quad,
\end{equation}
it can be incorporated in $h_{\pa}^{(0)}$ itself.

Therefore, the impact factors in Eq.~(\ref{hak}) assume the form
\begin{equation}
 h_{\pa}(\kk^2)=h_{\pa}^{(0)}(\kk;\as(\kk^2))\left(1+\frac{\as}{2\pi}\,\K\right)
 +\frac{\as}{2\pi}\left[h_{\pq}^{(0)}\frac{\tilde{P}_{\pq\pa}}{\e}
 +h_{\pg}^{(0)}\frac{\tilde{P}_{\pg\pa}}{\e}\right]\left(\frac{\kk^2}{\mu^2}
 \right)^{\e}    \label{impatto}
\end{equation}
and thus satisfy the DGLAP equations with splitting functions
$\tilde{P}_{\pc\pa}(\om=0)$.

After factorization of the collinear singularities contained in the latter, the only
finite renormalization is the one implied by the factor
$\ds{\left(1+\frac{\as}{2\pi}\,\K\right)}$, which is universal, i.e. independent of the
parton type.

Several comments are in order. First, the collinear singularities in $h_{\pa}$ do not
contain the $\sim 1/\om$ terms of the splitting functions, which have been subtracted
out in the leading term. In this context, we remark the difference between the
double-$\kk$ cross section defined here through $\kk$-factorization, and the
double-mini-jet inclusive cross section defined by Mueller and Navelet~\cite{MuNa87}.

In the latter case the cross section is inclusive over all fragmentation products of the
incoming partons which are not identified, and measures the gluonic $\kk$'s only because
of collinear strong ordering.  Therefore, it factorizes the gluon distribution density
in the parton, rather than the impact factor, with all its collinear singularities
included and, furthermore, can be applied only at leading $\ln s$ level because of
strong ordering.

On the other hand, in our case the definition of the $\kk$-dependent cross section is
more precise, but can be done theoretically rather than experimentally, because one
needs not only to identify all fragmentation products, but also to subtract out the
central region tail --- which appears to be hard to do experimentally. Because of this
difference, the collinear singularities to be factored out are different in the two
cases.

As a second point we remark that a universal renormalization, with the same $\K$
coefficient, holds also for the soft part of the one-loop time-like splitting
functions~\cite{FuPe80}.  This part has a next-to-leading $1/(1-z_1)$ singularity, the
leading one having a logarithmic factor, and is therefore analogous to the impact
factor, which corresponds to the NL$x$ constant piece in the high-energy limit.

The above analogy is perhaps a hint \cite{KoKo96} to explain the universality found
here. However our result does not seem to follow in a clearcut way, because of the
difference between collinear and high-energy factorization pointed out before.

%
\section{Which of the energy-scales?\labe{s:wes}}

The problem of choosing the energy scale $s_0(\kk,\kk_0)$ is not concerned only with the 
collinear behaviour of the impact factors, but also with the validity of the NL$x$
factorization formula (\ref{fatt1loop}) {\em to all orders} in $\as$ and with the
definition of the GGF.

One could ask whether Eq.~(\ref{fatt1loop}) really holds to all orders when $s_0=s_M$ is 
adopted as energy-scale. If not, is it possible to properly define the impact factors
and the GGF in such a way that, for a different choice of $s_0$ the factorization
formula is valid?

These questions naturally arise when noticing that the scale $s_M$ of Eq.~(\ref{fatlog})
has the defect of not being factorized in its $\ku,\kd$ dependence. This fact is argued
to contradict multi-Regge factorization of production amplitudes \cite{BrWe75}, which
implies short-range correlations of the $\kk$'s. Furthermore, the use of $O(2,1)$
variables~\cite{CiDeMi69} implies that Regge behaviour in the energy $s$ should involve
the boost (in the following $k_i\dug|\kk_i|$, $q\dug|\qq|$)
\begin{equation}
 \cosh(\zeta)=\frac{s}{2k_1k_2}\quad,  \label{spinta}
\end{equation}
thus suggesting the choice $s_0=k_1k_2$.

We are going to show that, if the factorized scale $s_0=k_1k_2$ is chosen, then at
1-loop level the definition of the GGF has to be modified with respect to
Eq.~(\ref{d:GGF}) by introducing two $\om$-independent additional kernel.

The 1-loop cross section restricted to half the phase space $z_1<q/\sqrt s$ can be
represented, apart from the $\pb$ impact factor and the Regge-gluon trajectory, by the
effective fragmentation vertex
\begin{align}\nonumber
 \int_{q/\sqrt s}^1\dif z_1\;F_{\pa}(z_1,\ku,\qq)&=h_{\pa}^{(0)}(\ku){\ab\over\pie\qq^2}
 \left[\ln{\sqrt s\over\Max(k_1,q)}+\fin_{\pa}(\ku,\qq)\right]\\ \lab{nasceH}
 &=h_{\pa}^{(0)}(\ku){\ab\over\pie\qq^2}\left[\ln{\sqrt s\over k_1}
 -\Th_{q\,k_1}\ln{q\over k_1}+\fin_{\pa}(\ku,\qq)\right]
\end{align}
($\Th_{q\,k_1}$ being a short hand notation for $\Th(q-k_1)$) where the constant (in $s$)
function $\fin_{\pa}$ vanishes for $\qq=0$ and is related to the 1-loop impact factor by
\begin{equation}\lab{r:fin}
 \underset{\text{fixed }\kd}{\int}\du\;h_{\pa}^{(0)}(\ku){\ab\over\pie\qq^2}
 \fin_{\pa}(\ku,\qq)=\ab h_{\pa}^{(1)}(\kd)\;.
\end{equation}
By adding the second half of the phase space, which is simply obtained by interchanging
$\pa\lra\pb$ and $1\lra2$, we reproduce the leading logarithmic term $\ln s/k_1k_2$ with
the desired energy scale, the $\fin$'s provides the impact factor corrections and two
new terms appear, contributing to the 1-loop cross section as~\cite{Cia98}

\begin{align}\lab{sez1loopH}
 &h_{\pa}^{(0)}(\ku)\Big[\ab H_L(\ku,\kd)+\ab H_R(\ku,\kd)\Big]h_{\pb}^{(0)}(\kd)\\
 \lab{d:nuclimp}
 H&\dug H_R(\ku,\kd)=H_L(\kd,\ku)=-{\ab\over\pie\qq^2}\Th_{q\,k_2}\ln{q\over k_2}
\end{align}
According to Eq.~(\ref{espfatt}), the {\em impact kernels} $H_L$ and $H_R$ cannot enter
the leading kernel $K^{(0)}$, furnishing a subleading contribution and, at the same
time, they cannot be absorbed in the impact factors, because of their unacceptable
collinear behaviour
\begin{equation}\lab{intH}
 \int\du\;h_{\pa}^{(0)}(\ku)H(\ku,\kd)=h_{\pa}^{(0)}(\kd)\left({\kd^2\over\mu^2}\right)^\e
 {1\over2\e^2\Ga(1+\e)} +\ord(\e)
\end{equation}
as the double $\e$-pole points out.

Therefore, the kernels $H_L$ and $H_R$ have to be included in the GGF as subleading
contributions in the impact points: the equation~\cite{Cia98}
\begin{equation}\lab{d:nuovaGGF}
 \G_\om=(\id+\ab H_L)[\om-\K]^{-1}(\id+\ab H_R)\;,
\end{equation}
to be interpreted at operatorial level, replaces Eq.~(\ref{d:GGF}).

To summarize, if we adopt a collinear safe energy-scale, meaning that such a choice
provides automatically IR finite impact factors, the impact kernels can be set equal to
zero, but in this case the energy term in Eq.~(\ref{fatt1loop}) is not factorized in its 
$\kk,\kk_0$ dependence; on the other hand, if we want to use a factorized energy-scale,
we have to introduce the impact kernels~(\ref{d:nuclimp}), in order to retain IR finite
impact factors. Of course, the final result is the same.

\section{Two-loop analysis\labe{s:2la}}

The calculation of NL$x$ correction to the BFKL equation has required a large
theoretical effort, the main part of which has been done by the Russian school and the
Florence group. Besides the difficulties of the technical calculation for evaluating
2-loops amplitudes, the determination of the NL$x$ part of the BFKL kernel involves
several conceptual problems mainly concerned with the identification and the subtraction 
of the leading term in a consistent way.

The previous section had the purpose of cleaning up the subtraction procedure, but it is
to be said that, chronologically, such an analysis was the last step but one in order to
derive the NL$x$ kernel. For its determination, in the ``simplest'' case of
parton-parton scattering, a 2-loops calculation is needed. In fact, by expanding
Eq.~(\ref{fatt1loop}) to 2-loops in NL$x$ approximation by using
Eqs.~(\ref{d:nuovaGGF},\ref{esph},\ref{espK}), we obtain, in addition to
Eq.~(\ref{espfatt}) plus the contribution of the impact kernels~(\ref{sez1loopH}),
\begin{align}\lab{esp2loop}
 {\dif\si^{(2)}_{\pa\pb}\over\dif\kk\,\dif\kk_0}&=\ab^2\left\{\ln^2{s\over s_0}\left[
 h_{\pa}^{(0)}K^{(0)}{}^2h_{\pb}^{(0)}\right]\right.\\ \nonumber
&\left.\quad+\ln{s\over s_0}\left[h_{\pa}^{(0)}K^{(0)}h_{\pb}^{(1)}+h_{\pa}^{(1)}
 K^{(0)}h_{\pb}^{(0)}+h_{\pa}^{(0)}\Big(K^{(1)}+H_LK^{(0)}+K^{(0)}H_R\Big)h_{\pb}^{(0)}
 \right]\right\}\;.
\end{align}
up to NNL$x$ terms.
It is evident that $K^{(1)}$ can be unambiguously extracted once the energy-scale $s_0$,
the 1-loop impact factor $h^{(1)}$ and the impact kernel $H_L,H_R$ are known.

\subsection{The cluster expansion\labe{ss:cluster}}

To obtain the 2-loops differential cross section in NL$x$ approximation, we have to
consider, besides real emission in MRK and L$x$ virtual corrections, also processes in
which the additional power of $\as$ is not compensated by a $\ln s$. This is accomplished 
by admitting that any (but only one) pair of the outgoing particles carry a small ---
i.e., not increasing with energy --- invariant mass. This is known as {\em
quasi-multi-Regge kinematics} (QMRK), in distinction to MRK where all the sub-energies of 
all pairs are large compared to the transverse momenta. At virtual level, we have to
consider corrections not including one $\ln s$.

After these premises, we are led to consider three kinds of processes: {\it(i)} elastic
diffusion, {\it (ii)} one particle and {\it(iii)} two particle production. In {\it(ii)}
and {\it(iii)} there can be a cluster of two particles with small invariant mass.

\subsubsection{Elastic diffusion\labe{sss:2pfs}}

\begin{figure}[t]
\centering
\includegraphics[width=15cm]{\fig 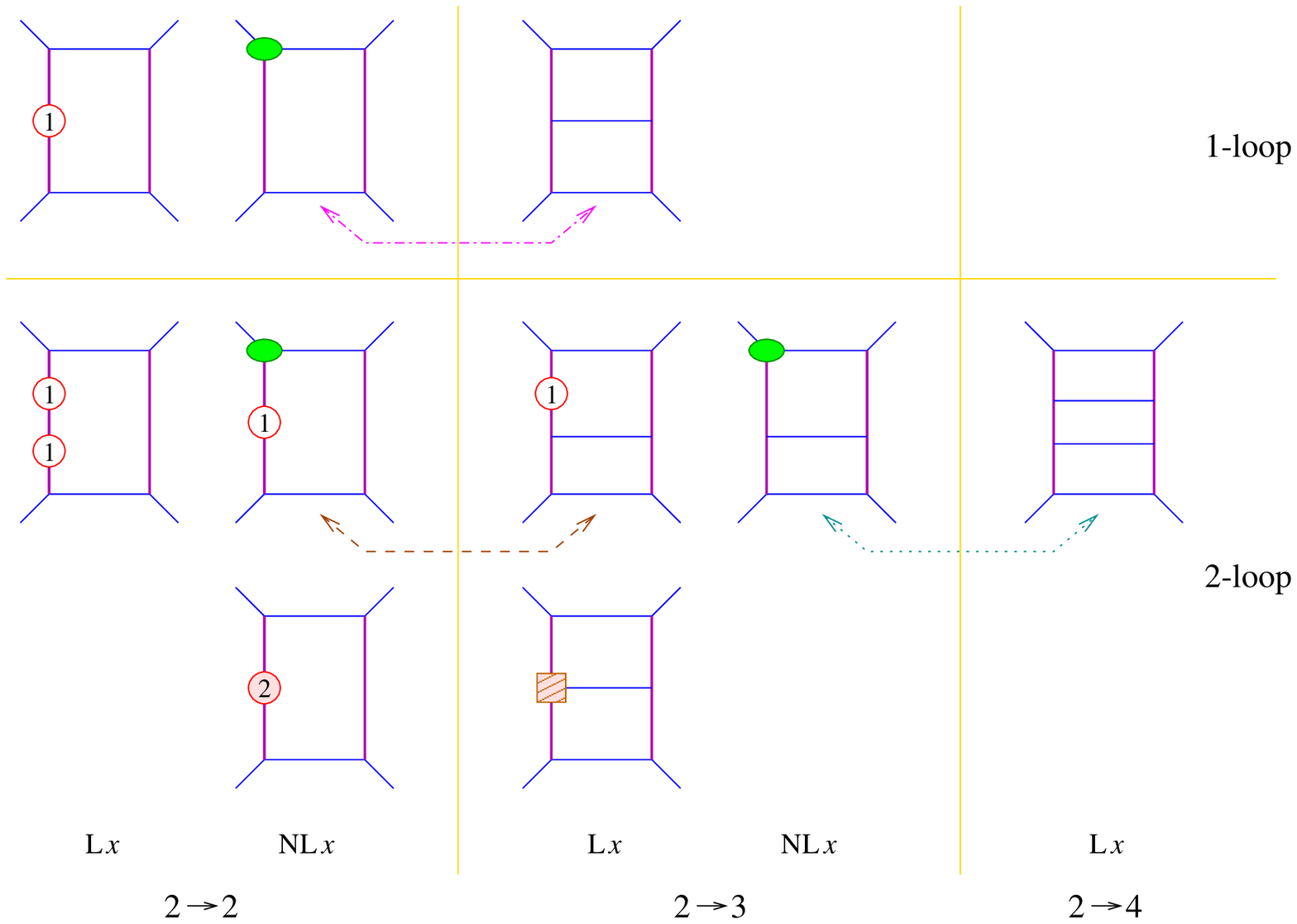}
\didascalia{Cluster expansion. The various class of cut diagrams contributing to the 1-loop
and 2-loops L$x$ + NL$x$ cross section are represented. Symmetric virtual correction
diagrams are understood. The dashed/dotted lines joins classes of diagrams that are
combined in order to form effective fragmentation vertices in the differential cross section.%
\labe{f:loop}} 
\end{figure}
The simplest process whose NL$x$ part of the kernel is identified is the elastic one
{\it(i)}. As Fig.~\ref{f:loop} shows, there are three different 2-loops structures
contributing at NL$x$ level: the leading gluon trajectory iteration, the product of
gluon trajectory and PPR vertex correction and the pure NL$x$ gluon trajectory. The
former, given in Eq.~(\ref{interc}), belongs to the iteration of the L$x$ kernel in
Eq.~(\ref{esp2loop}). Also the PPR corrections, defined as the non logarithmic
contribution to the 1-loop elastic amplitude has already been
derived~\cite{PPRcorr}. Therefore, the latter contribution, the NL$x$ gluon trajectory,
can be unambiguously defined, by observing that in the elastic process the only physical
scale is provided by the transferred momentum $-t\simeq\kk^2$. In dimensional
regularization $\Om^{(1)}$ reads~\cite{Regge2loop}
\begin{align}\nonumber
 \Om^{(1)}(\kk^2)&=\frac{1}{8\e^2}\bigg\{
 -\left(\frac{\kk^2}{\mu^2}\right)^\e\left({11N_c-2\Nf\over3N_c}-{11\pi^2\over18}
 \e^2\right)+\left(\frac{\kk^2}{\mu^2}\right)^{2\e}\left[\frac{11N_c-2\Nf}{6N_c}\right.\\
\lab{interc2loop}&\left.+\left(\frac{\pi^2}{6}-\frac{67}{18}+{5\Nf\over9N_c}\right)\e
 +\left(\frac{202}{27}-\frac{11\pi^2}{18}-\zeta(3)-{28\Nf\over27N_c}\right)\e^2\right]
 \bigg\}\;.
\end{align}

\subsubsection{Three-particle final states\labe{sss:3pfs}}

When considering one-particle production, the full amplitude with 1-loop corrections can 
be written~\cite{CaCi97b}
\begin{align}\lab{Mtre1loop}
&M_{2\to3}^{(1)}=M_{2\to3}^{(0)}\bigg\{1+\ab\bigg[\Pi_{\pa\pa}^{(+)}(\ku^2)
 +\Pi_{\pb\pb}^{(+)}(\kd^2)\nonumber\\
&+\frac{1}{2}\Om^{(0)}(\ku^2)\left(\ln\frac{s_1}{q^2}+\ln\frac{s_1}{k_1k_2}\right)
 +\frac{1}{2}\Om^{(0)}(\kd^2)\left(\ln\frac{s_2}{q^2}+\ln\frac{s_2}{k_1k_2}\right)
 \bigg]\bigg\}+\tilde{M}^{(1)}_{2\to3}
\end{align}
where $M_{2\to3}^{(0)}$ is given in Eq.~(\ref{Magb}), $s_1$ and $s_2$ are the
sub-energies variables defined in Eq.~(\ref{propregge}), $\Pi$ is the vertex corrections
of Eqs.~(\ref{corrvirt},\ref{corrqq},\ref{corrgg}) and $\tilde{M}^{(1)}_{2\to3}$ is the
reggeon-reggeon-gluon (RRG) vertex correction amplitude.

The RRG vertex corrections~\cite{RRGcorr} didn't take place in the 1-loop evaluation of
the cross section and their interference with the Born amplitude $M_{2\to3}^{(0)}$ give
a pure NL$x$ contribution to 2-loops. On the contrary, the trajectory and PPR
corrections are important in the 1-loop expansion (\ref{espfatt}) contributing to both
the L$x$ and NL$x$ term, as explained in Sec.~\ref{s:wes}. Therefore,
$\tilde{M}^{(1)}_{2\to3}$ provides a so called ``irreducible'' contribution to the NL$x$
kernel, whereas the remaining terms of $M_{2\to3}^{(1)}$ contribute to the NL$x$ kernel
after the subtractions of the leading term, of the impact factor corrections and of the
impact kernels.

The three ensuing structure contributing to the 2-loops cross section
($\Pi^{(+)},\Om^{(0)}$ and $\tilde{M}$) are drawn in Fig.~\ref{f:loop}. The one-gluon
irreducible contribution is given by~\cite{CaCi97b}
\begin{align}\lab{3irrid}
 {\dif\si^{(2,\text{irr})}_{2\to3}\over\du\,\dd}&=\ab^2 h_{\pa}^{(0)}\,K_{1\pg}^{(\text{irr})}
 \ln{s\over k_1k_2}\,h_{\pb}^{(0)}\;,\\ \nonumber
 K_{1\pg}^{(\text{irr})}&=\frac{1}{4}\left[-{2\pi\over\e\pie}\frac{(\qq^2/\mu^2)^\e}{\qq^2}
 \frac{\cos\pi\e}{\sin\pi\e}+\frac{11}{3}\left(\frac{1}{\e\pie\qq^2}+\frac{1}{\ku^2-\kd^2}
 \ln\frac{\ku^2}{\kd^2}\right)\right]\;,
\end{align}
where, for simplicity, an azimuthal average has been performed.

We now consider the one-particle emission with trajectory corrections diagrams together
with the elastic ones carrying one PPR and one trajectory correction, as the dashed line 
in Fig.~\ref{f:loop} indicates.

Let's restrict to particle emission in the positive rapidity half space $y>0$ and let's
take the elastic diagrams with PPR correction only in the upper line, as in
Fig.~\ref{f:3pfs}, the lower impact factor having been omitted. The first two
diagrams are nothing but the components of the effective fragmentation vertex
(cfr.\ Fig.~\ref{f:efffragvert}) times the leading virtual term $\Om^{(1)}\ln s$ in
Eq.~(\ref{Mtre1loop}) and yield
\begin{figure}[ht!]
\centering
\includegraphics[width=115mm]{\fig 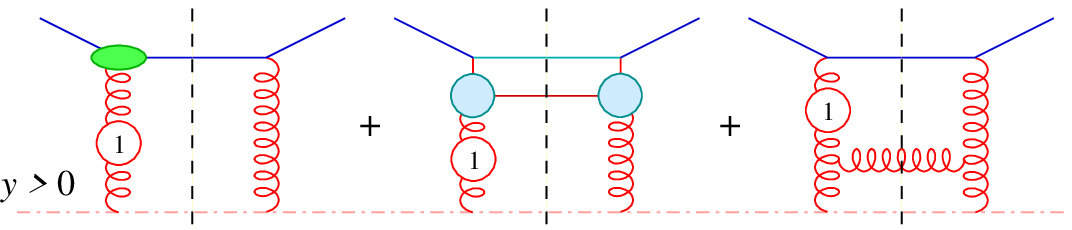}
\didascalia{Reducible contribution to the NL$x$ kernel from three-particle final states: PPR
corrections in the upper line (the symmetric virtual correction is understood) and real emission in the positive rapidity half space $y>0$ with 1-loop reggeon correction in the lower and upper propagator.%
\labe{f:3pfs}} 
\end{figure}
\begin{align}\nonumber
 {\dif\si^{(2)}_{2\to3}\over\du\,\dd}\Big|_{\text{\sl a+b}}^{y>0}&=h_{\pb}^{(0)}(\kd)
 \int_{q/\sqrt s}^1\dif z_1\;F_{\pa}(z_1,\ku,\qq){1\over2}\Om^{(0)}(\kd^2)\left[
 \ln{s_2\over q^2}+\ln{s_2\over k_1k_2}\right]\\ \label{Fasenza}
&=h_{\pb}^{(0)}(\kd)\Om^{(0)}(\kd^2)\int_{q/\sqrt s}^1\dif z_1\;F_{\pa}(z_1,\ku,\qq)
 \ln{zs\over q\sqrt{k_1k_2}}\;,
\end{align}
since $s_2\simeq z_1s$. It is easy to realize that we get $\ln s$ terms from both the
integration boundaries if you replace the NL$x$ fragmentation vertex with the leading
one
\begin{equation}\lab{vfeLx}
 F_{\pa}(z,\kk,\qq)\valutato_{\,\text{L}x}={1\over z}\,{\ab\over\pie\qq^2}
\end{equation}
and perform the $z$-integration. Therefore, the effective vertex $F_{\pa}$ provides an
important NL$x$ correction with respect to its L$x$ analogue.

On the other hand, the third diagram in Fig.~\ref{f:3pfs} yields
\begin{align}\nonumber
 {\dif\si^{(2)}_{2\to3}\over\du\,\dd}\Big|_{\text{\sl c}}^{y>0}&=h_{\pb}^{(0)}(\kd)
 \int_{q/\sqrt s}^1\dif z_1\;\widetilde{F}_{\pa}(z_1,\ku,\qq){1\over2}\Om^{(0)}(\ku^2)\left[
 \ln{s_1\over q^2}+\ln{s_1\over k_1k_2}\right]\\ \lab{Fatilde}
&=h_{\pb}^{(0)}(\kd)\Om^{(0)}(\ku^2)\int_{q/\sqrt s}^1\dif z_1\;
 \widetilde{F}_{\pa}(z_1,\ku,\qq)\ln{q\over z_1\sqrt{k_1k_2}}\;,
\end{align}
since $s_1\simeq q^2/z_1$. Here $\widetilde{F}_{\pa}$ is another effective fragmentation
vertex which, just like $F_{\pa}$, reduces to the RHS of Eq.~(\ref{vfeLx}) in the small
$z$-region and for all values of $z>q/\sqrt s$ in L$x$ approximation.

However, Eq.~(\ref{Fatilde}) provides a logarithmic contribution only in the vicinity of 
the lower integration boundary, and by making the substitution (\ref{vfeLx}) for
$\tilde{F}_{\pa}$, the error is only NNL$x$.

In this way, by taking the derivative of Eqs.~(\ref{Fasenza},\ref{Fatilde}) with respect 
to $\ln s$ and by using Eq.~(\ref{nasceH}), we can perform the $z_1$-integration exactly 
obtaining
\begin{align}\lab{derlogs3}
 {\de\over\de\ln s}{\dif\si^{(2)}_{2\to3}\over\du\,\dd}&\Big|_{\text{\sl a+b+c}}^{y>0}\\
 &=\left\{\left[\left({\omu\over2}+{3\omd\over2}\right)\ln{s\over k_1k_2}
 +\omd\ln{k_2\over k_1}+2\omd\fin_1\right]{\ab\over\pie\qq^2}
 +2\omd H_L\right\}h_{\pb}^{(0)}\nonumber
\end{align}
where we have set $\Om_i\dug\Om^{(0)}(\kk_i^2)$ and
$\fin_i\dug\fin_{\pa_i}(\kk_i,\qq)$. Including the other half phase space $y<0$
($\{1\lra2\},\{L\lra R\}$), inverting the $\ln s$ derivative and carrying out the
transverse integration of the $\fin$-terms in order to reproduce the impact factor
corrections, we end up with the ``reducible'' part of the one-particle emission cross
section~\cite{CaCi98}
\begin{align}\lab{3riduc}
 {\dif\si^{(2,\text{red})}_{2\to3}\over\du\,\dd}&=h_{\pa}^{(0)}\left\{(\omu+\omd){\ab\over
 \pie\qq^2}\ln^2{s\over k_1k_2}\right.\\
&\quad+\left.\left[(\omu-\omd){\ab\over\pie\qq^2}\ln{k_1\over k_2}+H_L 2\omd
 +2\omu H_R\right]\ln{s\over k_1k_2}\right\}h_{\pb}^{(0)}\nonumber\\ \nonumber
&\quad+\left[h_{\pa}^{(0)}2\omu{\ab\over\pie\qq^2}h_{\pb}^{(1)}+h_{\pa}^{(1)}
 {\ab\over\pie\qq^2}2\omd h_{\pb}^{(0)}\right]\ln{s\over k_1k_2}\;.
\end{align}

\subsubsection{Four-particle final states\labe{sss:4pfs}}

There remains to consider two-particle production at Born level (and the left over class 
of diagrams of the previous section). The contribution coming from a $\pq\bar{\pq}$ pair 
in the final state does not require any particular identification procedure, since it
doesn't appear at L$x$ level, giving therefore an irreducible contribution~\cite{QQ2loop}:
\begin{align}\lab{qqirrid}
 \left.{\dif\si^{(2)}_{2\to4}\over\du\,\dd}\right|_{\pq\bar{\pq}}&=\ab^2\,h_{\pa}^{(0)}\,
 K^{(\text{irr})}_{\pq\bar{\pq}}\ln{s\over k_1k_2}\,h_{\pb}^{(0)}\\ \lab{nucleoqq}
 K^{(\text{irr})}_{\pq\bar{\pq}}(\ku,\kd)&={\Nf\over6N_c}\left[\left(\ln{\qq^2\over\mu^2}
 -{5\over3}\right){1\over\qq^2}-{1\over\ku^2-\kd^2}\ln{\ku^2\over\kd^2}\right]\;.
\end{align}

The two-gluon production, on the contrary, gives L$x$ and NL$x$ contributions, and a
treatment similar to that of Sec.~\ref{sss:3pfs} has to be done. First of all, we
identify another irreducible contribution to the kernel, which consists in the
production of a gluon pair in the central region with a small invariant mass.
\begin{figure}[ht!]
\centering
\includegraphics[width=40mm]{\fig 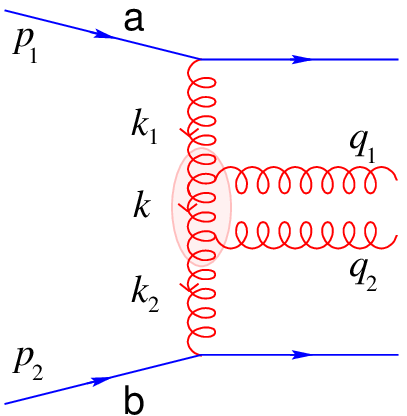}
\didascalia{Two-gluon emission in the central region in QMRK.
\labe{f:2glucentrali}} 
\end{figure}

To be explicit, according to Fig.~\ref{f:2glucentrali}, we parametrize the exchanged
momenta by
\begin{subequations}\labe{param2glu}
\begin{align}
 k_i^\mu&=z_ip_1^\mu+\bar{z}_ip_2^\mu+\kk_i^\mu\qquad(i=1,2)\;,\\
 k^\mu&=zp_1^\mu+\bar{z}p_2^\mu+\kk^\mu
\end{align}
\end{subequations}
so that the interesting kinematic region corresponds to the QMRK
\begin{equation}\lab{QMRK2g}
 1\gg z_1\sim z\gg z_2\simeq{\kk_2^2\over s}\;.
\end{equation}
The irreducible contribution to the $2\to4$ cross section is then defined by subtracting, in
the region (\ref{QMRK2g}), the leading term:
\begin{equation}\lab{d:24irr}
 {\dif\si^{(2,\text{irr})}_{2\to4}\over\du\dd\dif z_1\dif z}=
 {\dif\si^{(2,\text{tot})}_{2\to4}\over\du\dd\dif z_1\dif z}-h_{\pa}^{(0)}(\ku)
 h_{\pb}^{(0)}(\kd){1\over z_1}{1\over z}\int\dk\;
 {\ab\over\pie\qq_1^2}{\ab\over\pie\qq_2^2}\;.
\end{equation}
The irreducible contribution to the cross section is~\cite{GG2loop,CaCi97b}
\begin{equation}\lab{4irrid}
 {\dif\si^{(2,\text{irr})}_{2\to4}\over\du\,\dd}=\ab^2 h_{\pa}^{(0)}\,K_{2\pg}^{(\text{irr})}
 \ln{s\over k_1k_2}\,h_{\pb}^{(0)}
\end{equation}
in terms of the two-gluon irreducible kernel~\cite{GG2loop,CaCi97b}
\begin{align}
 K_{2\pg}^{(\text{irr})}&=\frac{1}{4}\bigg\{{2\Ga^2(1+\e)\over\e\Ga(1+2\e)\pie}
 \frac{(\qq^2/\mu^2)^\e}{\qq^2}\left[\frac{1}{\e}-\frac{11}{6}+\left(\frac{67}{18}
 -\frac{\pi^2}{2}\right)\e-\left(\frac{202}{27}-7\zeta(3)\right)\e^2\right]\nonumber\\
 &-H_{\text{coll}}(\ku,\kd)+\tilde{H}(\ku,\kd)\bigg\}\;,
\end{align}
where we have introduced the ``collinear'' kernel
\begin{align}
&H_{\text{coll}}(\ku,\kd)=\frac{1}{32\pi}\left[2\left(\frac{1}{\ku^2}+\frac{1}{\kd^2}\right)+
\left(\frac{1}{\kd^2}-\frac{1}{\ku^2}\right)\ln\frac{\ku^2}{\kd^2}+
\left(118-\frac{\ku^2}{\kd^2}-\frac{\kd^2}{\ku^2}\right)\times\right.\nonumber\\
&\left.\times\frac{1}{\sqrt{\ku^2\kd^2}}\left(\ln\frac{\ku^2}{\kd^2}
\tan^{-1}\frac{|\kd|}{|\ku|}+\im\Li\left(\ui\frac{|\kd|}{|\ku|}
\right)\right)\right]-\frac{\pi^2}{3k^2_>},
\end{align}
in which an azimuthal average has been performed, and the dilogarithmic one
\begin{align}
&\tilde{H}(\ku,\kd)=-\frac{\pi}{3k_>^2}+
\frac{2\qq\cdot(\ku+\kd)}{\pi\qq^2(\ku+\kd)^2}
\left[\ln\frac{\ku^2}{\kd^2}\ln\frac{\ku^2\kd^2}{(\ku^2+\kd^2)^2}+
\right.\nonumber\\
&\left.+\Li\left(1-\frac{\qq^2}{\ku^2}\right)-\Li\left(1-\frac{\qq^2}{\kd^2}\right)
+\Li\left(-\frac{\kd^2}{\ku^2}\right)-\Li\left(-\frac{\ku^2}{\kd^2}\right)
\right]+\nonumber\\
&+{2\over\pi}\left[\int_0^1\frac{dt}{(\ku-t\kd)^2}\left(\frac{\kd\cdot\qq}{\qq^2}
-\frac{\kd^2\qq\cdot(\ku+\kd)}{\qq^2(\ku+\kd)^2}(1+t)\right)
\ln\frac{t(1-t)\kd^2}{\ku^2(1-t)+\qq^2 t}+\right.\nonumber\\
&+(\ku\longleftrightarrow-\kd)\bigg].
\end{align}
where $k_>^2\dug\Max(\ku^2,\kd^2)$.

At this point, to complete the calculation, it remains to evaluate the subtracted term
in the RHS of Eq.~(\ref{d:24irr}) to NL$x$ accuracy. To do this, we add to it the
one-particle emission diagrams with PPR corrections (see the dotted line in
Fig.~\ref{f:loop}) and perform the following analysis: let $q_1$ be the gluon with the
largest rapidity and define
\begin{equation}\lab{rapgluoni}
 y_i\dug\ln z_i{\sqrt s\over q_i}\quad,\quad y_i\in[-Y_i,Y_i]\quad,\quad
 Y_i\dug\ln{\sqrt s\over q_i}\;.
\end{equation}
Working at fixed transverse momenta, we split the phase space into three regions:
\begin{align*}\hspace{5cm}
 &A:&0&<y_2<y_1<Y_1\;,&\hspace{5cm}\\
 &B:&-Y_2&<y_2<0<y_1<Y_1\;,&\\
 &C:&-Y_2&<y_2<y_1<0\;.&
\end{align*}
\begin{figure}[ht!]
\centering
\includegraphics[width=12cm]{\fig 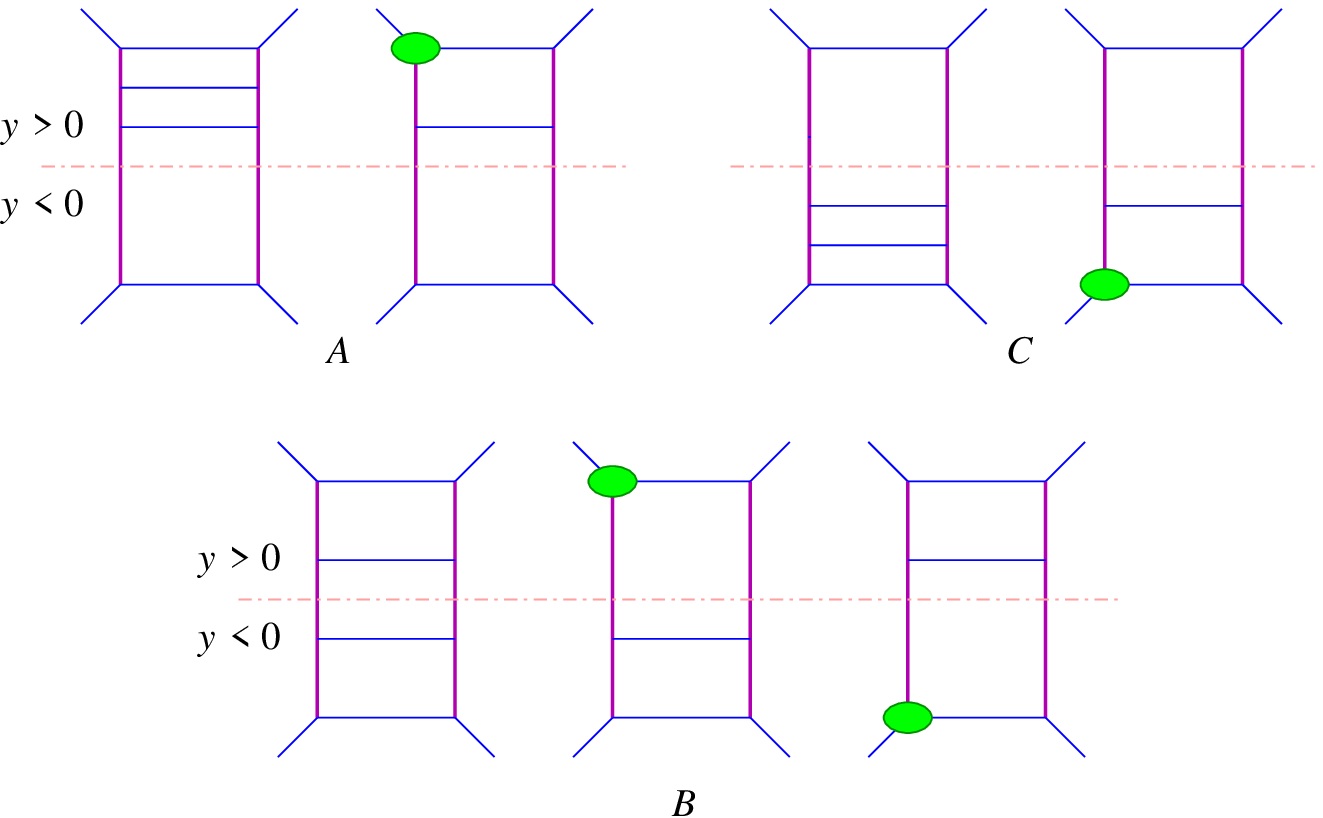}
\didascalia{How $2\to3$ and $2\to4$ diagrams are combined in order to reconstruct the
effective fragmentation vertices. Symmetric virtual corrections diagrams are understood.%
\labe{f:ABC}} 
\end{figure}
Then we add the one-particle emission diagrams with PPR corrections that we distinguish
into four types --- according to the sign of the rapidity of the emitted particle and to
the (upper or lower) leg which the PPR correction is attached to --- and we associate
them to the three regions $A,B$ and $C$ as in Fig.~\ref{f:ABC}. Repeating the analysis
of Sec.~\ref{sss:3pfs} we reproduce the effective fragmentation vertex in the relevant
phase space regions, and remembering that, when $y_1\simeq y_2$ the subtracted term is
exactly given by the last one in Eq.~(\ref{d:24irr}), we are allowed to write, at NL$x$
accuracy,
\begin{align}\lab{rhoABC}
  {\dif\si^{(2,\text{red})}_{2\to4}\over\du\,\dd}&=h_{\pa}^{(0)}(\ku)h_{\pb}^{(0)}(\kd)
 \int\dk\;{\ab\over\pie\qq_1^2}{\ab\over\pie\qq_2^2}\big(\rho_A+\rho_B+\rho_C)\;,\\
\intertext{where}\nonumber
 \rho_A&=\int_{q_1/\sqrt s}^1{\dif z_1\over z_1}\;\FF_{\pa}(z_1,\ku,\qq_1)
 \int_{q_2/\sqrt s}^{\Min\(1,{q_2\over q_1}z_1\)}{\dif z\over z}\;,\\ \nonumber
 \rho_B&=\int_{q_1/\sqrt s}^1{\dif z_1\over z_1}\;\FF_{\pa}(z_1,\ku,\qq_1)
 \int_{q_2/\sqrt s}^1{\dif\bar{z}_2\over\bar{z}_2}\;\FF_{\pb}(\bar{z}_2,\kd,\qq_2)\;,\\
 \nonumber
 \rho_C&=\int_{q_2/\sqrt s}^1{\dif\bar{z}_2\over\bar{z}_2}\;\FF_{\pb}(\bar{z}_2,\kd,\qq_2)
 \int_{q_1/\sqrt s}^{\Min\(1,{q_1\over q_2}\bar{z}_2\)}{\dif\bar{z}\over\bar{z}}
 =\rho_A|_{1\lra2}^{\pa\lra\pb}
\end{align}
and the ``reduced'' fragmentation vertex $\FF_{\pa}$ is
\begin{equation}\lab{d:vferid}
 F_{\pa}(z,\kk,\qq)\ugd h_{\pa}^{(0)}(\kk){\ab\over\pie\qq^2}\FF_{\pa}(z,\kk,\qq)\;.
\end{equation}
By deriving with respect to $\ln s$, performing the integrals and inserting the results
in Eq.~(\ref{rhoABC}), we obtain~\cite{CaCi98}
\begin{align}\nonumber
  {\dif\si^{(2,\text{red})}_{2\to4}\over\du\,\dd}&=h_{\pa}^{(0)}h_{\pb}^{(0)}\left\{
 {1\over2}\ln^2{s\over k_1k_2}\;{\ab\over\pie\qq_1^2}\circ{\ab\over\pie\qq_2^2}
\ln{s\over k_1k_2}\left[H_L\circ{\ab\over\pie\qq_2^2}+{\ab\over\pie\qq_1^2}
 \circ H_R\right]\right\}\\ \lab{4riduc}
 &\quad+\ln{s\over k_1k_2}\left[h_{\pa}^{(0)}{\ab\over\pie\qq_1^2}h_{\pb}^{(1)}+
 h_{\pa}^{(1)}{\ab\over\pie\qq_2^2}h_{\pb}^{(0)}\right]\;,
\end{align}
the small circle $\circ$ denoting composition of operators, i.e., a $\kk$-integration.

\subsection{Next-to-leading BFKL kernel\labe{ss:nlBFKL}}

The left over elastic contributions (the first and the third diagram in the $2\to2$,
2-loops group of Fig.~\ref{f:loop}) contribute to the cross section with
\begin{equation}\lab{2tutta}
  {\dif\si^{(2)}_{2\to2}\over\du\,\dd}=h_{\pa}^{(0)}h_{\pb}^{(0)}\d(\ku-\kd)
 \left\{{1\over2}\big(2\omu2\omd\big)\ln^2{s\over k_1k_2}+2\Om^{(1)}(\ku^2)
 \ln{s\over k_1k_2}\right\}\;.
\end{equation}

At least we are at the final step: by adding
Eqs.~(\ref{3irrid},\ref{3riduc},\ref{qqirrid},\ref{4irrid},\ref{4riduc},\ref{2tutta})
and identifying the corresponding terms in the expansion (\ref{esp2loop}), we are able
to single out the NL$x$ BFKL kernel at energy-scale $s_0=k_1k_2$~\cite{FaLi98,CaCi98}:
\begin{align}\nonumber
 K^{(1)}&=2\Om^{(1)}(\ku^2)\d(\ku-\kd)+K_{1\pg}^{(\text{irr})}+K_{\pq\bar{\pq}}^{(\text{irr})}+
 K_{2\pg}^{(\text{irr})}+{\omu-\omd\over(\ku-\kd)^2}\ln{k_1\over k_2}\\ \lab{Kuno}
 &=-b_0\ln{\ku^2\over\mu^2}\,K^{(0)}+K^{(1)}_{\text{s.i.}}
\end{align}
where in the second line we have explicitly shown the scale-invariance violating term and
$K^{(1)}_{\text{s.i.}}$ defines the scale-invariant part of the NL$x$ kernel. Note that
the coefficient of the $\ln\mu^2$ term is precisely the leading kernel with the
$\be$-function coefficient $b_0$ given in Eq.~(\ref{bzero}). Therefore it can be
interpreted as a running factor and we can express the total L$x$ + NL$x$ kernel in the
form
\begin{subequations}\labe{nucleomobile}
\begin{align}
 \K&=\ab(\mu^2)\left[\(1-b_0\as(\mu^2)\ln{\ku^2\over\mu^2}\)\right]K^{(0)}+
 \ab(\mu^2)K^{(1)}_{\text{s.i.}}\\
 &=\ab(\ku^2)\big[K^{(0)}+\ab(\mu^2)K^{(1)}_{\text{s.i.}}\big]+\ord(\as^3)
\end{align}
\end{subequations}
Factorizing the running coupling at the scale $\ku^2$ is an asymmetric
procedure, but is convenient for the discussion of the non-scale-invariant
BFKL equation.
Using a different scale (e.g., $\as(k_>^2)$) implies changing $K^{(1)}_{\text{s.i.}}$
so as to leave the total NL kernel invariant (see below). 

\subsection{Eigenvalues of the NL$x$ kernel\labe{ss:enlk}}

In order to investigate the physical features emerging from the NL$x$ kernel, we
determine the eigenvalues of the scale-invariant part by applying $\K$ to the
(spherically symmetric) eigenfunctions of the leading kernel:
\begin{equation}\lab{d:avlNL}
 \ab(\kk^2)\!\int\!\dif\kk'\(K^{(0)}(\kk,\kk')+\ab K^{(1)}_{\text{s.i.}}(\kk,\kk')\)
 (\kk'{}^2)^{\ga-1}=\ab(\kk^2)\left[\chi^{(0)}(\ga)
 +\ab\chi^{(1)}(\ga)\right](\kk'{}^2)^{\ga-1}\,.
\end{equation}
The leading eigenvalue $\chi^{(0)}$ is the same of Eq.~(\ref{avlK0}) with $m=0$:
\begin{equation}\lab{avlK0m0}
 \chi^{(0)}(\ga)=2\psi(1)-\psi(\ga)-\psi(1-\ga)\;,
\end{equation}
while the eigenvalue function $\chi^{(1)}$ of the next-to-leading kernel
$K^{(1)}_{\text{s.i.}}$ ($\equiv K^{(1)}$ from now on) is the sum of a gluonic part
\begin{align}\lab{avlglu}
 \chi^{(1,\pg)}(\ga)=&-{1\over2}\bigg[{11\over12}\(\chi^{(0)\,2}(\ga)+\chi^{(0)}{}'(\ga)\)\bigg]
 -{1\over4}\chi^{(0)}{}''(\ga)\\ \nonumber
&-\left({\pi\over\sin\pi\ga}\right)^2{\cos\pi\ga\over(1-2\ga)}
 \left({11\over12}+{\ga(1-\ga)\over36(1+2\ga)(1-{2\over3}\ga)}\right)\\ \nonumber
&+\left({67\over36}-{\pi^2\over12}\right)\chi^{(0)}(\ga)
 +{3\over2}\,\zeta(3)+{\pi^3\over4\sin\pi\ga}-\Phi(\ga)\;,\\ \lab{d:Phi}
\Phi(\ga)\dug\sum_{n=0}^{\infty}&(-)^n\left[{\psi(n+1+\ga)-
 \psi(1)\over(n+\ga)^2}+{\psi(n+2-\ga)-\psi(1)\over(n+1-\ga)^2}\right]\;,
\end{align}
and a quark --- $\Nf$-dependent --- one
\begin{align}\lab{avlqua}
 \chi^{(1,\pq)}(\ga)={\Nf\over6N_c}\bigg[&{1\over2}
 \(\chi^{(0)\,2}(\ga)+\chi^{(0)}{}'(\ga)\)-{5\over3}\chi^{(0)}(\ga)\\ \nonumber
&-{1\over N_c^2}\left({\pi\over\sin\pi\ga}\right)^2
 {\cos\pi\ga\over1-2\ga}\,{1+{3\over2}\ga(1-\ga)\over(1+2\ga)(1-{2\over3}\ga)}\bigg]\;.
\end{align}

We want to stress that the particular form of the NL$x$ kernel, and hence the NL$x$
eigenvalue, depends on the choice of both the scale of $\ab$ and the scale of the energy 
$s_0$. For instance, if we adopt $k_>^2$ as the argument of $\ab$, in Eq.~(\ref{Kuno}) 
we should replace
\begin{equation}\lab{camscalab}
 b\ln{k_1^2\over\mu^2}K^{(0)}=-b\ln{k_>^2\over\mu^2}K^{(0)}-b\Th_{k_2\,k_1}\ln
 {k_2^2\over k_1^2}K^{(0)}\;,
\end{equation}
so that the NL$x$ kernel undergoes a shift
\begin{equation}\lab{diffKnl}
 \De K^{(1)}=-b\Th_{k_2\,k_1}\ln{k_2^2\over k_1^2}K^{(0)}=-b{\Th_{k_2\,k_1}\over
 |\ku-\kd|^2}\ln{k_2^2\over k_1^2}
\end{equation}
and, correspondingly, the NL$x$ eigenvalue function acquires the additional term
\begin{equation}\lab{diffavlNL}
 \De\chi^{(1)}(\ga)=-b\psi'(1-\ga)
\end{equation}
which symmetrizes completely itself.

In DIS the energy dependence of the SF is commonly expressed by means of the Bjorken
variable $x\simeq Q^2/s$, $s$ being the photon-proton CM energy squared. In the limit of 
$Q^2\gg\kk_0^2$, where $\kk_0^2$ is the typical transverse momentum scale of the
proton's constituents, the correct energy-scale is $s_0=k^2\sim Q^2$. The shift in the
NL$x$ kernel corresponding to the change $s_0=kk_0\to k^2$ can be accomplished by a
simple trick. If we denote with $H(\ga)$ the eigenvalue function of the impact kernel
$H$ of Eq.~(\ref{d:nuclimp})
\begin{equation}\lab{avlH}
 H(\ga)=-{1\over2}\sum_{n=0}^{\infty}\left({\Ga(\ga+n)\over\Ga(\ga)n!}\right)^2{1\over(\ga+n)^2}
 \stackrel{\ga\to0}{=}-{1\over2\ga^2}+\ord(\ga^2)\;,
\end{equation}
the GGF (actually, its scale-invariant part) has the representation
(cfr.\ Eq.~(\ref{rapspeG}) with $m=0$)
\begin{subequations}\labe{rapGsi}
\begin{align}
 \G(s,k,k_0)&=\int\difo\difg\;\left({s\over kk_0}\right)^\om g_\om(\ga)\left({k^2\over k_0^2}
 \right)^\ga{1\over\pi k^2}\\
 &=\int\difo\difg\;\left({s\over k^2}\right)^\om g_\om(\ga)\left({k^2\over k_0^2}
 \right)^{\ga+\ho}{1\over\pi k^2}\\ \intertext{where}
 g_\om(\ga)&={1+\ab[H(\ga)+H(1-\ga)]\over\om-\ab\left[\chi^{(0)}(\ga)+\ab\chi^{(1)}(\ga)
 \right]}\qquad(s_0=kk_0)\;.
\end{align}
\end{subequations}
Therefore the energy-scale change $kk_0\to k^2$ is equivalent to replacing $\ga$ with
$\ga+\ho$. By shifting $\ga$ in $g_\om$ and then expanding in $\om$ up to NL$x$ level,
the GGF Mellin transform at the ``upper'' energy-scale $s_0=k^2$ reads
\begin{equation}\lab{gup}
 g^{[u]}_\om(\ga)={1+\ab[H(\ga)-\half\chi^{(0)}{}'(\ga)+H(1-\ga)]\over\om-\ab\left[\chi^{(0)}(\ga)
 +\ab\chi^{(1)}(\ga)-\half\ab\chi^{(0)}\chi^{(0)}{}'\right]}\qquad(s_0=k^2)\;.
\end{equation}
This means that the shift in the kernel's eigenvalue function is
\begin{equation}\lab{Dchienergia}
 \De\chi^{(1)}=-{1\over2}\chi^{(0)}\chi^{(0)}{}'\stackrel{\ga\to0}{=}{1\over2\ga^3}-\zeta(3)
 +\ord(\ga^2)\;.
\end{equation}
The shift (\ref{Dchienergia}) cancels the cubic singularity (and the $\zeta(3)$ term) of
\begin{align}\lab{svilchi1}
 \chi^{(1)}(\ga)\stackrel{\ga\to0}{=}&-{1\over2\ga^3}+{A_1\over\ga^2}+{A_2\over\ga}
 +A_3+\zeta(3)+\ord(\ga)\;,\\ \nonumber
 A_1&=-\left({11\over12}+{\Nf\over6N_c^3}\right)\;,\\ \nonumber
 A_2&=-{\Nf\over6N_c}\left({5\over3}+{13\over6N_c^2}\right)\;,\\ \nonumber
 A_3&=-{1\over4}\left[{395\over27}-2\zeta(3)-{11\pi^2\over18}+{\Nf\over N_c^3}
 \left({71\over27}-{\pi^2\over9}\right)\right]\;,
\end{align}
providing the right collinear behaviour for $k^2\gg k_0^2$, as explained in the next
section. Furthermore, by Eq.~(\ref{avlH}), the numerator in Eq.~(\ref{gup}) becomes
regular for $\ga=0$ too, where it takes the value $1+\ab H(1)=1-\ab\psi'(1)$, which
renormalizes the impact factors. This confirms the
importance of factorizing $H$ in the GGF in Eq.~(\ref{d:nuovaGGF}).

\section{NL$x$ resummed anomalous dimension\labe{s:NLrad}}

In order to determine the resummed anomalous dimension in NL$x$ approximation, we have
to check whether the NL$x$ BFKL equation is consistent with the RG equation. The running
of the coupling spoils the arguments of Sec.~\ref{ss:rad} and a new analysis is needed.

The choice of $k^2$ as the scale of $\as$ is suitable for that analysis. In fact, the
inhomogeneous BFKL equation for the unintegrated gluon density (\ref{Fbfkl}) reads
\begin{equation}\lab{FbfklNL}
 \om\left[\F_\om(\kk)-\F_\om^{(0)}(\kk)\right]={1\over b\ln{k^2\over\La^2}}\int\dif\kk'\;
 K(\kk,\kk')\,\F_\om(\kk')\quad,\quad b\dug{\pi\over N_c}b\;,
\end{equation}
where we have explicitly written the perturbative expression of the running coupling
$\ab(k^2)$ and the scale-invariant L$x$ + NL$x$ kernel $K=K^{(0)}+\ab K^{(1)}$. In
$\ga$-space, Eq.~(\ref{FbfklNL}) assumes a simpler form: with the definition
(\ref{MellC}) and by observing that $b\ln k^2/\La^2\to-b\de_\ga$ and
$K\to\chi=\chi^{(0)}+\ab\chi^{(1)}$, we obtain
\begin{equation}\lab{eqdifF}
 -b\om[\F_\om(\ga)-\F_\om^{(0)}(\ga)]'=\chi(\ga)\F_\om(\ga)\quad,\quad\F'\equiv\de_\ga\F\;,
\end{equation}
which is a first order linear differential equation for $\F_\om$ admitting the formal
solution~\cite{GrLeRy83}
\begin{subequations}\labe{solform}
\begin{align}
 \F_\om(\ga)&=\esp{-{X(\ga)\over b\om}}\int_\infty^\ga\dif\ga'\;\esp{X(\ga')\over b\om}
 \F_\om^{(0)}{}'(\ga')\;,\\
 \F_\om(\kk)&={1\over\pi k^2}\int\difg\;\left({k^2\over\La^2}\right)^\ga\F_\om(\ga)\;,
\end{align}
\end{subequations}
where $X$ is any primitive of $\chi$.

It has been argued~\cite{CoKw89,CaCi96b} that, in the limit $k^2\gg k_0^2$, the GGF
assume the factorized form
\begin{equation}\lab{Gfatt}
 \G_\om(\kk,\kk_0)=\fpe_\om(\kk)\gnp_\om(\kk_0)
 \times\left[1+\ord\({k_0^2\over k^2}\)\right]
\end{equation}
where $\fpe$ is purely perturbative while $\gnp$ is determined by the properties of non 
perturbative physics. After integration with the bare $\kk_0$-dependent gluon
distribution (Eq.~(\ref{fattDGNI})), the unintegrated gluon density $\F_\om(\kk)$
inherits the factorized form
\begin{equation}\lab{Ffatt}
 \F_\om(\kk)=\fpe_\om(\kk)\fnp_\om+\text{higher twist}\;.
\end{equation}
A comparison with Eqs.~(\ref{solform}) led us to identify the non perturbative factor
\begin{equation}\lab{d:Fnp}
 \fnp_\om\simeq{1\over2}\left(\int_{\infty-\ui\e}^0+\int_{\infty+\ui\e}^0\right)\dif\ga'\;
 \esp{X(\ga')\over b\om}\F_\om^{(0)}{}'(\ga')\;,
\end{equation}
where we have set the upper integration limit $\ga\simeq0$ (see below). The perturbative 
factor, on the other hand, carries the whole $\kk$-dependence of $\F_\om$ and is given
by
\begin{equation}\lab{d:Fpert}
 \fpe_\om(\kk)={1\over\pi k^2}\int\difg\;\left({k^2\over\La^2}\right)^\ga
\esp{-{X(\ga)\over b\om}}\;.
\end{equation}
In the anomalous dimension limit $k^2\gg\La^2$, Eq.~(\ref{d:Fpert}) can be evaluated by
the saddle point method, which yields~\cite{CaCi97}
\begin{equation}\lab{pusepert}
 \fpe_\om(\kk)\simeq{1\over k^2}\sqrt{b\om\over2\pi|\chi'(\gb)|}
 \exp\int^t\gb_\om(\tau)\,\dif\tau\qquad\(t\dug\ln{k^2\over\La^2}\)\;,
\end{equation}
where $\gb=\gb_\om(t)=\de_t[\gb t-X(\gb)/b\om]$ is the position of the saddle point
determined by the condition
\begin{equation}\lab{d:gabarra}
 {\om\over\ab(t)}=b\om t=\chi^{(0)}(\gb)+\ab\chi^{(1)}(\gb)\;.
\end{equation}
The representation (\ref{pusepert}) is valid only in the anomalous dimension regime
\begin{equation}\lab{regdimanom}
 b\om t\gg\chi_m\dug\underset{\ga\in]0,1[}{\Min} \chi(\ga)\;.
\end{equation}
If the above limitation were not satisfied, the representation (\ref{pusepert}) would
breaks down, due to large $\ga$-fluctuations.

Finally, we consider the singlet%
\footnote{In the NL$x$ approximation the solution of the
BFKL equation, taking contributions also from quark intermediate states, has to be
identified with the eigenvector of the singlet sector. In fact the anomalous dimension
in Eq.~(\ref{svdimanomNL}) corresponds to the larger eigenvalue of the singlet anomalous
dimension matrix.}
density (\ref{d:DGI})
\begin{align}\nonumber
 f_\om^{(+)}(Q^2)&=\int\difg\left({Q^2\over\La^2}\right)^\ga{\F_\om(\ga)\over\ga}\\
\lab{DGInl}&={1\over\gb}\sqrt{b\om\over2\pi|\chi'(\gb)|}\exp\int^{t_Q}\gb_\om(\tau)\,
 \dif\tau\qquad\(t_Q\dug\ln{Q^2\over\La^2}\)\;.
\end{align}
The coefficient function in front of the exponential term is simply a constant in the
$Q^2\to0$ ($\gb\to0$) limit and the singlet density satisfies
\begin{equation}\lab{evolDGI}
 {\dif\over\dif\ln Q^2}f_\om^{(+)}(Q^2)=\gb_\om(Q^2)f_\om^{(+)}(Q^2)\;,
\end{equation}
consistently with the RG equation (\ref{EDind}). Therefore, we interpret the function
$\gb_\om$ implicitly defined by (\ref{d:gabarra}) as the (highest eigenvalue of the)
resummed anomalous dimension in the singlet sector.

The structure of the NL$x$ eigenvalue function at coupling-scale and energy-scale $k^2$
is (cfr.\ Eqs.~(\ref{svilchi1},\ref{Dchienergia}))
\begin{align}\lab{struchi1}
 \chi^{(1)}(\ga)&={A_1\over\ga^2}+{A_2\over\ga}+A_3+\ord(\ga)
\end{align}
This imply that, up to 3-loops, the NL$x$ anomalous dimension has the expansion
\begin{align}\lab{d:gNL}
 \gb_\om(\ab)&=\gl\({\ab\over\om}\)+\ab\gnl\({\ab\over\om}\)
 \qquad\qquad\left(\gnl={\chi^{(1)}(\gl)\over-\chi^{(0)}{}'(\gl)}\right)\\ \lab{svdimanomNL}
 &={\ab\over\om}+\ab\left(A_1+A_2{\ab\over\om}+A_3{\ab^2\over\om^2}\right)+\ord(\ab^4)
\end{align}
which can be checked to be consistent with the known expression up to $\ord(\as^2)$.
It is apparent that, in order to recover
the expansion (\ref{svdimanomNL}), the double pole of $\chi^{(1)}$, and hence the shift
to scale $s_0=k^2$, plays a crucial role: if a cubic singularity were been present, the
anomalous dimension would assume the non realistic form $\gb\simeq\ab^{2/3}/\om^{1/3}$.

\section{Pomeron: perturbative versus non perturbative features\labe{s:psh}}

The study of the $\om$-singularities of the GGF in the NL$x$ approximation shows a
completely different scenario with respect to the L$x$ one. A first point concerns the
running of the coupling, which changes drastically the spectral properties of the
GGF. In particular, the regularization of $\as(Q^2)$ to values of $Q^2$ around and below 
the Landau pole $Q^2=\La^2$ affects the position and the strength of the singularities of
$\G_\om$~\cite{CaCi96b}.

The second point, to which we would address the reader's attention, concerns the nature
of those singularities. Note that both the perturbative and the non perturbative factors
in Eq.~(\ref{Gfatt}) carry an $\om$-dependence and could contribute to the singular
behaviour of the GGF.

The questions then arise:
\begin{itemize}
\item are both $\fpe_\om$ and $\gnp_\om$ singular in $\om$?
\item if so, which of them carries the rightmost $\om$-singularity?
\end{itemize}
The structure of the $\om$-singular points and their dependence on the particular
regularization of the IR coupling has been studied in some
models~\cite{CaCi96b,CiCoSa99b} whose evolution equation resembles in their main
features the BFKL running coupling one. It turns out that the rightmost singularity of
the GGF (the pomeron $\op$) is a really non perturbative phenomenon, which is very much
dependent on the behaviour of the strong coupling in the soft region $k^2\sim\La^2$. For
that reason, it should be related to the universal singurarity of soft physics, also.
\begin{figure}[hb!]
\centering
\includegraphics[width=10cm]{\fig 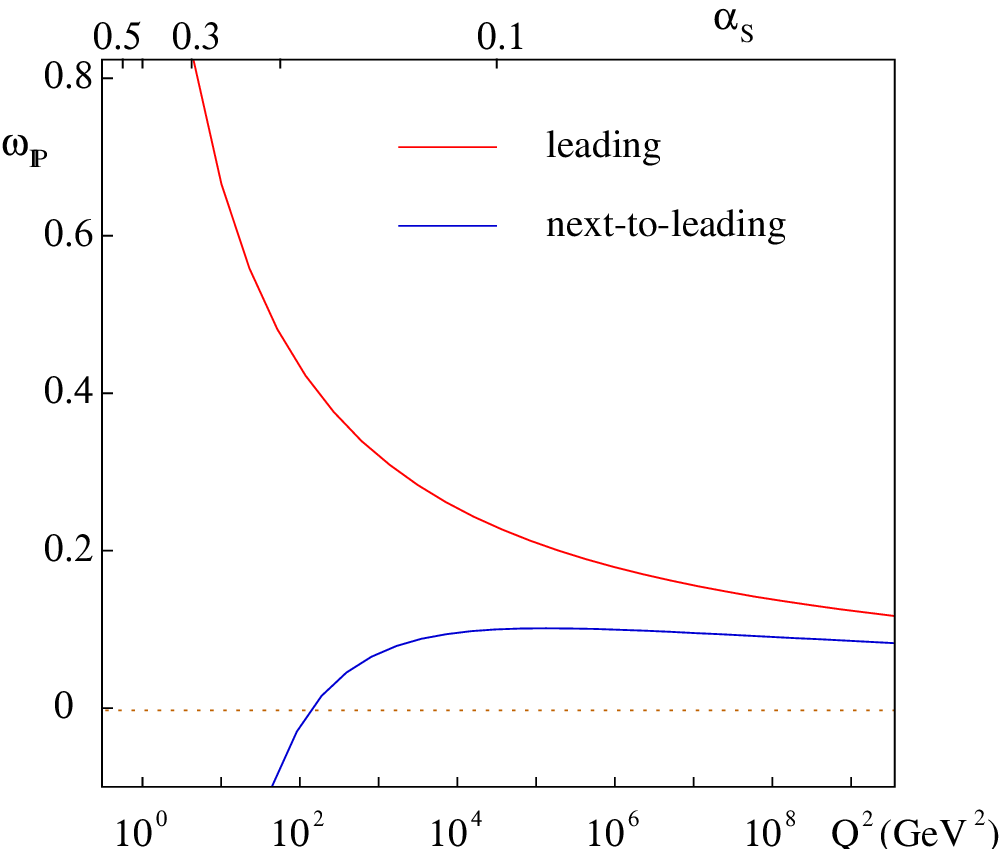}
\didascalia{The BFKL hard pomeron $\op(\as)$ in leading and next-to-leading
	approximation.\labe{f:pomBFKL}} 
\end{figure}

As far as the perturbative factor is concerned, the integral representation
(\ref{d:Fpert}) has an essential singularity at $\om=0$. However, the anomalous dimension 
representation (\ref{pusepert}), depending on $\om$ through $\gb_\om$, is singular in
correspondence of the branch cut $\op(\ab)$ of $\om\mapsto\gb_\om(\ab)$ given by
\begin{align}\nonumber
 \op(\ab)\dug&\ab\chi_m\simeq\ab[\chi^{(0)}\({1\over2}\)+\ab\chi^{(1)}\({1\over2}\)\\
 \lab{d:pomduro}=&\op^{(\text{L}x)}(\ab)[1-6.47\ab]\;.
\end{align}
Note that the value  of $\op(\ab)$ is not affected by the choice of the energy-scale
within the factorized class $s_0=k^pk_0^{1-p}$, since the shift $\De\chi^{(1)}(1/2)$ is
proportional to $\chi^{(0)}{}'(1/2)=0$ (cfr.\ Eq.~(\ref{Dchienergia})).

Due to the formal analogy of the definitions~(\ref{d:pomduro}) and (\ref{d:pomLx}),
$\op(\ab)$ is often called the {\em hard pomeron} singularity. To be fair, however, only 
the hard nature of $\op(\ab)$ is a fact. In fact we cannot guarantee it to be the
rightmost singularity of the GGF. In addition, there may be sound doubts it to be a true 
singularity. In fact, $\op(\ab)$ signals the breakdown of the anomalous dimension
representation (\ref{pusepert}) where the saddle point estimates does no more work.

Nevertheless, even if $\op(\ab)$ doesn't correspond to a singular behaviour of $\G_\om$,
but rather to a singularity of the anomalous dimension, it may be related to a
power-like behaviour in an intermediate small-$x$ moderate-$Q^2$ region. This assertion
is confirmed by studies of some simplified models \cite{CaCi96b,CiCoSa99b} one of which
we shall deal with in Chap.~\ref{c:cm}.

The last term in the RHS of Eq.~(\ref{d:pomduro}) emphasizes the large and negative
corrections stemming from the NL$x$ BFKL kernel. In Fig.~\ref{f:pomBFKL} you can see
that the maximum value of $\op(\ab)\simeq0.11$ is reached for $\ab\simeq0.08$ and in the 
range of HERA data $Q^2\simeq1\div10^4\GeV^2$, i.e., $\ab\simeq0.1\div0.4$, the
NL$x$ hard pomeron is very small and even negative, in deep disagreement with the data
(see Figs.~\ref{f:esponcresc}).
\begin{figure}[ht!]
\centering
\includegraphics[width=10cm]{\fig 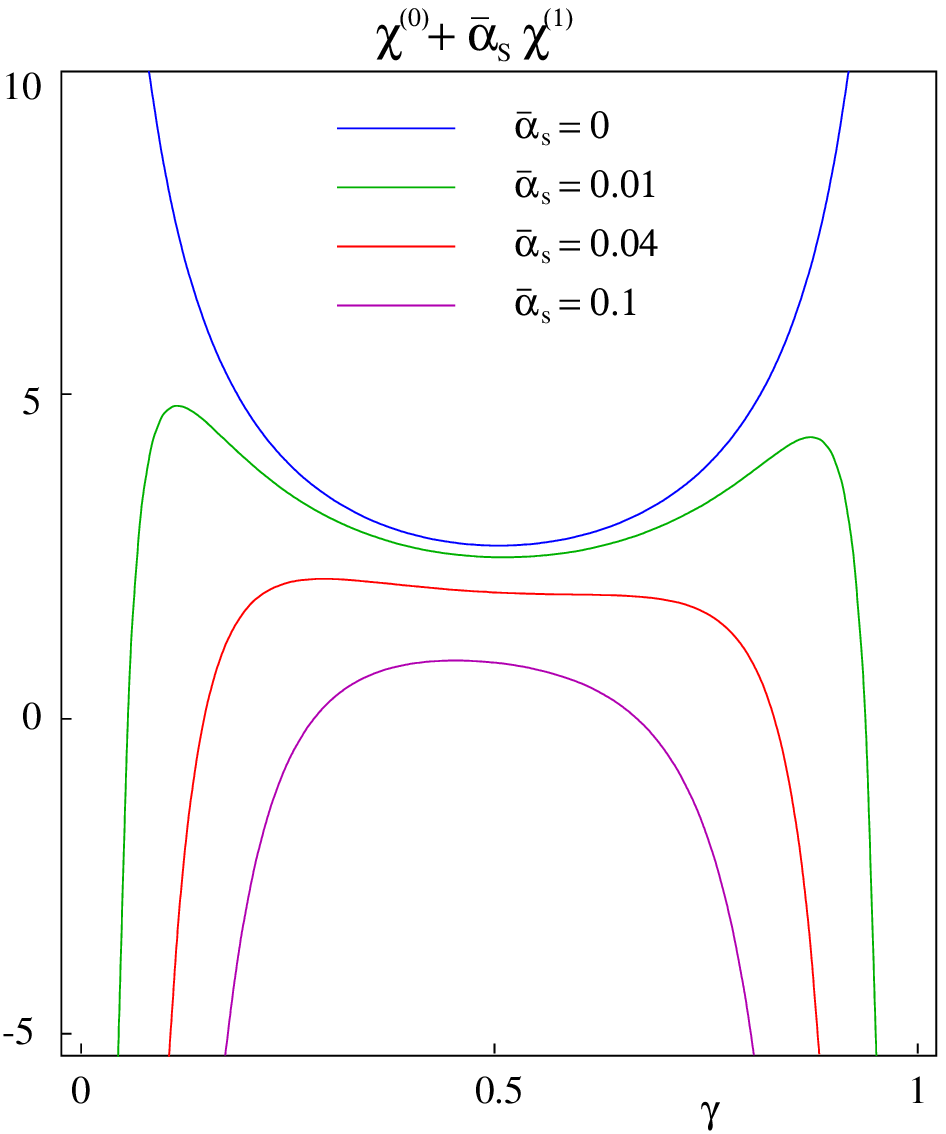}
\didascalia{The BFKL eigenvalue function for different values of $\ab$ and $\Nf=0$. It
is apparent the change in shape and the disappearance of the minimum for $\ab>0.04$.
\labe{f:avlNL}} 
\end{figure}

The situation is even worse if we look for the solution of Eq.~(\ref{d:pomduro}) for
finite (non vanishing) values of $\ab$:
\begin{equation}\lab{opduro}
 \op(\ab)=\underset{\ga\in]0,1[}{\Min} \ab[\chi^{(0)}(\ga)+\ab\chi^{(1)}(\ga)]\;.
\end{equation}
The plot of the eigenvalue function in Fig.~\ref{f:avlNL} shows that a stable minimum
for $\chi$ exists only for very small values of $\ab<0.04$, whereas for $\ab>0.04$ the
shape of $\chi$ is the opposite of the L$x$ eigenvalue. Among the physical consequences
of this pathological behaviour there is the loss of positivity for the total cross
section, as it was pointed out in Refs.~\cite{Ros98,Lev98}.

Does a reliable phenomenological analysis require the high energy $\ln1/x$ resummation
in NNL$x$ approximation? A simple arithmetic-geometric extrapolation based on
the L$x$ and NL$x$ computation-time would suggest something like $20\div100$ years of
work. But it seems that the whole $\ln1/x$ BFKL hierarchy has slow or even bad
convergence problems. Rather than a 3-loops calculation it would be better to add some
physical information, some consistency constraints with well established physics...

\chapter{Improvement of the small-$\boldsymbol{x}$ equation by RG analysis%
\labe{c:rga}}

After many years of calculations and expectations, the NL$x$ corrections to the kernel
of the BFKL equation turned out to be so large as to question the very meaning of the
high energy expansion (\ref{fatt1loop}), rising the compelling question of how to
improve it.

Exact higher order calculations would be for sure a formidable task, both for their size
and because they mix with unitarity effects~\cite{MuBaLiWu94} and thus cannot be
described within the BFKL equation alone. Furthermore, there seems to be serious
problems with the convergence of the kernel, as we will show in the next section.

The main issue is to understand the reasons of the pathological behaviour of the NL$x$
corrections. Just like for the problems concerning the subtraction of the leading terms
from the reducible contributions of the NL$x$ kernel, the discussion of the large NL$x$
correction is fundamentally related to the issue of the choice of scales.

\section{Origin of the double logarithms\labe{s:odl}}

In a high energy scattering process of two objects with transverse scales $\kk^2$ and
$\kk_0^2$, the L$x$ BFKL equation resums all the leading logarithmic terms of the cross
section of the form
\begin{equation}\lab{termdomi}
 \left(\ab\,K^{(0)}\,\ln{s\over s_0}\right)\quad,\quad n\geq0
\end{equation}
and we have already remarked that, in L$x$ approximation, the choice of $s_0$ is
immaterial.

In the NL$x$ approximation, the energy-scale is intimately connected to the form of the
NL$x$ kernel (Sec.~\ref{ss:nlBFKL}) and also to the structure of the GGF
(Sec.~\ref{s:wes}). However, there is not a particular scale which seems to be
preferable, rather there are various candidates according to the regime being
investigated. We have already discussed the collinear-safe scale $s_M$ and the
factorized Regge-motivated ones $kk_0$, $k^2$. Nevertheless, in the situation where
$k^2\gg k_0^2$, the DGLAP approach (Chap.~\ref{c:pthp}) tells us that each power of
$\ab$ is accompanied by a large $\ln k^2/k_0^2$ and that the correct scaling variable,
carrying the whole $s$-dependence, is given by $x\simeq k^2/s$. In practice, the cross
section has terms like
\begin{equation}\lab{singlog}
 {1\over k^2}\left(\ab\,\ln{k^2\over k_0^2}\,\ln{s\over k^2}\right)\qquad (k^2\gg k_0^2)\;.
\end{equation}
Now, if we rewrite the general term (\ref{singlog}) by adopting $kk_0$ as the scale of
the energy, double logarithms ($\ln^2$) of the transverse momenta appear in the
perturbative series:
\begin{equation}\lab{doppilog}
 (\ref{singlog})={1\over k^2}\sum_{m=0}^n{n\choose m}(-2)^{-m}\left(\ab\ln^2
 {k^2\over k_0^2}\right)^m\left(\ab\ln{s\over kk_0}\ln{k^2\over k_0^2}\right)^{n-m}\;.
\end{equation}
For instance, the $\ord(\ab^2)$ cross section at L$x$ level reads (see
Eq.~(\ref{esp2loop}))
\begin{equation}\lab{sig2loopLx}
 {\dif\si^{(2,\text{L}x)}_{\pa\pb}\over\dif\kk\,\dif\kk_0}=\ab^2h_{\pa}^{(0)}(\kk)
 h_{\pb}^{(0)}(\kk_0)\ln^2{s\over s_0}K^{(0)}{}^2(\kk,\kk_0)\;.
\end{equation}
In the collinear limit $\kk\gg\kk_0$ we have $s_0=k^2$ and
\begin{align}\nonumber
 K^{(0)}{}^2(\kk,\kk_0)&\simeq\int\dif\kk'\;{1\over\pi(\kk-\kk')^2}
 {1\over\pi(\kk'-\kk_0)^2}\\ \lab{K0quadcoll}
&\simeq{1\over k^2}\left[\ln{k^2\over k_0^2}+\ord(1)\right]\;,
\end{align}
which shows the single logarithmic term and, in Mellin $\ga$-space, gives rise to a
double pole $\ab^2/\ga^2$.

By shifting the energy-scale $s_0=k^2\to kk_0$, we get
\begin{equation}\lab{doplogalf2}
 K^{(0)}{}^2\ln^2{s\over k^2}\simeq{1\over k^2}
 \ln{k^2\over k_0^2}\ln^2{s\over k^2}\;\longrightarrow\;{1\over k^2}
 \ln{k^2\over k_0^2}\ln^2{s\over kk_0}-{1\over k^2}
 \ln^2{k^2\over k_0^2}\ln{s\over kk_0}+\text{NNL}x\;.
\end{equation}
The L$x$ term, i.e., the coefficient of $\ln^2s$, remains unchanged whilst the
NL$x$ term acquires the double logarithmic contribution $K^{(0)}\ln^2({k^2/k_0^2})$
whose eigenvalue gives rise to a cubic pole $\ab^2/\ga^3$.

In general, the appropriate DGLAP single logs corresponds to $\ab^n/\ga^k$ and
$\ab^n/(1-\ga)^k$ poles with $k\leq n+1$. On the contrary, the presence of double logs
involves poles with $n+1<k\leq2n+1$.

We conclude with Salam~\cite{Sa98} that the $\ab$-expansion of the kernel in conjunction
with a non collinear-safe energy scale, generate double logs, i.e., strong collinear
singularities, in the lower order kernels.

\section{A toy kernel\labe{s:giocattolo}}

Now, with the aid of a toy model~\cite{Sa98}, we want to show that those strong
collinear singularities are responsible of the instability of the finite order
$\ab\ln1/x$ truncation. We can get an idea of that mechanism by considering the Lund
model~\cite{AnGuSa96}, consisting in a small-$x$ evolution equation for the unintegrated
gluon density --- derived in the context of the linked-dipole-chain (LCD) model ---
similar to the BFKL one but differing in the collinear region. The relevant equation can
be written
\begin{equation}\lab{eqlund}
 \om\F_\om=\F^{(0)}_\om+\K_\om\F_\om\;,
\end{equation}
where $\om$ is the variable conjugate to $x=k^2/s$ --- so that we are considering
$s_0=k^2$ --- and the kernel $\K_\om$ admits the usual power-like eigenfunctions
(\ref{afzK0}) and its eigenvalue is
\begin{equation}\lab{avllund}
 \ab\chi(\ga,\om)=\ab[2\psi(1)-\psi(\ga)-\psi(1-\ga+\om)]\;.
\end{equation}
In complete analogy with the method of solution explained in the context of the BFKL
equation, the anomalous dimension $\gb_\om(\ab)$ is determined by the equation
\begin{equation}\lab{dimlund}
 \om=\ab\chi(\gb,\om)
\end{equation}
or, equivalently,
\begin{equation}\lab{chiefflund}
 \om=\ab\chi_{\eff}(\gb,\ab)\;,
\end{equation}
where $\chi_{\eff}(\ga,\ab)$ is implicitly defined by the solution of Eq.~(\ref{dimlund})
once $\ga$ and $\ab$ have been fixed.

In the DGLAP limit, corresponding to the region close to $\ga=0$, the effective
eigenvalue function has only simple poles $1/\ga$ to all orders in $\ab$:
\begin{equation}\lab{sveff1}
 \chi_{\eff}(\ga,\ab)={1\over\ga}\;[1+\ord(\ab)]+\ord(\ga^0)\qquad(s_0=k^2)\;.
\end{equation}
In the opposite DGLAP limit, $k_0^2\gg k^2$, the relevant region of $\ga$ is close to
$\ga=1$. Making the transformation to the relevant Bjorken scale $s_0=k_0^2$, using
$\ga\to\ga+\om$ (cfr.\ Secs.~\ref{ss:enlk} and \ref{s:isxe}), there is only the pole
$1/(1-\ga)$:
\begin{equation}\lab{sveff2}
 \chi_{\eff}(\ga,\ab)={1\over1-\ga}\;[1+\ord(\ab)]+\ord\((1-\ga)^0\)\qquad(s_0=k_0^2)\;.
\end{equation}
So one is completely free of double transverse logarithms in both DGLAP limits.

In terms of the symmetric energy-scale $s_0=kk_0$, Eq.~(\ref{avllund}) becomes
($\ga\to\ga+\ho$)
\begin{equation}\lab{avl0simm}
 \chi(\ga,\om)=2\psi(1)-\psi(\ga+\ho)-\psi(1-\ga+\ho)\;.
\end{equation}
The $\ab$-expansion of the ensuing effective eigenvalue function
\begin{subequations}\labe{svilsimm}
\begin{align}\chi_{\eff}(\ga,\ab)&=\sum_{n=0}^\infty\ab^n\chi_{\eff}^{(n)}(\ga)\;,\\
 \chi_{\eff}^{(0)}(\ga)&=2\psi(1)-\psi(\ga)-\psi(1-\ga)\;,\\
 \chi_{\eff}^{(1)}(\ga)&=-{1\over2}\chi_{\eff}^{(0)}(\ga)[\psi'(\ga)+\psi'(1-\ga)]\;,\\
 \chi_{\eff}^{(2)}(\ga)&=-{1\over2}\chi_{\eff}^{(1)}(\ga)[\psi'(\ga)+\psi'(1-\ga)]-{1\over8}
 \chi_{\eff}^{(0)}(\ga)^2[\psi''(\ga)+\psi''(1-\ga)]\;,
\end{align}
\end{subequations}
signals the presence of double logarithms from NL$x$ level on, as one can see from
\begin{subequations}\labe{poligrossi}
\begin{align}
 \chi_{\eff}^{(1)}(\ga)&=-{1\over2\ga^3}+\ord\({1\over\ga}\)+\{\ga\lra1-\ga\}\;,\\
 \chi_{\eff}^{(2)}(\ga)&=+{1\over4\ga^5}+\ord\({1\over\ga^3}\)+\{\ga\lra1-\ga\}\;.
\end{align}
\end{subequations}
Note that the most divergent part of $\chi_{\eff}^{(1)}$ in the regions of $\ga$ close to 0 and
close to 1 is exactly the same of the NL$x$ BFKL eigenvalue (at scale $s_0=kk_0$)!
Therefore, as previously remarked, the cubic poles of Eq.~(\ref{d:avlNL}) are a
consequence or the fact that the high-energy-factorization scale $kk_0$ is not collinear 
safe, i.e., is different from the Bjorken scale $k_>^2$.
\begin{figure}[hb!]
\begin{center}
\includegraphics[width=120mm]{\fig 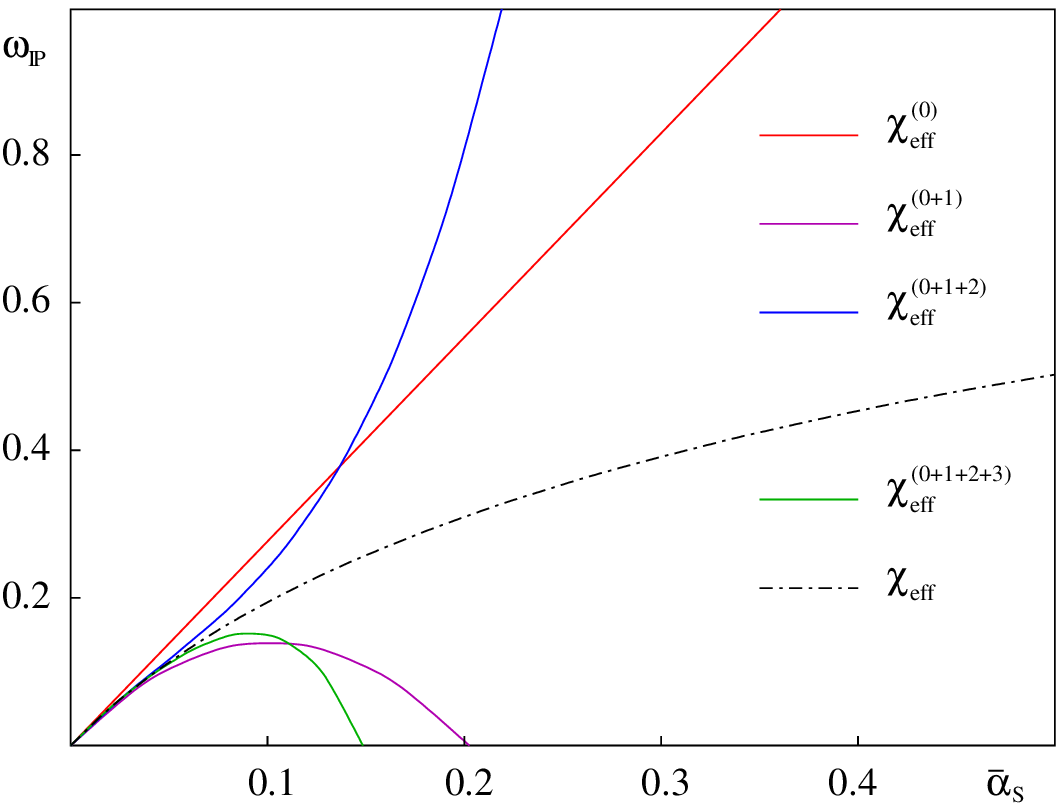}
    \didascalia{The toy-kernel pomeron singlularity as a function of $\ab$,
      at leading, NL$x$, NNL$x$ , N$^3$L$x$ and all orders.\labe{f:toy}}
  \end{center}
\end{figure}

If we look at the pomeron singularity $\op(\ab)=\ab\chi_{\eff}(1/2)$ of the toy-kernel in
Fig.~\ref{f:toy}, we observe an alternate behaviour of the perturbative estimates, whose
convergence is rather slow and limited to very low values of $\ab$ in the lower
approximations. For realistic values of $\ab$, say 0.2, the fixed-order expansion are
quite unreliable, and going from NL$x$ to NNL$x$ does not bring one any closer to the
all-order result except for $\ab\lesssim0.1$.

To summarize, what one learns from this toy kernel is that the double transverse
logarithms lead to a very poor convergence of the BFKL kernel at subleading orders and
in particular the features of the analytic structure of the kernel and of the physical
observables after the resummation are quite unrelated to the NL$x$ ones. Therefore,
essential ingredients of any improvement of the BFKL approach should be the correct
treatment of the collinear behaviour, as predicted by the RG, and the resummation of the 
corresponding collinear singularities to all orders.

\section{Improved small-$\boldsymbol x$ kernel\labe{s:isxe}}

In small-$x$ physics, Regge theory and the RG come to a sort of clash, which provides
non trivial consistency requirements. The general form of the RG improved small-$x$
kernel is constrained by {\it(i)} the exact L$x$ and NL$x$ calculation and {\it(ii)}
the collinear singulatity structure which we are going to specify. The latter requires to 
study the kernel in the collinear limits $k\gg k_0$ and $k\ll k_0$ where the relevant
energy-scales are $k^2$ and $k_0^2$ respectively. It is also useful to consider the
symmetric high-energy-factorization scale $kk_0$.

It is easy to see that, under energy-scale transformations of that kind, the GGF in
Eq.~(\ref{fatt1loop}) transforms according to
\begin{equation}\lab{trasimG}
 \Gs_\om(\kk,\kk_0)=\left({k_0\over k}\right)^\om\Gu_\om(\kk,\kk_0)
 =\left({k\over k_0}\right)^\om\Gl_\om(\kk,\kk_0)\;,
\end{equation}
where the symmetric ``$s$'', upper ``$u$'' and lower ``$l$'' suffixes refer to the
energy-scales $kk_0$, $k^2$ and $k_0^2$. The similarity transformations (\ref{trasimG})
apply straightforwardly to the small-$x$ kernel defined by
\begin{equation}\lab{d:sxK}
\G_\om=[\om-\K_\om]^{-1}\;,
\end{equation}
which we should consider $\om$-dependent~\cite{CiCo98b}:
\begin{equation}\lab{trasimK}
 \Ks_\om(\kk,\kk_0)=\left({k_0\over k}\right)^\om\Ku_\om(\kk,\kk_0)
 =\left({k\over k_0}\right)^\om\Kl_\om(\kk,\kk_0)\;.
\end{equation}
Note that in the definition (\ref{d:sxK}) of our improved small-$x$ kernel, the GGF does
not contain any impact kernel.

We have already expressed the opinion that $\ab$ is not a suitable expansion parameter
for the kernel. The introduction of the $\om$-dependence of $\K_\om$ will reveal crucial 
in order to derive a novel small-$x$ expansion which automatically resums the collinear
singularities.

From a perturbative point of view, however, we observe that, in general, the
renormalized kernel $\K_\om(\kk,\kk';\mu^2,\ab(\mu^2))$ for non vanishing values of
its arguments $\kk$ and $\kk'$, is IR finite (see Secs.~\ref{ss:lxBFKLe} and
\ref{ss:enlk}) and RG invariant, so that it can be expanded as a power series in
$\ab(k^2)$ with scale-invariant coefficient kernels~\cite{CiCo98b}
\begin{equation}\lab{serieKom}
 \K_\om(\kk,\kk';\mu^2,\ab(\mu^2))=\sum_{n=0}^\infty\left[\ab(k^2)\right]^{n+1}
 K_\om^{(n)}(\kk,\kk')\;.
\end{equation}
In other words, the only non scale-invariant source is the running coupling, which in
Eq.~(\ref{serieKom}) has been evaluated, for simplicity, at scale $k^2$. Different
coupling-scales correspond to different, but always scale-invariant, coefficient kernels.

\section{Form of the collinear singularities\labe{s:fcs}}

We have shown in Sec.~\ref{s:NLrad} (see also Refs.~\cite{CoKw89,CaCi96b}) that the BFKL
equation satisfies RG factorization in an asymptotic way. The asymptotic form of the GGF
for $t\dug\ln k^2/\La^2\gg t_0\dug\ln k_0^2/\La^2$ is given by
\begin{equation}\lab{gasint}
 k^2\Gu_\om(\kk,\kk_0)\simeq C_\om(\ab(t))\Big[\exp\int_{t_0}^t\ga^+_\om
 (\ab(\tau))\;\dif\tau\Big]\gnp_\om(t_0)\;,
\end{equation}
where $\ga^+_\om$ is the larger eigenvalue of the singlet anomalous dimension
matrix, defined by the saddle point condition (\ref{d:gabarra}).

Therefore, the RG invariant kernel in the LHS of Eq.~(\ref{serieKom}) acquires collinear 
singurarities for $k'/k\to0$ ($k/k'\to0$), which corresponds to strong ordering of the
transverse momenta in the direction of the ``upper'' Bjorken scale $k^2$ (``lower'' scale
$k_0^2$). Such singularities are due to the non singular part $\tilde{\ga}$ of the
singlet anomalous dimension~(\ref{svdimanomNL}) 
which, neglecting the (small) $\pq\bar{\pq}$ contribution, is
\begin{align}\lab{nsdimanom}
 \tilde{\ga}_\om&=\gagg_\om-{\ab\over\om}=\ab A_1(\om)+\ab^2A_2(\om)+\cdots\;,\\
 A_1(\om)&=-{11\over12}+\ord(\om)\quad,\quad A_2(\om)=0+\ord(\om)\;,\nonumber
\end{align}
the singular part being taken into account by the BFKL iteration itself.
It follows~\cite{CiCo98b} that, for $k\gg k'$,
\begin{align}\lab{kupper}
 \Ku_{\om}(\kk,\kk')&\simeq{\ab(t)\over k^2}\;\exp\int_{t'}^t
 \tilde{\ga}_\om(\as(\tau))\;\dif\tau\qquad(t\gg t')\\ \nonumber
 &={\ab(t)\over k^2}\left(1-b\ab(t)\ln{k^2\over k'{}^2}\right)^
 {-{A_1(\om)\over b}}\;.
\end{align}

Expanding Eq.~(\ref{kupper}) in $\ab(t)$ and comparing with the
general definition (\ref{serieKom}), leads to the identification of the
kernels $\overset{\text{\tiny u}}{K}{}_\om^{(n)}$ in the collinear limit, whose eigenvalue
functions turn out to have the singularities
\begin{equation}
 \chiu_\om^{(n)}(\ga)\simeq{1\cdot A_1(A_1+b)\cdots(A_1+(n-1)b)\over
 \ga^{n+1}}\qquad(\ga\ll1)\;, \label{chiupper}
\end{equation}
which correspond to single logarithmic scaling violations for $k\gg k_0$.  A similar
reasoning yields the collinear behavior of $\Kl_\om$ in the opposite strong
ordering region $k'\gg k$
\begin{align}\lab{klower}
 \Kl_{\om}(\kk,\kk')&\simeq{\ab(t')\over k'{}^2}\;\exp\int_t^{t'}
 \tilde{\ga}_\om(\as(\tau))\;\dif\tau\qquad(t'\gg t)\\ \nonumber
 &={\ab(t)\over k'{}^2}\left(1-b\ab(t)\ln{k'{}^2\over k^2}\right)^
 {{A_1(\om)\over b}-1}
\end{align}
and to the singularities
\begin{equation}
 \chil_\om^{(n)}(\ga)\simeq{1\cdot(A_1-b)\cdots(A_1-nb)\over
 (1-\ga)^{n+1}}\qquad(1-\ga\ll1)\;.\label{chilower}
\end{equation}

However, the similarity relations (\ref{trasimK}) connect the kernels $\Ku$ and
$\Kl$. Therefore $\Ku$ has the singularities (\ref{chilower}) shifted at $\ga=1+\om$
also, and similarly $\Kl$ has the singularities (\ref{chiupper}) shifted at
$\ga=-\om$. As a consequence, the symmetric kernel $\Ks$ --- for the energy-scale
$s_0=kk_0$ --- has both kinds of singularities shifted by $\pm\om/2$:
\begin{subequations}\labe{chin}
\begin{align}
 \chis_\om^{(n)}(\ga)&\simeq{1\cdot A_1(A_1+b)\cdots(A_1+(n-1)b)\over
 (\ga+\ho)^{n+1}}\qquad(\ga\simeq-\ho)\;,\\
 &\simeq{1\cdot(A_1-b)(A_1-2b)\cdots(A_1-nb)\over
 (1-\ga+\ho)^{n+1}}\qquad(\ga\simeq1+\ho)\;.
\end{align}
\end{subequations}
Note the $b$-dependent asymmetry of the singularities in Eq.~(\ref{chin}) under the
$\ga\lra1-\ga$ transformation. It is due to the fact that the expansion
(\ref{serieKom}) involves $\ab(t)$ (and not $\ab(t')$). Of course, the kernel $\K_{\om}$
itself must be symmetric under $t\lra t'$ exchange, so that expressing $\ab(t')$
in terms of $\ab(t)$
\begin{equation}
 \ab(t')={\ab(t)\over1-b\ab(t)\ln\ds{k^2\over k'{}^2}}
\end{equation}
leads to the symmetry constraints (in the following we omit the suffix ``$s$'')
\begin{equation}\lab{vincolo}
 \chi_\om^{(n)}(\ga)=\sum_{m\leq n}{n\choose m}(-b\de_{\ga})^{n-m}\chi_\om^{(m)}(1-\ga)\;.
\end{equation}
It is straightforward to check by the binomial identity
\begin{equation}
 {r+n\choose n}=\sum_{m=0}^n{r\choose m}{n\choose m}
\end{equation}
that the symmetry constraints (\ref{vincolo}) are indeed satisfied by
Eq.~(\ref{chin}). In particular we must have
\begin{equation}\lab{simmetria}
 \chi_\om^{(0)}(1-\ga)=\chi_\om^{(0)}(\ga)\quad,\quad
 \chi_\om^{(1)}(1-\ga)=\chi_\om^{(1)}(\ga)+b\chi_\om^{(0)}{}'(\ga)\;,
\end{equation}
showing that the antisymmetric part of $\achi_\om^{(1)}(\ga)$ is
$-{b\over2}\achi_\om^{(0)}{}'(\ga)$.

The coefficient kernels $K_\om^{(n)}$ take up collinear singularities not only from the
non singular part of the gluon anomalous dimension $\tilde{\ga}^{\pg\pg}$, but also from
$\pq\bar{\pq}$ states which are coupled to it in the one-loop gluon/quark-sea anomalous
dimension matrix
\begin{equation}\lab{andimmat}
 \tilde{\gga}_\om\dug\gga_\om-{\ab\over\om}{0\quad0\choose\ts{C_F\over C_A}
 \;1}=\ab\boldsymbol{A}(\om)\;.
\end{equation}
Although the inclusion of the two-channel evolution (\ref{andimmat}) involves the
collinear problem conceptually, the numerical effect of the quark-sea contribution to
the gluon anomalous dimension is pretty small~\cite{CaCi97,CiCoSa99}. Therefore, in the
following, we will restrict our analysis to the gluon sector only.

\subsection{Form of the leading coefficient kernel}

Having determined the RG constraints, we are going to develop a procedure for defining
the improved small-$x$ kernel at NL$x$ accuracy. The improved coefficient kernels
$K_\om^{(n)}$ are constructed by requiring that
\begin{itemize}
\item[{\it(1)}] the Green's function $\G_{\om}$ reproduce the known NL$x$ calculations;
\item[{\it(2)}] the collinear singularities be as in Eq.~(\ref{chin}).
\end{itemize}

In order to implement condition {\it(1)} we have first to relate the $\om$-dependent
formulation of $\G_{\om}$ in Eq.(\ref{d:sxK}) to the customary expression of the BFKL
kernel at NL$x$ level
\begin{equation}\lab{ksuom}
 \K=\ab K^{(0)}+\ab^2K^{(1)}+\cdots\;.
\end{equation}
The $\om$-dependent formulation of Eq.~(\ref{serieKom}) yields instead the NL$x$ expansion
\begin{align}\lab{komsuom}
 \K_\om&=\ab K_0^{(0)}+\ab\om K^{(0,1)}+\ab^2K_0^{(1)}+\cdots\;,\\
 &K_\om^{(n)}\dug K_0^{(n)}+\om K^{(n,1)}+\cdots\,, \nonumber
\end{align}
which is actually more general than Eq.~(\ref{ksuom}) because the $\ab\om$ term, coming
from the $\om$-expansion of $K_\om^{(0)}$, is a possible NL$x$ contribution too.

Now it turns out that, at NL$x$ level, the formulation (\ref{komsuom}) reduces to the
one in (\ref{ksuom}), provided the impact kernels $H_L$ and $H_R$ of
Eq.~(\ref{d:nuovaGGF}) are taken into account~\cite{CiCoSa99}. In fact, by using the expansion
(\ref{komsuom}) and simple operator identities, we can write
\begin{equation}\lab{operatori}
 [\om-\K_\om]^{-1}=\left(1-\ab K^{(0,1)}\right)^{-{1\over2}}
 \left[\om-(\ab K^{(0)}+\ab^2K^{(1)}+\cdots)\right]^{-1}
 \left(1-\ab K^{(0,1)}\right)^{-{1\over2}}
\end{equation}
provided we set
\begin{subequations}
\begin{align}\lab{identifico0}
 K^{(0)}&=K_0^{(0)}\;,\\ \lab{identifico1}
 K^{(1)}&=K_0^{(1)}+K_0^{(0)}K^{(0,1)}\;.
\end{align}
\end{subequations}
Eqs.~(\ref{operatori}) and (\ref{d:nuovaGGF}) show that the two formulations above
differ by just a redefinition of the impact kernels, while Eq.~(\ref{identifico1})
means that $K_0^{(1)}$ is given by $K^{(1)}$, after subtraction of the term already
accounted for in the $\om$-dependence of $K_\om^{(0)}$.

The simplest way to define a L$x$ improved coefficient kernel fulfilling
Eqs.~(\ref{identifico0}) and (\ref{chin})$|_{n=0}$ is to displace the poles of
$\achi^{(0)}_\om$ in Eq.~(\ref{avlK0m0}) by shifting the arguments of the $\psi$-functions,
as suggested in Ref.~\cite{Sa98}:
\begin{align}\lab{lund}
 \chi_\om^{(0)}(\ga)&=2\psi(1)-\psi(\ga+\ho)-\psi(1-\ga+\ho)\\
 &=\chi^{(0)}(\ga)-{1\over2}\om{\pi^2\over\sin^2\pi\ga}+\cdots\;.\nonumber
\end{align}
The kernel $K_\om^{(0)}$, corresponding to Eq.~(\ref{lund}) is that occurring in the Lund
model~\cite{AnGuSa96} and is given by
\begin{equation}\lab{soglia}
 K_\om^{(0)}(\kk,\kk')=K^{(0)}(\kk,\kk')\left({k_<\over k_>}\right)^{\om}\;,
\end{equation}
where $k_>\dug\Max(k,k')$ and $k_<\dug\Min(k,k')$.  It is thus related to the
customary leading kernel $K^{(0)}$ by the ``threshold factor'' $(k_</k_>)^{\om}$. This
means that the $s$-dependence provided by its inverse Mellin transform is~\cite{CiCoSa99}
\begin{equation}\lab{sogliola}
 K^{(0)}(s;\kk,\kk')\equiv\int{\dif\om\over2\pi\ui}\left({s\over kk'}\right)^{\om}
 {1\over\om}K_\om^{(0)}(\kk,\kk')=K^{(0)}(\kk,\kk')\,\Th(s-k_>^2)\;.
\end{equation}

Can one justify the form of the kernel (\ref{soglia}) ``a priori''?  From the point of
view of the RG improved equation, any kernel which ($i$) reduces to $K^{(0)}$ in the
$\om\to0$ limit and ($ii$) has the leading simple poles of Eq.~(\ref{chin}) for $n=0$,
is an acceptable starting point. An alternative choice of this kind will differ from
$K_\om^{(0)}$ by a NL$x$ kernel without $\ga=0$ or $\ga=1$ singularities. The resulting
ambiguity can thus be reabsorbed by a proper subtraction in the NL$x$ coefficient
kernel.

Nevertheless, the threshold interpretation of Eqs.~(\ref{soglia}) and (\ref{sogliola})
is appealing. For instance, the first iteration of such a kernel provides the expression
\begin{align}\lab{itero}
 K^{(0)\,2}(s;\ku,\kd)=&\int{\dif\om\over2\pi\ui}\left({s\over k_1k_2}\right)^{\om}
 \left({1\over\om}K_\om^{(0)}\right)^2\\
 =&\int\dif\kk\;K^{(0)}(\ku,\kk)\left(\ln{s\over k_1k_2}-
 \eta(k_1,k)-\eta(k_2,k)\right)K^{(0)}(\kk,\kd)\nonumber
\end{align}
where
\begin{equation}
 \cosh\eta(k_i,k)\equiv{k^2+k_i^2\over2kk_i}\,. \label{eta}
\end{equation}
The threshold condition implied by Eq.~(\ref{itero})
\begin{equation}
 {s\over2k_1k_2}=\cosh\eta>\cosh\(\eta(k_1,k)+\eta(k,k_2)\) \label{toller}
\end{equation}
is reminiscent of phase space in Toller variables~\cite{Tol65} and may be
regarded as an alternative way of stating coherence effects~\cite{CCFM88}, as implied
in the original version of the Lund model itself.

Whether or not such hints will eventually provide a more direct justification of
$K_\om^{(0)}$, the fact remains that Eq.~(\ref{lund}) resums the $\om$-dependence of the
$\ga$-singularities, and thus provides the correct singularities of the scale-dependent
terms of the NL kernel. Therefore, it is a good starting point, yielding NL
contributions which are smoother than those in the $\as(t)$-expansion, as we now
discuss.

\subsection{Form of the next-to-leading contribution}\label{ss:fnlc}

The NL$x$ improved coefficient kernel is constrained by Eq.~(\ref{chin})$|_{n=1}$ and by
Eq.~(\ref{identifico1}) which yields the $\om=0$ limit of the eigenvalue function
\begin{equation}\lab{chiom0}
 \chi_{\om=0}^{(1)}(\ga)=\chi^{(1)}(\ga)+{1\over2}\chi^{(0)}(\ga)
 {\pi^2\over\sin^2\pi\ga}\;.
\end{equation}

The subtraction term so obtained is important because it has cubic poles at $\ga=0,1$
which cancel the corresponding ones occurring in $\chi^{(1)}(\ga)$. Furthermore, the
impact kernels of Eq.~(\ref{operatori}) have quadratic poles which similarly
account for the ones occurring in $H_L$ and $H_R$. This means that the
remaining contributions are, in both cases, much smoother in the $\om$-dependent
formulation.

In order to implement condition {\it(2)} on $\achi_\om^{(1)}$, we note that the $\om=0$
limit (\ref{chiom0}) still contains double and single poles at $\ga=0,1$, which should
be shifted according to Eq.~(\ref{chin}). By neglecting the (small) $\pq\bar{\pq}$
contributions, the explicit form of Eq.~(\ref{chiom0}), following from
Eq.~(\ref{avlglu}) for the energy-scale $s_0=kk_0$, is~\cite{CiCoSa99}
\begin{align}
 \chi_{\om=0}^{(1)}(\ga)=&-{1\over2}\left({11\over12}\(\chi^{(0)\,2}(\ga)+\chi^{(0)}{}'(\ga)\)
 \right)+\left[-{1\over4}\,\chi^{(0)}{}''(\ga)+{1\over2}
 \chi^{(0)}(\ga){\pi^2\over\sin^2\pi\ga}\right]+\nonumber\\
&-{1\over4}\left\{\!
 \left({\pi\over\sin\pi\ga}\right)^2\!\!{\cos\pi\ga\over3(1-2\ga)}\left(11+
 {\ga(1-\ga)\over(1+2\ga)(3-2\ga)}\right)\right\}+\nonumber\\ \lab{chi1}
&+\left({67\over36}-{\pi^2\over12}\right)\!
 \chi^{(0)}(\ga)+{3\over2}\,\zeta(3)+{\pi^3\over4\sin\pi\ga}-\Phi(\ga)\;,
\end{align}
where $\Phi(\ga)$ has been defined in Eq.~(\ref{d:Phi}).  Here we have singled out some
singular terms which have a natural physical interpretation, namely the running coupling
terms (in round brackets), the energy-scale-dependent terms (in square brackets) and the
collinear terms (in curly brackets).

The running coupling terms have a double pole at $\ga=1$ only, and account for the
asymmetric part of $\chi^{(1)}$ (given by $-{b\over2}\chi^{(0)}{}'$) which provides the
$b$-dependent double pole on $\achi_\om^{(1)}$ in Eq.~(\ref{chin}). The collinear terms
have symmetric double poles with residue $A_1(\om=0)$, in accordance with
Eq.~(\ref{chin}) also. Both types of singularities can be shifted by adding a NNL$x$ term,
vanishing in the $\om=0$ limit, which we take to be
\begin{equation}\lab{shift}
 A_1(\om)\psi'(\ga+\ts{\om\over2})-A_1(0)\psi'(\ga)+\(A_1(\om)-b\)
 \psi'(1-\ga+\ts{\om\over2})-\(A_1(0)-b\)\psi'(1-\ga)\;.
\end{equation}
This term incorporates the $\om$-dependence of the one-loop anomalous dimension
(\ref{nsdimanom}) too.

The energy-scale-dependent term in square brackets contains the subtraction
(\ref{chiom0}) and has, therefore, simple poles at $\ga=0,1$ only, which we can shift by
adding the contribution%
\footnote{Of course, such simple poles, which are dependent on the choice
   (\ref{lund}) of $\achi_\om^{(0)}$, do not occur --- by construction --- in
   the NL$x$ eigenvalue function $\chi^{(1)}(\ga)$. They are just part of the
   NNL$x$ ambiguity of our resummation scheme, whose size is evaluated in
   Sec.~\ref{ss:reschem}.}
\begin{equation}\lab{add}
 {\pi^2\over6}\(\chi_0^{\om}(\ga)-\chi_0(\ga)\)\,.
\end{equation}
By then collecting Eqs.~(\ref{chiom0}), (\ref{shift}) and (\ref{add}) we obtain the
final eigenvalue function
\begin{equation}\lab{chi1om}
 \chi_1^{\om}(\ga)\equiv\tilde{\chi}_1(\ga)+A_1(\om)\psi'(\ga+\ho)
 +\(A_1(\om)-b\)\psi'(1-\ga+\ho)+{\pi^2\over6}\chi_0^{\om}(\ga)\;,
\end{equation}
where
\begin{equation}
 \tilde{\chi}^{(1)}(\ga)\equiv\chi^{(1)}(\ga)+{1\over2}\chi^{(0)}(\ga)
 {\pi^2\over\sin^2\pi\ga}-{\pi^2\over6}\chi^{(0)}(\ga)-A_1(0)\psi'(\ga)
 -\(A_1(0)-b\)\psi'(1-\ga)      \label{chitilde}
\end{equation}
is a symmetric function without $\ga=0$ or $\ga=1$ singularities at all.  The
expression (\ref{chi1om}) satisfies in addition the symmetry constraints
(\ref{simmetria}), having antisymmetric part $-{b\over2}\achi_\om^{(0)}{}'$.

Of course, there is some ambiguity involved in the choice of the subtraction terms
(\ref{shift}, \ref{add}), which boils down to the possibility of adding to
(\ref{chitilde}) a term, vanishing in the $\om=0$ limit, and having only higher twist
$\ga$-singularities, around $\ga=-1,-2,\cdots$ and $\ga=2,3,\cdots$. We will discuss
that ambiguity in Sec.~\ref{ss:reschem}.

\subsection{Numerical importance of collinear effects at NLO\labe{sec:collval}}

\begin{figure}[ht!]
\centering
\includegraphics[width=10cm]{\fig 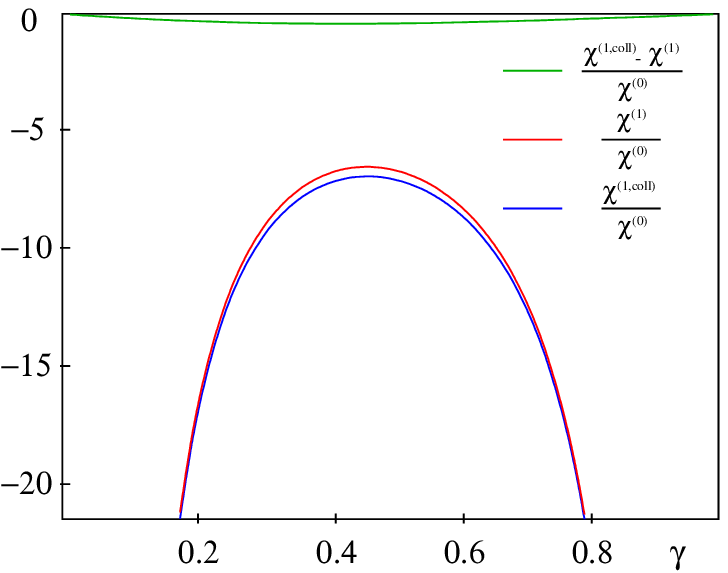}
 \didascalia{A comparison of the collinearly-enhanced (double and triple poles only)
    part of the NL$x$ corrections with the full NL$x$ corrections;
    $\Nf=0$.\labe{fig:chi1c}}
\end{figure}

Above we have given the general form for the collinear singularities of the kernel at
all orders. It is of interest to consider at NL$x$ level just how much of the full corrections
come from these collinearly enhanced terms. Accordingly we look at the part of the NL$x$
corrections which contains just double and triple poles, $\chi^{(1)}_{\text{coll}}$:
\begin{equation}
 \chi^{(1)}_{\text{coll}} =\frac{A_1}{\ga^2}+\frac{A_1-b}{(1-\ga)^2}
 -\frac{1}{2\ga^3}-\frac{1}{2(1-\ga)^3}\;.
\end{equation}
This is compared with the full $\chi^{(1)}$ in Fig.~\ref{fig:chi1c}, where we have
plotted their ratios to $\chi^{(0)}$. The remarkable observation is that over a range of
$\ga$, the collinear approximation reproduces the true corrections to within $7\%$.  It
is obviously impossible to say whether this is true at higher orders as well.  However
the fact that the study of collinear terms has such predictive power at NL$x$ is a
non-trivial point in favour of our resummation approach.

\section{Factorization of non-perturbative effects\labe{s:fnpe}}

The next step is to obtain the improved GGF (\ref{d:sxK}), i.e., to solve
\begin{equation}\lab{g}
  \om\G_{\om}(\kk,\kk_0)=\d^2(\kk-\kk_0)+\int\dif\kk'\;
 \K_{\om}(\kk,\kk')\G_{\om}(\kk',\kk_0)\;.
\end{equation}
where an extention of the representation (\ref{serieKom}) in the region around the
Landau pole $k^2\simeq\La^2$ ($t=0$) is needed. However, for perturbation theory to be
applicable, the non perturbative effects of such region should be factored out, as is
predicted by the RG, and has been argued for at L$x$ and NL$x$
level~\cite{CoKw89,CaCi96b}.

We assume that, by a suitable regularization of $\as(t)$ around
the Landau pole, $\K_{\om}$ can be defined as a hermitian operator
bounded from above in an $\L^2$ Hilbert space, with a continuum (or
possibly discrete) spectrum $\Sp(\K_\om)\subset\;]-\infty,\mp(\om)]$. Typical
regularizations of this kind may
\begin{itemize}
\item[(a)] cut-off $\as(t)$ below some value $t=\tb>0$:
         \begin{equation}\lab{cutoff}
          \ab(t)={1\over bt}\Th(t-\tb)\;;
         \end{equation}
\item[(b)] freeze it in the form
\begin{equation}\lab{freeze}
 \ab(t)={1\over bt}\Th(t-\tb)+{1\over b\tb}\Th(\tb-t)\;,
\end{equation}
possibly with some smoothing out around the cusp.
\end{itemize}
In such a framework, a formal solution for the Green's function $\G_{\om}$ is given by
the spectral representation
\begin{equation}\lab{rapspet}
 \G_{\om}(\kk,\kk_0)=\int_{-\infty}^{\mp(\om)}{\dif\mu\over\pi}\;
 {\afz_{\om}^{\mu}(\kk)\afz_{\om}^{\mu\,*}(\kk_0)\over\om-\mu}
\end{equation}
in terms of a complete and orthonormal set of (real) eigenfunctions
$\afz_{\om}^{\mu}:\mu\in\Sp(\K_\om)$
\begin{equation}\lab{autof}
 \K_{\om}\afz_{\om}^{\mu}=\mu\afz_{\om}^{\mu}\;.
\end{equation}

Let's refer for definiteness to the frozen-$\ab(t)$ regularization (\ref{freeze}), which
allows a simple classification of the eigenfunctions $\afz_{\om}^{\mu}(\kk)$ of
Eq.~(\ref{autof}), according to their behavior for $t\to-\infty$. In fact, since
$\ab(t)$ is fixed for $t<\tb$, in the $t\to-\infty$ limit the eigenfunctions must behave 
just like the scale-invariant ones
\begin{equation}\lab{afz}
 (k^2)^{\ga(\mu)-1}={1\over k}\esp{\ui\nu(\mu)t}\quad,\quad\ga={1\over2}+\ui\nu\;.
\end{equation}
Because of the symmetry property of the kernel with respect to the exchange of its
arguments $\kk\lra\kk'$, to each eigenvalue $\mu$ there corresponds two opposite values
$\pm\nu(\mu)$ and we can decompose the eigenfunctions $\afz_{\om}^{\mu}(\kk)$ as
\begin{align}\lab{onde}
 \afz^{\mu}(\kk)&\underset{\qquad\;}{=}\;{1\over2\ui}\left(E^{\nu(\mu)}(\kk)-
 E^{-\nu(\mu)}(\kk)\right)\\
 &\underset{t\to-\infty}{\simeq}\;{1\over2\ui k}\left(c(\nu)\esp{\ui\nu(\mu)t}
 -c^*(\nu)\esp{-\ui\nu(\mu)t}\right)\nonumber
\end{align}
for suitable functions $E^{\nu}(\kk)$ having a plane-wave asymptotic behavior for large
and negative $t$ (the $\om$ index has been dropped).

The precise superposition coefficient $c(\nu)$ of left- and right-moving waves occurring
in Eq.~(\ref{onde}) is determined by the condition that $\afz^{\mu}(\kk)$ be regular for
$t\to+\infty$, i.e., be vanishing at least as rapidly as $1/k$, so as to allow an $\L^2$
normalization.

If we assume $E^{\nu(\mu)}$ and $E^{-\nu(\mu)}$ to be boundary values of an imaginary
analytic function $\afzt_\om^\mu(\kk)$ of $\mu$, whose branch cut lies along the
spectrum (see Fig.~\ref{f:rapspe}), we can rewrite the spectral
representation~(\ref{rapspet}) as a contour integral
\begin{equation}\lab{intcont}
 \G_{\om}(\kk,\kk_0)=\int_{C(\om)}{\dif\mu\over2\pi\ui}\;
 {\afz_{\om}^{\mu}(\kk)\afzt_{\om}^\mu(\kk_0)\over\om-\mu}\,,
\end{equation}
encircling the spectrum $\Sp(\K_\om)$. By distorting the $\mu$-contour (because
$\afzt^{\mu}$ is well behaved, for $\re(\ui\nu)>0$) and by picking up the residue at
the $\mu=\om$ pole, we end up with the factorized expression~\cite{CiCoSa99}
\begin{equation}\lab{fattor}
 \G_\om(\kk,\kk_0)=\afz_\om(\kk)\afzt_\om(\kk_0)\qquad(k^2\gg k_0^2) 
\end{equation}
where $\afz_\om(\kk)\dug\afz_\om^\om(\kk)$ and the same for $\afzt$.
This procedure can be carried through provided $k/k_0$ is large enough for the decrease
of $\afz^{\om}$ to compensate the increase of $E^{\nu(\om)}$.
\begin{figure}[ht!]
\centering
\includegraphics[width=66mm]{\fig 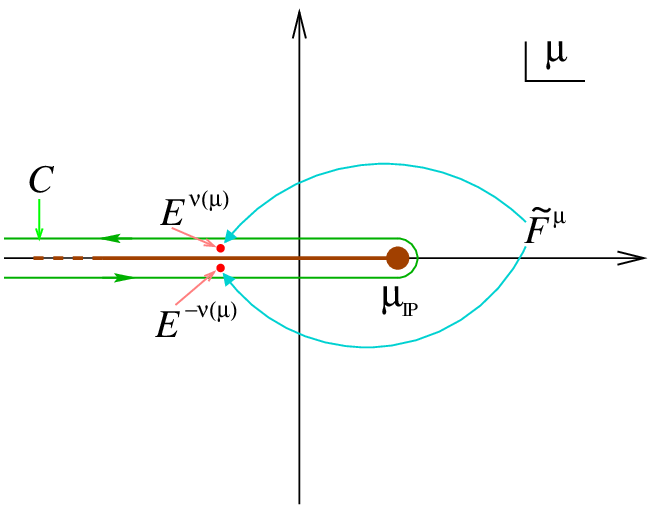}
 \didascalia{The function $\afzt^\mu$ is defined as that analytic function of $\mu$ whose 
limit on its branch cut --- corresponding to the spectrum of $\K$ --- is given by
$E^{\nu(\mu)}$ ($E^{-\nu(\mu)}$) over (under) the cut. In this way the real
integration in Eq.~(\ref{rapspet}) can be replaced by a complex integration along the
contour $C(\om)$ of Eq.~(\ref{intcont}).\labe{f:rapspe}}
\end{figure}

The plausibility argument above is further supported by the explicit model we will give
in Chap.~\ref{c:cm} for arbitrary values of $t$ and $t_0$, and hints at the general
validity of Eq.~(\ref{fattor}). Therefore, for $k\gg k_0$, the Green's function is
asymptotically proportional --- up to higher twist contributions --- to the regular (for
$t\to\infty$) solution $\afz_{\om}(\kk)$ of the homogeneous BFKL equation (\ref{autof})
with eigenvalue $\mu=\om$:
\begin{equation}\lab{eqx}
 \om\afz_\om=\K_\om\afz_\om\;,
\end{equation}
which becomes the basic quantity to be found.

\section{The small-$\boldsymbol\om$ expansion\labe{s:soe}}

Having determined the factorization property of the GGF in Eq.~(\ref{fattor}), we are
now in a position to determine its perturbative (large-$\kk$) behaviour determined by
the small-$x$ equation (\ref{eqx}) whose kernel we know at NL$x$ level.
In this section the scale of the energy $s_0$ is arbitrary, but one have to keep in mind 
that the resulting quantities depend on the scale chosen.

As usual it is better to work in $\ga$-space, where we represent the eigenfunctions
$\afz_\om^\mu$ in a form%
\footnote{The canonical dimension of the eigenfunctions is $[\afz]=-1$.}
hinted by Eq.~(\ref{d:Fpert}):
\begin{equation}\lab{rappres}
 \afz_\om^\mu(\kk)={1\over k}\intmel\difg\exp
 \left\{(\ga-\half) t-{1\over b\mu}X_\om(\ga,\mu)\right\}\;,
\end{equation}
where $X_\om(\ga,\mu)$ is to be found by solving Eq.~(\ref{autof}).

By using the above representation for $\afz_\om^\mu$ and
keeping the general expression (\ref{serieKom}) for the kernel, the eigenvalue equation
(\ref{autof}) becomes
\begin{equation}\lab{azker}
 0=[\K_\om-\mu]\afz_\om^\mu(\kk)={1\over k}\int\difg\;\esp{(\ga-\half) t-{1\over b\mu}
 X_\om(\ga,\mu)}\left[\sum_{n=0}^{\infty}\ab(t)^{n+1}\chi_\om^{(n)}(\ga)-\mu\right]\;.
\end{equation}
We assume that in the small-$\mu$, large-$t$ regime
\begin{equation}\lab{regime}
 bt\gtrsim{1\over\mu}\gtrsim{1\over\om}\gg1\;,
\end{equation}
--- resembling the anomalous dimension regime (\ref{regdimanom}) --- the integral
(\ref{azker}) is dominated by a stable saddle point
$\gb_\om^\mu(\ab(t))$ determined by the condition
\begin{subequations}\labe{puntos}
\begin{align}
 &\de_{\ga}\left\{(\ga-\half)t-{1\over b\mu}X_\om(\ga,\mu)\right\}_{\ga=\gb}=0
 \quad\iff\quad b\mu t=
 \de_\ga X_\om^\mu(\gb)\;,\\
 &\de_\ga^2X_\om(\gb,\mu)<0\;.
\end{align}
\end{subequations}
which need to be checked ``a posteriori''.

By expanding in $\ga$ the exponent and the eigenvalue functions $\achi_\om^{(n)}$ around
$\ga=\gb$ and by keeping the ensuing fluctuations up to the relevant order, as explained 
in Ref.~\cite{CiCoSa99}, Eq.~(\ref{azker}) provides relations between the eigenvalues 
of the improved coefficient kernels and the $\ga$-derivatives of $X_\om(\ga,\mu)$ which
allows us to write the latter in terms of the former. More precisely, in the regime
(\ref{regime}), we can express
\begin{equation}\lab{chimu}
 \chi_\om(\ga,\mu)\dug \de_\ga X_\om(\ga,\mu)\;,
\end{equation}
by an expansion in $\mu$ whose coefficients are the eigenvalues $\achi_\om^{(n)}$ and
their $\ga$-derivatives:
\begin{equation}\lab{mues}
 \chi_{\om}(\ga,\mu)=\eta_\om^{(0)}(\ga)+\mu\,\eta_\om^{(1)}(\ga)+\mu^2
 \,\eta_\om^{(2)}(\ga)+\cdots\;,
\end{equation}
where the $\mu$-expansion coefficient $\eta_\om^{(i)}$ could be derived, in principle, to 
all orders, and up to the third order (NNNL$x$) reads~\cite{CiCoSa99}
\begin{subequations}\labe{ete}
\begin{align}
 \eta_\om^{(0)}&=\chi_\om^{(0)}\;,\\
 \eta_\om^{(1)}&={\achi_\om^{(1)}\over\achi_\om^{(0)}}\;,\\
 \eta_\om^{(2)}&={1\over\achi_\om^{(0)}}\Bigg[{\achi_\om^{(2)}\over\achi_\om^{(0)}
 }+b\left({\achi_\om^{(1)}\over\achi_\om^{(0)}}\right)^{\prime}-\left({
 \achi_\om^{(1)}\over\achi_\om^{(0)}}\right)^2\Bigg]\;,\\
 \eta_\om^{(3)}&={1\over\achi_\om^{(0)}}\left[{\achi_\om^{(3)}\over\achi_\om^{(0)\,2}}
 +{b\over\achi_\om^{(0)}}\left({\achi_\om^{(2)}\over\achi_\om^{(0)}}\right)
 ^{\prime}-{\achi_\om^{(1)}\achi_\om^{(2)}\over\achi_\om^{(0)\,3}}+b\,
 \eta_\om^{(2)}{}'-2\,\eta_\om^{(1)}\eta_\om^{(2)}\right]\;.
\end{align}
\end{subequations}
In the $\mu=\om$ case, Eq.~(\ref{mues}) becomes an $\om$-expansion~\cite{CiCo98b} for
\begin{equation}\lab{omegaes}
 \chi_\om(\ga)\dug\chi_\om(\ga,\om)=\chi_\om^{(0)}(\ga)+\om{\achi_\om^{(1)}(\ga)
 \over\achi_\om^{(0)}(\ga)}+\text{NNL}x\;,
\end{equation}
which provides, through Eqs.~(\ref{chimu}), (\ref{rappres}) and (\ref{fattor}), the
perturbative expression for the GGF we were looking for.

The analogy of Eqs.~(\ref{puntos})$|_{\mu=\om}\;\lra\;$(\ref{d:gabarra}),
(\ref{rappres})$\;\lra\;$(\ref{d:Fpert}) and the L$x$ truncation of Eq.~(\ref{omegaes})
leads to the identification of $\chi_\om(\ga)$ with the {\em effective eigenvalue
function} of the improved kernel $\K_\om$.
\begin{figure}[ht!]
\centering
\includegraphics[width=9cm]{\fig 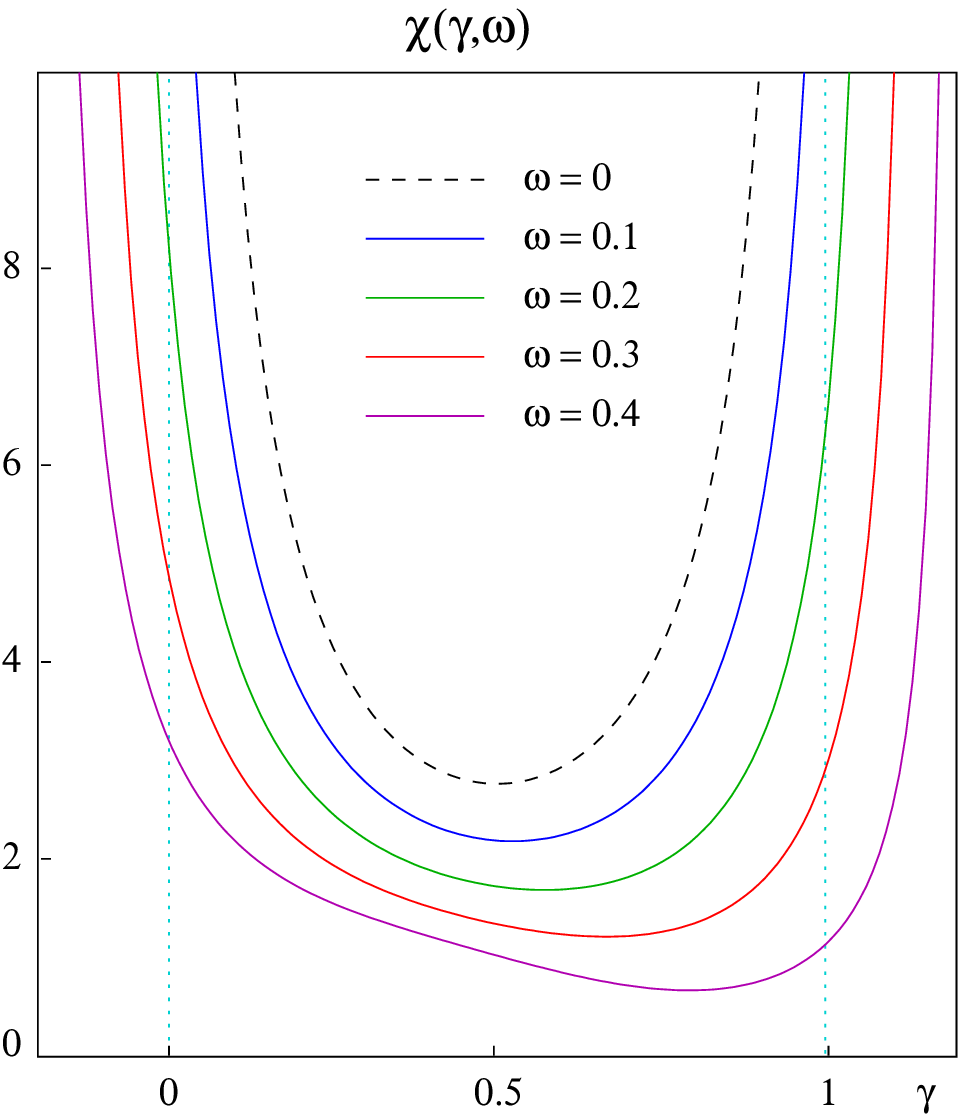}
 \didascalia{The NL$x$ truncation (\ref{omegaes}) of the effective eigenvalue function
	$\chi_\om(\ga)$; $\Nf=0$.\labe{f:avleff}}
\end{figure}

Before entering the details of the determination of the improved anomalous dimension and
of the hard pomeron, we would like to spend a few words about the major features of the
improved eigenvalue $\achi_\om(\ga)$. In Fig.~\ref{f:avleff} you can see the effective
eigenvalue function for various values of the new expansion parameter $\om$. The
stability of the shape, i.e., the presence of a minimum even for sizeable values of
$\om$ is apparent, at variance with Fig.~\ref{f:avlNL}. The shift of the poles is
evident too, providing the consistency with the RG. The asymmetric shape is due to our
(asymmetric) choice of the coupling-scale $k^2$.

\section{Improved anomalous dimension\labe{s:iad}}

In order to study the anomalous dimension limit $Q^2\gg\La^2$ of the integrated gluon
density, let's set the energy scale for the GGF to $s_0=k^2$.  Due to the validity of RG
factorization in the large-$t$ limit of Eq.~(\ref{fattor}), we can state that the gluon
density $f^{(\pg/H)}_\om(Q^2)$ in the hadron $H$ has a universal
$t_{\!_{Q}}\dug\ln Q^2/\La^2$ dependence
\begin{equation}\lab{fatdenglu}
 f^{(\pg/H)}_\om(Q^2)=g_\om(t_{\!_{Q}})\,\tilde{g}_\om^{(H)}\;,
\end{equation}
where the $H$-dependent coefficient arise from Eq.~(\ref{fattDGNI})
\begin{equation}\lab{coefNP}
 \tilde{g}_\om^{(H)}=\int\dk_0\;\afzt_\om(\kk_0)\,f_\om^{(\pg/H)}(\kk_0)
\end{equation}
and is in general non perturbative, while the perturbative one is independent on the
probe and is determined by
\begin{equation}\lab{coefP}
 g_\om(t_{\!_Q})=\int^{Q^2}\dk\;\afz_\om(\kk)=\int\difg\;{1\over\ga}\exp\left\{
 \ga t_{\!_Q}-{1\over b\om}\overset{u}{X}{}_\om(\ga)\right\}
\end{equation}
where we have specified the function $X_\om$ at the ``upper'' energy-scale $k^2$.

The asymptotic behaviour of Eq.~(\ref{coefP}) in the RG regime can be found from the
saddle point (\ref{puntos}) which, thanks to the identity
\begin{equation}\lab{idgabarra}
 \gb_\om t-X_\om(\gb)=\int^t\gb_\om(\ab(\tau))\;\dif\tau+\text{const}
\end{equation}
(check it by differentiating with respect to $t$) yields the result
\begin{equation}\lab{gdiman}
 g_{\om}(t_{\!_Q})\simeq\left({1\over\gb_\om}\sqrt{b\om\over2\pi|
 \chiu_\om'(\gb)|}+\cdots\right)\exp\int^{t_{\!_Q}}\gb_\om(\ab(\tau))\;\dif\tau\;.
\end{equation}
The coefficient in front, coming from the saddle point fluctuations, has been evaluated
at NL level only. If we work at NL level, the saddle point approximation (\ref{gdiman})
is enough, and provides the effective anomalous dimension~\cite{CiCoSa99}
\begin{align}\nonumber
 \ga^{\eff}_\om(\ab(t))\dug{\dif\over\dif t}\ln g_\om(t)
&=\gb_\om-{\dif\over\dif t}\ln\left[\gb_\om
 \sqrt{|\chiu_\om'(\gb_\om)|}\right]\\ \label{gaeff}
&=\gb_\om-{b\om\over\chiu'_\om(\gb_\om)}\left({1\over\gb_\om}+
 {\chiu''_\om(\gb_\om)\over2\chiu'_\om(\gb_\om)}\right)+\cdots
\end{align}
where in the last equality use have been made of the relation
\begin{equation}\lab{dergaba}
 b\om t=\chi_\om(\gb_\om)\quad\imp\quad b\om=\chi'_\om(\gb_\om){\dif\over\dif t}
 \gb_\om(\ab(t))\;.
\end{equation}
\begin{figure}[ht!]
\centering
\includegraphics[width=12cm]{\fig 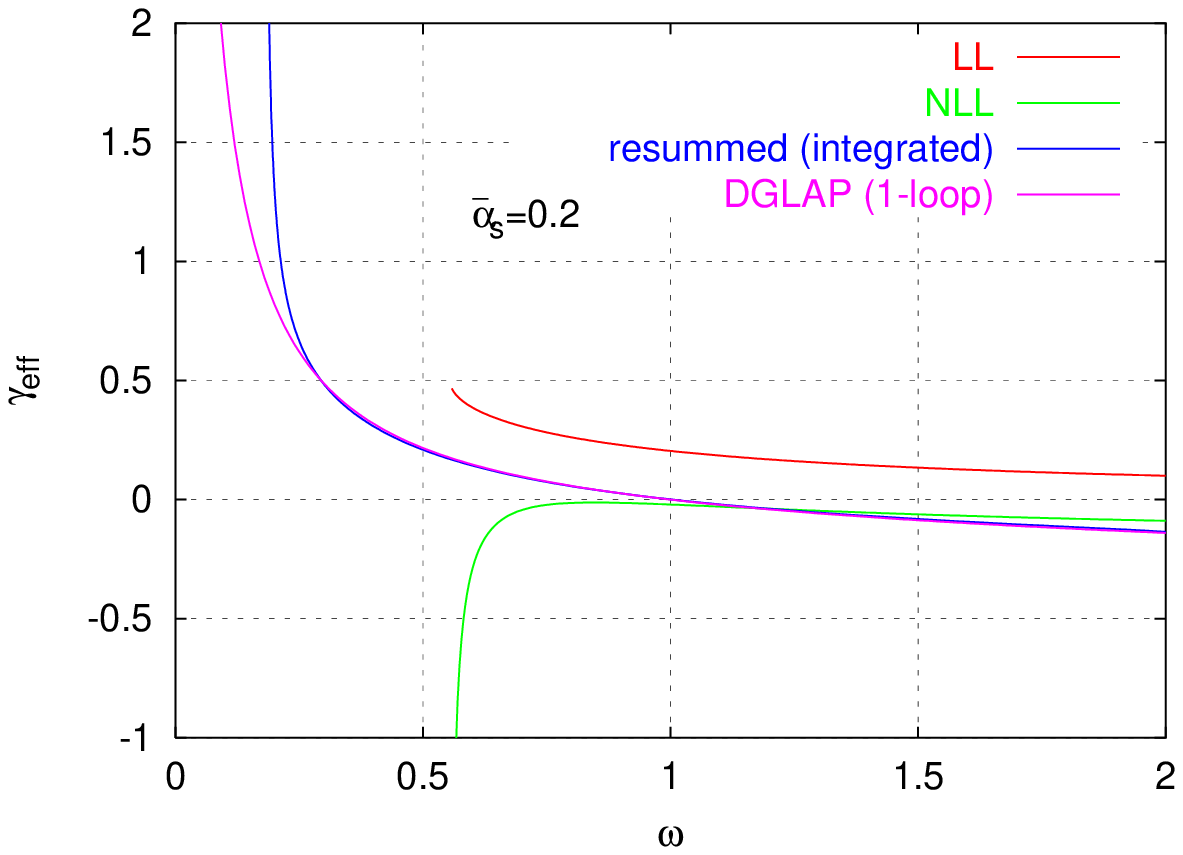}
\didascalia{The anomalous dimension in various approximations.\labe{f:anomdim}}
\end{figure}

Fig.~\ref{f:anomdim} shows the (purely gluonic) anomalous dimension as a function of
$\om$ for $\ab=0.2$. The L$x$ anomalous dimension is just $\gl=\chi^{(0)\,-1}(\om/\ab)$
and has the familiar branch-cut at $\om=4\ln2 \ab$. The NL$x$ anomalous dimension
(\ref{d:gNL}) has a divergent structure around the same point as $\gl$. The resummed
result, defined in Eqs.~(\ref{coefP}) and (\ref{gaeff}), shows a divergence at a much
lower $\om$, defined by $\om_c(t)$, corresponding to the vanishing of the gluon density
\begin{equation}\lab{omcrit}
  g_{\om_c(t)}(t)=0\;,
\end{equation}
thus extending the resummed anomalous dimension beyond the spurious singularity of the
BFKL truncation. What
is particularly remarkable is the similarity to the DGLAP result until
very close to the divergence. The momentum sum rule is automatically
conserved: for $\om=1$ we have $\ga^{\eff}_1=0$. Note however that, since $\om_c(t)$
comes from a zero of $g_{\om}(t)$, the $\om=\om_c(t)$ singularity does
not transfer to $g_{\om}(t)$ itself. Furthermore, the singularity of the anomalous
dimension representation (\ref{gdiman}), coming from the failure of the saddle point
expansion at the value $\om=\om_s(t)$ such that
\begin{equation}\lab{omsp}
 \chi_{\om_s}'(\gb_{\om_s}(t))=0\;,
\end{equation}
is different from $\om_c(t)$.

\section{Improved hard pomeron\labe{s:ihp}}
\begin{figure}[ht!]
\centering
\includegraphics[width=12cm]{\fig 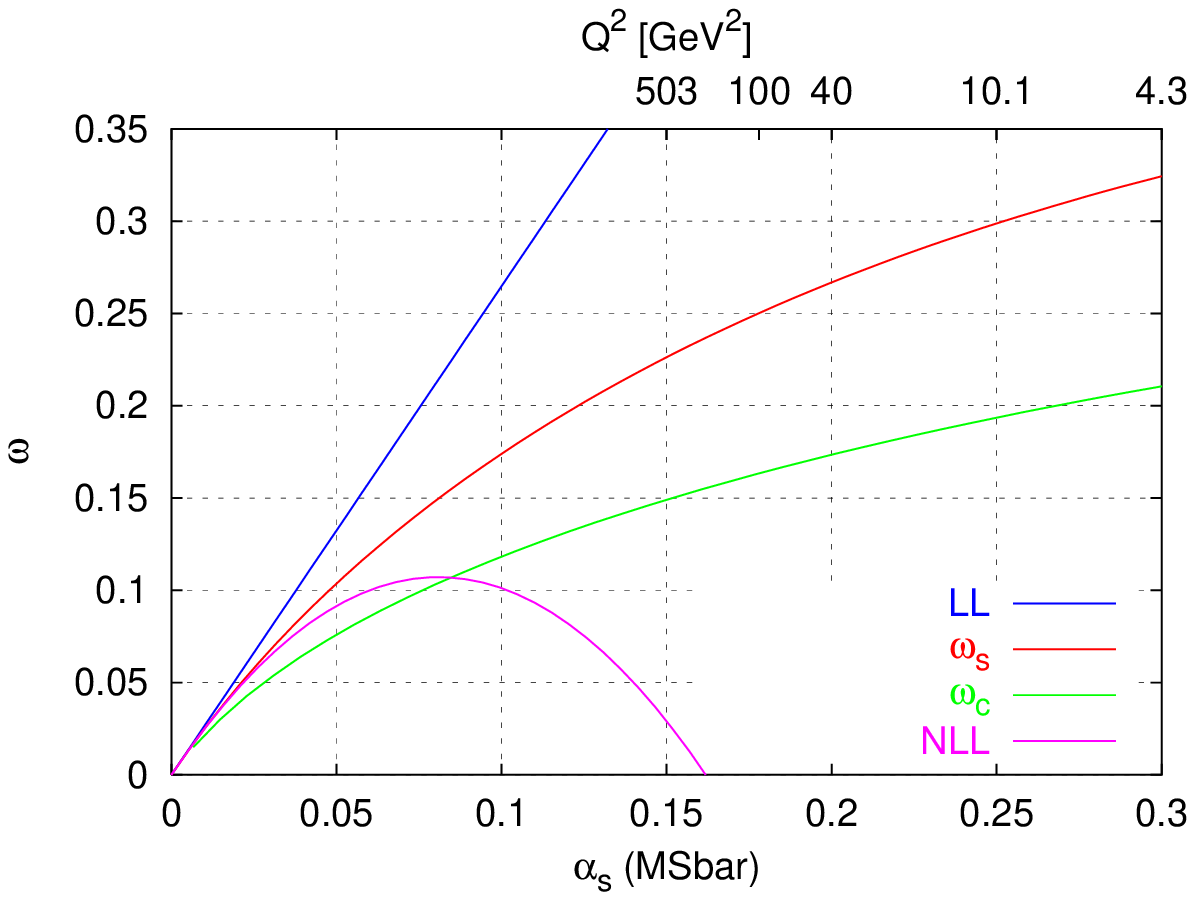}
\didascalia{$\om_c$ and $\om_s$ as a function of $\as$ for the BFKL kernel with
	$\Nf=0$.\labe{f:omcoms}}
\end{figure}

The values of the exponents $\om_c$ and $\om_s$ as a function of $\as$
(and $Q^2$), are shown in Fig.~\ref{f:omcoms} and compared with the
L$x$ and pure NL$x$ results. It is apparent that the improved equation
provides sensible predictions even for sizeable values of $\as$. The above difference should not be too confusing. The exponent
$\om_s(t)$ signals the breakdown of the formal small-$x$ expansion of
the anomalous dimension of Eq.~(\ref{gaeff}), due to infinite
saddle-point fluctuations, while $\om_c(t)$ tells us the position of
the singularity of the resummed anomalous dimension. Their difference
arises from their different definitions, not from some instability of
our approach.

What are the roles of the two hard pomeron estimates $\om_c$ and $\om_s$? What, if any,
their relation to the pomeron singularity $\op$, the leading singularity of the
GGF? We are not in a position to say a lot about those questions yet. Probably there is
an intermediate small-$x$ moderate-$Q^2$ regime where the perturbative physics
dominates, where $\om_c$ or $\om_s$ could be the effective small-$x$ growth
exponent. The non perturbative soft pomeron, determined by the spectral condition
$\op=\mp(\op)$, should however govern the very low-$x$ region.

In order to gain some hints on these issues, we have devised a simplified
model~\cite{CiCoSa99b} which allows one to treat in a semi-analytic way the whole
$t$-space dependence and thus to estimete the (in)dependence of some physical quantities
on the strong coupling region and also to clarify possible transition mechanisms between
various $x$-regimes. But this is the subject of the last chapter.

\section{Estimate of the NL$\boldsymbol x$ truncation error\labe{s:eoe}}

What is the error that we make in the NL truncation of the RG improved equation? Our
claim is that, in the improved formulation, based on the $\om$-expansion
(\ref{omegaes}), this error is smaller than in the formal NL expansion in $\as(t)$. Let
us in fact estimate the remaining terms in Eq.~(\ref{mues}). According to
Eq.~(\ref{chin}) further subleading eigenvalue functions contain at least higher order
collinear poles which contribute to $\eta_\om^{(2)}$, $\eta_\om^{(3)}$ and so on. A
first observation is that, even if $\achi_\om^{(n)}$ has ($n+1$)-th order poles, the
$\eta_\om^{(n)}:n\geq0$ have at most {\em simple} poles, due to the powers of
$\chi_\om^{(0)}$ in the denominator, roughly due to the replacement
$\ab(t)\sim\om/\achi_\om^{(0)}$. Therefore, their contribution cannot be too big, even
for small values of $\ga=\ord(\om)$.

Furthermore, one can check that, if $\pq\bar{\pq}$ contributions are neglected, the
leading collinear poles actually {\em cancel out} in the expansions (\ref{ete}) of
$\eta_\om^{(2)}$, $\eta_\om^{(3)}$, \dots around both $\ga=0$ and $\ga=1$.  The
mechanism of this cancellation is explained in Ref.~\cite{CiCoSa99}.

From a physical point of view, it is not possible for simple poles to survive in
$\eta_\om^{(n)}:n\geq2$, because, when replaced in the saddle point condition
(\ref{dergaba}), they would provide $\om^n:n\geq2$ corrections to the 1-loop anomalous
dimensions which cannot possibly be there. In fact, the full anomalous dimension is
accounted for by Eqs.~(\ref{omegaes},\ref{puntos}) as follows
\begin{equation}\lab{adsemp}
 b\om t\simeq{1\over\gb_\om}+{A_1(\om)\over\gb_\om}\quad\imp\quad
 \gb_\om=\ab\left({1\over\om}+A_1(\om)\right)\;,
\end{equation}
where we have taken the small-$\ga$ limit of the collinear safe
eigenvalue function $\chiu_\om(\ga)$.

We therefore conclude that, in the purely gluonic case, the NL
$\om$-expansion (\ref{puntos}) takes into account the collinear behavior
to all-orders, and that no further resummation is needed.

\section{Stability\labe{s:stab}}

The effective eigenvalue $\achi_\om(\ga)$ reveals its stability in shape providing a
stable minimum and thus a reliable framework for small-$x$ physics. Nevertheless, there
are some ambiguities, one of which is proper of the present resummation method and
concerns the pole-shifting procedure of Sec.~\ref{s:fcs}, others are typical of the
perturbative treatments and concern the renormalization scheme prescription and the
renormalization scale choice.

The original L$x$+NL$x$ formalism suffered from considerable instabilities under
renormalization group scale and scheme changes~\cite{BFKLP99}. An important
characteristic of any resummed approach is that it should be relatively insensitive to
such changes, and generally stable.

\subsection{Renormalization scale\labe{ss:Rschema}}
 
Note first that in our approach the renormalization scale only enters
through the RG invariant $\La$ parameter of $t=\ln k^2/\La^2$ (Eqs.~(\ref{serieKom}) and
(\ref{coefP})). It is then easy to see that the physical results are
$\La$-independent. A redefinition of $\La$ is essentially a shift in
$t$, say by an amount $\De t$.  There is a corresponding modification
of $\chi_\om^{(1)}$, $\chi_\om^{(2)}$, \dots by the amounts
\begin{equation}\lab{deltachi}
  \chi_\om^{(1)}\to\chi_\om^{(1)}+b\De t\chi_\om^{(0)}\quad,\quad
 \chi_\om^{(2)}\to\chi_\om^{(2)}+2b\De t\chi_\om^{(1)}+b^2(\De t)^2
 \chi_\om^{(0)}\quad,\quad\dots\;.
\end{equation}
In the $\gamma$-representation (\ref{rappres}), this corresponds to a
modification of $X_\om$ by an amount $b\mu\ga\De t$. In fact the transformation
(\ref{deltachi}) changes the coefficient $\eta^{(1)}_\om$ only, the remaining ones
$\eta^{(2)}_\om$, $\eta^{(3)}_\om$,\dots being left invariant. This change exactly
cancels the modification of $t$ itself:
\begin{equation}
  \exp\left\{\ga t-{1\over b\mu}X_\om(\ga,\mu)\right\}\to\exp\left\{\ga(t+\De t)
 -{1\over b\mu}(X_\om(\ga,\mu)+b\mu\ga\De t)\right\}\;,
\end{equation}
thus implying that the physical results are independent of the $\La$-parameter
choice~\cite{CiCoSa99}.  This automatic resummation of the renormalization scale
alleviates the need for techniques such as BLM resummation~\cite{BrLeMa83}, advocated
for example in~\cite{BFKLP99}, which show a strong renormalization scheme dependence.

\subsection{Renormalization scale\labe{ss:Rscala}}

The issue of renormalization scheme dependence is in fact closely related.
Consider a scheme $S$ related to the $\overline{MS}$ scheme by 
\begin{equation}\lab{schmchng}
  \as^{(S)} = \as^{(\MSbar)} + T \as^2\,,
\end{equation}
with an appropriate modification of $\chi_\om^{(1)}$.  Except for terms of $\ord(\as^3)$
and higher, this is identical to a renormalization scale change. Indeed if one defines
the scheme change by a modification of $\La$ then renormalization scheme changes behave
exactly as renormalization scale changes, and so have no effect on the answer.  Using
instead \ref{schmchng} there is some residual dependence on the scheme at $\ord(\as^3)$,
but as one can see in Fig.~\ref{f:scheme} for the $\Upsilon$ scheme, which has
$T\simeq1.17$ (for $\Nf=0$), the effect of the change of scheme is small.
\begin{figure}[ht!]
\centering
\includegraphics[width=12cm]{\fig 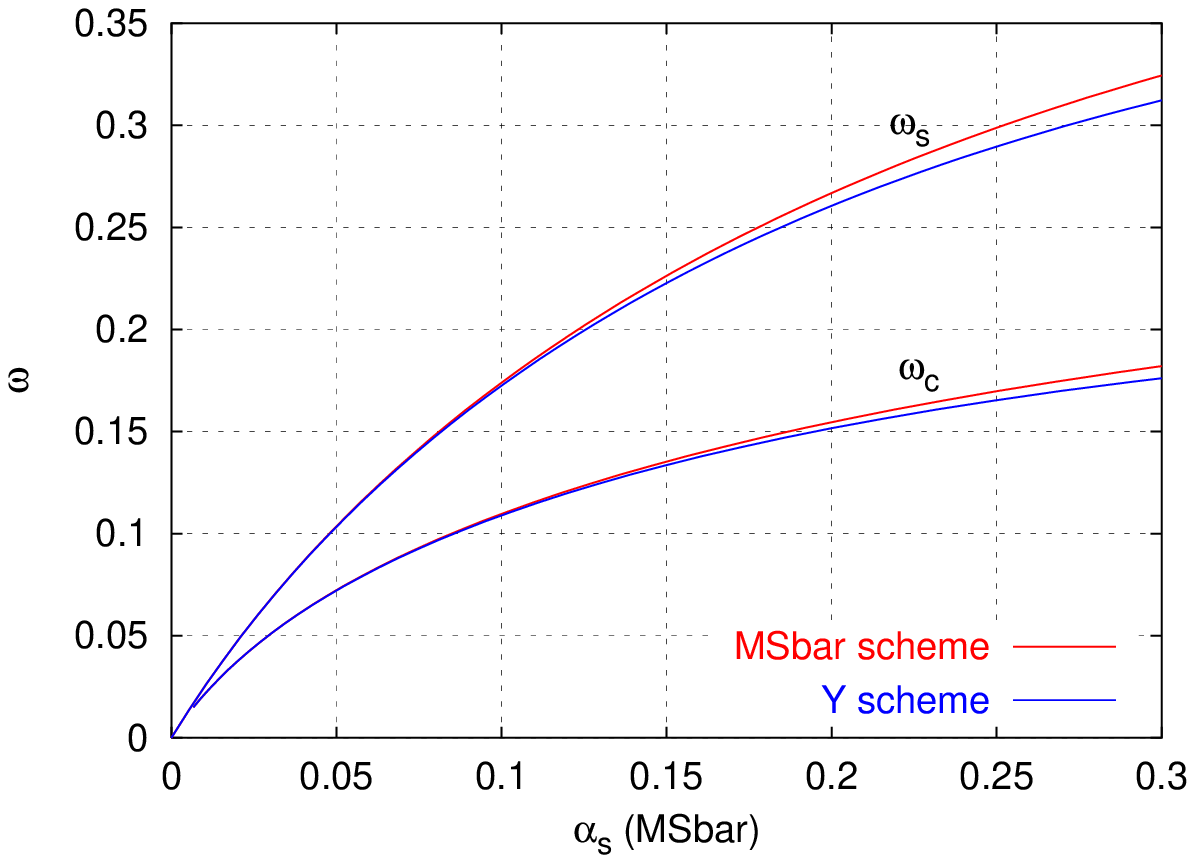}
\didascalia{Renormalization scheme uncertainty of the two exponents;
      $\overline{MS}$ scheme and $\Upsilon$ scheme; $\as$ is always
      shown in the $\overline{MS}$ scheme, and is connected to the
      $\Upsilon$ scheme value of $\as$ via (\ref{schmchng}).\labe{f:scheme}}
\end{figure}

\subsection{Resummation scheme\labe{ss:reschem}}

In resumming the double transverse logarithms (energy-scale terms), there is some
freedom in one's choice of how to shift the poles around $\ga=0$ and $\ga=1$. In a
similar manner to what was done in~\cite{Sa98} we consider two choices. The one
explicitly discussed in this chapter can be summarized as
\begin{equation*}
  \psi^{(n-1)}(\ga) \to \psi^{(n-1)}(\ga + \ho)\;,
\end{equation*}
with an equivalent procedure around $\ga=1$. We refer to this as
resummation scheme~(a).  An alternative possibility is (scheme (b))
\begin{equation*}
  \frac{1}{\ga^n} \to \frac{1}{(\ga + \ho)^n}\;.
\end{equation*}
For instance, in this second scheme, the L$x$ improved eigenfunction would be
\begin{equation}\lab{chi0alt}
 \chi^{(0)}_\om(\ga)=\chi^{(0)}(\ga)-\frac1{\ga}-\frac1{1-\ga}+\frac1{\ga+\ho}
 +\frac1{1-\ga+\ho}\;.
\end{equation}
A comparison of these two resummation schemes~\cite{CiCoSa99} is given in
Fig.~\ref{f:resumscheme} and the difference between them is again reasonably small.
\begin{figure}[ht!]
\centering
\includegraphics[width=12cm]{\fig 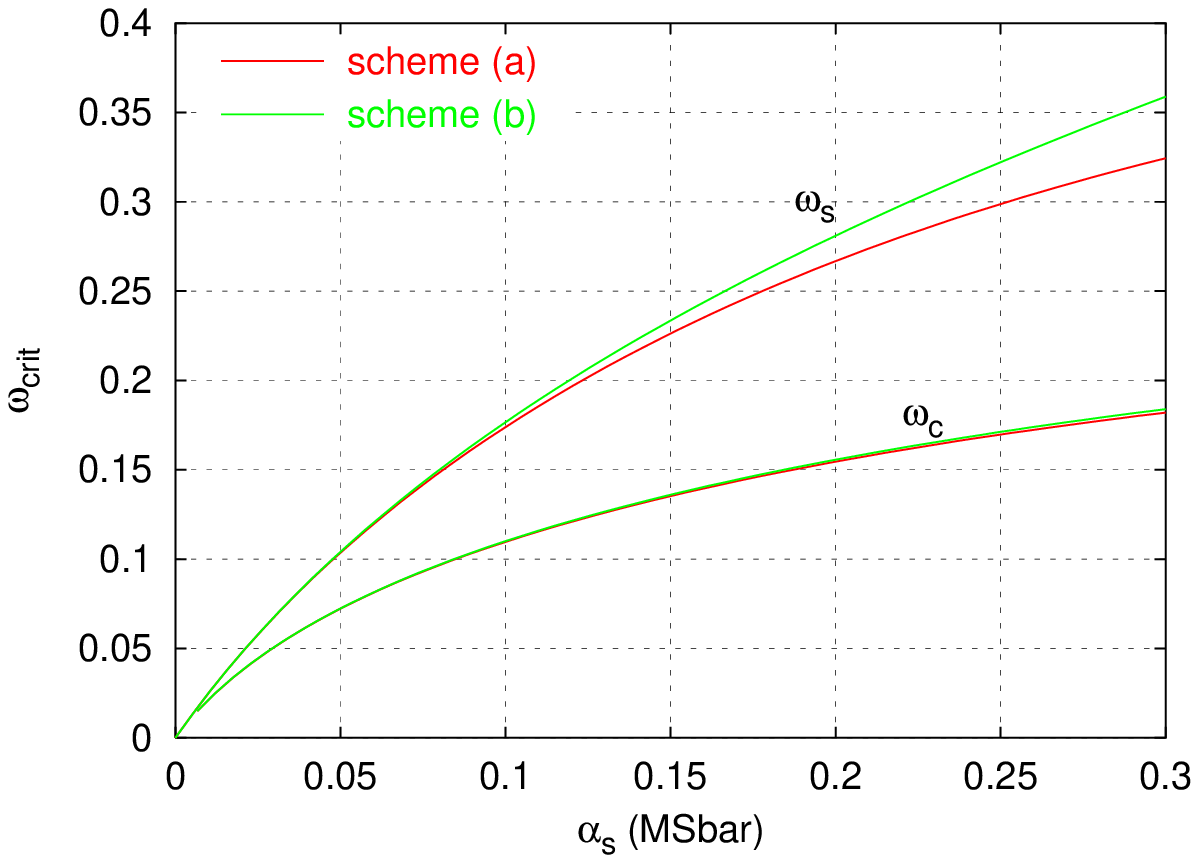}
\didascalia{Resummation scheme uncertainty of the two critical exponents $\om_s$ and
$\om_c$.\labe{f:resumscheme}}
\end{figure}

Aside from the explicit renormalization-scale independence, the
stability of our approach is connected with the resummation of the
collinear poles, for both the double-log, energy-scale dependent terms
(the $1/\ga^3$ and $1/(1-\ga)^3$ poles at NL$x$ order) and for the single-log
ones of Eq.~(\ref{adsemp}).

\chapter{The collinear model\labe{c:cm}}

Let's briefly review the results of the previous chapters. The study of perturbative
high energy processes requires resummation techniques in order to correctly take into
account the energy dependence. The BFKL approach provides such a framework, but exhibits
an unstable behaviour as soon as the coupling exceeds very small values. It has been
shown that the implementation of RG constrained collinear physics cures the BFKL
hierarchy pathologies and the use of $\om$ instead of $\as$ as expansion parameter
furnishes a stable framework where to study small-$x$ physics.

However, the improved formulation of Chap.~\ref{c:rga} still leaves some open questions,
concerning in particular the relative importance of perturbative versus non perturbative
effects and the identification of the physical observables. Of course a lot of work has
to be done yet in checking the validity of the factorization properties (\ref{fattor})
of the GGF and on the novel small-$x$ equation (\ref{azker}) determining the
perturbative solution.

For those purposes, we have devised a simple, but powerful tool for studying the problem
of small-$x$ physics that we call {\em the collinear model} --- namely a model where all
and only the collinearly-enhanced physics is correctly included, in particular the full
dependence on the one-loop running coupling, the splitting functions, and the so-called
energy-scale terms (the ones in square brackets in Eq.~(\ref{chi1}).  The model has the
properties of correctly reproducing the one-loop renormalization group results and of
being symmetric (given a problem with two transverse scales, its results do not depend
on which of the scales is larger). These are properties desired from the resummation of
the NLL corrections in the case of the full BFKL kernel.

While it does not correctly resum the series of leading and subleading logarithms of $s$
(i.e.\ the non-collinear part of the problem), in that region it does have a structure
which qualitatively is very similar to that of the BFKL equation (cfr.\ Fig.~\ref{f:confronto})
and can usefully serve as a model. In contrast to the BFKL equation, it is very easily
soluble, as a Schr\"odinger-like problem.

\section{Definition of the model\labe{s:dom}}

Our model~\cite{CiCoSa99b} is defined in such a way to reproduce the DGLAP limits for
branchings of gluons with ordered transverse momenta, and the anti-DGLAP limit for
branchings with anti-ordered transverse momenta. In practice the kernel of this model
extends the collinear asymptotic form (\ref{kupper}) of the small-$x$ kernel down to
$t\dug\ln k^2/\La^2\geq t'\dug\ln k'{}^2/\La^2$ while in the opposite region $t\leq t'$
it is defined by Eq.~(\ref{klower}). It is clear that the two expressions refer to the
``upper'' $k^2$ and ``lower'' $k_0^2$ energy-scales of the GGF respectively.  By taking
into account the energy-scale transformation relations (\ref{trasimK}) the symmetric
expression of the collinear kernel reads
\begin{align}\lab{eq:kernel} 
 \KK_\om(t,t')&\dug kk'\Ks_\om(k,k')\\
&=\ab(t)\exp\left\{-{1+\om\over2}(t-t')+A_1(\om)\int_{t'}^t\ab(\tau)\,
 \dif\tau\right\}\Th(t-t')+(t\leftrightarrow t')\;,\nonumber
\end{align}
$A_1(\om)$ being the non singular part of the gluon anomalous dimension
(\ref{nsdimanom}) at 1-loop.

Therefore the kernel for a single small-$x$ branching actually resums many branchings,
of which the last (and only the last) is governed by the $1/\om$ part of the splitting
function and is taken into account in the iteration of the kernel giving the GGF
\begin{equation}\lab{GGFcoll}
 \G_\om={1\over\om}\left[\id-{1\over\om}\K_\om\right]^{-1}
\end{equation}
of the model.

The collinear properties of the kernel (\ref{eq:kernel}) can also be seen in
$\ga$-space, by the expansion
\begin{equation*}
  \K_{\om}(k,k')=\sum_{n=0}^{\infty}[\ab(k^2)]^{n+1}\,K_\om^{(n)}(k,k')\;,
\end{equation*}
where the scale-invariant kernels $K_\om^{(n)}$ have eigenvalue functions
$\achi_\om^{(n)}(\ga)$ given by
\begin{equation}\lab{chincol}
 \chi_\om^{(n)}(\ga)={1\cdot A_1(A_1+b)\cdots(A_1+(n-1)b)\over
 (\ga+\ts{1\over2}\om)^{n+1}}+{1\cdot(A_1-b)(A_1-2b)\cdots(A_1-nb)\over
 (1-\ga+\ts{1\over2}\om)^{n+1}}\;.
\end{equation}
The leading order eigenvalue (cfr.~Eq.~(\ref{identifico0}))
\begin{equation}\lab{chi0col}
 \chi^{(0)}(\ga)\dug\chi_{\om=0}^{(0)}(\ga)={1\over\ga}+{1\over1-\ga}
\end{equation}
differs from BFKL quite significantly numerically, but retains a very similar structure
--- a saddle point at $\ga=1/2$ (see Fig.~\ref{f:confronto}{\sl a}), implying a power
growth of the cross section, and diffusion.
\begin{figure}[ht!]
\centering
\includegraphics[width=15cm]{\fig 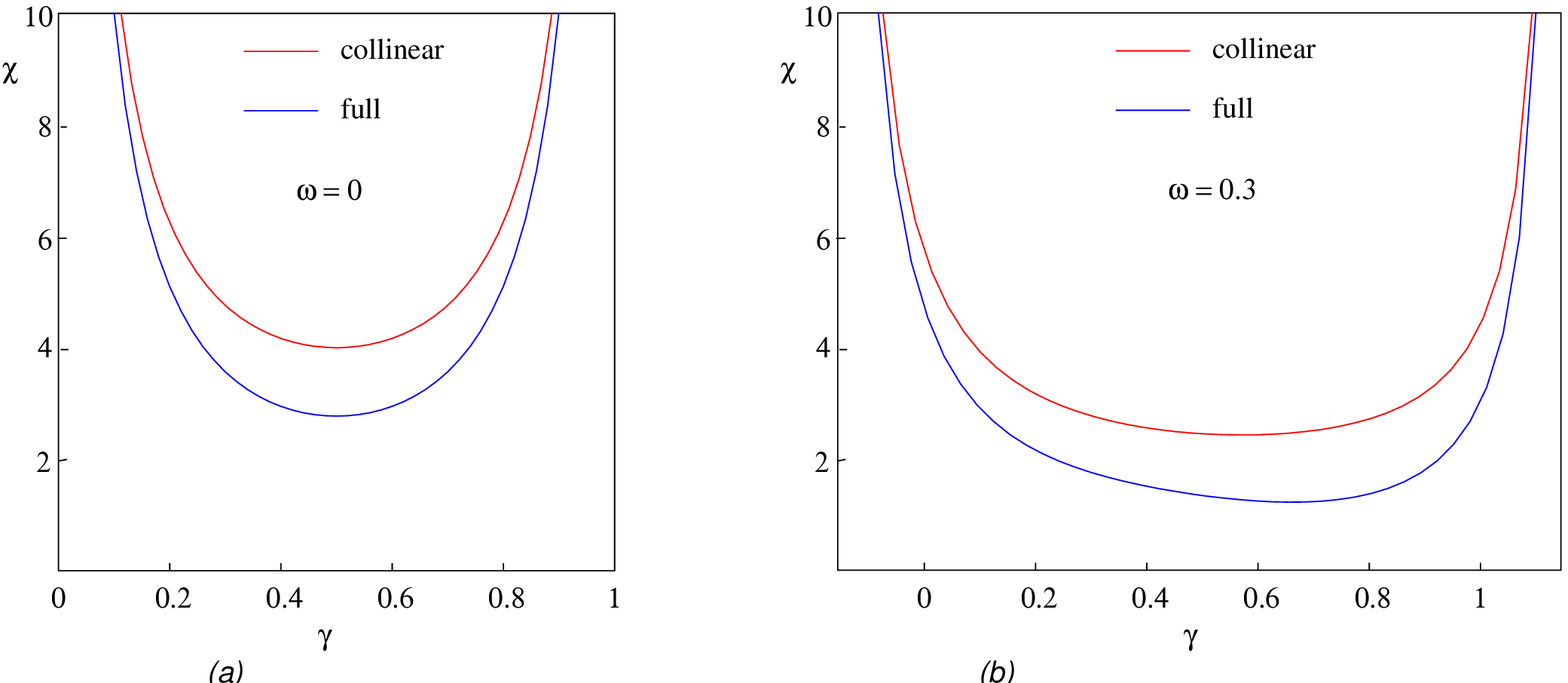}
\put(-112,-7){({\sl a})}\put(-33,-7){({\sl b})}
\didascalia{Comparison of the effective eigenvalue functions of the collinear model and
of the full small-$x$ kernel of Chap.~\ref{c:rga}. In ({\sl a}) the leading ($\om=0$)
ones and in ({\sl b}) the L$x$+NL$x$ ones for $\om=0.3$.\labe{f:confronto}}
\end{figure}
The NL$x$ eigenvalue (cfr.~Eq.~(\ref{identifico1}))
\begin{equation}\lab{chi1col}
 \chi^{(1)}(\ga)\dug\chi^{(1)}_{\om=0}(\ga)+\chi^{(0)}_{\om=0}\,\de_\om\chi^{(1)}_\om
 |_{\om=0}=\frac{A_1}{\ga^2}+\frac{A_1-b}{(1-\ga)^2}-\frac{\chi^{(0)}(\ga)}{2}
 \left(\frac{1}{\ga^2}+\frac{1}{(1-\ga)^2}\right)
\end{equation}
turns out to be very close, even numerically, to the full BFKL NLO kernel, reproducing
it to better than $7\%$ over the whole range of $\ga$ from $0$ to $1$ (cfr.\
Fig.~\ref{fig:chi1c}). This confirms that collinearly enhanced effects dominate the
NL$x$ kernel.

\section{Differential equation formulations\labe{s:difeqform}}

\subsection{First order formulation\labe{ss:1of}}

The main advantage of our collinear kernel from the point of view of this article is its
relative simplicity. Specifically it can be written in factorized form:
\begin{equation}\lab{eq:kernelfacts}
 \KK_\om(t,t') = U_\om(t) \,V_\om(t')\, \Theta(t-t') + U_\om(t') \,V_\om(t)\, \Theta(t'-t)\,,
\end{equation}
where (the $\om$-dependence will be understood)
\begin{subequations}\labe{d:UV}
\begin{align}
 U(t) &=\ab(t)\exp\left\{-{1+\om\over2}t+A_1(\om)
  \int^t\ab(\tau)\,\dif\tau\right\}\,,\\
 V(t) &=\exp\left\{{1+\om\over2}t-A_1(\om)\int^t\ab(\tau)
 \,\dif\tau\right\}\,.
\end{align}
\end{subequations}
This allows us to recast the homogeneous BFKL equation
\begin{equation}\lab{eq:homogen}
 \om\cF(t)\equiv\om k\F_\om(k) = U(t) \int_{-\infty}^t\dif t'\;V(t') \,\cF(t') + 
           V(t) \int^{ \infty}_t\dif t'\;U(t') \,\cF(t')
\end{equation}
as a differential equation. Dividing $\cF$ into two parts,
\begin{subequations}\labe{eq:ABFdef}
  \begin{align}\lab{eq:Adef}
    \cFL(t) &= U(t) \int_{-\infty}^t\dif t'\;V(t') \,\cF(t')\\ \lab{eq:Bdef}
    \cFH(t) &=   V(t) \int^{ \infty}_t\dif t'\;U(t') \,\cF(t')\\
    \om\cF(t) &= \cFL(t) + \cFH(t)
  \end{align}
\end{subequations}
and taking the derivative leads to a pair of coupled differential equations:
\begin{subequations}\labe{eq:DEUV}
  \begin{align}
    \frac{\dif\cFL}{\dif t} &= \frac{U'}{U} \cFL + U V \cF\,,\\
    \frac{\dif\cFH}{\dif t} &= \frac{V'}{V} \cFH - U V \cF\,.
  \end{align}
\end{subequations}
For the specific kernel (\ref{eq:kernel}) that we consider, we have~\cite{CiCoSa99b}
\begin{subequations}\labe{eq:DE}
\begin{align}
  \frac{\dif\cFL}{\dif t} &= \left(-\frac{1+\om}{2} +
    A_1\ab + \frac{\ab'}{\ab}\right)\cFL +\ab\cF \\
 \frac{\dif\cFH}{\dif t} &=\left(\frac{1+\om}{2}-A_1\ab\right)\cFH-\ab\cF
\end{align}
\end{subequations}
Since we have two coupled equations, there are two independent solutions. Examining the
equation for large and positive $t$, where $\ab$ is small, one sees that they can be
classified as a regular solution
\begin{equation}\lab{eq:FRbigt}
  \cF_R \sim \exp\left(-\frac{1+\om}{2}t\right) ,
\end{equation}
which is dominated by $\cFL$, and an irregular solution 
\begin{equation}\lab{eq:FIbigt}
  \cF_I \sim \exp\left(\frac{1+\om}{2}t\right) ,
\end{equation}
dominated by $\cFH$.

\subsection{Second order formulation}\labe{ss:2of}

The coupled set of differential equations (\ref{eq:DEUV}) can be recast in the form of a
simple second order equation for $\cF$. In fact, by using (\ref{eq:ABFdef}), we can first
rewrite (\ref{eq:DEUV}) in the form
\begin{equation}\lab{eq:2ndorda}
 \om \cF = \left[ \left(\partial_t - \frac{U'}{U}\right)^{-1}
 -\left(\partial_t - \frac{V'}{V}\right)^{-1}\right]UV \cF\,.
\end{equation}
Then, in order to eliminate the resolvents appearing in (\ref{eq:2ndorda}) we introduce
the operator
\begin{align}\lab{eq:oper}
  \Dt &= \left(\partial_t + \frac{U'}{U} - \frac{w'}{w}\right)
            \left(\partial_t - \frac{U'}{U}\right)
      = \left(\partial_t + \frac{V'}{V} - \frac{w'}{w}\right)
            \left(\partial_t - \frac{V'}{V}\right)\\ \nonumber
 &=\de^2-{w'\over w}\;\de+{U'V''-U''V'\over w}\;,
\end{align}
and the wronskian
\begin{equation}\lab{eq:wronskian}
  w(t) =W[U,V]\equiv UV'-U'V=\ab(t)\left(1 + \om - 2A_1\ab -
    \frac{\ab'}{\ab} \right)\,.
\end{equation}
By applying $\Dt$ to (\ref{eq:2ndorda}) we finally obtain~\cite{CiCoSa99b}
\begin{equation}\lab{eq:2ndord}
  \om\Dt\cF=\left(\frac{U'}{U}-\frac{V'}{V}\right)UV\cF=-w(t)\cF,
\end{equation}
which is the second order formulation that we were looking for.

Eq.~(\ref{eq:2ndord}) can be recast in normal form by the similarity transformation
\begin{equation}\lab{eq:simtrans}
  \cF(t)={\rm const}\cdot\sqrt{w(t)}\,f(t)\,,
\end{equation}
which leads to a Schr\"odinger type equation,
\begin{subequations}\labe{eq:schr}
\begin{align}\lab{schr}
  (-\partial_t^2 + V_{\eff})\,f&= 0\\ \lab{veff}
  V_{\eff} &=\frac{1}{4}\left(\frac{w'}{w}\right)^2-\frac12
 \left(\frac{w'}{w}\right)'-\frac{w}{\om}+\frac{U''V'-V''U'}{w}\;.
\end{align}
\end{subequations}
Note that the above derivation is valid for any form of the running coupling $\ab(t)$
which extrapolates the perturbative form $(bt)^{-1}$ into the strong coupling region
around the Landau pole $t=0$.

In what follows we consider various regularizations of the Landau pole, in particular
the cut-off and the frozen $\as$ cases defined in Eqs.~(\ref{cutoff}) and
(\ref{freeze}). It is obvious that such different regularizations will change the form
of the potential (Fig.~\ref{f:potenziale}) and thus the boundary conditions on $\cF$ (or
$f$) coming from the strong coupling region.  A similar Schr\"odinger formulation was
found~\cite{CaCi96b} for the (Airy) diffusion model~\cite{GrLeRy83,CoKw89} with running
coupling, with a potential which roughly corresponds to the bottom of the well in
Fig.~\ref{f:potenziale}.
\begin{figure}[ht!]
\centering
\includegraphics[width=10cm]{\fig 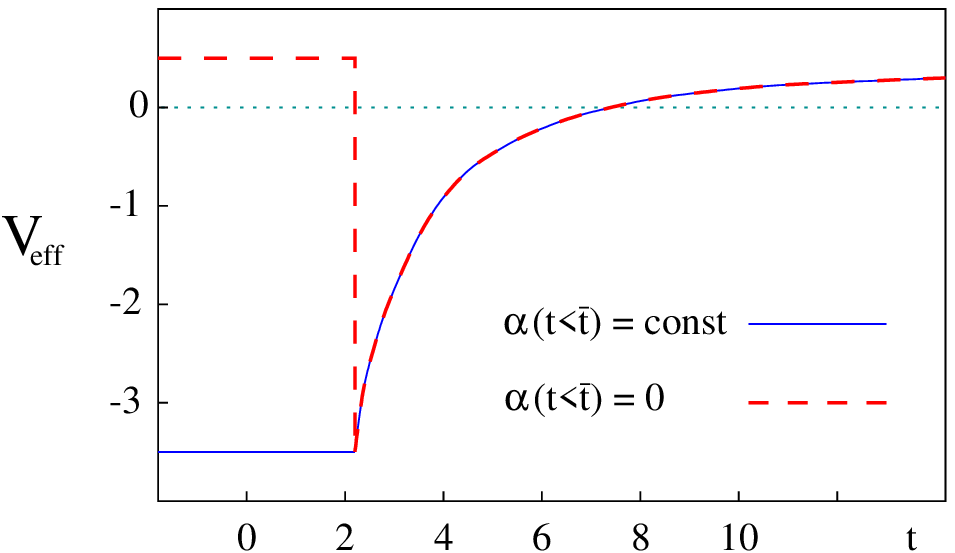}
\didascalia{Qualitative $t$-dependence of the effective potential for
        the regularizations of type (a) and (b) of the coupling strength.\labe{f:potenziale}}
\end{figure}
\\ \indent We consider in particular the solution $\cF_R(t)$ ($\cF_L(t)$) of the homogeneous
equation (\ref{eq:2ndord}) which is regular for $t \to +\infty$ ($t\to-\infty$). If both
conditions are satisfied, $\cF_R=\cF_L$ is an eigenfunction of the BFKL equation. The
pomeron singularity $\om=\op$ is the maximum value of $\om$ for which this occurs
(ground state).

If $\om > \op$, $\cF_R\neq\cF_L$ and the two solutions have rather different
features. Due to the locality of the differential equation, $\cF_R$ is
\emph{independent} of the regularization in the region $t>\tb$. On the other hand,
$\cF_L$ will be dependent on the behaviour of $\ab(t)$ for $t<\tb$. For instance, in the
case of $\ab(t)$ being frozen (\ref{freeze}), $\cF_L$ has the exponential behaviour
\begin{equation}\lab{fsek}
  \cF_L(t)\sim\esp{\kappa t}\,,\qquad(t<\tb)\,,
\end{equation}
where $\kappa=\sqrt{V_{\eff}(t<\tb)}$ is found from (\ref{veff}) to be
\begin{equation}\lab{kapa}
  \kappa^2 = \left[ \frac12(1+\om) - A_1\ab(\tb)\right]
             \left[ \frac12(1+\om) - A_1\ab(\tb)-\frac{2\ab(\tb)}{\om}\right].
\end{equation}

\subsection{Factorization of non-perturbative effects\labe{s:fact}}

The basic tool for describing BFKL evolution is the Green's function
$\GG_\om(t,t_0)\equiv kk_0\G_\om(k,k_0)$, which satisfies the inhomogeneous small-$x$
equation
\begin{equation}\lab{eq:green}
  \om \GG_\om(t,t_0)=\delta(t-t_0)+\KK_\om\otimes\GG_\om(t,t_0),
\end{equation}
and is supposed to be well-behaved for $t,t_0\to\pm \infty$. The problem of
factorization is the question of the (in)dependence on the non-perturbative
strong-coupling region.

For $t\neq t_0$, $\GG_\om$ satisfies the same differential equation as $\cF$ and is thus
a superposition of two independent solutions. The large-$t$ behaviour implies a
regularity condition and suggests the expression
\begin{equation}\lab{eq:fact}
  \GG_\om(t,t_0) = \cF_R(t)\cF_L(t_0)\Th(t-t_0)+\cF_L(t)\cF_R(t_0)
 \Th(t_0-t)\quad,\quad(t\neq t_0)
\end{equation}

Actually (\ref{eq:fact}) is a rigorous consequence of the second-order formulation. In
fact, $\GG_\om$ satisfies the differential equation
\begin{equation}\lab{edg}
 \left(\Dt+{w\over\om}\right)\GG_\om={1\over\om}\Dt\d(t-t_0)\,.
\end{equation}
With a little thought, one can realize that $\GG_\om$ must contain a delta function term
in the form
\begin{equation}\lab{scomp}
 \GG_\om(t,t_0)={1\over\om}\d(t-t_0)+\hat{\GG}_\om(t,t_0)
\end{equation}
and hence
\begin{equation}\lab{edhg}
 \left(\Dt+{w\over\om}\right)\hat{\GG}_\om=-{w\over\om^2}\d(t-t_0)
\end{equation}
showing that $\hat{\GG}$ is continuous function at $t=t_0$ with discontinuous
derivative. Eq.~(\ref{edhg}) is an inhomogeneous Schr\"odinger type equation with a
delta source, and its solution is just like the RHS of Eq.~(\ref{eq:fact}) also for
$t=t_0$. We conclude that
\begin{equation}\lab{gdelta}
  \GG(t,t_0)={1\over\om}\d(t-t_0)+\cF_R(t)\cF_L(t_0)\Th(t-t_0)+
 \cF_L(t)\cF_R(t_0)\Th(t_0-t)
\end{equation}
with the normalization
\begin{equation}\lab{fh}
  \cF_{R,L}(t)=\frac{\sqrt{w(t)}}{\om}f_{R,L}(t)\quad,\quad
  W[f_R,f_L]\dug f_R f'_L - f_R'f_L = 1\,.
\end{equation}
The main consequence of (\ref{eq:fact}) is that the regularization dependence is
factorized away in $\cF_L$, whenever $t$ or $t_0$ are large enough. This happens in
particular in the collinear limit $t-t_0\gg 1$ relevant for structure functions.

\section{Solutions: analytical features\labe{s:af}}

The collinear model just defined can be solved in principle as a Schr\"odinger problem
by known analytical and numerical techniques and for both types of solutions occurring
in the Green's function (\ref{eq:fact}) (i.e., the left-regular and the right-regular
ones).

The right-regular solution $\cF_R$ is, for large $t$, perturbative, i.e., independent of the
potential in the strong coupling region $t\leq\tb$, while the left-regular one $\cF_L$
is dependent on the strong coupling. The left-regular solution $\cF_L$ can be written as
a superposition of the right-regular solution $\cF_R$ and an other generic (linearly
independent on $\cF_R$) solution $\cF_I$ (which is necessarily right-irregular, i.e.,
$\cF_I\to\infty$ for $t\to+\infty$)%
\footnote{The right-irregular solution $\cF_I$ has to be specified by suitable boundary
conditions for $t\to+\infty$, thus it is independent of the strong coupling region. This 
is the reason why $\cF_I$ is useful for parametrizing the left-regular function $\cF_L$.}.
By normalizing $\cF_I$ in such a way that
\begin{equation}\lab{normcond}
 W[f_R,f_I]=1\quad,\quad\cF_I(t)\ugd{\sqrt{w(t)}\over\om}\,f_I(t)\;,
\end{equation}
we have
\begin{equation}\lab{fleft}
 \cF_L(t)=\cF_I(t)+S(\om)\cF_R(t)\;.
\end{equation}

The coefficient $S(\om)$ is formally found from the matching conditions of both sides of
Eq.~(\ref{fleft}) and their first derivatives at $t=\tb$:

\begin{equation}\lab{esse}
 S(\om)={\cF_I(\tb^+)\over\cF_R(\tb^+)}\cdot
 {\L_I(\om)-\L_L(\om)\over\L_L(\om)-\L_R(\om)}\quad,\quad
 \L_i\dug{\cF'_i(\tb^+)\over\cF_i(\tb^+)}\;.
\end{equation}
The coefficient $\L_L$ is provided by the strong coupling boundary conditions. In the
two cases of regularization that we are considering we have
\begin{align}\lab{elleLf}
 &\left.\L_L\right|_{\rm freezing}=\kw(\om)\\ \lab{elleLc}
 &\left.\L_L\right|_{\rm cutoff}=\left[\left({1+\om\over2}-A_1\ab(\tb)\right)
 \left(1-{\ab(\tb)\over\om(1+\om)}\right)+{\ab'(\tb)\over\om(1+\om)}
 \right]\left(1+{\ab(\tb)\over\om(1+\om)}\right)^{-1}\,,
\end{align}
where $\ab(\tb)$ is the $t\to\tb^+$ limit of $\ab(t)$.

Let us remark that, while $\cF_I$ and $\cF_R$ are expected to have an essential
singularity at $\om=0$ only, (cfr.\ Sec.~\ref{s:heb}), the coefficient $S(\om)$ displays
the leading part of the $\om$-spectrum in $\re\om>0$, and in particular the leading
Pomeron singularity $\op$. In fact, if $\om\to\om_0\in\Sp(\K_\om)$, then%
\footnote{This argument apply in the case of $\om_0$ in the discrete spectrum of
$\K_\om$. If $\om_0$ belongs to the continuous spectrum, then $S(\om)$ is discontinuous
at $\om=\om_0$, the singularity being a branch cut.}
$\cF_L\to\text{const}\times\cF_R$ and
the normalization condition (\ref{normcond}) can be satisfied only if $S(\om)\to\infty$.

The soft pomeron $\op$ is the greatest $\om$-singularity of $S(\om)$ and in the
frozen-$\as$ regularization it corresponds to the endpoint of the continuous spectrum.
In the cut-off case, we expect also a discrete component in the spectrum (since there is
a potential well, see Fig.~\ref{f:potenziale}) with poles at $\om$-values such that
$\L_L(\om)=\L_R(\om)$, both being analytic functions in $\re\om>0$. Mixed cases are
also possible, depending on the detailed shape of the potential in the strong coupling
region.

Eq.~(\ref{fleft}) allows us to rewrite the Green's function for $t>t_0$ and $\om>\op$ in
terms of $S(\om)$ in Eq.~(\ref{esse}) as follows:
\begin{equation}\lab{fattg}
 \GG_\om(t,t_0)=\cF_R(t)\left[\cF_I(t_0)+S(\om)\cF_R(t_0)\right]\,.
\end{equation}
If both $t,t_0\gg1$, but $t-t_0=\ord(1)$, Eq.~(\ref{fattg}) is dominated by the first
term, the second being suppressed exponentially in $t_0$ because the regular solution is
much smaller than the irregular one for $t_0\gg1$. This term is, on the other hand,
defined by boundary conditions for $t,t_0\to+\infty$ only, and is therefore independent
of the strong coupling region. Its analytical and numerical form will be discussed in
more detail in the following.

The second term in Eq.~(\ref{fattg}) carries the regularization dependence and contains
the leading $\om$-singularities, in particular $\op$ (Ref.~\cite{CaCi97b}). In $Y\dug\log
(s/kk_0)$ space, the sum in Eq.~(\ref{fattg}) defines two asymptotic regimes, as we
shall see.

\section{Strong-coupling features\labe{s:scf}}

Within the collinear model we can study the spectrum of $\K_\om$,
which provides the $\om$-singularities of the Green's function embodied in $S(\om)$, and in turn determine the high-energy
behaviour of the cross section.

The leading $\om$-singularity is the Pomeron $\op$, i.e., the maximum
$\om$ value for which a zero energy bound state for the potential (\ref{veff}) is
present. The pomeron properties depends on the coupling regularization. For smooth and
non negative couplings the effective potential is a regular function bounded from below
and increasing with $\om$ so that there exists a value of $\om=\op>0$ such that for
$\om>\op$ there are no more zero energy solutions. This demonstrates that the spectrum of 
the collinear kernel is bounded from above ($\Sp(\K_\om)\subset\;]-\infty,\op]$) for any
regularization ($\op$ depends however on the regularization!).

In the following, we
consider the two cases of freezing (\ref{freeze}) and cutting-off (\ref{cutoff})
$\ab(t)$. The first case is fairly simple, because for $t<\tb$ the effective potential
is constant and at its minimum value. Therefore $\op$, the upper limit of the continuous 
spectrum, given by $V^{\eff}_\om(t<\tb)-0$, is determined by the condition
\begin{equation}\lab{pomfrez}
 \op={4\ab(\tb)\over1+\op-2\ab(\tb)A_1(\op)}\;.
\end{equation}
This is to be compared with the saddle point definition of the hard
Pomeron (\ref{omsp}) $\om_s(\tb)$, which is obtained by
minimizing the solution (as a function of $\ga$) of
\begin{equation}\lab{oms}
  \om_s =\ab\left(\chi_0^{\om_s}(\ga) + \om_s
  \frac{\chi_1^{\om_s}}{\chi_0^{\om_s}}\right)\,.
\end{equation}
Taking the $b=0$ form for $\chi_1^{\om}$ the minimum is at $\ga=1/2$,
and $\om_s$ is given by
\begin{equation}
  \om_s = \frac{4\ab}{1 + \om_s - 2\ab A_1}\,.
\end{equation}
which is identical to the true $\op$.

The second critical exponent $\om_c(t)$ introduced in Sec.~\ref{s:iad}, corresponds to
the rightmost singularity of the anomalous dimension. In the collinear model the
integrated gluon distribution can be defined as
\begin{equation}\lab{eq:igluondef}
 g_\om(t_{\!_Q})\dug\int_{-\infty}^{t_{\!_Q}}\dif t\;\exp\left\{{1+\om\over2}t+A_1
 \int_t^{t_{\!_Q}}\ab(\tau)\;\dif\tau\right\}\cF_\om(t)={\cFL(t)\over\ab(t)}
\end{equation}
because $g_\om(t)$ can be shown to satisfy, in the collinear limit, the usual DGLAP
equation with anomalous dimension $(1/\om+A_1)\ab$. The singularity of the anomalous
dimension (at energy scale $k^2$)
\begin{equation}\lab{eq:intgam}
 \ga^{\eff}_\om\(\ab(t)\)\dug{\dif\over\dif t}\ln g_\om(t)
\end{equation}
is at the point $\om=\om_c(t)$ where $g_\om(t)$ goes to zero, i.e.,
where $\cFL(t)$ goes to zero.

It is easier to discuss $\om_c(t)$ in the cut-off case, in which $\ab(t)=0$ for $t<\tb$.
In this case (\ref{eq:igluondef}) allows us to write
\begin{equation}\lab{FaaB}
 \om\cF(t)=\cFL(t)+\cFH(t)=\ab(t)g(t)+\cFH(t)\;.
\end{equation}
By substituting (\ref{FaaB}) in the first order differential equation (\ref{eq:DE}) it
is not difficult to show that both $g(t)$ and $\cFH(t)$ are continuous at $t=\tb$ and
exponentially behaved for $t<\tb$: $g\sim\cFH\sim\exp\(\half(1+\om)t\)$. It follows
that
\begin{equation}\lab{condracc}
 {\cFL(\tb^+)\over\ab(\tb^+)}=g(\tb)={\cFH(\tb)\over\om(1+\om)}\;.
\end{equation}
For $\om=\op$, $\cF_{\op}(t)$ corresponds to the ground state of the system and have no
zeroes:
\begin{equation}\lab{nonzeroF}
 0\neq\cF_{\op}(\tb^+)={1\over\op}\(\cFL(\tb^+)+\cFH(\tb)\)={1\over\op}\left[
 {\ab(\tb^+)\over\op(1+\op)}+1\right]\cFH(\tb)\;,
\end{equation}
hence $\cFH(\tb)\neq0$ and also $g_{\op}(\tb)\neq0$. In order to have $\cFL(\tb^+)=0$
one has to increase the depth of the potential well, i.e., to lower $\om<\op$. This
means that $\om_c(\tb)<\op$.

The relationships just found
($\op^{\rm freezing}=\om_s(\tb),\;\op^{\rm cutoff}>\om_c(\tb)$)
represent two extreme cases of the boundary condition dependence of
$\op$. If the coupling $\ab(t)>0$ is positive but has intermediate
size and shape for $t<\tb$, we expect in general that
\begin{equation}\lab{teww}
 \om_c(\tb)<\op<\om_s(\tb)\;.
\end{equation}

\section{Perturbative regime: $\!\boldsymbol\om$-expansion and $\!$WKB $\!$limit%
\labe{s:oewl}}

Approximate homogeneous solutions in the large-$t$ region can be found, as in the full
small-$x$ equation of Chap.~\ref{c:rga}, by the method of the $\ga$-representation
(\ref{rappres}) and $\om$-expansion (\ref{omegaes}). The regular solution is
approximated, for $\om\ll1$, by the expression:
\begin{equation}\lab{rapintf}
  \cF_R(t)\simeq \int\difg\;\exp\left[(\ga-\half)t -
    \frac{X^\om}{b\om}\right]\,,
\end{equation}
with 
\begin{equation}\lab{XNL}
  \de_\ga X^\om \equiv \chi(\ga,\om) = \chi^\om_0(\ga) + \om
  \frac{\chi^\om_0(\ga)}{\chi^\om_1(\ga)}+\ord(\om^2)
\end{equation}
where, for the collinear model,
\begin{subequations}\labe{chicol}
  \begin{align}
    \chi^\om_0(\ga) &= \frac1{\ga + \half \om}  +  \frac1{1-\ga +
      \half\om},\\
    \chi^\om_1(\ga) &= \frac{A_1}{(\ga + \half \om)^2}  +  \frac{A_1 -
      b}{(1-\ga + \half\om)^2}.
  \end{align}
\end{subequations}
In Chap.~\ref{c:rga} this representation was extensively studied, and shown to be a solution
of the problem up to next-to-leading order in $\om$ and to all orders for the collinear
structure. But this was only guaranteed to work true for small values of $\om$ (whereas
for the continuation with the DGLAP anomalous dimensions it is useful to be able to
access high $\om$ as well). Also there was no way of determining the coefficient of any
higher-order error introduced by the procedure.  A comparison of this representation
with the exact solution, as is possible in the collinear model, is therefore important
(cfr.\ Sec.~\ref{ss:rrs}).

The expressions (\ref{rapintf}--\ref{chicol}) can be further specialized in the
large-$t$ limit, where (\ref{rapintf}) is dominated by a saddle point at
$\ga=\gb_\om(t):b\om t=\chi(\gb,\om)$ and yields
\begin{equation}\lab{pscol}
 \cF_R(t)=\sqrt{b\om\over2\pi|\chi_\om'(\gb)|}\exp\int^t\gb(\tau)\;\dif\tau\;.
\end{equation}
We can obtain the large-$t$ behaviour of the right-regular solution by means of the WKB
ap\-proximation for solving the differential equation (\ref{eq:schr}). In fact, on the
basis of Eqs.~(\ref{eq:schr}) and (\ref{fh}), one can prove the asymptotic expansion
\begin{equation}\lab{wkb}
 \cF_R(t)={1\over\om}\sqrt{w(t)\over2\,\kw(t)}\exp\left\{-\int_{t_\kw}^t\kw(\tau)
 \,\dif\tau\right\}\times\left[1+\ord\({1\over t}\)\right]\,,
\end{equation}
where $\kw(t)$ is defined in terms of the effective potential (\ref{veff}) as
\begin{equation}\lab{Kwkb}
 \kw^2(t)\equiv V_{\eff}(t)=\frac12\left(1+\om -{2A_1-b\over bt}\right)
             \left[\frac12\left(1+\om-{2A_1-b\over bt}\right) -
             \frac{2}{b\om t}\right]+\ord\({1\over t^2}\)
\end{equation}
and $t_\kw$ is the zero of the WKB momentum: $\kw(t_\kw)=0$.

Eq.~(\ref{pscol}) and (\ref{Kwkb}) are asymptotic representations of the same function
and are related by
\begin{equation}\label{gb}
 \gb=\frac12\left(1+{1\over t}\right)-\kw(t)+\ord\({1\over t^2}\)\;.
\end{equation}
It is interesting to note that in the collinear model, because of the differential
equation (\ref{eq:schr}), the present method yields the irregular solution also, which
is obtained by just changing the sign of the WKB momentum $\kw(t)$, i.e.,
\begin{equation}\lab{wkbirr}
 \cF_I(t)={1\over\om}\sqrt{w(t)\over2\,\kw(t)}\exp\left\{
 +\int_{t_\kw}^t\kw(\tau)\,\dif\tau\right\}\,.
\end{equation}
This is useful for evaluating the Green's function according to Eq.~(\ref{fattg}).

\section{Test of the $\boldsymbol\om$-expansion\labe{s:toe}}

\subsection{Right-regular solution\labe{ss:rrs}}

The possibility of evaluating (at least numerically) the GGF of the collinear model to
arbitrary precision, allows us to test the validity of the $\om$-expansion approximation 
method introduced in Sec.~\ref{s:soe} by comparing the exact right-regular solution of
the differential equation (\ref{eq:DE}), or equivalently (\ref{eq:schr}), with the
$\ga$-representation (\ref{rapintf}) where $X_\om(\ga)$ is determined by the NL$x$
truncation of the $\om$-expansion (\ref{XNL}). Fig.~\ref{f:forma} shows the exact and
approximated function $\cF_R(t)$ for $\om=0.15$. One sees that for larger values of $t$,
the $\om$-expansion is in good agreement with the exact solution, while for smaller $t$,
where the solutions oscillate the results from the $\om$-expansion are slightly out of
phase with the exact solution. In general we are interested in the behaviour to the
right of the rightmost zero.
\begin{figure}[ht!]
\centering
\includegraphics[width=10cm]{\fig 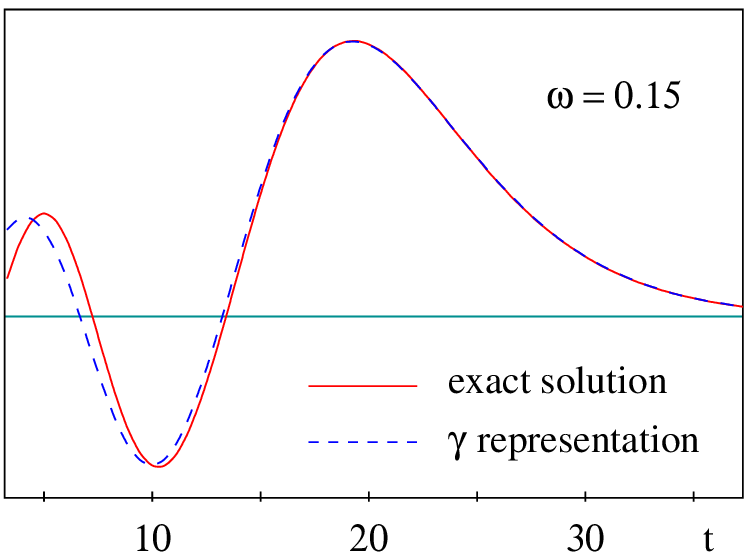}
\didascalia{Right-regular solution from the explicit solution of the differential
equation and from the $\om$-expansion in arbitrary normalization; shown for
$\om=0.15$.\labe{f:forma}}
\end{figure}

\subsection{Critical exponent\labe{ss:crex}}

One quantitative test of the $\om$-expansion concerns the critical exponent $\om_c$,
i.e., the value of $\om$ at which the effective anomalous dimension of
Eq.~(\ref{eq:intgam}) diverges. We recall that this is connected with the position of
the rightmost zero of the regular solution: $\cF_{\om_c}(t)=0$. The $\om$-expansion
determination of the position of this zero, i.e., the zero of $\cF_R(t)$ determined by
means of the $\ga$-representation (\ref{rapintf}) and the $\om$-expansion (\ref{XNL}),
involves a small error which we call $\delta\om_c$.  Fig.~\ref{f:omtdiff} shows
$\delta\om_c/\om_c$ as a function of $\om_c$, and we see that the relative error on
$\om_c$ goes roughly as $\om^2$, or equivalently as $\ab^2$.  This corresponds to a
NNL$x$ correction and is beyond our level of approximation. We note also that even for
relatively large values of $\om\sim0.3$, the relative correction remains of the order of
$5\%$ which is quite acceptable. In other words the NNL$x$ correction that arises is not
accompanied by a large coefficient.
\begin{figure}[ht!]
\centering
\includegraphics[width=12cm]{\fig 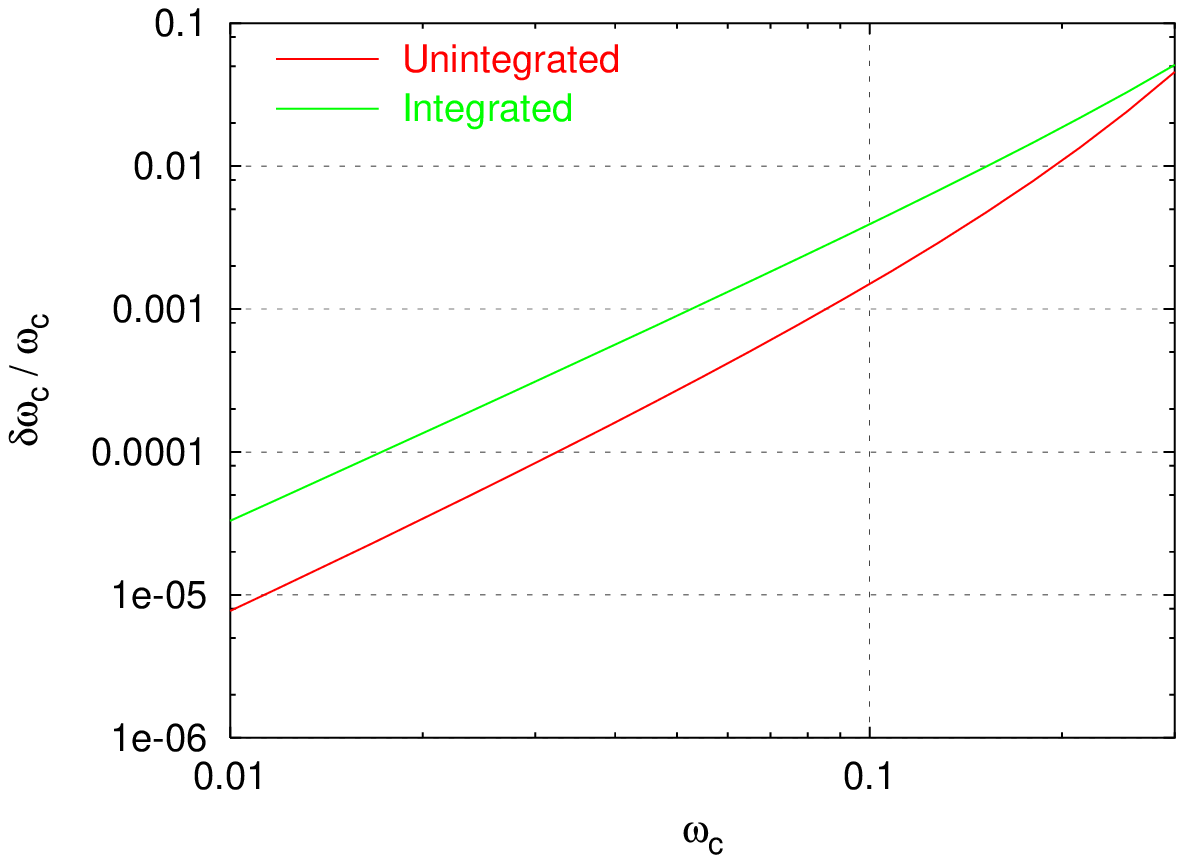}
\didascalia{The error, $\delta \om_c$ in the determination of $\om_c$
      within the $\om$-expansion. Shown for both the unintegrated
      and integrated solutions.\labe{f:omtdiff}}
\end{figure}

\subsection{Anomalous dimension\labe{ss:andimcol}}

Our second quantitative test of the $\om$-expansion concerns the anomalous dimension
$\ga_\om$, as defined in Eq.~\eqref{eq:intgam}. The error in the $\om$-expansion result
for the anomalous dimension, $\delta \ga$, is plotted in Fig.~\ref{f:dgam} as a
function of $\ab$ for two values of $\om$.  Let us first concentrate on the region for
$\ab < 0.01$. We see that the error is roughly independent of $\om$, and proportional
to $\ab^2$. In other words the difference between the exact result and the
$\om$-expansion is a term of $\ord(\ab^2)$. We recall that the terms that we wish to
include properly are the leading terms $(\ab/\om)^n$, the NL terms $\ab(\ab/\om)^n$
and the collinear terms $(\ab/\om) \om^n$. A first correction of $\ord(\ab^2)$ is
consistent with all these terms having been correctly included.

For $\ab > 0.01$ we see that the error in the anomalous dimension has a more
complicated behaviour: it changes sign (the dip) and then diverges, at the same point as
the divergence in the anomalous dimension itself (solid curve): this is just a
consequence of the position of the divergence of the anomalous dimension from the
$\om$-expansion being slightly different from the true position
\begin{figure}[ht!]
\centering
\includegraphics[width=12cm]{\fig 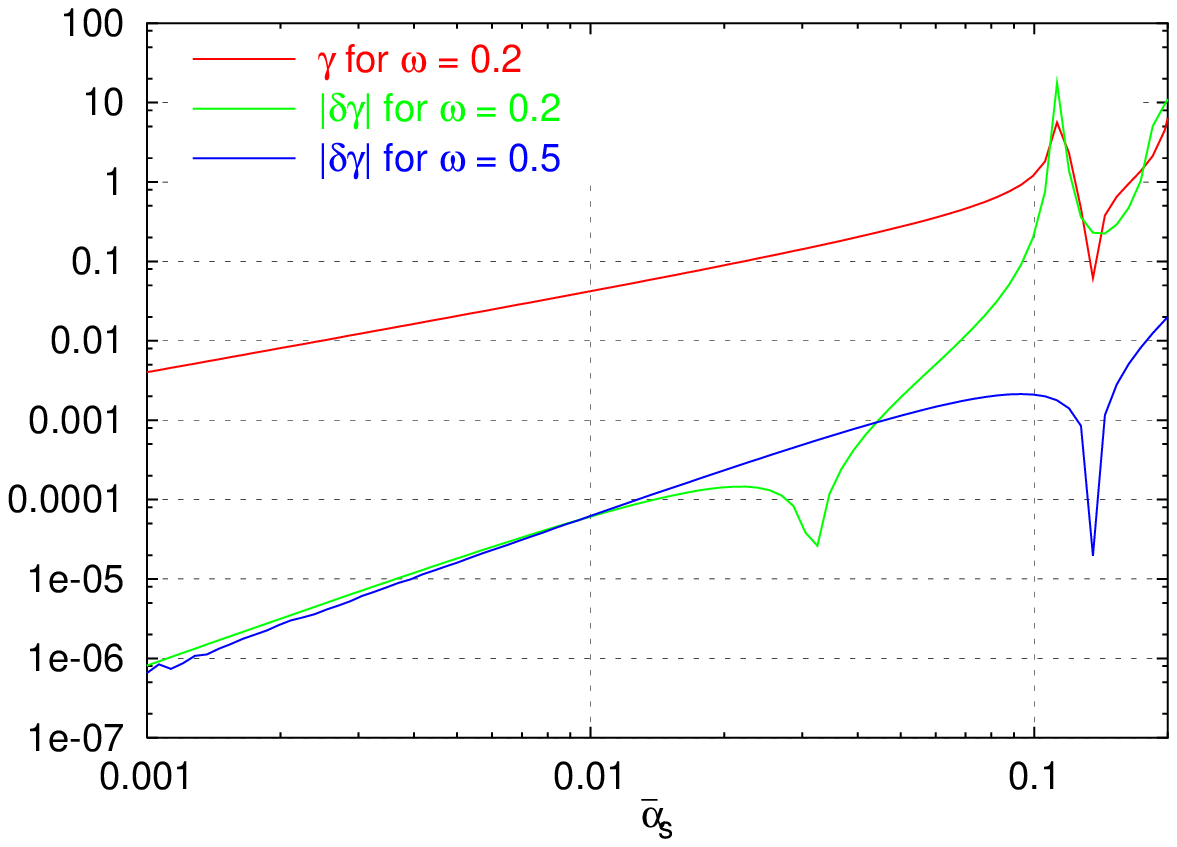}
\didascalia{Error in the effective anomalous dimension of Eq.~(\ref{eq:intgam}) from the
	$\om$-expansion plotted against $\as(t)$.\labe{f:dgam}}
\end{figure}

\section{High energy behaviour\labe{s:heb}}

It is widely believed that a two-scale process --- described by a
small-$x$ equation of BFKL type --- is perturbative for large enough
$t$ and $t_0$, while it becomes a strong coupling process if the
energy is so large as to allow diffusion to small values of $t\simeq0$
($k^2\simeq\La^2$)~\cite{BaLo93,FoRo97}.

In the collinear model the Green's function has the explicit
expression (\ref{fattg}), in which the strong-coupling information is
clearly embodied in the reflection coefficient $S(\om)$. Therefore 
it allows a direct study of the relative importance of the
``perturbative'' part $\cF_R\otimes\cF_I$ and of the ``strong-coupling''
part $S\,\cF_R\otimes\cF_R$, induced by diffusion through the boundary
conditions at $t=\tb$.

For large $t$ and $t_0$, the factorized form (\ref{eq:fact}) of $\hat{\GG}_\om$ can be
approximated by the WKB expression of Eq.~(\ref{wkb},\ref{wkbirr}), that we study in the
special case $t=t_0$, so that the GGF in $Y\dug\ln s/kk_0$ space is
\begin{equation}\lab{gytt}
 \hat{\GG}(Y;t,t)\simeq\int\difo\;\esp{\om Y}{w(t)\over2\om^2\kw_\om(t)}
 \left[1+S(\om)\exp\left(-2\int_{t_\kw}^t\kw_\om(\tau)\,\dif\tau\right)\right]
\end{equation}
for large enough $t$. The (perturbative) first term is characterized by a saddle point
at $\om=\omb(t)>\om_s(t)>0$ placed between the vertical asymptote of $1/\kw_\om(t)$ at
$\om\simeq\om_s(t)$ on the left and the fast rising exponent $\esp{\om Y}$ on the right.
Since around $\om_s$ we have $\kw_\om^2(t)\simeq\text{const}(t)\;[\om-\om_s(t)]$, the
saddle point condition reads
\begin{equation}\lab{Y}
 \left.Y=\frac12{\de\over\de\om}\log\kw^2_\om\right|_{\om=\omb}\simeq
 {1\over2(\omb-\om_s)}\;,
\end{equation}
hence
\begin{equation}\lab{ombarra}
 \omb(Y,t)\simeq\om_s(t)+{1\over2Y}\;.
\end{equation}
The saddle point (\ref{ombarra}) is self consistent provided the WKB approximation is
valid, i.e., the variation of the WKB momentum $\kw(t)$ is small on the typical
wave-length of the solution:
\begin{subequations}\labe{valPS}
\begin{equation}
 {\de_t\kw\over\kw}\Big/{2\pi\over\kw}\ll1\quad\iff\quad2Y\ll\left({2\over b}
 \right)^{2\over3}\om_s(t)^{-{5\over3}}\sim\ab(t)^{-{5\over3}}
\end{equation}
and provided $Y$ is not too small so that the saddle point is not too shallow:
\begin{equation}
 {1\over2Y}\ll\om_s(t)\sim\ab(t)\;.
\end{equation}
\end{subequations}
The saddle point estimate is finally performed by means of the integral
\begin{equation}
 \int_{\epsilon-\ui\infty}^{\epsilon+\ui\infty}
 {\dif x\over2\ui\sqrt{\pi x}}\esp{x}=1\,,\quad(x\equiv(\om-\om_s)Y)
\end{equation}
rather than by na\"\i ve quadratic fluctuations around $x=1/2$ and yields~\cite{CiCoSa99b}
\begin{equation}\lab{GYpert}
 \hat{\GG}^{(\text{pert})}(Y;t,t)\simeq{1\over\sqrt{2\pi\chi_m''\ab Y}}\,\esp{\om_s(t)Y}\times
 \left[1+\ord(\as^5Y^3)\right]\quad,\quad
 \chi_m''(t)\dug\underset{\ga}{\Min}\,\chi\(\ga,\om_s(t)\)\;.
\end{equation}

The contribution of the (non perturbative) second term of Eq.~(\ref{gytt}) is evaluated
with the residue method
\begin{equation}\lab{GYnp}
 \hat{\GG}^{(\text{nopt})}(Y;t,t)=\sum_{\om_i\in\{\text{poles}\}}\esp{\om_i Y}R_i
 \esp{-2\int_{t_\kw}^t\kw_{\om_i}(\tau)\,\dif\tau}\;.
\end{equation}
\begin{figure}[ht!]
\centering
\includegraphics[width=9cm]{\fig 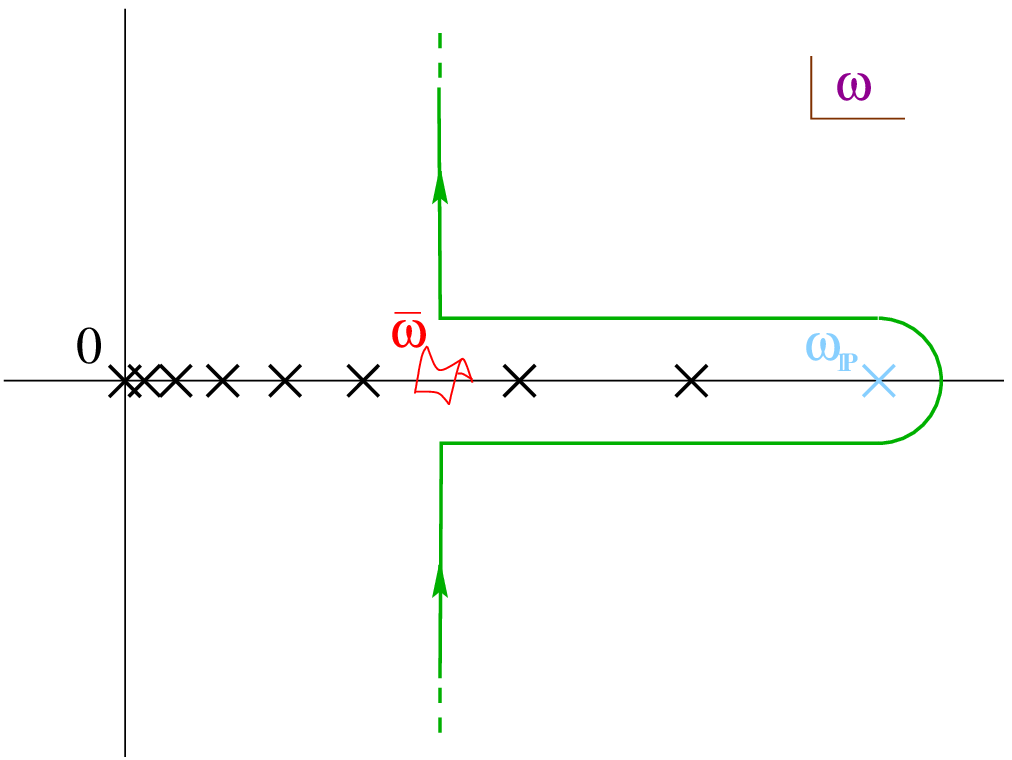}
\didascalia{The contour of integration for the Green's function $\hat{\GG}_\om(Y;t,t)$,
displaying the pole contribution of the non perturbative part and the background
integral parallel to the imaginary axis relevant for the perturbative part of the GGF.
\labe{f:contour}}
\end{figure}
Retaining only the leading pole $\op$ and observing that 
$\kw_{\op}(\tau)\simeq(1+\op)/2+\ord(1/\tau)$, we end up with the result~\cite{CiCoSa99b}
\begin{equation}\lab{res}
 \hat{\GG}(Y;t,t)\simeq{1\over\sqrt{2\pi\chi_m''\ab Y}}\,\esp{\om_s(t)Y}\times
 \left[1+\ord(\as^5Y^3)\right]+\esp{\op Y}R_{\mathbb{P}}\esp{-(1+\op)t}\,.
\end{equation}
showing that in the well-defined intermediate regime (\ref{valPS}) the exponent
$\om_s(t)$ appears to be an observable quantity. Increasing $Y$ the GGF enters a second, 
non perturbative regime where the second term of Eq.~(\ref{res}) dominates on the first
one, i.e.,
\begin{equation}\lab{d:Yt}
 Y>Y_t\dug{1+\op\over\op-\om_s(t)}\,t\;.
\end{equation}

Two remarks are in order.
\\ $\bullet$ It is possible to calculate the higher order corrections to
Eq.~(\ref{GYpert})) as shown in Ref.~\cite{CiCoSa99b}:
\begin{equation}\lab{subcor}
 \left[1+\ord(\as^5Y^3)\right]\quad\longrightarrow\quad\sum_{n=0}^\infty
  \frac{1}{n!}\left(\frac{\om_s^2\chi_m'' Y^3}{24 b t^3}\right)^n 
  =  \exp\left(\frac{\chi_m^2\chi_m''b^2\,\ab^5Y^3}{24}\right)\;.
\end{equation}
It turns out that these corrections coincide with the ``non-Regge correction'' to the
``Regge exponent'' $\om_s(t)Y$ found by other authors~\cite{KoMu98,Lev98} in different but
related context . We notice, however, that the true Regge contribution is the second
term in Eq.~(\ref{res}), which is of strong-coupling type, with a $t$-independent and
eventually leading exponent $\op$. The perturbative part, which dominates in the
large-$t$ limit, comes from the background integral and has no reason to be Regge
behaved.
\\ $\bullet$ The quantity $\om_c(t)$ --- the formal anomalous dimension singularity in the
definition (\ref{eq:intgam}) --- does not directly appear in the $Y$ dependence, because
the oscillating behaviour of the $\cF$'s, which is relevant for increasing values of
$Y$, is masked by the onset of $\op$ dominance for $Y>Y_t$.

We conclude that the two-scale Green's function shows a perturbative (non-Regge) regime
where the exponent $\om_s(t)Y$ shows up with calculable corrections
(Eq.~(\ref{subcor})), provided the parameter $\ab^5Y^3$ is small. Even before the latter
gets large, at $Y\gtrsim Y_t$ the Pomeron-dominated regime takes over, characterized by
the regular solution, which is confined to the strong-coupling region of small $t$'s.

\section{The tunneling mechanism\labe{s:tunn}}

The transition mechanism between the perturbative and non perturbative regimes described 
in the previous section reveals completely new features with respect to the well-known
diffusion one~\cite{FoRo97}, as we will see soon. We are going to present some numerical results, the
analytical description being rather hard.

Note first that several aspects of the two regimes are due to just running $\as$
effects. The latter enters at two levels: first, because of the ``attraction'' in the
potential well of Fig.~\ref{f:potenziale}, it modifies the traditional fixed-$\as$ kind
of diffusion, weighting it towards lower transverse scales. Secondly, because of the
strong-coupling boundary condition, it also introduces a qualitatively new kind of
diffusion, perhaps more properly referred to as `tunneling': namely there is a certain
$Y = Y_t$, defined in Eq.~(\ref{d:Yt}), at which the non-perturbative pomeron suddenly
takes over and beyond which the Green's function is dominated by the regular solution,
and thus confined to small $t$-values.

This `tunneling' phenomenon is qualitatively different from diffusion, in so far as
there is not a gradual decrease in the relevant scale for the evolution, until
non-perturbative scales become important, but rather there is a point beyond which low
scales suddenly dominate, without intermediate scales ever having been relevant.

To study these effects at a qualitative level it suffices to consider
a very simplified version of the collinear model: one which retains
only the running of the coupling, but not the $\om$-shifts of the
$\ga=0,1$ poles, nor the $A_1$ component of the NL
corrections~\cite{CiCoSa99b}. We then examine the solution to 
\begin{equation}
  \GG(Y,t,t_0) = \delta(t-t_0) + \int_0^Y\dif y\;K\otimes\GG(y)\,,
\end{equation}
for this simplified kernel and we consider the effective exponent of
the evolution
\begin{equation}
  \om_{\eff}=\frac{\dif}{\dif Y}\ln\hat{\GG}(Y,t,t)
\end{equation}
as a function of $Y$, where $\hat{\GG}(Y,t,t_0)$ is simply $\GG(Y,t,t_0)$ with
$\d(t-t_0)$ subtracted, in analogy with Eq.~\eqref{scomp}.
\begin{figure}[ht!]
\centering
\includegraphics[width=12cm]{\fig 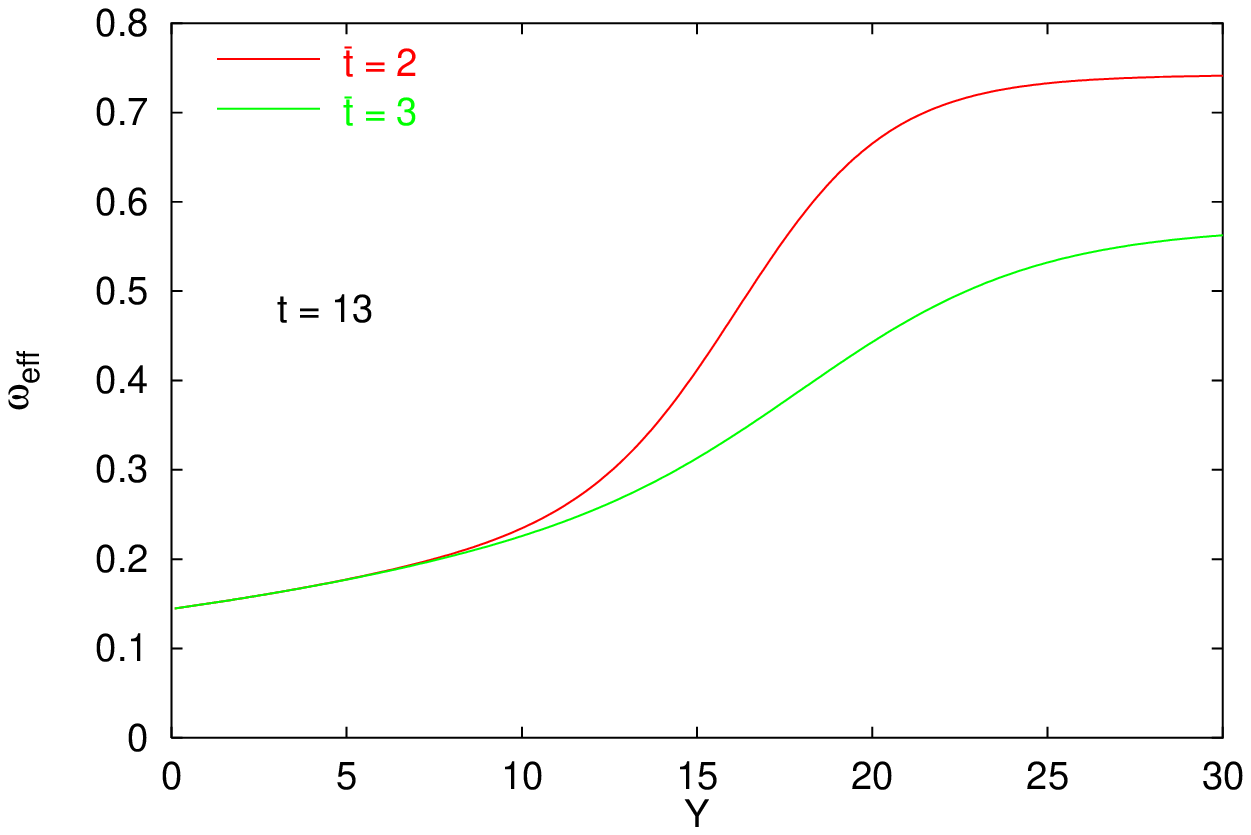}
\didascalia{The effective intercept $\om_{\eff}$ as a function of $Y$ for $t=13$ and two
	different values of the infra-red cutoff $\tb$. \labe{f:expstep}}
\end{figure}
Fig.~\ref{f:expstep} illustrates the basic behaviour of $\om_{\eff}$ for rather extreme
kinematics --- not intended to be phenomenologically relevant, but rather to show
clearly the relevant features. Two values of the infra-red cutoff are considered. What
is seen is that the exponent at first increases slowly and smoothly, and then at a
certain threshold $Y$ increases rapidly towards $\op$. The saturation of $\op$ occurs
later in $Y$ for increased $\tb$ (i.e., decreased $\op$) as expected from
\eqref{d:Yt}. Traditional smooth diffusion into the infra-red would have led to the
opposite behaviour, namely the higher $\tb$ case (lower $\op$) being saturated first.

In order to have a closer look to the underlying physical mechanism, we analyse the
$Y$-dependence of the so called Bartels' cigar consisting in the contour plot of the
function
\begin{equation}\lab{sigaro}
 {\GG(Y-y;t,t')\GG(y;t',t_0)\over\GG(Y;t,t_0)}\;.
\end{equation}
Once one has fixed $Y$, $t$ and $t_0$, the $y$-section of Eq.~(\ref{sigaro}) as a
function of $t'$ describes the relative importance of the intermediate states of
different transverse momentum $k'=\esp{t'/2}$ in the evolution of $\GG$ from 0 to $Y$.
\begin{figure}[p!]
\centering
\includegraphics[height=13cm,width=14cm]{\fig 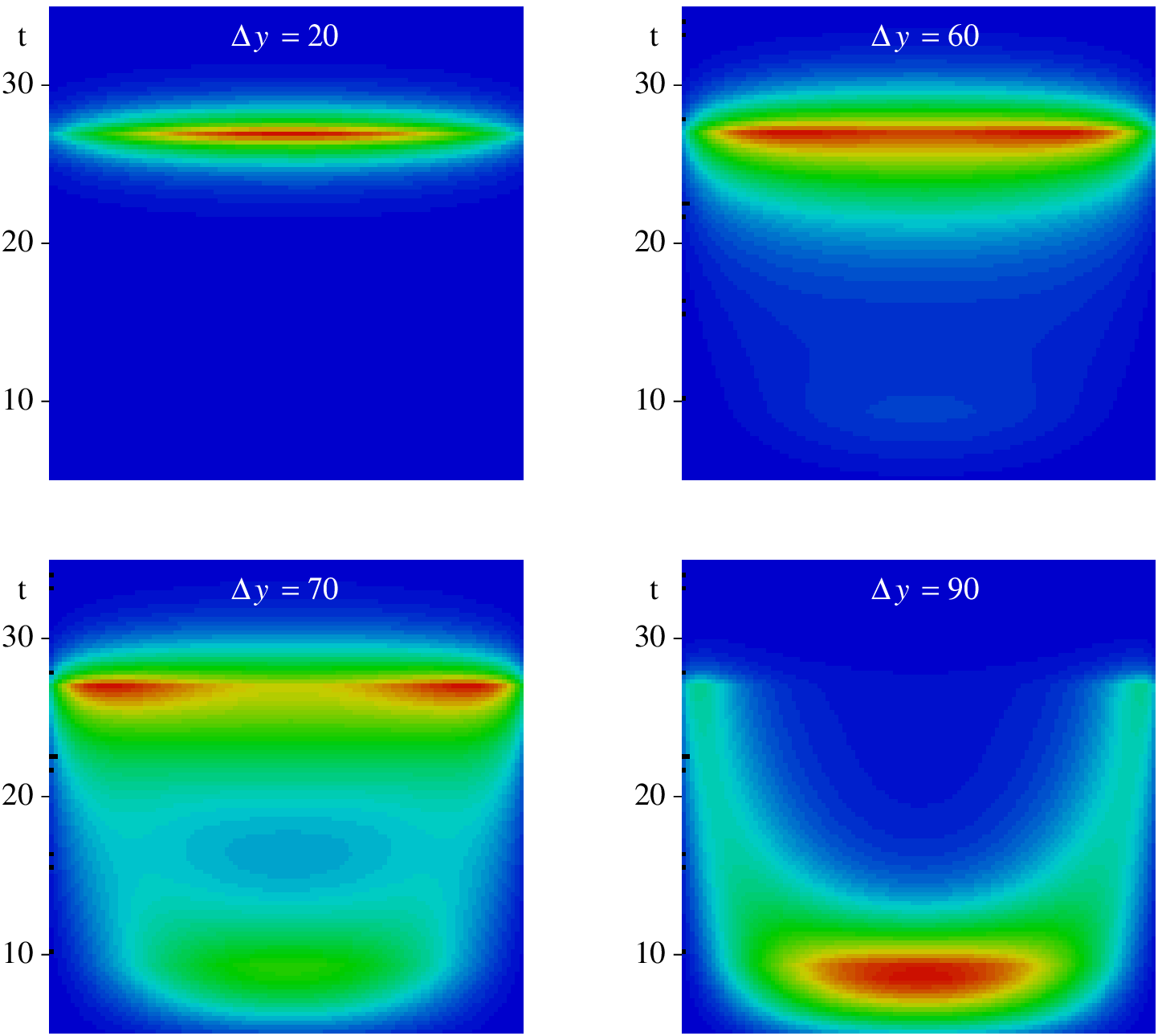}
\didascalia{Domains of the dominant paths in transverse momentum $t'$ and rapidity $y$ 
space for the evolution of the GGF in $Y$ between two hard and equal scales $t=t_0$. In
the first plots the relevant paths stays in the perturbative region, their spread
increasing with $Y$. The last plots show the sharp transition of the paths in the non
perturbative region corresponding to the pomeron dominated cross section.
\labe{f:sigari}}
\end{figure}

Let us start with a not too large rapidity interval $Y$ and consider for simplicity
$t_0=t\gg1$. The first plot of Fig.~\ref{f:sigari} shows that the domain of the dominant 
paths in transverse momentum-rapidity space, represented by the light coloured region,
is restricted around the value of $t$ of the initial and final state, with a small
spread. Increasing the value $Y$, the cigar gets thicker and thicker. In normal
diffusion, the reach and the broadening of the cigar in the non perturbative region is
gradual, according to its width $\sim\sqrt{\ab(t)Y}$.

In the collinear model instead, we observe (Fig.~\ref{f:sigari}) a jump (tunneling) in
the IR region of an increasing fraction of paths as $Y$ gets larger, until all the
relevant intermediate states belong to the strong coupling domain, thus providing the
non perturbative regime mentioned in Sec.~\ref{s:heb}.

An impressing picture of the tunneling phenomenon can be obtained by plotting
the distribution of transverse momenta in the middle of the evolution interval
$Y/2$: Fig.~\ref{f:anatra} displays the contour plot of the function
\begin{equation}\lab{anatra}
 {|\GG(Y/2;t,t')|^2\over\GG(Y;t,t)}
\end{equation}
in $t'$ and $Y/2$. In the left region (small evolution) the transverse momenta are
spread around $t$ in the perturbative region and the small-$x$ effective growth exponent
\begin{equation}\lab{d:omeff}
 \om_{\eff}(Y,t)\dug{\dif\ln\GG(Y;t,t)\over\dif Y}
\end{equation}
is given by the saddle point one $\om_s(t)$ plus the ($Y$-derivative of the) corrections 
(\ref{subcor}). For $Y>Y_t$ tunneling is at work and the small-$x$ exponent is given by
$\op$.
\begin{figure}[ht!]
\centering
\includegraphics[width=10cm]{\fig 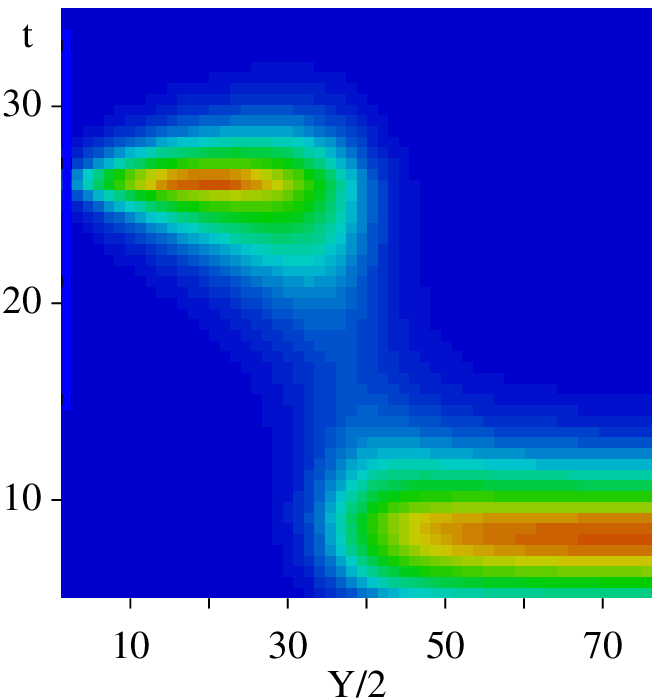}\\[5mm]
\hspace{-20mm}\includegraphics[width=10cm]{\fig 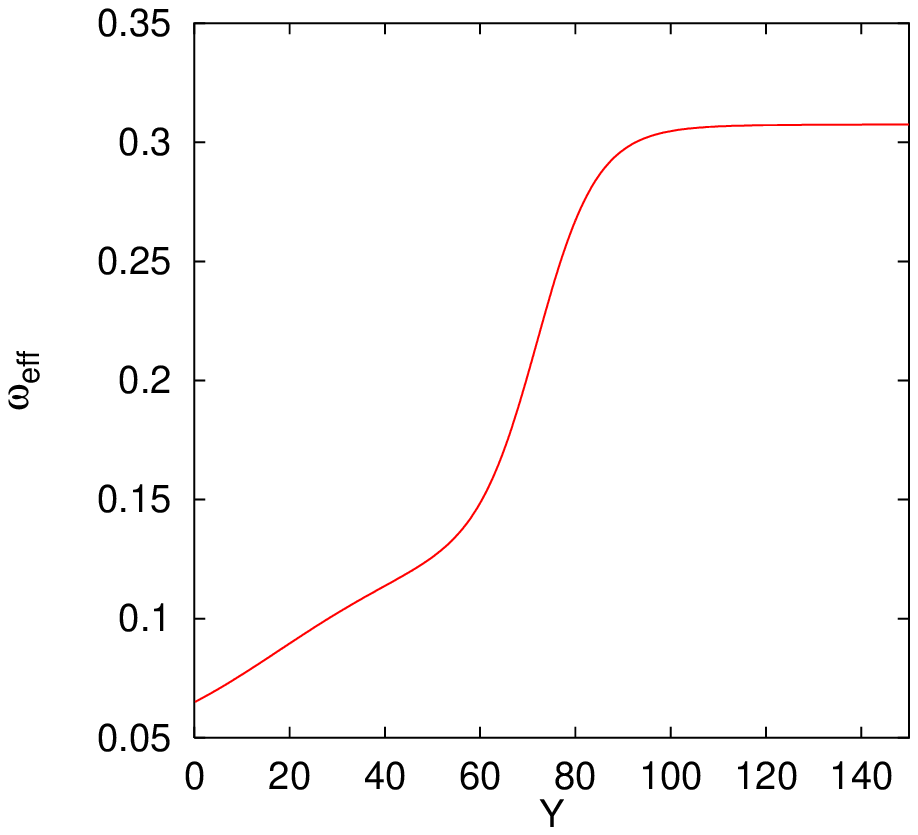}
\didascalia{Middle section of the Bartels' cigars of Fig.~\ref{f:sigari} for continuous
value of $Y$ and the corresponding small-$x$ effective growth exponent\labe{f:anatra}}
\end{figure}

To conclude this chapter, we stress that all the estimates based on the collinear model
are by no means intended to be quantitative results. This model reveals itself useful in 
order to justify some of the assumptions adopted in the treatment of the complete
small-$x$ kernel such as the factorization properties and the $\om$-expansion, and to
clarify the role of the critical exponents $\om_s(t)$, $\om_c(t)$ and $\op$.

In particular, for two-scales processes, we have elucidated a completely new transition
mechanism into the non perturbative domain which, for large enough energies
$Y\gtrsim t(1+\op)/(\op-\om_s(t))$ and for sizeable values of $\op-\om_s(t)$ may be
spectacular.

\chapter{Conclusive remarks\labe{c:cr}}

This last and short chapter summarizes the basic results of the work presented in this
thesis and indicates the main open questions and directions for future studies as well.

We have investigated the properties of QCD in the kinematic region of small-$x$ with
particular attention to deeply inelastic scattering of electrons on hadrons and in
particular to the physics at HERA. The most precise experimental results in the
small-$x$ domain concerns the measurements of the $F_2$ structure function which clearly 
exhibits a marked rise (Fig.~\ref{f:crescita}) towards low values of $x$ in a wide range of $Q^2$. Even though
the standard DGLAP	evolution equations are able to describe the whole set of data,
nevertheless they have to be supplied with an initial condition (see
Eq.~(\ref{inputpert})) for the partonic distribution functions which cannot be explained 
within the DGLAP approach itself. Furthermore, there is a variety of two-scale
processes, e.g.\ $\ga^*\ga^*$ scattering and forward-jets, whose $x$-behaviour is
determined independently of $Q^2$-evolution which is suppressed.

The natural framework where to study the small-$x$ behaviour of those inclusive
processes is the BFKL one, where the large logarithmic coefficients of the perturbative
series have been resummed at leading order and, recently, also at next-to-leading
order. Even though the BFKL equation provides the right power-like shape for the
small-$x$ growth of structure functions, nevertheless the quantitative estimate of the
growth exponent is absolutely not compatible with its measurements
(Fig.~\ref{f:esponcresc}). This is true both at leading order --- which overestimates
the small-$x$ rise --- and in the NL$x$ approximation, where the corrections are so
large and negative that, for the accessible values of $Q^2$, the $x$-behaviour changes
its slope (Fig.~\ref{f:pomBFKL}).

The analysis performed in Secs.~\ref{s:odl} and \ref{s:giocattolo} have shown that the
responsibles of such large corrections and of the ensuing instability of the BFKL
hierarchy are the big collinear contributions to the coefficient kernels in the BFKL
expansion. More precisely, the coefficient kernels $K^{(n)}(\kk,\kk')$ of
Eq.~(\ref{espK}) are affected by
single and double logarithms of $k^2/ k'{}^2$ in the collinear limits
$k^2\gg k'{}^2$ and $k^2\ll k'{}^2$. The single logarithms are essentially those
predicted by the renormalization group equation which are responsibles for the
logarithmic scaling violations in $Q^2$ for the partonic densities. The double
logarithms, on the other hand, are due to the mismatch between the high energy
factorization scale $s_0=kk_0$ --- which is supposed to be the correct scale which
allows the factorization formula (\ref{fatt1loop}) --- and the Bjorken scale
$s_B=\Max(k^2,k_0^2)$ providing the relevant scaling variable $x_B\simeq s_B/s$ in the
collinear limits $k^2\gg k_0^2$ and $k^2\ll k_0^2$.

Those large logarithms are unavoidable in the $\as$-expansion of the BFKL kernel
(\ref{espK}) and the corresponding series is oscillatory and unstable for not very small
values of $\as$. In addition, the total cross section stemming from the na\"\i ve NL$x$
truncation of that series may turn out to be negative in the very large $s$-limit! A
resummation of the large collinear contributions mentioned before is therefore
mandatory.

By a RG analysis of the collinear singularities of the GGF (\ref{gasint}) and hence of
the kernel (\ref{kupper},\ref{klower}) it is possible to determine the coefficients of
single and double logarithms of the transverse momenta $\kk,\kk_0$ to all orders in
$\as\ln1/x$. The resummation of those logarithms can be accomplished by a suitable shift
of the $\ga$-singularities of the eigenvalue functions $\achi^{(n)}_\om(\ga)$ of the
improved coefficient kernels of Eq.~(\ref{serieKom}) as indicated in
Eqs.~(\ref{chin}). By then incorporating the constraints of the known L$x$ and NL$x$
BFKL eigenvalue functions, we are able to construct --- up to NNL$x$ corrections --- the
resummed L$x$ and NL$x$ eigenvalues, given in Eqs.~(\ref{lund}) and (\ref{chi1om}).

Having determined the collinear-resummed L$x$ and NL$x$ kernels, it remains the question 
of determining the improved GGF, i.e., the solution of Eq.~(\ref{g}). The technical
difficulty in solving such an equation is mainly concerned with the running of
$\as(k^2)$ and in particular with the singularity of its perturbative expression at the
Landau point $k^2=\La^2$. In practice, a suitable regularization of $\as(k^2)$ has to be
adopted.

At this point we proceed in two directions: the first one consists in
considering the large $t=\ln k^2/\La^2$ regime ($t\gg t_0$) where perturbative (regularization
independent) features and strong coupling (regularization dependent) ones are argued to
factorize and therefore can be considered independently. The second one is to
define a simplified small-$x$ model which can be exactly solved (at least
numerically) where the qualitative features of the real case can be studied as well as
the issue of factorization.

The complete small-$x$ equation has been studied in Chap.~\ref{c:rga} where, in the regime
$k^2\gg k_0^2$, the GGF is argued to assume the factorized form (\ref{fattor}) in terms
of a perturbative factor $\afz_\om(\kk)$ which obeys the homogeneous equation
(\ref{eqx}). The solution of the latter has been given in terms of the
$\ga$-representation (\ref{rappres}) and of the $\om$-expansion (\ref{omegaes}).

The merit of the $\om$-expansion is to provide an effective eigenvalue function
$\achi_\om(\ga)$ whose shape is stable even for sizeable values of the expansion
parameter $\om$ (Fig.~\ref{f:avleff}) and whose error stemming from neglecting NNL$x$
contributions is much smaller than the corresponding NNL$x$ truncation of the BFKL
kernel in the $\as$-expansion. As far as the stability of the shape is concerned, this
means that a minimum in the function $\ga\mapsto\achi_\om(\ga)$ is always present, even
for large values of $\om$. Because of that, a stable saddle point exists, preventing us
from negative cross sections and providing a small-$x$ growth exponent $\om_s(t)$ ---
defined in Eq.~(\ref{omsp}) and plotted in Fig.~\ref{f:omcoms} --- which should be
observable in some intermediate-$x$ regime (see Sec.~\ref{s:heb}). In the range
$Q^2\simeq4\div40\GeV^2$, i.e., $\as\simeq0.2\div0.3$, the critical
exponent $\om_s(t)\simeq0.27\div0.32$ is not far from the corresponding values of the
HERA data.

It is to be noted that, while $\om_s(t)$ decreases for large $t$, the trend of the
experimental growth exponent is to increas with $Q^2$. This is due to the scaling
violations, which the DGLAP approach takes into account by resumming the $\ln Q^2$
contributions which are not directly incorporated in $\om_s(t)$. A more direct
observation of the critical exponent $\om_s(t)$ is expected from two-scale processes
where the $Q^2$-evolution is less important (cfr.\ beginning of Chap.~\ref{c:NLbfkl}).

The error coming from the NL truncation in the $\om$-expansion is uniformly (i.e.,
independently of $\ga$) $\ord(\om^2)$, the neglected coefficients having no $\ga=0$
nor $\ga=1$ singularities at all. This error is therefore of the same size as the
ambiguity in the definition of $\achi_1^{\om}$. The corresponding error in the saddle
point condition (\ref{puntos}) is a roughly $\ga$-independent change of scale
$\De(bt)=\ord(\om)$,
affecting the anomalous dimension with a relative uncertainty $\De\as/\as\sim\as\om$
while for the critical exponent $\De\om_s(t)\sim\dot{\om}_s(t)\De t$. The important
thing to note is that all these errors are NNL$x$ and that their coefficients do not
show a strong $t$-dependence.

In connection with the ambiguities of the finite order truncation, one can consider also 
the renormalization scale and scheme dependence as well as the resummation scheme
dependence. In Sec.~\ref{s:stab} it is shown that such dependences are very small and do 
not affect considerably the critical $\om$-exponents $\om_s$ and $\om_c$.

The collinear model presented in Chap.~\ref{c:cm} allows us to investigate in more
detail the physical consequences of the collinear-resummed kernel. First of all, the
model offers a method of analysis of non perturbative effects on the observable quantities, in
particular on the very small-$x$ growth exponent (Sec.~\ref{s:scf}). The main results of 
that analysis is that, in two-scale processes, for large enough values of the external
scales $k^2$ and $k_0^2$ and for not too large values of the energy $s$, there exists a
regime governed by perturbative physics where the effective small-$x$ growth exponent is
given by $\om_s(t)$. A preliminary phenomenological analysis seems to indicate that
the $x$-growth in $\ga^*\ga^*$ and forward-jet processes are
similar, and consistent with the $\om_s$ value quoted before.  For increasing values of the energy, there is a transition point beyond
which most of the exchanged gluons lie in the strong coupling region and the growth
exponent is given by the ``pomeron'' which is of non perturbative nature. This
transition mechanism does not take place continuously like ``diffusion''. For this
reason, we refer to it as ``tunneling''.

The study of the RG improved small-$x$ equation has just begun and leaves many open
questions yet, which hopefully will find an answer after further investigation. Just to
mention some of them, there is to clarify the role of the critical exponent $\om_c(t)$
--- the true singularity of the resummed anomalous dimension. A second point concerns
the evaluation of higher twist contributions stemming from poles in the effective
eigenvalue functions at values of $\ga<-\om/2$ and $\ga>1+\om/2$. A preliminary study of
such corrections can be done in the collinear model by suitably modifying the
pole-structure of the eigenvalue. A major task concerns the implementation of unitarity
which however should involve the study of strong coupling effects.

To conclude, small-$x$ physics emerges from our work as a rather unique framework where
the perturbative and strong coupling features play a complementary role. I think we have 
shown that by combining $\ln1/x$ and $\ln Q^2$ evolutions we can progress a great deal
towards understanding the perturbative aspects of the problem. Further progress will
require a deeper understanding of non perturbative physics.
\appendix
\chapter{The Structure Functions in DIS\labe{a:sf}}

According to the definition of the kinematics variables of DIS in Sec.~{\ref{s:dis} (see
Fig.~\ref{f:dis}),the transition matrix element is given, in the lowest order of the EM
interaction, by
\begin{equation}\lab{emdis}
 \vm{l'\,X|T|l\,H}=\overline{u}_{\si'}(p_1')\,\ui e\ga^\mu\,u_\si(p_1)\,
 {-\ui\over q^2}\,\vm{X|(-e)J_\mu(0)|p_2,\la}\;,
\end{equation}
where $\si'$, $\si$ and $\la$ are the spin components of the scattered 
electron, initial electron and target proton respectively, $e$ is the electronic
EM charge and $J_\mu(x)$ is the quark part of the EM current.

The inclusive unpolarized differential cross section (with respect to
the outgoing electron) is obtained by multiplying (\ref{emdis}) by the
appropriate phase space factor and reads
\begin{equation}\lab{sdd}
 p_1'{}^0\,{\dif\si\over\dif^3\vec{p}_1{}\!'}={2\over s}\,{\al^2\over Q^4}\,L^{\mu\nu}
 W_{\mu\nu}\;,
\end{equation}
where we have defined the leptonic tensor
\begin{subequations}\labe{d:tenslep}
\begin{align}
 L^{\mu\nu}(p_1,q)\dug&{1\over2}\sum_{\si,\si'}\Big(\overline{u}_{\si'}(p_1')
 \ga^\mu u_\si(p_1)\Big)^*\Big(\overline{u}_{\si'}(p_1')\ga^\nu u_\si(p_1)
 \Big)\\
 =&2\left(p_1^\mu p_1'{}^\nu+p_1^\nu p_1'{}^\mu\right)+q^2 g^{\mu\nu}
\end{align}
\end{subequations}
and the hadronic tensor
\begin{subequations}\labe{d:tensadr}
\begin{align}
 W_{\mu\nu}(p_2,q)\dug&{1\over4\pi}\sum_X(2\pi)^4\d^4(p_2+q-p_X)\,
 {1\over2}\sum_\la\vm{p_2,\la|J_\mu(0)|X}\vm{X|J_\nu(0)|p_2,\la}\\ \lab{wcomm}
 =&{1\over4\pi}\int\dif^4z\;\esp{\ui qz}\,{1\over2}\sum_\la
 \vm{p_2,\la|[J_\mu(z),J_\nu(0)]|p_2,\la}\;.
\end{align}
\end{subequations}
Because of current conservation at the hadronic vertex,
$\de_\mu J^\mu(z)|X\rangle=0$ and hence
$q^\mu W_{\mu\nu}=0=q^\nu W_{\mu\nu}$. The most general form of the
hadronic tensor fulfilling the requirements of Lorentz covariance,
space-inversion and time-reversal conservation and current
conservation is ($p_2q\equiv p_2\cdot q$)
\begin{equation}\lab{d:SF}
 W^{\mu\nu}=F_1\,\left({q^\mu q^\nu\over q^2}-g^{\mu\nu}\right)+
 F_2\,{1\over p_2q}\left(p_2^\mu-{p_2q\over q^2}q^\mu\right)
 \left(p_2^\nu-{p_2q\over q^2}q^\nu\right)\;,
\end{equation}
where the Lorentz invariant coefficient $F_i(x,Q^2):i=1,2$ are called
{\em structure functions} of the proton. It is worthwile to stress that the
structure functions depends only on the Lorentz invariants of the lower vertex (the blob) in
Fig.~\ref{f:dis}, i.e., on $x$ and $Q^2$.

Keeping $p_2$ and $q$ fixed, let's introduce the longitudinal polarization unit vector
\begin{equation}\lab{d:vetlong}
 \po_L^\mu(q)\dug{Q\over p_2q}\left[1+{M^2Q^2\over(p_2q)^2}\right]^{-\half}
 \left(p_2^\mu-{p_2q\over q^2}q^\mu\right)\;,\quad
 q\cdot\po_L=0\;,\quad\po_L\cdot\po_L=1
\end{equation}
and a couple of transverse unit vectors $\po_1$ and $\po_2$ such
that
\begin{equation*}
 q\cdot\po_i=0\;,\quad\po_L\cdot\po_i=0\;,\quad\po_i\cdot\po_i=-1
 \;,\quad\po_1\cdot\po_2=0\;,\quad(i=1,2)\;.
\end{equation*}
The sets of vectors $\{\po_1,\po_2\}$ and $\{\po_L,\po_1,\po_2\}$ form
pseudo-orthonormal basis for the subspaces $<p_2,q>^\perp$ and
$<q>^\perp$ respectively. Defining the projectors
\begin{align}\lab{proiettori}
 \Pi^{\mu\nu}_L&\dug\po_L^\mu\po_L^\nu\\
 \Pi^{\mu\nu}_T&\dug-(\poc{}_1^\mu\po_1^\nu+\poc{}_2^\mu\po_2^\nu)\\
 \Pi^{\mu\nu}_\perp&\dug\Pi^{\mu\nu}_L+\Pi^{\mu\nu}_T=
 g^{\mu\nu}-{q^\mu q^\nu\over q^2}
\end{align}
in the longitudinal, transverse and orthogonal (to $q^\mu$) space
respectively, we can decompose the hadronic tensor as
\begin{subequations}\labe{tensadr}
\begin{align}
 W^{\mu\nu}&=-F_1\,\Pi_\perp^{\mu\nu}+F_2\,{p_2q\over
 Q^2}\left(1+{Q^2\over \nu^2}\right)\Pi_L^{\mu\nu}\\
 &=-F_1\,\Pi_T^{\mu\nu}+{F_L\over2x}\,\Pi_L^{\mu\nu}\\
 &={1\over2x}\left[F_L\,\Pi_\perp^{\mu\nu}-\left(1+{Q^2\over\nu^2}
 \right)F_2\,\Pi_T^{\mu\nu}\right]\;,
\end{align}
\end{subequations}
where the longitudinal structure function is given by
\begin{equation}\lab{d:longSF}
 F_L(x,Q^2)\dug\left(1+{Q^2\over\nu^2}\right)F_2-2xF_1\;.
\end{equation}

If we consider virtual photon absorption by a proton\footnote{Dealing
with an off-shell incoming particle, a particular flux convention has
to be adopted.}, the ensuing cross section involves just the hadronic
tensor (\ref{d:tensadr}) and the former assumes the form
\begin{equation}
 \si(\ga^*\,H)\propto\poc{}^\mu\po^\nu\,W_{\mu\nu}\;,
\end{equation}
$\po^\mu$ being the polarization vector of the virtual photon. It is
easy then to extract any of the structure functions simply by choosing 
the appropriate vector $\po^\mu$. If we parametrize the generic polarization vector
$\po^\mu=\la\po_L+\tau\po_T$ where $\po_T$ is a unitary transverse vector, then 
$\poc{}^\mu\po^\nu\,W_{\mu\nu}$ gives
\begin{itemize}
\item $F_1$ for $|\la|=0$ and $|\tau|=1$;
\item $F_L$ for $|\la|=1$ and $|\tau|=0$;
\item $F_2$ for $|\la|=|\tau|=2x/(1+{Q^2\over\nu^2})$.
\end{itemize}

\section{$\boldsymbol{M,m\to0}$ limit}

For high values of the energy $\sqrt{s}$ and of the momentum transfer $Q$ with respect to the
typical hadronic masses, the latter can safely be neglected, together with the electronic mass which 
is even smaller. In this case $\nu^{-1}=0$ and for the relevant kinematical variables holds
\begin{equation}\lab{qxys}
 Q^2=xys\;.
\end{equation}
In terms of these variables and of the azimuthal angle $\phi$ in the CM frame, the phase space
element in Eq.~(\ref{sdd}) becomes
\begin{equation*}
 p_1'{}^0\,{\dif\over\dif^3\vec{p}_1{}\!'}={2sx^2\over Q^2}{\dif\over\dif x\,\dif Q^2\,\dif\phi}
\end{equation*} 
so that we can write the differential cross section with respect to $x$, $Q^2$ and $\phi$ at fixed
$s$ as
\begin{subequations}\lab{sezurtoSF}
\begin{align}
 {\dif\si\over\dif x\,\dif Q^2\,\dif\phi}&={Q^2\over2x^2s}(\ref{sdd})={\al^2\over x^2s^2Q^2}
 L^{\mu\nu}W_{\mu\nu}\nonumber\\
 &={2\al^2\over xQ^4}\left\{xy^2F_1+(1-y)F_2\right\}\\
 &={\al^2\over xQ^4}\left\{[1+(1-y)^2]F_2-y^2F_L\right\}
\end{align}
\end{subequations}
where $\al\dug e^2/4\pi$ is the EM fine structure constant. Note that in this massless case
$F_L=F_2-2xF_1$.
\chapter{Dimensional regularization\labe{a:regdim}}

The dimensional regularization technique consists in performing the loop and phase space 
(divergent) integrals of Feynman diagrams and squared matrix elements in a modified
space-time of generic dimension $D\neq4$, where such integrals exist. The ensuing
expressions are then analytically continued for complex values of $D$.

For our purposes, the main modifications with respect to the $D=4$ case concern the
phase space measure $\dif\phi^{(n)}$ and the expression of the coupling $\as$.

The generally adopted convention for momentum integration is to replace the volume
element as
\begin{equation}\lab{modvol}
 {\dif^4k\over(2\pi)^4}\quad\longrightarrow\quad{\dif^D k\over(2\pi)^D}\;.
\end{equation}
Introducing the (half) extra-dimension parameter $\e:D\ugd4+2\e$, the phase space
measure (\ref{sf2n})
\begin{equation}\lab{sf2nE}
 \dif\phi^{(2+n)}={1\over(2s)^2}{1\over(4\pi)^n}\prod_{i=1}^{n}{\dif z_i\over z_i}
 \prod_{j=1}^{n+1}{\dif\kk_j\over(2\pi)^2}\quad\longrightarrow\quad
 {1\over(2s)^2}{1\over(4\pi)^n}\prod_{i=1}^{n}{\dif z_i\over z_i}\prod_{j=1}^{n+1}
 {\dif\kk_j\over(2\pi)^{2+2\e}}\;,
\end{equation}
i.e., only the $(2\pi)^2$ denominator under $\dif\kk_j$ is modified.

In addition, in going from L$x$ to NL$x$ approximation, also the longitudinal measure is
modified. In particular the two-body phase space reads
\begin{equation}\lab{sfdueE}	       
 \dif\phi^{(2)}={1\over4s^2}{\dif\kk\over(2\pi)^{2+2\e}}
\end{equation}
and the three-body one is
\begin{equation}\lab{sftreE}
 \dif\phi^{(3)}={1\over16\pi s^2}{\dif z_1\over z_1(1-z_1)}
 {\dif\ku\dif\kd\over(2\pi)^{4+4\e}}\;.
\end{equation}

In $4+2\e$ dimensions the QCD coupling constant $g$ acquire a mass-dimension $[g]=-\e$,
therefore a massive parameter $\mu$ has to be introduced, so to consider $g\mu^\e$ as the
effective dimensionless coupling. The convention generally adopted in the papers
reporting the calculations of the NL$x$ BFKL kernel is to use the $\overline{MS}$
renormalization scheme and to define the dimensionless colour strength as
\begin{equation}\lab{asE}
 \as\dug{g^2\mu^{2\e}\Ga(1-\e)\over(4\pi)^{1+\e}}\;,
\end{equation}
which reduces to the customary expression in the $\e\to0$ limit. Accordingly, we have
adopted the following definition for the Born impact factor:
\begin{equation}\lab{hzeroE}
 h_{\pa}^{(0)}(\kk)\dug{2^{1+\e}C_{\pa}\as\over\sqrt{N_c^2-1}\,\Ga(1-\e)}\,
 {1\over\mu^{2\e}\kk^2}\;.
\end{equation}

As transverse measure we keep the natural one
\begin{equation}\lab{dkE}
 \dif\kk\dug\dif^{2+2\e}\kk\quad;\quad\d(\kk)\dug\d^{2+2\e}(\kk)\;.
\end{equation}
\chapter{Integral representations for functions in transverse momentum space\labe{a:irtms}}

\section{The leading BFKL eigenvalue function\labe{a:avlbfkl}}

Given a set $\Om$ of complex functions defined on a vectorial space $V$, an operator
$\K$ acting on $\Om$ is said to be {\em rotationally invariant} if it commutes with all the
rotation operators on $V$. If $\K(\kk,\kk')$ is the kernel of the linear integral
operator $\K$, then $\K$ is rotationally invariant if and only if
\begin{equation}\lab{d:invrot}
 \K(R\kk,R\kk')=\K(\kk,\kk')\quad,\quad\kk,\kk'\in V
\end{equation}
where $R$ is the generic rotation operator in $V$.

Let $S_\la$ denote the scaling operator on $\Om$ defined by
\begin{equation}\lab{d:opscal}
 [S_\la f](\kk)\dug f(\la\kk)\quad,\quad f\in\Om\;.
\end{equation}
An operator $\K$ is said to be {\em scale invariant} if it commutes with $S_\la$ for all 
$\la>0$. In a $d$-dimensional vector space $V$, the integral operator $\K$ is scale
invariant if and only if
\begin{equation}\lab{d:invscal}
 \K(\la\kk,\la\kk')=\la^{-d}\K(\kk,\kk')\;.
\end{equation}

Under rather general conditions, if the operator $\K$ is rotationally invariant, then it
admits a complete set of eigenfunction $\{f_j:j\in J\}$ of the form
\begin{equation}\lab{afzrot}
 f_j(\kk)=\Psi_j(\kk^2) Y_j(\phi)
\end{equation}
where $\phi$ denotes the set of angular variables and $Y_j$ is a spherical harmonic in
$d$ dimensions. Furthermore, if $\K$ is also scale-invariant, the radial eigenfunctions
$\Psi$ can only be powers of the variable $\kk^2$.

The BFKL operator $\K$ acts on functions defined in the $d=2$ dimensional transverse
space $V$; the only angular variable is the azimuthal angle $\phi$. The action of $\K$ on 
a generic function $f\in\Om$ can be explicitly written as
\begin{equation}\lab{azioneK}
 [\K f](\kk)=\ab\int\dif\kk'{1\over\pi(\kk-\kk')^2}
 \left[f(\kk')-{\kk^2f(\kk)\over2\kk'{}^2}\right]
\end{equation}
the first and second term corresponding to the real and virtual part of the kernel
respectively.  It is easy to check that, for fixed coupling $\ab$, the BFKL kernel
defines a rotationally and scale invariant operator.

Since in two dimensions spherical harmonics corresponds to complex exponentials
\begin{equation}\lab{espcom}
 Y_m(\phi)={\esp{\ui m\phi}\over\sqrt\pi}
\end{equation}
the eigenfunctions of $\K$ are of the form
\begin{equation}\lab{d:afzbfkl}
 f_{\ga,m}(\kk)=\kk^{\ga-1}{\esp{\ui m\phi}\over\sqrt\pi}
\end{equation}
and a complete set is obtained by allowing $m\in\Z$ and $\ga=1/2+\ui\nu:\nu\in\R$.
The eigenfunctions (\ref{d:afzbfkl}) are normalized according to
\begin{equation}\lab{normafz}
 \d^2(\kk-\kk_0)=\sum_{m\in\Z}\intmel\difg\;f_{\ga,m}(\kk)f^*_{\ga,m}(\kk_0)
\end{equation}
and satisfy the orthonormality conditions
\begin{equation}\lab{ortoafz}
 \int\dif\kk\;f^*_{\ga,m}(\kk)f_{\ga',m'}(\kk)=2\pi\d_{mm'}\,\d(\nu-\nu')\quad,
 \quad\ga=\half+\ui\nu\quad,\quad\ga'=\half+\ui\nu'\;.
\end{equation}

The IR finiteness of the kernel can be checked directly on the eigenfunctions. For
simplicity we restrict ourselves to the $m=0$ case. By using the dimensional
regularization in $d=2+2\e$ and the formula
\begin{equation}\lab{formulagamma}
 \int{\dif^{2+2\e}\kk\over\pi^{1+\e}}\;{1\over[\kk^2]^{1-\al}[(\qq-\kk)^2]^{1-\be}}=
 {\Ga(1-\al-\be-\e)\Ga(\al+\e)\Ga(\be+\e)\over\Ga(\al+\be+2\e)\Ga(1-\al)\Ga(1-\be)}
 (\qq^2)^{\al+\be+\e-1}
\end{equation}
we get:
\begin{align}\lab{avlreal}
 \int{\dif^{2+2\e}\kk'\over\pi^{\e}}\;\K^{(\text{R})}(\kk,\kk')(\kk'{}^2)^{\ga-1}
 &=\left[{1\over\e}+\psi(1)-\psi(\ga)-\psi(1-\ga)+\ord(\e)\right]\ab(\kk^2)^{\e+\ga-1}\\
\lab{avlvirt}\int{\dif^{2+2\e}\kk'\over\pi^{\e}}\;\K^{(\text{V})}(\kk,\kk')(\kk'{}^2)^{\ga-1}
 &=\left[-{1\over\e}+\psi(1)+\ord(\e)\right]\ab(\kk^2)^{\e+\ga-1}
\end{align}
where the real and virtual BFKL kernel are defined in Eqs.~(\ref{d:nucleiBFKL}) and
$\psi\dug\Ga'/\Ga$ is the logarithmic derivative of the gamma function. It is evident
the singular behaviour of both the real and virtual terms in $d=2$ dimension
corresponding to the poles in $\e$. However, when adding the two contributions, the
poles cancel and in the $\e\to0$ limit we recover the $m=0$ eigenvalue function
$\chi(\ga)\dug\chi(\ga,0)=2\psi(1)-\psi(\ga)-\psi(1-\ga)$. When acting on the general
eigenfunction the kernel yields
\begin{align}\lab{avlm}
 &\int\dif\kk\;\K(\kk,\kk')\,f_{\ga,m}(\kk')=\ab\chi(\ga,m)\,f_{\ga,m}(\kk)\\
 &\chi(\ga,m)=2\psi(1)-\psi\(\ga+{|m|\over2}\)-\psi\(1-\ga+{|m|\over2}\)\;.
\end{align}

\section{Mellin representation in transverse momentum space\labe{a:mellin}}

The BFKL kernel can be expressed in terms of its eigenfunctions (\ref{d:afzbfkl}) and its
eigenvalues (\ref{avlm}) by means of the spectral representation
\begin{equation}\lab{rapspeK}
 \K(\kk,\kk')=\sum_{m\in\Z}\intmel\difg\;f_{\ga,m}(\kk)\ab\chi(\ga,m)f^*_{\ga,m}(\kk)\;
\end{equation}
In particular, the azimuthal average of the kernel, defined by
\begin{equation}\lab{medazimK}
 \K(\kk^2,\kk'{}^2)\dug\int{\dif\te\over2\pi}\;\K(\kk,\kk')\quad,
 \quad\te=\widehat{\kk\kk'}
\end{equation}
has $(\kk^2)^{\ga-1}$ as eigenfunctions and $\chi(\ga)\dug\chi(\ga,0)$ as eigenvalue
function. Thanks to Eq.~(\ref{rapspeK}), it can be given the integral representation
\begin{align}\lab{intrapK}
 \K(\kk^2,\kk'{}^2)&={1\over\pi\kk^2}\intmel\difg\;\left({\kk^2\over\kk'{}^2}\right)^\ga
 \ab\chi(\ga)\;,\\ \lab{chiKaz}
 \ab\chi(\ga)&=\int\dif\kk\;\left({\kk^2\over\kk'{}^2}\right)^{-\ga}\K(\kk^2,\kk'{}^2)\;,
\end{align}
showing that $\chi(\ga)$ is nothing but the Mellin transform of the azimuthally averaged 
kernel $\K$. This is due to the fact that the radial eigenfunctions of $\K$ are powers
of $\kk^2$.

The Mellin transformation allows to diagonalize convolutions of function in both
longitudinal and transverse space, such as the high energy factorization formula
(\ref{ffae}). By defining
\begin{subequations}\labe{mellinomga}
\begin{align}
 F_{2,\om}(\ga)&\dug\int_0^\infty{\dif Q^2\over Q^2}\;\left({Q^2\over\La^2}\right)^{-\ga}
 \int_0^1{\dif x\over x}\;x^\om\,F_2(x,Q^2)\\
 \sip_\om(\ga)&\dug\int_0^\infty{\dif\rho\over\rho}\;\rho^{-\ga}\int_0^1
 {\dif\xi\over\xi}\;\xi^\om\,\sip(\xi,\rho)\quad,\quad\xi={x\over z}\quad,\quad
 \rho={Q^2\over\La^2}\\ \lab{MellC}
 \F_\om(\ga)&\dug\int\dif\kk\;\left({\kk^2\over\La^2}\right)^{-\ga}\int_0^1{\dif z\over z}
 \;z^\om\,\F(z,\kk)
\end{align}
\end{subequations}
the mentioned factorization formula can be written in the simple algebraic form
(\ref{ffaega}) or as a $\kk$-integral if the Mellin transform is carried out only in the 
longitudinal variables $x$ and $z$.

The Mellin transforms (\ref{mellinomga}) can be inverted by using
\begin{subequations}\labe{invmellinomga}
\begin{align}
 F_2(x,Q^2)&=\intmel\difg\;\left({Q^2\over\La^2}\right)^\ga\int\difo\;
 x^{-\om}\,F_{2,\om}(\ga)\\
 \sip\({x\over z},{Q^2\over\La^2}\)&=\intmel\difg\;\left({Q^2\over\La^2}\right)^\ga
 \int\difo\;\left({x\over z}\right)^{-\om}\,\sip_\om(\ga)\\
 \F(z,\kk)&={1\over\pi\kk^2}\intmel\difg\;\left({\kk^2\over\La^2}\right)^\ga\int\difo\;
 z^{-\om}\,F_\om(\ga)
\end{align}
\end{subequations}
where the integration in the complex $\om$-plane is parallel to the imaginary axis and
to the right of all the singularities of the integrating function.

\end{document}